\documentclass[aps,pre,reprint,tightenlines,twocolumn,showpacs,showkeys]{revtex4}
\usepackage[dvips]{graphicx}
\usepackage[dvips]{color}
\usepackage{hyperref,breakurl,amsmath,amssymb,amsbsy,esint,eepic,tipa,pst-node}

\newcommand{\ii}{\textrm{i}}
\newcommand{\ee}{\textrm{e}}
\newcommand{\dd}{\textrm{d}}
\newcommand{\Tr}{\textrm{Tr}}
\newcommand{\bTr}{\textrm{bTr}}
\newcommand{\Det}{\textrm{Det}}
\newcommand{\TT}{\textrm{T}}
\newcommand{\diag}{\textrm{diag}}
\newcommand{\erfc}{\textrm{erfc}}
\newcommand{\re}{\textrm{Re}}
\newcommand{\im}{\textrm{Im}}
\newcommand{\csch}{\textrm{csch}}
\newcommand{\dupl}{\textrm{dupl.}}
\newcommand{\lin}{\textrm{lin.}}
\newcommand{\Id}{\mathbf{1}}
\newcommand{\Zero}{\mathbf{0}}
\newcommand{\la}{\left\langle}
\newcommand{\ra}{\right\rangle}

\begin{document}

\title{Hermitian and non-Hermitian covariance estimators for multivariate Gaussian and non-Gaussian assets from random matrix theory}

\author{Andrzej \surname{Jarosz}}
\email{jedrekjarosz@gmail.com}
\affiliation{The Henryk Niewodnicza\'{n}ski Institute of Nuclear Physics, Polish Academy of Sciences, Radzikowskiego 152, 31-342 Krak\'{o}w, Poland}

\topmargin=0cm
\allowdisplaybreaks[4]

\begin{abstract}
The random matrix theory method of planar Gaussian diagrammatic expansion is applied to find the mean spectral density of the Hermitian equal-time and non-Hermitian time-lagged cross-covariance estimators, firstly in the form of master equations for the most general multivariate Gaussian system, secondly for seven particular toy models of the true covariance function. For the simplest one of these models, the existing result is shown to be incorrect and the right one is presented, moreover its generalizations are accomplished to the exponentially-weighted moving average estimator as well as two non-Gaussian distributions, Student t and free L\'{e}vy. The paper revolves around applications to financial complex systems, and the results constitute a sensitive probe of the true correlations present there.
\end{abstract}


\maketitle

%%%%%%%%%%%%%%%%%%%%%%%%%%%%%%%%%%%%%%%%%%%%%%%%%%%%%%%%%%%%%%%%%%%%%%
%%%%%%%%%%%%%%%%%%%%%%%%%%%%%%%%%%%%%%%%%%%%%%%%%%%%%%%%%%%%%%%%%%%%%%
%%%%%%%%%%%%%%%%%%%%%%%%%%%%%%%%%%%%%%%%%%%%%%%%%%%%%%%%%%%%%%%%%%%%%%

\section{Introduction}
\label{s:Introduction}

An important problem in various fields of science featuring systems of a large number of time-dependent objects is to unravel correlations between these objects, either at the same or at distinct time moments (Sec.~\ref{sss:GaussianCovarianceFunctions}). This knowledge may be gained from the past realized values of the correlations, although it is obscured by a measurement noise caused by an insufficient amount of information in any such time series (Sec.~\ref{sss:MeasurementNoise}). This paper discusses how to account for this noise in applications to models of complex systems which exhibit properties such as sectors of increased cross-correlations (Sec.~\ref{sss:IndustrialSectors}) or vector autoregression type of temporal dynamics (Sec.~\ref{sss:VAR}).

Often in the below discussion the reference point will be financial markets---systems which are very complex, relevant for everyday life, accessible to empirical study and with the underlying mechanisms full of question marks---however, the results are beneficial for all kinds of complex systems, such as in brain science~\cite{KwapienDrozdzIoannides2000} or meteorology~\cite{MayyaAmritkar2006}.

%%%%%%%%%%%%%%%%%%%%%%%%%%%%%%%%%%%%%%%%%%%%%%%%%%%%%%%%%%%%%%%%%%%%%%
%%%%%%%%%%%%%%%%%%%%%%%%%%%%%%%%%%%%%%%%%%%%%%%%%%%%%%%%%%%%%%%%%%%%%%

\subsection{Gaussian covariance functions and their historical estimation}
\label{ss:GaussianCovarianceFunctionsAndTheirHistoricalEstimation}

%%%%%%%%%%%%%%%%%%%%%%%%%%%%%%%%%%%%%%%%%%%%%%%%%%%%%%%%%%%%%%%%%%%%%%

\subsubsection{Gaussian covariance functions}
\label{sss:GaussianCovarianceFunctions}

\emph{Random matrix theory approach to complex systems.} Consider a system of $N$ real variables (labeled by $i = 1 , 2 , \ldots , N$), and each one of them is observed throughout consecutive time moments $a = 1 , 2 , \ldots , T$ (separated by some time $\delta t$); let the value of object $i$ at time $a$ be denoted by \smash{$R_{i a}$}, all of which constitute an $N \times T$ matrix $\mathbf{R}$. For such a large complex system, a standard approach is to replace complexity by randomness, i.e., assume that $\mathbf{R}$ is a random matrix, endowed with some joint probability density function (JPDF) $\mathcal{P} ( \mathbf{R} )$, and investigate its statistical properties (here: the mean density of its eigenvalues), which then should help understand some of the generic features of the underlying complex system. Since there are two discrete dimensions here---spatial and temporal---random matrix theory (RMT) is a natural framework for doing that.

\emph{Gaussian approximation.} In most of the paper $\mathcal{P} ( \mathbf{R} )$ is chosen Gaussian of zero mean. This is an important restriction because the density of eigenvalues is not universal, i.e., it depends on the specific form of $\mathcal{P} ( \mathbf{R} )$. Even worse, for complex financial systems this JPDF is typically far from Gaussian (cf.~App.~\ref{a:NonGaussianProbabilityDistributionsForFinancialInstruments} for an introduction to non-Gaussian effects on financial markets).

The Gaussian approximation is dictated by analytical difficulties---it leads to complicated solutions yet much more tractable than in a non-Gaussian case---hence, it is a natural place to begin the analysis. Furthermore, it does provide a basis for non-Gaussian extensions, e.g., to the Student t-distribution through random volatility models, in which one considers the variances to be random variables (cf.~App.~\ref{aa:RandomVolatilityModels} for an introduction and Sec.~\ref{sss:TM1TLCEStudent} for an application), or to the free L\'{e}vy distribution through free probability calculus (cf.~Secs.~\ref{sss:RotationalSymmetry}, \ref{sss:TM1TLCELevy}).

\emph{Covariance functions.} A real multivariate Gaussian distribution of zero mean is fully described by a two-point covariance function,
\begin{equation}\label{eq:TrueRealCovarianceFunctionDefinition}
\la R_{i a} R_{j b} \ra ,
\end{equation}
where the brackets stand for the averaging over $\mathcal{P} ( \mathbf{R} )$. The attention in the literature has been mainly on the equal-time ($b = a$) covariances, described by an $N \times N$ real symmetric matrix \smash{$C_{i j} \equiv \langle R_{i a} R_{j a} \rangle$}, where stationarity (independence of $a$) has been assumed. This paper revolves around a more involved object, the time-lagged ($t \equiv b - a \neq 0$) covariances, described by an $N \times N$ real time-dependent matrix \smash{$C_{i j} ( t ) \equiv \langle R_{i a} R_{j , a + t} \rangle$}, obeying \smash{$\mathbf{C} ( t ) = \mathbf{C} ( - t )^{\TT}$}, where translational symmetry in time (over the considered period $T$) has again been assumed.

%%%%%%%%%%%%%%%%%%%%%%%%%%%%%%%%%%%%%%%%%%%%%%%%%%%%%%%%%%%%%%%%%%%%%%

\subsubsection{Measurement noise}
\label{sss:MeasurementNoise}

\emph{Unbiased estimators.} A basic way to experimentally measure the above covariance matrices for a given complex system is to take a historical time series of some length $T$ of the values \smash{$R_{i a}$}, compute the realized covariances \smash{$R_{i a} R_{j a}$} or \smash{$R_{i a} R_{j , a + t}$}, and average them over the past time moments $a$---this results in unbiased estimators which in the matrix notation read
\begin{subequations}
\begin{align}
\mathbf{c} &\equiv \frac{1}{T} \mathbf{R} \mathbf{R}^{\TT} ,\label{eq:RealcDefinition}\\
\mathbf{c} ( t ) &\equiv \frac{1}{T - t} \mathbf{R} \mathbf{D} ( t ) \mathbf{R}^{\TT} , \quad D_{a b} ( t ) \equiv \delta_{a + t , b} .\label{eq:RealctDefinition}
\end{align}
\end{subequations}

\emph{Weighted estimators and the EWMA.} Other choices are also in use, e.g., ``weighted estimators'' in which the past realized returns \smash{$R_{i a}$} are multiplied by some real and positive weights \smash{$w_{a}$}, (i) the smaller the older the data (i.e., this sequence increases with $a$) which reflects the fact that older measurements are less relevant for today estimation of the covariance, and (ii) obeying the sum rule \smash{$\frac{1}{T} \sum_{a = 1}^{T} w_{a}^{2} = 1$} to make this case consistent with [(\ref{eq:RealcDefinition}), (\ref{eq:RealctDefinition})]. Denoting \smash{$W_{a b} \equiv w_{a} \delta_{a b}$}, this prescription is equivalent to $\mathbf{R} \to \mathbf{R} \mathbf{W}$, i.e.,
\begin{subequations}
\begin{align}
\mathbf{c}^{\textrm{weighted}} &\equiv \frac{1}{T} \mathbf{R} \mathbf{W}^{2} \mathbf{R}^{\TT} , \label{eq:RealWeightedcDefinition}\\
\mathbf{c}^{\textrm{weighted}} ( t ) &\equiv \frac{1}{T - t} \mathbf{R} \mathbf{W} \mathbf{D} ( t ) \mathbf{W} \mathbf{R}^{\TT} .\label{eq:RealWeightedctDefinition}
\end{align}
\end{subequations}
In yet other words, this is equivalent to the standard estimators [(\ref{eq:RealcDefinition}), (\ref{eq:RealctDefinition})] with a modified true covariance, \smash{$\langle R_{i a} R_{j b} \rangle \to w_{a} w_{b} \langle R_{i a} R_{j b} \rangle$}.

A typical example is the ``exponentially-weighted moving average'' (EWMA),
\begin{equation}\label{eq:EWMADefinition}
w_{a}^{2} \equiv T \frac{1 - \kappa}{\kappa^{- T} - 1} \kappa^{- a} ,
\end{equation}
with just one parameter, $\kappa \in [ 0 , 1 ]$, defining the exponential suppression of the older data with the characteristic time \smash{$\tau_{\textrm{EWMA}} = - 1 / \log \kappa$}. Estimator [(\ref{eq:RealWeightedcDefinition}), (\ref{eq:EWMADefinition})] (with $\kappa = 0.94$) is widespread in financial industry through the RiskMetrics~1994 methodology of risk assessment~\cite{RiskMetrics1996,MinaXiao2001}.

\emph{Measurement noise and thermodynamic limit.} A fundamental obstacle for practical usage of the estimators is that a description by a finite time series will necessarily incur inaccuracies---since $N$ time series of length $T$ (i.e., $N T$ quantities) are used to estimate the \smash{$\sim N^{2}$} independent elements of the ``true'' covariance matrix, the error committed (the ``noise-to-signal ratio'') will be governed by the ``rectangularity ratio'' of $\mathbf{R}$, $r \equiv N / T$. In other words, $\mathbf{c}$ and $\mathbf{c} ( t )$ are consistent estimators of $\mathbf{C}$ and $\mathbf{C} ( t )$, respectively, i.e., they tend to the true values as $r \to 0$ (i.e., for $T$ large compared to $N$, which is a standard limit in statistics). On the other hand, a more practically relevant regime is when both $N$ and $T$ are large and of comparable magnitude---such as several hundred financial assets sampled daily over a few years---which is modeled by the ``thermodynamic limit,'' being a standard limit in RMT,
\begin{equation}\label{eq:ThermodynamicLimit}
N , T \to \infty , \quad r = \frac{N}{T} = \textrm{finite} .
\end{equation}
In this case, the estimators reproduce the true values of the covariances but with additional marring with the ``measurement noise,'' the more pronounced the larger $r$.

One might immediately propose to further and further increase the observation frequency (i.e., decrease $\delta t$, i.e., increase $T$, i.e., decrease $r$) in order to suppress the noise, but unfortunately this procedure is limited by the Epps effect (cf.~the end of~App.~\ref{aaa:MultivariateRandomVolatilityModels})---the very object one is measuring, the covariance matrix, changes with the observation frequency above its certain level.

Remark that the thermodynamic limit should be combined with the following limits:

(i) The RMT methods used in this paper are valid only provided that
\begin{equation}\label{eq:SmallTimeLagLimit}
t \ll T .
\end{equation}
Otherwise, finite-size effects would become relevant, as illustrated in Fig.~\ref{fig:TM2aTLCETwoTests} (a).

(ii) For the EWMA estimators [(\ref{eq:RealWeightedcDefinition}), (\ref{eq:RealWeightedctDefinition})], one should additionally take
\begin{equation}\label{eq:EWMALimit}
\kappa \to 1^{-} , \quad \vartheta \equiv T ( 1 - \kappa ) = \textrm{finite} ,
\end{equation}
since then the characteristic time \smash{$\tau_{\textrm{EWMA}} \sim T / \vartheta$} stretches over the scale of the entire time series. Unfortunately, the limits (\ref{eq:SmallTimeLagLimit}) and (\ref{eq:EWMALimit}) are incompatible with each other due to \smash{$t \ll \tau_{\textrm{EWMA}}$}, thus the theoretical results will reproduce Monte Carlo simulations with some discrepancy, albeit not very big, especially well inside the bulk [cf.~Fig.~\ref{fig:TM1TLCEEWMA} (a)].

\emph{Noise cleaning of the equal-time covariance estimator.} A basic way to observe the measurement noise is by comparing the eigenvalues of the equal-time covariance matrix $\mathbf{C}$ and its estimator $\mathbf{c}$. They are important because they represent uncorrelated causes (``principal components,'' ``explicative factors'') for the correlations of the objects \smash{$R_{i a}$} at the same moment of time (the temporal index $a$ will thus be skipped below); indeed, diagonalizing the (symmetric and positive-definite) matrix $\mathbf{C}$, \smash{$\mathbf{C} \mathbf{v}_{j} = \lambda_{j} \mathbf{v}_{j}$}, yields a linear decomposition, \smash{$R_{i} = \sum_{j = 1}^{N} [ \mathbf{v}_{j} ]_{i} e_{j}$}, into uncorrelated factors of variances given by the eigenvalues, \smash{$\langle e_{j} e_{k} \rangle = \lambda_{j} \delta_{j k}$}. This procedure is known as the ``principal component analysis'' (PCA). A financial interpretation is that the eigenvectors describe uncorrelated portfolios (investments of \smash{$[ \mathbf{v}_{j} ]_{i}$} part of one's wealth into asset $i$) while the corresponding eigenvalues quantify their risks (this is important for the Markowitz risk assessment method~\cite{Markowitz1952}).

The simplest situation is \smash{$\mathbf{C} = \sigma^{2} \Id_{N}$}, in which case the mean spectrum of the estimator $\mathbf{c}$ (which is known as the ``Wishart random matrix''~\cite{Wishart1928}) is given by the famous ``Mar\v{c}enko-Pastur (MP) distribution''~\cite{MarcenkoPastur1967} (cf.~App.~\ref{aa:TM1ETCE} for derivation). In other words, the single eigenvalue \smash{$\lambda = \sigma^{2}$} is ``smeared'' by the measurement noise over the interval \smash{$[ x_{-} , x_{+} ]$}, where \smash{$x_{\pm} = \sigma^{2} ( 1 \pm \sqrt{r} )^{2}$}, with a density (\ref{eq:TM1ETCEEq03}).

In the seminal papers~\cite{LalouxCizeauBouchaudPotters1999,PlerouGopikrishnanRosenowAmaralStanley1999} (the results below come from~\cite{LalouxCizeauBouchaudPotters1999}; cf.~Fig.~1 there), the empirical spectral density of the estimator $\mathbf{c}$ derived from $N = 406$ financial returns of the S\&P500 index, sampled over $T = 1309$ days (i.e., $r \approx 0.31$), has been fitted by the MP distribution. A fit with \smash{$\sigma^{2} \approx 0.74$} happened to well describe most of the spectrum, except two regions, comprised of c.a.~$6 \%$ of the eigenvalues: (i) One small peak around a very big eigenvalue \smash{$\lambda_{1}$} (c.a.~$25$ times larger than \smash{$x_{+}$}). (ii) Several well-separated peaks just above \smash{$x_{+}$}. Moreover, the eigenvector \smash{$[ \mathbf{v}_{1} ]_{i} \approx 1 / \sqrt{N}$}, which represents an approximately equal investment in all the assets---a very risky portfolio, depending on the behavior of the market as a whole, and thus called the ``market mode.'' In the first approximation, \smash{$R_{i} \approx [ \mathbf{v}_{1} ]_{i} e_{1}$}, i.e., just one factor, the market, governs the behavior of all the assets. Furthermore, the eigenvectors corresponding to the intermediate peaks describe portfolios of strongly correlated assets from industrial sectors. To summarize, the bulk of the spectrum may be thought to represent a pure noise, whereas the spikes which leak out of it carry relevant information about the true structure of equal-time correlations.

This use of the MP distribution for discerning the true information from noise has led to a number of ``cleaning schemes'' devised to construct a covariance matrix \smash{$\widetilde{\mathbf{C}}$} which better reflects experimental data. For instance, in the ``eigenvalue clipping'' method~\cite{LalouxCizeauPottersBouchaud2000}, \smash{$\widetilde{\mathbf{C}}$} consists of the empirical eigenvalues lying above \smash{$x_{+}$} and their eigenvectors, while the eigenvalues below the upper MP edge, supposedly originating purely from the measurement noise, are all replaced by a common value \smash{$\lambda_{\textrm{noise}}$} such that the trace of the covariance matrix is unaltered. This scheme has been argued~\cite{BouchaudPotters2009} to slightly outperform classical cleaning methods such as the ``shrinkage algorithm''~\cite{LedoitWolf2004}.

These discoveries commenced a series of applications of RMT to econophysics. In particular, an important research program is to construct more realistic ``priors'' $\mathbf{C}$ which---unlike the trivial case above---already incorporate some features of the market, then perform the PCA of the estimator $\mathbf{c}$, and compare it with empirical data. More generally, one should assume a certain JPDF $\mathcal{P} ( \mathbf{R} ; \{ \textrm{parameters} \} )$, dependent on parameters which describe some aspects of the market dynamics, and compare the spectra of both estimators $\mathbf{c}$ and $\mathbf{c} ( t )$ with empirical data, thereby assessing the parameters of the selected JPDF. In particular, quite naturally, it will be demonstrated that true correlations between different assets but at the same moment of time are more visible in the spectrum of $\mathbf{c}$, while $\mathbf{c} ( t )$ becomes very useful for investigating true time-lagged correlations. In the next Sec.~\ref{ss:ModelsOfSpatialAndTemporalCorrelations}, a few such forms of the JPDF used in this paper will be introduced.

%%%%%%%%%%%%%%%%%%%%%%%%%%%%%%%%%%%%%%%%%%%%%%%%%%%%%%%%%%%%%%%%%%%%%%
%%%%%%%%%%%%%%%%%%%%%%%%%%%%%%%%%%%%%%%%%%%%%%%%%%%%%%%%%%%%%%%%%%%%%%

\subsection{Models of spatial and temporal correlations}
\label{ss:ModelsOfSpatialAndTemporalCorrelations}

As explained above, a basic approach to distinguishing relevant information about the true correlations in a given complex system from the measurement noise inevitably present in any covariance estimator is by comparing the empirical spectrum of an estimator with its theoretical spectrum obtained from the assumption of no correlations; any difference between the two results then from the true correlations. This is Toy Model 1 (Sec.~\ref{ss:TM1}) below. A more advanced version of this program is to assume a certain parametric form of the true correlations, compute the theoretical spectrum of an estimator and compare it with an experiment---this has a potential to more accurately determine the parameters of the system. This Section introduces two important classes of models of the true correlations which will be used for this purpose in the paper.

%%%%%%%%%%%%%%%%%%%%%%%%%%%%%%%%%%%%%%%%%%%%%%%%%%%%%%%%%%%%%%%%%%%%%%

\subsubsection{Industrial sectors}
\label{sss:IndustrialSectors}

\emph{Factor models.} In order to construct a prior $\mathbf{C}$ which would reflect the spike structure visible in the spectrum of $\mathbf{c}$, a simple way is through the ``factor models'' [``factor component analysis'' (FCA)].

For instance, a ``one-factor model'' [``market model,'' ``capital asset pricing model'' (CAPM)]~\cite{Sharpe1964}, sufficient to describe the market mode, assumes that each asset \smash{$R_{i}$} (the temporal index $a$ is omitted here) follows a certain factor \smash{$\phi_{0}$} (the ``market''; a random variable with volatility $\Sigma$) with strength \smash{$\beta_{i}$} (the ``market beta''), i.e., \smash{$R_{i} = \beta_{i} \phi_{0} + \epsilon_{i}$}, where the \smash{$\epsilon_{i}$} (the ``idiosyncratic noises,'' with volatilities \smash{$\sigma_{i}$} of comparable magnitude) are uncorrelated with \smash{$\phi_{0}$} and each other. The cross-covariance matrix reads, \smash{$C_{i j} = \Sigma^{2} \beta_{i} \beta_{j} + \sigma_{i}^{2} \delta_{i j}$}, and indeed has one large ($\sim N$) eigenvalue plus a sea of $( N - 1 )$ small ones.

More generally, a ``$K$-factor model''~\cite{Noh2000,PappPafkaNowakKondor2005,BrinnerConnor2008} further supposes that the idiosyncratic noises depend on $K$ hidden factors, corresponding to the industrial sectors, \smash{$\epsilon_{i} = \sum_{\alpha = 1}^{K} \beta_{i \alpha} \phi_{\alpha} + \widetilde{\epsilon}_{i}$}, where the industrial factors and the new idiosyncratic noises have volatilities \smash{$\Sigma_{\alpha}$} and \smash{$\sigma_{i}$}, respectively, and are all uncorrelated. The spectrum of the cross-covariance matrix, \smash{$C_{i j} =$} \smash{$\Sigma^{2} \beta_{i} \beta_{j} +$} \smash{$\sum_{\alpha = 1}^{K} \Sigma_{\alpha}^{2} \beta_{i \alpha} \beta_{j \alpha} +$} \smash{$\sigma_{i}^{2} \delta_{i j}$}, may be shown to contain not only the market mode but also the $K$ sectorial modes.

Other constructions along these lines exist, e.g., the ``hierarchically-nested factor models''~\cite{TumminelloLilloMantegna2007}. Regardless of the model, if the assets are Gaussian and there are no temporal correlations, the matrix $\mathbf{C}$ can simply be considered diagonal, with an appropriate structure of its eigenvalues.

\emph{Power-law-distributed $\mathbf{C}$.} Another model of $\mathbf{C}$ has been suggested~\cite{BouchaudPotters2009} to even better reflect the coexistence of small and large industrial sectors~\cite{BurdaJurkiewicz2004,MalevergneSornette2004}---a hierarchical, power-law distribution of the sector sizes~\cite{Marsili2002},
\begin{equation}\label{eq:PowerLawCDefinition}
\rho_{\mathbf{C}} ( \lambda ) = \frac{\mu ( \mu - 1 )^{\mu} \left( 1 - \lambda_{\min} \right)^{\mu}}{\left( \lambda - 1 + \mu \left( 1 - \lambda_{\min} \right) \right)^{\mu + 1}} \Theta \left( \lambda - \lambda_{\min} \right)
\end{equation}
[this is the density of eigenvalues of $\mathbf{C}$ defined in (\ref{eq:HermitianMSDDefinition}); it is normalized and yields \smash{$\frac{1}{N} \Tr \mathbf{C} = 1$}], which is described by two parameters, a slope $\mu > 1$ and the smallest sector size \smash{$\lambda_{\min} \in ( 0 , 1 - 1 / \mu )$} [$\Theta ( \cdot )$ is the Heaviside theta (step) function].

In~\cite{BouchaudPotters2009} (cf.~Fig.~2 there), $N = 500$ U.S. stocks has been observed over $T = 1000$ days (i.e., $r = 0.5$), and shown that the empirical spectrum of $\mathbf{c}$, after removing the market mode, is very well fitted with the Wishart distribution with (\ref{eq:PowerLawCDefinition}), where $\mu = 2$ and \smash{$\lambda_{\min} = 0.35$}.

This article investigates the above situation in two versions: (i) $\mathbf{C}$ has $K$ distinct eigenvalues (Toy Model 2a, Sec.~\ref{sss:TM2a}); (ii) $\mathbf{C}$ has power-law-distributed eigenvalues (\ref{eq:PowerLawCDefinition}) with $\mu = 2$ (Toy Model 2b, Sec.~\ref{sss:TM2b}).

%%%%%%%%%%%%%%%%%%%%%%%%%%%%%%%%%%%%%%%%%%%%%%%%%%%%%%%%%%%%%%%%%%%%%%

\subsubsection{Vector autoregression models}
\label{sss:VAR}

\emph{VAR.} A simple yet profound model for the evolution of statistical variables e.g. in macro-economy (cf.~the European Central Bank's ``Smets-Wouters model''~\cite{SmetsWouters2002}) or in finances (cf.~the ``linear causal influence models''~\cite{PottersBouchaudLaloux2005}) is ``vector autoregression'' (VAR) in which the value of any variable depends linearly on the values of some other variables in some previous times, plus an idiosyncratic noise (the ``error term'') \smash{$e_{i a}$} which, in anticipation of the applications below, is chosen complex Gaussian of zero mean and the nonzero covariance \smash{$\langle e_{i a} \overline{e_{j b}} \rangle \equiv C^{\textrm{id.}}_{i j} ( b - a )$},
\begin{equation}\label{eq:VARDefinition}
R_{i a} = e_{i a} + \sum_{j = 1}^{N} \sum_{b < a} K_{i j} ( b - a ) R_{j b} .
\end{equation}
[The summation over time $b$ is assumed to stretch from $- \infty$, i.e., this is far from the beginning of the dynamics. The ``influence kernel'' \smash{$K_{i j} ( c \geq 0 ) \equiv 0$}.] This recurrence equation can be readily solved,
\begin{equation}
\begin{split}\label{eq:VARExplicitly}
R_{i a} &=\\
&= \sum_{n \geq 0} \sum_{a_{1} , \ldots , a_{n} < 0} \sum_{j = 1}^{N} \cdot\\
&\cdot \left[ \mathbf{K} ( a_{1} ) \mathbf{K} ( a_{2} ) \ldots \mathbf{K} ( a_{n} ) \right]_{i j} e_{j , a + a_{1} + \ldots + a_{n}} ,
\end{split}
\end{equation}
where certain constraints must be imposed on the kernel to make this sum convergent (to be specified below on simplified examples). Consequently, the variables are complex Gaussian of zero mean and the covariance function translationally symmetric in time (i.e., dependent on $c = b - a$) which reads
\begin{equation}
\begin{split}\label{eq:VARCovarianceFunction}
\mathbf{C} ( c ) &=\\
&= \sum_{n , m \geq 0} \sum_{\substack{a_{1} , \ldots , a_{n} < 0\\b_{1} , \ldots , b_{m} < 0}} \cdot\\
&\cdot \mathbf{K} ( a_{1} ) \mathbf{K} ( a_{2} ) \ldots \mathbf{K} ( a_{n} ) \cdot\\
&\cdot \mathbf{C}^{\textrm{id.}} \left( \sum_{m^{\prime} = 1}^{m} b_{m^{\prime}} - \sum_{n^{\prime} = 1}^{n} a_{n^{\prime}} + c \right) \cdot\\
&\cdot \mathbf{K} ( b_{m} )^{\dagger} \mathbf{K} ( b_{m - 1} )^{\dagger} \ldots \mathbf{K} ( b_{1} )^{\dagger} .
\end{split}
\end{equation}

\emph{SVAR(1).} In this paper only quite drastically simplified versions of the VAR model will be considered: (i) The temporal dependence in (\ref{eq:VARDefinition}) is only on the previous moment, \smash{$\mathbf{K} ( c ) = \mathbf{B} \delta_{c , - 1}$}, for some $N \times N$ matrix $\mathbf{B}$ [this is abbreviated VAR(1)]. (ii) The covariance function of the error terms is trivial in the temporal sector, \smash{$\mathbf{C}^{\textrm{id.}} ( b - a ) = \mathbf{C}^{\textrm{id.}} \delta_{a b}$}; moreover, the spatial covariance matrix will be assumed diagonal, \smash{$C^{\textrm{id.}}_{i j} = ( \sigma_{i}^{\textrm{id.}} )^{2} \delta_{i j}$} [this is called ``structural VAR'' (SVAR)]. Under these circumstances, the covariance function decays exponentially with the time lag,
\begin{equation}\label{eq:SVAR1CovarianceFunction}
\mathbf{C} ( c ) = \left\{ \begin{array}{ll} \mathbf{F} \mathbf{B}^{\dagger , c} , & \textrm{for } c \geq 0 , \\ \mathbf{B}^{| c |} \mathbf{F} , & \textrm{for } c < 0 , \end{array} \right.
\end{equation}
where for short, \smash{$\mathbf{F} \equiv \sum_{n \geq 0} \mathbf{B}^{n} \mathbf{C}^{\textrm{id.}} \mathbf{B}^{\dagger , n}$}.

\emph{$\mathbf{B}$ diagonal.} The matrix $\mathbf{B}$ will first be chosen diagonal, \smash{$B_{i j} = \beta_{i} \delta_{i j}$}, which means that there is no correlation between different variables. Then \smash{$F_{i j} = f_{i} \delta_{i j}$}, where for short, \smash{$f_{i} \equiv ( \sigma_{i}^{\textrm{id.}} )^{2} / ( 1 - | \beta_{i} |^{2} )$}, and there must be \smash{$| \beta_{i} | < 1$} for all $i$. Hence, the covariance function becomes
\begin{equation}\label{eq:SVAR1BDiagonalCovarianceFunction}
C_{i j} ( c ) = f_{i} \left| \beta_{i} \right|^{| c |} \ee^{- \ii \, \textrm{arg} ( \beta_{i} ) c} \delta_{i j} ,
\end{equation}
i.e., it is diagonal with each term consisting of a distinct (i) proportionality factor \smash{$f_{i}$}; (ii) exponential decay with the time lag governed by \smash{$| \beta_{i} |$}; (iii) sinusoidal modulation with the time lag governed by \smash{$\textrm{arg} ( \beta_{i} )$} (however, the \smash{$\beta_{i}$} will henceforth be always assumed real, so no oscillations).

Three versions of this model will be investigated in this article: (i) all the \smash{$\beta_{i}$} equal to each other and \smash{$\sigma_{i}^{\textrm{id.}}$} equal to each other (Toy Model 3, Sec.~\ref{ss:TM3}); (ii) all the \smash{$\beta_{i}$} equal to each other but arbitrary \smash{$\sigma_{i}^{\textrm{id.}}$} (Toy Model 4a, Sec.~\ref{sss:TM4a}); (iii) arbitrary \smash{$\beta_{i}$} and \smash{$\sigma_{i}^{\textrm{id.}}$} (Toy Model 4b, Sec.~\ref{sss:TM4b}).

\emph{$\mathbf{B}$ with a market mode.} In order to incorporate correlations between different variables, a more involved non-diagonal structure of $\mathbf{B}$ is required. For instance, to account for the market mode (cf.~Sec.~\ref{sss:IndustrialSectors}), one may assume~\footnote{This model has been suggested to me by Zdzis{\l}aw Burda, and initially analyzed with Zdzis{\l}aw Burda, Giacomo Livan and Artur \'{S}wi\textpolhook{e}ch.} that there is one asset (the ``market'') which depends only on itself, \smash{$R_{1 a} = e_{1 a} + \alpha R_{1 , a - 1}$}, while any other asset depends on the market and on itself, \smash{$R_{i a} = e_{i a} + \beta R_{1 , a - 1} + \gamma R_{i , a - 1}$}, for $i > 1$, where the beta coefficients $\alpha \in ( 0 , 1 )$, $\beta \in \mathbb{R}$, $\gamma \in ( 0 , 1 )$ are identical for all the assets. Consequently, the nonzero matrix elements of $\mathbf{B}$ read, \smash{$B_{1 1} = \alpha$}, \smash{$B_{i 1} = \beta$}, \smash{$B_{i i} = \gamma$}, for $i > 1$ (i.e., the diagonal and first column). The covariance function thus becomes
\begin{subequations}
\begin{align}
&C_{1 1} ( c ) = \frac{1}{1 - \alpha^{2}} \alpha^{| c |} ,\label{eq:SVAR1BMarketCovarianceFunction1}\\
&C_{i i} ( c ) = \frac{\beta^{2}}{( 1 - \alpha^{2} ) ( \alpha - \gamma ) ( 1 - \alpha \gamma )} \alpha^{1 + | c |} +\nonumber\\
&\quad\quad\quad + \frac{1}{1 - \gamma^{2}} \left( 1 - \frac{\beta^{2} \gamma}{( \alpha - \gamma ) ( 1 - \alpha \gamma )} \right) \gamma^{| c |} ,\label{eq:SVAR1BMarketCovarianceFunction2}\\
&\left. \begin{array}{ll} C_{1 i} ( c ) , & \textrm{for } c \geq 0 , \\ C_{i 1} ( c ) , & \textrm{for } c < 0 \end{array} \right\} =\nonumber\\
&\quad\quad\quad = \frac{\beta}{\alpha - \gamma} \left( \frac{1}{1 - \alpha^{2}} \alpha^{| c |} - \frac{1}{1 - \alpha \gamma} \gamma^{| c |} \right) ,\label{eq:SVAR1BMarketCovarianceFunction3}\\
&\left. \begin{array}{ll} C_{i 1} ( c ) , & \textrm{for } c \geq 0 , \\ C_{1 i} ( c ) , & \textrm{for } c < 0 \end{array} \right\} = \frac{\beta}{( 1 - \alpha^{2} ) ( 1 - \alpha \gamma )} \alpha^{1 + | c |} ,\label{eq:SVAR1BMarketCovarianceFunction4}\\
&C_{i j} ( c ) = \frac{\beta^{2}}{( \alpha - \gamma ) ( 1 - \alpha \gamma )} \cdot \nonumber\\
&\quad\quad\quad \cdot \left( \frac{1}{1 - \alpha^{2}} \alpha^{1 + | c |} - \frac{1}{1 - \gamma^{2}} \gamma^{1 + | c |} \right) ,\label{eq:SVAR1BMarketCovarianceFunction5}
\end{align}
\end{subequations}
for $i , j > 1$ and $i \neq j$ (i.e., five distinct matrix entries; the first row and column interchange upon $c \to - c$). It has the following (time-dependent) eigenvalues: (i) an $( N - 2 )$-fold degenerate eigenvalue,
\begin{equation}\label{eq:SVAR1BMarketCovarianceFunctionEigenvalue1}
\lambda_{1} ( c ) \equiv \frac{1}{1 - \gamma^{2}} \gamma^{| c |} ,
\end{equation}
describing the self-interaction of the assets; (ii) two eigenvalues solving the quadratic equation
\begin{equation}
\begin{split}\label{eq:SVAR1BMarketCovarianceFunctionEigenvalue2}
&( \alpha - \gamma ) \Big( ( 1 - \alpha \gamma )^{2} + ( N - 1 ) \beta^{2} \Big) ( \alpha \gamma )^{| c |} -\\
&- \bigg( \left( 1 - \gamma^{2} \right) \Big( ( \alpha - \gamma ) ( 1 - \alpha \gamma ) + ( N - 1 ) \beta^{2} \alpha \Big) \alpha^{| c |} +\\
&+ \left( 1 - \alpha^{2} \right) \Big( ( \alpha - \gamma ) ( 1 - \alpha \gamma ) - ( N - 1 ) \beta^{2} \gamma \Big) \gamma^{| c |} \bigg) \cdot\\
&\cdot ( 1 - \alpha \gamma ) \lambda_{2 , 3} ( c ) +\\
&+ \left( 1 - \alpha^{2} \right) \left( 1 - \gamma^{2} \right) ( \alpha - \gamma ) ( 1 - \alpha \gamma )^{2} \lambda_{2 , 3} ( c )^{2} = 0
\end{split}
\end{equation}
[one of them is large as $\textrm{O} ( N )$], describing the interactions of the market with itself and individual assets with the market. This is the content of Toy Model 4c, initially analyzed in Sec.~\ref{sss:TM4c}.

%%%%%%%%%%%%%%%%%%%%%%%%%%%%%%%%%%%%%%%%%%%%%%%%%%%%%%%%%%%%%%%%%%%%%%
%%%%%%%%%%%%%%%%%%%%%%%%%%%%%%%%%%%%%%%%%%%%%%%%%%%%%%%%%%%%%%%%%%%%%%
%%%%%%%%%%%%%%%%%%%%%%%%%%%%%%%%%%%%%%%%%%%%%%%%%%%%%%%%%%%%%%%%%%%%%%

\section{Master equations}
\label{s:MasterEquations}

%%%%%%%%%%%%%%%%%%%%%%%%%%%%%%%%%%%%%%%%%%%%%%%%%%%%%%%%%%%%%%%%%%%%%%
%%%%%%%%%%%%%%%%%%%%%%%%%%%%%%%%%%%%%%%%%%%%%%%%%%%%%%%%%%%%%%%%%%%%%%

\subsection{Outline}
\label{ss:MasterEquationsOutline}

%%%%%%%%%%%%%%%%%%%%%%%%%%%%%%%%%%%%%%%%%%%%%%%%%%%%%%%%%%%%%%%%%%%%%%

\subsubsection{Random matrix models}
\label{sss:RandomMatrixModels}

\begin{table*}
\begin{tabular}[t]{|p{0.09\textwidth}||p{0.09\textwidth}|p{0.09\textwidth}|p{0.09\textwidth}|p{0.09\textwidth}|p{0.09\textwidth}|p{0.09\textwidth}|p{0.09\textwidth}|p{0.09\textwidth}|p{0.09\textwidth}|}
\hline
& Case 1 [(\ref{eq:ArbitraryCovarianceFunctionDefinition1}), (\ref{eq:ArbitraryCovarianceFunctionDefinition2})] & \multicolumn{5}{l|}{Case 2 (\ref{eq:Case2Definition})} & Case 3 (\ref{eq:Case3Definition}) & \multicolumn{2}{l|}{Case 2 + 3 (\ref{eq:Case2Plus3Definition})} \\
\cline{3-7}\cline{9-10}
& & $\mathbf{C}$, $\mathbf{A}$ & \smash{$\mathbf{C} = \Id_{N}$} $\mathbf{A} = \textrm{ diag}$ & \multicolumn{3}{l|}{\smash{$\mathbf{A} = \Id_{T}$}} & & $\mathbf{C}$, $\mathbf{A}$ & \smash{$\mathbf{C} = \Id_{N}$}\\
\cline{5-7}
& & & & Gaussian & free L\'{e}vy & free L\'{e}vy, $\beta = 0$ & & &\\
\hline\hline
ETCE & [(\ref{eq:Case1ETCEEq06a}), (\ref{eq:Case1ETCEEq06b}), (\ref{eq:Case1ETCEEq08a}), (\ref{eq:Case1ETCEEq08b}), (\ref{eq:Case1ETCEEq09}), (\ref{eq:Case1ETCEEq03})] & \multicolumn{3}{l|}{(\ref{eq:Case2ETCEEq06}) (recalculation)} & [(\ref{eq:Case2ETCEEq08a}), (\ref{eq:Case2ETCEEq08b})] & (\ref{eq:Case2ETCEEq10}) & [(\ref{eq:Case3ETCEEq01a})-(\ref{eq:Case3ETCEEq02b}), (\ref{eq:Case1ETCEEq09}), (\ref{eq:Case1ETCEEq03})] & --- & ---\\
\hline
TLCE & [(\ref{eq:Case1TLCEEq05a})-(\ref{eq:Case1TLCEEq05h}), (\ref{eq:Case1TLCEEq07a}), (\ref{eq:Case1TLCEEq07b}), (\ref{eq:Case1ETCEEq09}), (\ref{eq:Case1ETCEEq03})] & [(\ref{eq:Case2TLCEEq01a})-(\ref{eq:Case2TLCEEq02b}), (\ref{eq:Case1TLCEEq07a}), (\ref{eq:Case1TLCEEq07b}), (\ref{eq:Case1ETCEEq09}), (\ref{eq:Case1ETCEEq03})] & (\ref{eq:Case2C1ADiagonalTLCEMasterEq}), borderline: [(\ref{eq:Case2C1ADiagonalTLCERExt}), (\ref{eq:Case2C1ADiagonalTLCERInt})] & (\ref{eq:Case2A1TLCEMasterEq}), borderline: [(\ref{eq:Case2A1TLCERExt}), (\ref{eq:Case2A1TLCERInt})] & [(\ref{eq:Case2A1TLCELevyEq01a}), (\ref{eq:Case2A1TLCELevyEq01b})] & (\ref{eq:Case2A1TLCELevyEq02}) & [(\ref{eq:Case3TLCEEq01a})-(\ref{eq:Case3TLCEEq04}), (\ref{eq:Case1ETCEEq09}), (\ref{eq:Case1ETCEEq03})] & [(\ref{eq:Case2Plus3TLCEEq11a})-(\ref{eq:Case2Plus3TLCEEq11c}), (\ref{eq:Case2Plus3TLCEEq06a}), (\ref{eq:Case2Plus3TLCEEq06b}), (\ref{eq:Case2Plus3TLCEEq04}), (\ref{eq:Case2Plus3TLCEEq10a}), (\ref{eq:Case2Plus3TLCEEq10b}), (\ref{eq:Case2Plus3TLCEEq08}), (\ref{eq:Case2Plus3TLCEEq15})] & [(\ref{eq:Case2Plus3TLCEC1Eq02a})-(\ref{eq:Case2Plus3TLCEC1Eq03})]\\
\hline
\end{tabular}
\caption{Summary of the general master equations obtained in Sec.~\ref{s:MasterEquations}. The only recalculated result is for the ETCE in Case 2; all the rest is new.}
\label{tab:MasterEquations}
\end{table*}

The goal of this article is a basic study of the covariance estimators [(\ref{eq:RealcDefinition}), (\ref{eq:RealctDefinition})] and [(\ref{eq:RealWeightedcDefinition}), (\ref{eq:RealWeightedctDefinition})]. Actually, a few more assumptions about them will be made (without changing the notation) which do not influence the end results at the leading order in the thermodynamic limit [(\ref{eq:ThermodynamicLimit}), (\ref{eq:SmallTimeLagLimit})], but which greatly simplify the calculations: (i) The returns are complex instead of real. (ii) The indices in the Kronecker delta in the delay matrix (\ref{eq:RealctDefinition}) are understood modulo $T$, i.e., it is unitary. (iii) The normalization $1 / ( T - t )$ in (\ref{eq:RealctDefinition}) is replaced by $1 / T$. Points (ii) and (iii) are easily justified by $t \ll T$; an explanation concerning the crucial point (i) is offered in~Sec.~\ref{sss:ComplexVsReal}. To summarize, the following two $N \times N$ random matrix models are investigated:

(i) The (Hermitian) ``equal-time covariance estimator'' (ETCE),
\begin{equation}\label{eq:cDefinition}
\mathbf{c} \equiv \frac{1}{T} \mathbf{R} \mathbf{R}^{\dagger} .
\end{equation}

(ii) The (non-Hermitian) ``time-lagged covariance estimator'' (TLCE),
\begin{equation}\label{eq:ctDefinition}
\mathbf{c} ( t ) \equiv \frac{1}{T} \mathbf{R} \mathbf{D} ( t ) \mathbf{R}^{\dagger} , \quad D_{a b} ( t ) \equiv \delta_{( a + t ) \textrm{ mod } T , b} .
\end{equation}
[An even more general model $\mathbf{b}$ (\ref{eq:Case1TLCEEq01}), which encompasses the weighted TLCE (\ref{eq:RealWeightedctDefinition}), is also touched upon in Secs.~\ref{sss:Case1TLCE} and~\ref{sss:Case2TLCE}.]

In these definitions, the $N \times T$ return matrix $\mathbf{R}$ consists of complex Gaussian random numbers of zero mean and the covariance function of the form,
\begin{subequations}
\begin{align}
\la R_{i a} \overline{R_{j b}} \ra &= \mathcal{C}_{i j , a b} ,\label{eq:ArbitraryCovarianceFunctionDefinition1}\\
\la R_{i a} R_{j b} \ra = \la \overline{R_{i a}} \overline{R_{j b}} \ra &= 0 .\label{eq:ArbitraryCovarianceFunctionDefinition2}
\end{align}
\end{subequations}

%%%%%%%%%%%%%%%%%%%%%%%%%%%%%%%%%%%%%%%%%%%%%%%%%%%%%%%%%%%%%%%%%%%%%%

\subsubsection{Mean spectral density}
\label{sss:MSD}

\emph{Basic quantities.} One quantity only will be calculated for the ETCE and TLCE---their ``mean spectral density'' (MSD) [(\ref{eq:HermitianMSDDefinition}), (\ref{eq:NonHermitianMSDDefinition})]. It is introduced in App.~\ref{aaa:HermitianGreenFunction} (Hermitian case) and App.~\ref{aaa:NonHermitianGreenFunctions} (non-Hermitian case), where it is also encoded in the form of more convenient objects---the Green function [(\ref{eq:HolomorphicGreenFunctionDefinition}), (\ref{eq:NonHolomorphicGreenFunctionDefinition})] or the $M$-transform [(\ref{eq:HolomorphicMTransformDefinition}), (\ref{eq:NonHolomorphicMTransformDefinition})].

\emph{Master equations for various true covariances.} The rest of this Section is devoted to deriving the ``master equations'' obeyed by these objects for several cases of the true covariance function (\ref{eq:ArbitraryCovarianceFunctionDefinition1}): (i) Case 1: Arbitrary \smash{$\mathcal{C}_{i j , a b}$} (Sec.~\ref{ss:Case1}). (ii) Case 2: Factorized \smash{$\mathcal{C}_{i j , a b} = C_{i j} A_{b a}$} (Sec.~\ref{ss:Case2}). (iii) Case 3: Translationally symmetric in time \smash{$\mathcal{C}_{i j , a b} = C_{i j} ( b - a )$} (Sec.~\ref{ss:Case3}). (iv) Case 2 + 3: Factorized and translationally symmetric in time \smash{$\mathcal{C}_{i j , a b} = C_{i j} A ( b - a )$} (Sec.~\ref{ss:Case2Plus3TLCE}).

All these results are summarized in Tab.~\ref{tab:MasterEquations}. They are new discoveries except for the ETCE in Case 2 (Sec.~\ref{sss:Case2ETCE}), which is presented for (i) completeness; (ii) comparison of the derivation process with the TLCE in the same case (Sec.~\ref{sss:Case2TLCE}); (iii) an extension to the free L\'{e}vy distribution of the returns (in Sec.~\ref{sss:Case2ETCE} for the ETCE, then in Sec.~\ref{sss:RotationalSymmetry} for the TLCE).

\emph{Application to toy models.} These general equations are subsequently (Sec.~\ref{s:TM}) solved for four specific Toy Models (1, 2a, 2b, 3) of the true covariance function, and commented on for three other more involved Toy Models (4a, 4b, 4c); they have already been mentioned in Sec.~\ref{ss:ModelsOfSpatialAndTemporalCorrelations}. Also, it is explained how they can be a basis for extensions to non-Gaussian probability distributions of the returns: Student (Sec.~\ref{sss:TM1TLCEStudent}) and free L\'{e}vy (Sec.~\ref{sss:TM1TLCELevy}), as well as the EWMA estimator (Sec.~\ref{sss:TM1TLCEEWMA}).

\emph{Techniques of derivation.} The method used to obtain the master equations is the planar diagrammatic expansion and Dyson-Schwinger (DS) equations. All its necessary concepts are introduced in a self-contained way in App.~\ref{aaa:DSAndGUE} (Hermitian case) and App.~\ref{aaa:DSAndGinUE} (non-Hermitian case). Occasionally (Sec.~\ref{sss:RotationalSymmetry}), the ``$N$-transform conjecture''---a mathematical hypothesis relating the eigenvalues and singular values of a non-Hermitian random matrix whose mean spectrum possesses rotational symmetry around zero---is alternatively used which greatly simplifies calculations for the relevant models.

\emph{Universality issues.} Remark also that the MSD is not universal---it depends on the particular probability distribution of the returns, and not only on the symmetries of the problem---but nonetheless it is useful in applications e.g. to finances (cf.~Sec.~\ref{sss:MeasurementNoise}). However, for the models with a rotationally symmetric mean spectrum, a universal form-factor is proposed and tested (Secs.~\ref{sss:TM1TLCEErfc} and~\ref{sss:TM2a}) which describes the steep decline of the MSD close to the borderline of its domain.

%%%%%%%%%%%%%%%%%%%%%%%%%%%%%%%%%%%%%%%%%%%%%%%%%%%%%%%%%%%%%%%%%%%%%%
%%%%%%%%%%%%%%%%%%%%%%%%%%%%%%%%%%%%%%%%%%%%%%%%%%%%%%%%%%%%%%%%%%%%%%

\subsection{Case 1: True covariances arbitrary}
\label{ss:Case1}

In this Section, the general structure of the true covariance function [(\ref{eq:ArbitraryCovarianceFunctionDefinition1}), (\ref{eq:ArbitraryCovarianceFunctionDefinition2})] is assumed, and the technique of Secs.~\ref{aaa:DSAndGUE} and~\ref{aaa:DSAndGinUE} applied to the ETCE and TLCE [actually, its generalization (\ref{eq:Case1TLCEEq01})], respectively.

%%%%%%%%%%%%%%%%%%%%%%%%%%%%%%%%%%%%%%%%%%%%%%%%%%%%%%%%%%%%%%%%%%%%%%

\subsubsection{Equal-time covariance estimator}
\label{sss:Case1ETCE}

\emph{Step 0:} Prior to that, remark that $\mathbf{c}$ is a product of two Gaussian random matrices, \smash{$\frac{1}{\sqrt{T}} \mathbf{R}$} and \smash{$\frac{1}{\sqrt{T}} \mathbf{R}^{\dagger}$} (the factor $1 / T$ could be split between the two terms in a different way, but with such a choice, they are of the same large-$T$ order), while the method is used most easily just to Gaussian matrices. To remove this obstacle, a linearization procedure has been proposed~\cite{GudowskaNowakJanikJurkiewiczNowak2003}, which introduces the larger matrix,
\begin{equation}\label{eq:Case1ETCEEq01}
\mathbf{c}^{\lin} \equiv \left( \begin{array}{cc} \Zero_{N} & \frac{1}{\sqrt{T}} \mathbf{R} \\ \frac{1}{\sqrt{T}} \mathbf{R}^{\dagger} & \Zero_{T} \end{array} \right) .
\end{equation}
Its basic property is that upon squaring,
\begin{equation}\label{eq:Case1ETCEEq02}
\left( \mathbf{c}^{\lin} \right)^{2} \equiv \left( \begin{array}{cc} \frac{1}{T} \mathbf{R} \mathbf{R}^{\dagger} & \Zero \\ \Zero & \frac{1}{T} \mathbf{R}^{\dagger} \mathbf{R} \end{array} \right) ,
\end{equation}
it yields a block-diagonal matrix with the blocks being the two possible products of the two terms in question, which means that \smash{$\mathbf{c}^{\lin}$} has the same non-zero eigenvalues as $\mathbf{c}$, counted twice (i.e., $2 \min ( N , T )$), plus $| N - T |$ zero modes. In other words, the spectrum of \smash{$\mathbf{c}^{\lin}$} consists of $\min ( N , T )$ $2$-valued square roots of the eigenvalues of $\mathbf{c}$, plus $| N - T |$ zero modes. This translates into a close relationship between the holomorphic $M$-transforms of the two random matrices,
\begin{equation}\label{eq:Case1ETCEEq03}
M_{\mathbf{c}^{\lin}} ( z ) = \frac{2 r}{1 + r} M_{\mathbf{c}} ( z^{2} ) .
\end{equation}
The cost of this linearization is the increase in the dimension of the matrix from $N \times N$ to $( N + T ) \times ( N + T )$. The four blocks of \smash{$\mathbf{c}^{\lin}$} (and of analogous matrices henceforth) will be denoted by $N N$, $N T$, $T N$, $T T$ (from left to right and top to bottom), according to their dimensions. For instance, for the Green function matrix (\ref{eq:HolomorphicGreenFunctionMatrixDefinition}),
\begin{equation}\label{eq:Case1ETCEEq04}
\mathbf{G}_{\mathbf{c}^{\lin}} ( z ) = \left( \begin{array}{cc} \mathbf{G}^{N N} & \mathbf{G}^{N T} \\ \mathbf{G}^{T N} & \mathbf{G}^{T T} \end{array} \right) ,
\end{equation}
and similarly for the self-energy matrix (cf.~the discussion of the 1LI diagrams in App.~\ref{aaa:DSAndGUE}).

\emph{Step 1:} The nonzero matrix elements of \smash{$\mathbf{c}^{\lin}$} are all Gaussian random numbers of zero mean, hence the full information about them is encoded in the propagators, and the nonzero ones are directly inferred from (\ref{eq:ArbitraryCovarianceFunctionDefinition1}) to be
\begin{equation}\label{eq:Case1ETCEEq05}
\la [ \mathbf{c}^{\lin} ]^{N T}_{i a} [ \mathbf{c}^{\lin} ]^{T N}_{b j} \ra = \frac{1}{T} \mathcal{C}_{i j , a b} .
\end{equation}

\emph{Step 2:} Knowing the propagators, one may explicitly write the second DS equation (\ref{eq:DS2}),
\begin{subequations}
\begin{align}
\Sigma^{N N}_{i j} &= \frac{1}{T} \sum_{a , b = 1}^{T} \mathcal{C}_{i j , a b} G^{T T}_{a b} ,\label{eq:Case1ETCEEq06a}\\
\Sigma^{T T}_{b a} &= \frac{1}{T} \sum_{i , j = 1}^{T} \mathcal{C}_{i j , a b} G^{N N}_{j i} ,\label{eq:Case1ETCEEq06b}
\end{align}
\end{subequations}
and \smash{$\boldsymbol{\Sigma}^{N T} = \boldsymbol{\Sigma}^{T N} = \Zero$}.

\emph{Step 3:} Since the nonzero blocks of the self-energy matrix lie on its diagonal [(\ref{eq:Case1ETCEEq06a}), (\ref{eq:Case1ETCEEq06b})],
\begin{equation}\label{eq:Case1ETCEEq07}
\boldsymbol{\Sigma}_{\mathbf{c}^{\lin}} = \left( \begin{array}{cc} \boldsymbol{\Sigma}^{N N} & \Zero \\ \Zero & \boldsymbol{\Sigma}^{T T} \end{array} \right) ,
\end{equation}
the first DS equation (\ref{eq:DS1}) implies that the Green function matrix is also block-diagonal, with \smash{$\mathbf{G}^{N T} = \mathbf{G}^{T N} = \Zero$} and
\begin{subequations}
\begin{align}
\mathbf{G}^{N N} &= \frac{1}{z \Id_{N} - \boldsymbol{\Sigma}^{N N}} ,\label{eq:Case1ETCEEq08a}\\
\mathbf{G}^{T T} &= \frac{1}{z \Id_{T} - \boldsymbol{\Sigma}^{T T}} .\label{eq:Case1ETCEEq08b}
\end{align}
\end{subequations}
One may say that the first set of DS equations decouples into a spatial and temporal sector; of course, the two sectors are coupled by the second set of DS equations [(\ref{eq:Case1ETCEEq06a}), (\ref{eq:Case1ETCEEq06b})].

\emph{Step 4:} Eqs. [(\ref{eq:Case1ETCEEq06a}), (\ref{eq:Case1ETCEEq06b}), (\ref{eq:Case1ETCEEq08a}), (\ref{eq:Case1ETCEEq08b})] are the master equations for the four matrices, \smash{$\mathbf{G}^{N N}$}, \smash{$\mathbf{G}^{T T}$}, \smash{$\boldsymbol{\Sigma}^{N N}$}, \smash{$\boldsymbol{\Sigma}^{T T}$}. Their explicit solution may be attempted if the true covariance function \smash{$\mathcal{C}_{i j , a b}$} has been specified---six examples will be analyzed in Sec.~\ref{s:TM}. Once the two blocks of the Green function matrix are found, adding their traces yields the holomorphic Green function (\ref{eq:HolomorphicGreenFunctionDefinition}) of \smash{$\mathbf{c}^{\lin}$},
\begin{equation}\label{eq:Case1ETCEEq09}
G_{\mathbf{c}^{\lin}} ( z ) = \frac{1}{N + T} \left( \Tr \mathbf{G}^{N N} + \Tr \mathbf{G}^{T T} \right) ,
\end{equation}
which is then easily related to the holomorphic Green function of the estimator $\mathbf{c}$ [(\ref{eq:Case1ETCEEq03}), (\ref{eq:HolomorphicMTransformDefinition})], and thereby to its MSD (\ref{eq:MSDFromHolomorphicGreenFunction}).

%%%%%%%%%%%%%%%%%%%%%%%%%%%%%%%%%%%%%%%%%%%%%%%%%%%%%%%%%%%%%%%%%%%%%%

\subsubsection{Generalized time-lagged covariance estimator}
\label{sss:Case1TLCE}

Consider the following generalization of the TLCE,
\begin{equation}\label{eq:Case1TLCEEq01}
\mathbf{b} \equiv \frac{1}{T} \mathbf{R} \mathbf{E} \mathbf{R}^{\dagger} \mathbf{F} ,
\end{equation}
where $\mathbf{E}$ and $\mathbf{F}$ are arbitrary constant Hermitian or non-Hermitian matrices (of dimensions $T \times T$ and $N \times N$, respectively). [Setting $\mathbf{E} = \mathbf{D} ( t )$ and \smash{$\mathbf{F} = \Id_{N}$} reduces $\mathbf{b}$ to $\mathbf{c} ( t )$.]

\emph{Step 0:} Before that, the matrix product in (\ref{eq:Case1TLCEEq01}) should be linearized in order to obtain a Gaussian random matrix,
\begin{equation}\label{eq:Case1TLCEEq02}
\mathbf{b}^{\lin} \equiv \left( \begin{array}{cc} \Zero_{N} & \frac{1}{\sqrt{T}} \mathbf{R} \mathbf{E} \\ \frac{1}{\sqrt{T}} \mathbf{R}^{\dagger} \mathbf{F} & \Zero_{T} \end{array} \right) .
\end{equation}
Analogously to the Hermitian case, the nonholomorphic $M$-transforms of the original and linearized matrices remain in the simple relationship (\ref{eq:Case1ETCEEq03}), so it is enough to consider only the latter.

\emph{Step 1:} The non-Hermitian procedure differs from its Hermitian counterpart essentially by one step, which is ``duplication'' [(\ref{eq:MatrixValuedGreenFunctionMatrixDefinition1})-(\ref{eq:MatrixValuedGreenFunctionMatrixDefinition3})] of the random matrix in question. Combining the notation from \emph{Step 0} of Sec.~\ref{sss:Case1ETCE} and (\ref{eq:MatrixValuedGreenFunctionMatrixDefinition3}), the Green function matrix [(\ref{eq:MatrixValuedGreenFunctionMatrixDefinition1}), (\ref{eq:MatrixValuedGreenFunctionMatrixDefinition2})] will have the block structure
\begin{equation}\label{eq:Case1TLCEEq03}
\mathbf{G}_{\mathbf{b}^{\lin,\dupl}} ( z , \overline{z} ) = \left( \begin{array}{cc|cc} \mathbf{G}^{N N} & \mathbf{G}^{N T} & \mathbf{G}^{N \overline{N}} & \mathbf{G}^{N \overline{T}} \\ \mathbf{G}^{T N} & \mathbf{G}^{T T} & \mathbf{G}^{T \overline{N}} & \mathbf{G}^{T \overline{T}} \\ \hline \mathbf{G}^{\overline{N} N} & \mathbf{G}^{\overline{N} T} & \mathbf{G}^{\overline{N} \overline{N}} & \mathbf{G}^{\overline{N} \overline{T}} \\ \mathbf{G}^{\overline{T} N} & \mathbf{G}^{\overline{T} T} & \mathbf{G}^{\overline{T} \overline{N}} & \mathbf{G}^{\overline{T} \overline{T}} \end{array} \right) ,
\end{equation}
and similarly for the self-energy matrix.

\emph{Step 2:} The propagators (covariances) of the returns (\ref{eq:ArbitraryCovarianceFunctionDefinition1}) allow to easily deduce the propagators of the duplicated version (\ref{eq:MatrixValuedGreenFunctionMatrixDefinition2}) of \smash{$\mathbf{b}^{\lin}$},
\begin{subequations}
\begin{align}
\la [ \mathbf{b}^{\lin,\dupl} ]^{N T}_{i a} [ \mathbf{b}^{\lin,\dupl} ]^{T N}_{b j} \ra &= \frac{1}{T} \sum_{k = 1}^{N} \sum_{c = 1}^{T} E_{c a} F_{k j} \mathcal{C}_{i k , c b} ,\label{eq:Case1TLCEEq04a}\\
\la [ \mathbf{b}^{\lin,\dupl} ]^{N T}_{i a} [ \mathbf{b}^{\lin,\dupl} ]^{\overline{T} \overline{N}}_{\overline{b} \overline{j}} \ra &= \frac{1}{T} \sum_{c = 1}^{T} \sum_{d = 1}^{T} E_{c a} E^{\dagger}_{\overline{b} d} \mathcal{C}_{i \overline{j} , c d} ,\label{eq:Case1TLCEEq04b}\\
\la [ \mathbf{b}^{\lin,\dupl} ]^{\overline{N} \overline{T}}_{\overline{i} \overline{a}} [ \mathbf{b}^{\lin,\dupl} ]^{T N}_{b j} \ra &= \frac{1}{T} \sum_{k = 1}^{N} \sum_{l = 1}^{N} F^{\dagger}_{\overline{i} k} F_{l j} \mathcal{C}_{k l , \overline{a} b} ,\label{eq:Case1TLCEEq04c}\\
\la [ \mathbf{b}^{\lin,\dupl} ]^{\overline{N} \overline{T}}_{\overline{i} \overline{a}} [ \mathbf{b}^{\lin,\dupl} ]^{\overline{T} \overline{N}}_{\overline{b} \overline{j}} \ra &= \frac{1}{T} \sum_{k = 1}^{N} \sum_{c = 1}^{T} E^{\dagger}_{\overline{b} c} F^{\dagger}_{\overline{i} k} \mathcal{C}_{k \overline{j} , \overline{a} c} .\label{eq:Case1TLCEEq04d}
\end{align}
\end{subequations}
As explained, since the random matrix in question is Gaussian of zero mean, these propagators carry the entire information about it.

\emph{Step 3:} The second set of DS equations (\ref{eq:DS2}) acquires thus the form,
\begin{subequations}
\begin{align}
\Sigma^{N N}_{i j} &= \frac{1}{T} \sum_{a , b , c , k} E_{c a} F_{k j} \mathcal{C}_{i k , c b} G^{T T}_{a b} ,\label{eq:Case1TLCEEq05a}\\
\Sigma^{T T}_{b a} &= \frac{1}{T} \sum_{i , j , c , k} E_{c a} F_{k j} \mathcal{C}_{i k , c b} G^{N N}_{j i} ,\label{eq:Case1TLCEEq05b}\\
\Sigma^{N \overline{N}}_{i \overline{j}} &= \frac{1}{T} \sum_{a , \overline{b} , c , d} E_{c a} E^{\dagger}_{\overline{b} d} \mathcal{C}_{i \overline{j} , c d} G^{T \overline{T}}_{a \overline{b}} ,\label{eq:Case1TLCEEq05c}\\
\Sigma^{\overline{T} T}_{\overline{b} a} &= \frac{1}{T} \sum_{i , \overline{j} , c , d} E_{c a} E^{\dagger}_{\overline{b} d} \mathcal{C}_{i \overline{j} , c d} G^{\overline{N} N}_{\overline{j} i} ,\label{eq:Case1TLCEEq05d}\\
\Sigma^{\overline{N} N}_{\overline{i} j} &= \frac{1}{T} \sum_{\overline{a} , b , k , l} F^{\dagger}_{\overline{i} k} F_{l j} \mathcal{C}_{k l , \overline{a} b} G^{\overline{T} T}_{\overline{a} b} ,\label{eq:Case1TLCEEq05e}\\
\Sigma^{T \overline{T}}_{b \overline{a}} &= \frac{1}{T} \sum_{\overline{i} , j , k , l} F^{\dagger}_{\overline{i} k} F_{l j} \mathcal{C}_{k l , \overline{a} b} G^{N \overline{N}}_{j \overline{i}} ,\label{eq:Case1TLCEEq05f}\\
\Sigma^{\overline{N} \overline{N}}_{\overline{i} \overline{j}} &= \frac{1}{T} \sum_{\overline{a} , \overline{b} , k , c} E^{\dagger}_{\overline{b} c} F^{\dagger}_{\overline{i} k} \mathcal{C}_{k \overline{j} , \overline{a} c} G^{\overline{T} \overline{T}}_{\overline{a} \overline{b}} ,\label{eq:Case1TLCEEq05g}\\
\Sigma^{\overline{T} \overline{T}}_{\overline{b} \overline{a}} &= \frac{1}{T} \sum_{\overline{i} , \overline{j} , k , c} E^{\dagger}_{\overline{b} c} F^{\dagger}_{\overline{i} k} \mathcal{C}_{k \overline{j} , \overline{a} c} G^{\overline{N} \overline{N}}_{\overline{j} \overline{i}} ,\label{eq:Case1TLCEEq05h}
\end{align}
\end{subequations}
with the remaining blocks of the self-energy matrix zero.

\emph{Step 4:} The block structure of the self-energy matrix which emerges from [(\ref{eq:Case1TLCEEq05a})-(\ref{eq:Case1TLCEEq05h})] is such that each of its main four blocks (holomorphic-holomorphic, etc.) is block-diagonal,
\begin{equation}\label{eq:Case1TLCEEq06}
\boldsymbol{\Sigma}_{\mathbf{b}^{\lin,\dupl}} ( z , \overline{z} ) = \left( \begin{array}{cc|cc} \boldsymbol{\Sigma}^{N N} & \Zero & \boldsymbol{\Sigma}^{N \overline{N}} & \Zero \\ \Zero & \boldsymbol{\Sigma}^{T T} & \Zero & \boldsymbol{\Sigma}^{T \overline{T}} \\ \hline \boldsymbol{\Sigma}^{\overline{N} N} & \Zero & \boldsymbol{\Sigma}^{\overline{N} \overline{N}} & \Zero \\ \Zero & \boldsymbol{\Sigma}^{\overline{T} T} & \Zero & \boldsymbol{\Sigma}^{\overline{T} \overline{T}} \end{array} \right) ,
\end{equation}
and owing to it, the first set of DS equations (\ref{eq:DS1}) may be rewritten as two matrix equations (of dimensions $2 N \times 2 N$ and $2 T \times 2 T$, respectively),
\begin{subequations}
\begin{align}
\left( \begin{array}{c|c} \mathbf{G}^{N N} & \mathbf{G}^{N \overline{N}} \\ \hline \mathbf{G}^{\overline{N} N} & \mathbf{G}^{\overline{N} \overline{N}} \end{array} \right) &= \left( \begin{array}{c|c} z \Id_{N} - \mathbf{\Sigma}^{N N} & - \mathbf{\Sigma}^{N \overline{N}} \\ \hline - \mathbf{\Sigma}^{\overline{N} N} & \overline{z} \Id_{N} - \mathbf{\Sigma}^{\overline{N} \overline{N}} \end{array} \right)^{- 1} ,\label{eq:Case1TLCEEq07a}\\
\left( \begin{array}{c|c} \mathbf{G}^{T T} & \mathbf{G}^{T \overline{T}} \\ \hline \mathbf{G}^{\overline{T} T} & \mathbf{G}^{\overline{T} \overline{T}} \end{array} \right) &= \left( \begin{array}{c|c} z \Id_{T} - \mathbf{\Sigma}^{T T} & - \mathbf{\Sigma}^{T \overline{T}} \\ \hline - \mathbf{\Sigma}^{\overline{T} T} & \overline{z} \Id_{T} - \mathbf{\Sigma}^{\overline{T} \overline{T}} \end{array} \right)^{- 1} .\label{eq:Case1TLCEEq07b}
\end{align}
\end{subequations}
Analogously to the Hermitian case [(\ref{eq:Case1ETCEEq08a})-(\ref{eq:Case1ETCEEq08b})], the first set of DS equations decouples into a spatial and temporal sector.

\emph{Step 5:} As in the Hermitian case, the nonholomorphic Green function (\ref{eq:NonHolomorphicGreenFunctionDefinition}) of \smash{$\mathbf{b}^{\lin}$} is given by (\ref{eq:Case1ETCEEq09}), which can a priori be calculated once the master equations [(\ref{eq:Case1TLCEEq05a})-(\ref{eq:Case1TLCEEq05h}), (\ref{eq:Case1TLCEEq07a}), (\ref{eq:Case1TLCEEq07b})] have been solved. Using (\ref{eq:Case1ETCEEq03}) and [(\ref{eq:NonHolomorphicMTransformDefinition}), (\ref{eq:MSDFromNonHolomorphicGreenFunction})] leads then to the nonholomorphic Green function and MSD of $\mathbf{b}$.

As explained in App.~\ref{aaa:NonHermitianGreenFunctions}, another interesting part of the duplicated Green function matrix [(\ref{eq:MatrixValuedGreenFunctionMatrixDefinition1})-(\ref{eq:MatrixValuedGreenFunctionMatrixDefinition3})] is the order parameter---the negated product of the normalized traces of its off-diagonal blocks (\ref{eq:BDefinition}),
\begin{equation}
\begin{split}\label{eq:Case1TLCEEq08}
\mathcal{B}_{\mathbf{b}^{\lin}} ( z , \overline{z} ) = - & \frac{1}{N + T} \left( \Tr \mathbf{G}^{N \overline{N}} + \Tr \mathbf{G}^{T \overline{T}} \right) \cdot\\
&\cdot \frac{1}{N + T} \left( \Tr \mathbf{G}^{\overline{N} N} + \Tr \mathbf{G}^{\overline{T} T} \right) .
\end{split}
\end{equation}
Recall that it necessarily is a real and nonnegative number, which zeroes only on the borderline of the mean spectral domain, \smash{$\partial \mathcal{D}_{\mathbf{b}^{\lin}}$} (and \smash{$\partial \mathcal{D}_{\mathbf{b}}$}) (\ref{eq:BorderlineEquation}), thus providing a way for its analytical determination.

%%%%%%%%%%%%%%%%%%%%%%%%%%%%%%%%%%%%%%%%%%%%%%%%%%%%%%%%%%%%%%%%%%%%%%

\subsubsection{Complex instead of real assets do not influence the leading order}
\label{sss:ComplexVsReal}

An argument will now be given why complex instead of real assets lead to much simpler models and yet do not change the end results at the leading order in the thermodynamic limit (\ref{eq:ThermodynamicLimit}), as anticipated in~Sec.~\ref{sss:RandomMatrixModels}. Of course, the microscopic properties for these two classes are altogether different, however, they are not the subject of this article; for an advanced calculation of some of such properties for a Wishart matrix with real entries, cf.~e.g.~\cite{RecherKieburgGuhr2010-1,RecherKieburgGuhr2010-2}. The proof will be given only for $\mathbf{b}$, since for the ETCE it has already been presented in~\cite{BurdaJurkiewiczWaclaw2005-1}, albeit in a less general setting of the true covariance function factorized (Case 2 below)---it has been demonstrated that all the moments of the ETCE are identical at the leading order for complex or real assets.

If the random variables \smash{$R_{i a}$} were real, with the most general structure of the propagator (\ref{eq:TrueRealCovarianceFunctionDefinition}), then in addition to [(\ref{eq:Case1TLCEEq04a})-(\ref{eq:Case1TLCEEq04d})], there would be six nontrivial propagators of \smash{$\mathbf{b}^{\lin,\dupl}$}, namely between its blocks: $N T$ and \smash{$\overline{N} \overline{T}$}, $T N$ and \smash{$\overline{T} \overline{N}$}, as well as the self-propagators of $N T$, $T N$, \smash{$\overline{N} \overline{T}$}, \smash{$\overline{T} \overline{N}$}, e.g.,
\begin{equation}\label{eq:ComplexVsRealEq01}
\la [ \mathbf{b}^{\lin,\dupl} ]^{N T}_{i a} [ \mathbf{b}^{\lin,\dupl} ]^{N T}_{j b} \ra = \frac{1}{T} \sum_{c = 1}^{T} \sum_{d = 1}^{T} E_{c a} E_{d b} \mathcal{C}_{i j , c d} ,
\end{equation}
etc. This would lead to eight additional nontrivial blocks in the self-energy matrix, those which are zero in (\ref{eq:Case1TLCEEq06}), e.g.,
\begin{equation}\label{eq:ComplexVsRealEq02}
\Sigma^{N T}_{i a} = \frac{1}{T} \sum_{b , c , d , j} E_{c b} E_{d a} \mathcal{C}_{i j , c d} G^{T N}_{b j} ,
\end{equation}
etc. In other words, the solutions of all our models would become significantly harder.

However, these new blocks (\ref{eq:ComplexVsRealEq02}) are one order of $T$ smaller than [(\ref{eq:Case1TLCEEq05a})-(\ref{eq:Case1TLCEEq05h})], i.e., may be neglected in the thermodynamic limit. Indeed, if one follows the summation indices in (\ref{eq:ComplexVsRealEq02}), $j \to b \to c \to d$, and treats it like matrix multiplication, one finds that \smash{$\Sigma^{N T}_{i a}$} is like one matrix entry of the resulting product, i.e., of order $1 / T$. On the other hand, in e.g. (\ref{eq:Case1TLCEEq05a}) one has one matrix entry through the summation over $k$, plus a trace from $a \to b \to c$; this trace contributes an order of $T$, hence \smash{$\Sigma^{N N}_{i j}$} is of order $1$.

For instance, in the simplest case of $\mathbf{E} = \mathbf{D} ( t )$, \smash{$\mathbf{F} = \Id_{N}$} (the TLCE) and \smash{$\mathcal{C}_{i j , a b} = \sigma^{2} \delta_{i j} \delta_{a b}$} (no correlations; this is Toy Model 1 below, Sec.~\ref{ss:TM1}) and for real assets, the eight entries [(\ref{eq:Case1TLCEEq05a})-(\ref{eq:Case1TLCEEq05h})] of the self-energy matrix are identical as in the complex case and read
\begin{subequations}
\begin{align}
\boldsymbol{\Sigma}^{N N} &= \sigma^{2} \left( \frac{1}{T} \Tr \left( \mathbf{D} \mathbf{G}^{T T} \right) \right) \Id_{N} ,\label{eq:ComplexVsRealEq03a}\\
\boldsymbol{\Sigma}^{T T} &= \sigma^{2} \left( \frac{1}{T} \Tr \mathbf{G}^{N N} \right) \mathbf{D} ,\label{eq:ComplexVsRealEq03b}\\
\boldsymbol{\Sigma}^{N \overline{N}} &= \sigma^{2} \left( \frac{1}{T} \Tr \mathbf{G}^{T \overline{T}} \right) \Id_{N} ,\label{eq:ComplexVsRealEq03c}\\
\boldsymbol{\Sigma}^{\overline{T} T} &= \sigma^{2} \left( \frac{1}{T} \Tr \mathbf{G}^{\overline{N} N} \right) \Id_{T} ,\label{eq:ComplexVsRealEq03d}\\
\boldsymbol{\Sigma}^{\overline{N} N} &= \sigma^{2} \left( \frac{1}{T} \Tr \mathbf{G}^{\overline{T} T} \right) \Id_{N} ,\label{eq:ComplexVsRealEq03e}\\
\boldsymbol{\Sigma}^{T \overline{T}} &= \sigma^{2} \left( \frac{1}{T} \Tr \mathbf{G}^{N \overline{N}} \right) \Id_{T} ,\label{eq:ComplexVsRealEq03f}\\
\boldsymbol{\Sigma}^{\overline{N} \overline{N}} &= \sigma^{2} \left( \frac{1}{T} \Tr \left( \mathbf{D}^{\TT} \mathbf{G}^{\overline{T} \overline{T}} \right) \right) \Id_{N} ,\label{eq:ComplexVsRealEq03g}\\
\boldsymbol{\Sigma}^{\overline{T} \overline{T}} &= \sigma^{2} \left( \frac{1}{T} \Tr \mathbf{G}^{\overline{N} \overline{N}} \right) \mathbf{D}^{\TT} ,\label{eq:ComplexVsRealEq03h}
\end{align}
\end{subequations}
however, the eight other entries (\ref{eq:ComplexVsRealEq02}) are now nonzero,
\begin{subequations}
\begin{align}
\boldsymbol{\Sigma}^{N T} &= \sigma^{2} \frac{1}{T} \left( \mathbf{G}^{T N} \right)^{\TT} ,\label{eq:ComplexVsRealEq04a}\\
\boldsymbol{\Sigma}^{T N} &= \sigma^{2} \frac{1}{T} \left( \mathbf{G}^{N T} \right)^{\TT} ,\label{eq:ComplexVsRealEq04b}\\
\boldsymbol{\Sigma}^{N \overline{T}} &= \sigma^{2} \frac{1}{T} \left( \mathbf{D} \mathbf{G}^{T \overline{N}} \right)^{\TT} ,\label{eq:ComplexVsRealEq04c}\\
\boldsymbol{\Sigma}^{\overline{N} T} &= \sigma^{2} \frac{1}{T} \left( \mathbf{D}^{\TT} \mathbf{G}^{\overline{T} N} \right)^{\TT} ,\label{eq:ComplexVsRealEq04d}\\
\boldsymbol{\Sigma}^{T \overline{N}} &= \sigma^{2} \frac{1}{T} \left( \mathbf{G}^{N \overline{T}} \mathbf{D}^{\TT} \right)^{\TT} ,\label{eq:ComplexVsRealEq04e}\\
\boldsymbol{\Sigma}^{\overline{T} N} &= \sigma^{2} \frac{1}{T} \left( \mathbf{G}^{\overline{N} T} \mathbf{D} \right)^{\TT} ,\label{eq:ComplexVsRealEq04f}\\
\boldsymbol{\Sigma}^{\overline{N} \overline{T}} &= \sigma^{2} \frac{1}{T} \left( \mathbf{G}^{\overline{T} \overline{N}} \right)^{\TT} ,\label{eq:ComplexVsRealEq04g}\\
\boldsymbol{\Sigma}^{\overline{T} \overline{N}} &= \sigma^{2} \frac{1}{T} \left( \mathbf{G}^{\overline{N} \overline{T}} \right)^{\TT} ,\label{eq:ComplexVsRealEq04h}
\end{align}
\end{subequations}
but it is clear that they are one order of $T$ smaller than the other entries, and thus irrelevant in the thermodynamic limit.

%%%%%%%%%%%%%%%%%%%%%%%%%%%%%%%%%%%%%%%%%%%%%%%%%%%%%%%%%%%%%%%%%%%%%%
%%%%%%%%%%%%%%%%%%%%%%%%%%%%%%%%%%%%%%%%%%%%%%%%%%%%%%%%%%%%%%%%%%%%%%

\subsection{Case 2: True covariances factorized}
\label{ss:Case2}

In this Section, a special form of the true covariance function (\ref{eq:ArbitraryCovarianceFunctionDefinition1}) is investigated---factorized into a spatial and temporal part,
\begin{equation}\label{eq:Case2Definition}
\mathcal{C}_{i j , a b} = C_{i j} A_{b a} ,
\end{equation}
where $\mathbf{C}$ and $\mathbf{A}$ are some constant Hermitian matrices of dimensions $N \times N$ and $T \times T$, respectively, describing the decoupled spatial and temporal covariances. (Note the convention about the indices of $\mathbf{A}$.)

%%%%%%%%%%%%%%%%%%%%%%%%%%%%%%%%%%%%%%%%%%%%%%%%%%%%%%%%%%%%%%%%%%%%%%

\subsubsection{Equal-time covariance estimator}
\label{sss:Case2ETCE}

\emph{Planar diagrammatics derivation.} For the ETCE, the second set of DS equations [(\ref{eq:Case1ETCEEq06a}), (\ref{eq:Case1ETCEEq06b})] takes on the form
\begin{subequations}
\begin{align}
\boldsymbol{\Sigma}^{N N} &= \mathbf{C} \gamma ,\label{eq:Case2ETCEEq01a}\\
\boldsymbol{\Sigma}^{T T} &= \mathbf{A} g ,\label{eq:Case2ETCEEq01b}
\end{align}
\end{subequations}
where for short,
\begin{subequations}
\begin{align}
\gamma &\equiv \frac{1}{T} \Tr \left( \mathbf{A} \mathbf{G}^{T T} \right) ,\label{eq:Case2ETCEEq02a}\\
g &\equiv \frac{1}{T} \Tr \left( \mathbf{C} \mathbf{G}^{N N} \right) .\label{eq:Case2ETCEEq02b}
\end{align}
\end{subequations}
In other words, the matrix structure of the self-energy blocks is discovered---they are proportional to $\mathbf{C}$ or $\mathbf{A}$, respectively, with unknown proportionality constants $\gamma$, $g$, which depend on the blocks of the Green function matrix, and this by multiplying these blocks by $\mathbf{A}$ or $\mathbf{C}$, respectively, and taking the traces normalized by $1 / T$.

Plugging [(\ref{eq:Case2ETCEEq01a}), (\ref{eq:Case2ETCEEq01b})] into the first set of DS equations [(\ref{eq:Case1ETCEEq08a}), (\ref{eq:Case1ETCEEq08b})], and using the definition of the holomorphic Green function matrix (\ref{eq:HolomorphicGreenFunctionMatrixDefinition}), one obtains
\begin{subequations}
\begin{align}
\mathbf{G}^{N N} &= \frac{1}{z \Id_{N} - \mathbf{C} \gamma} = \frac{1}{\gamma} \mathbf{G}_{\mathbf{C}} \left( \frac{z}{\gamma} \right) ,\label{eq:Case2ETCEEq03a}\\
\mathbf{G}^{T T} &= \frac{1}{z \Id_{T} - \mathbf{A} g} = \frac{1}{g} \mathbf{G}_{\mathbf{A}} \left( \frac{z}{g} \right) .\label{eq:Case2ETCEEq03b}
\end{align}
\end{subequations}

Now, one should combine Eqs.~[(\ref{eq:Case2ETCEEq03a}), (\ref{eq:Case2ETCEEq03b})] with the definitions of $\gamma$, $g$ [(\ref{eq:Case2ETCEEq02a}), (\ref{eq:Case2ETCEEq02b})] in a twofold way:

Firstly, insert the former into the latter, then use the definitions of the holomorphic Green function (\ref{eq:HolomorphicGreenFunctionDefinition}) and $M$-transform (\ref{eq:HolomorphicMTransformDefinition}), which allow to rewrite \smash{$\frac{1}{T} \Tr ( \mathbf{A} \mathbf{G}^{T T} ) = \frac{1}{T} \Tr ( \mathbf{A} ( z \Id_{T} - \mathbf{A} g )^{- 1} ) = M_{\mathbf{A}} ( z / g ) / g$}, and analogously in the spatial sector. This yields two scalar equations for two scalar unknowns $\gamma$, $g$,
\begin{equation}\label{eq:Case2ETCEEq04}
\gamma g = M_{\mathbf{A}} \left( \frac{z}{g} \right) = r M_{\mathbf{C}} \left( \frac{z}{\gamma} \right) .
\end{equation}

Secondly, take the traces of [(\ref{eq:Case2ETCEEq03a}), (\ref{eq:Case2ETCEEq03b})] and substitute into [(\ref{eq:Case1ETCEEq09}), (\ref{eq:Case1ETCEEq03})], using also (\ref{eq:Case2ETCEEq04}). This implies that the holomorphic $M$-transform of the ETCE is expressed through $\gamma$, $g$ as
\begin{equation}\label{eq:Case2ETCEEq05}
M_{\mathbf{c}} \left( z^{2} \right) = \frac{1}{r} \gamma g .
\end{equation}

The master equations [(\ref{eq:Case2ETCEEq04}), (\ref{eq:Case2ETCEEq05})] may be rewritten in an even more concise way by exploiting the notion of the ``holomorphic $N$-transform'' which for any Hermitian random matrix $\mathbf{H}$ is the functional inverse of the holomorphic $M$-transform,
\begin{equation}\label{eq:HolomorphicNTransformDefinition}
M_{\mathbf{H}} \left( N_{\mathbf{H}} ( z ) \right) = N_{\mathbf{H}} \left( M_{\mathbf{H}} ( z ) \right) = z .
\end{equation}
This quickly leads to one simple equation for \smash{$M \equiv M_{\mathbf{c}} ( z )$},
\begin{equation}\label{eq:Case2ETCEEq06}
r M N_{\mathbf{A}} ( r M ) N_{\mathbf{C}} ( M ) = z .
\end{equation}
It is the only master equation presented in this paper which has already been known (cf.~e.g.~\cite{BurdaJurkiewiczWaclaw2005-1,BurdaGorlichJaroszJurkiewicz2004}).

\emph{Free probability derivation.} Equation (\ref{eq:Case2ETCEEq06}) can also be derived in another even simpler way~\cite{BurdaJaroszJurkiewiczNowakPappZahed2010}, based on free probability theory of Voiculescu and Speicher~\cite{VoiculescuDykemaNica1992,Speicher1994}. Even though this derivation is already known, it will now be sketched because of three reasons: (i) It is based on an application of the multiplication law (\ref{eq:FreeProbabilityMultiplicationLaw}) to a product of two Hermitian matrices, a procedure which may a priori not be valid. For the ETCE, it does work. However, an analogous reasoning may be applied to the TLCE as well and it is quite surprising to discover that it yields a wrong result, albeit only slightly (cf.~Sec.~\ref{sss:RotationalSymmetry}). (ii) It can easily be adjusted to the situation of the returns distributed according to the free L\'{e}vy law, which is a model of a non-Gaussian behavior of financial assets (cf.~App.~\ref{aaa:FreeLevyDistribution}). However, this extension has been accomplished~\cite{BurdaJurkiewiczNowakPappZahed2004} only for \smash{$\mathbf{C} = \Id_{N}$}, \smash{$\mathbf{A} = \Id_{T}$} and the free L\'{e}vy distribution with zero skewness. Therefore, it will in what follows be worked out for arbitrary $\mathbf{C}$, $\mathbf{A}$ and the L\'{e}vy parameters $\alpha$, $\beta$, $\gamma$ (except the case $\alpha = 1$ with $\beta \neq 0$). (iii) This result will in turn constitute a basis for a derivation of the master equation for the TLCE and free L\'{e}vy assets (cf.~Sec.~\ref{sss:RotationalSymmetry}).

In order to rederive (\ref{eq:Case2ETCEEq06}), recall the ``multiplication law'' of free probability. If \smash{$\mathbf{U}_{1}$} and \smash{$\mathbf{U}_{2}$} are two free unitary random matrices (``freeness'' is a generalization of the notion of statistical independence to noncommuting objects; here it may be thought of simply as independence), then the MSD of their product is obtained by multiplying their $N$-transforms,
\begin{equation}\label{eq:FreeProbabilityMultiplicationLaw}
N_{\mathbf{U}_{1} \mathbf{U}_{2}} ( z ) = \frac{z}{z + 1} N_{\mathbf{U}_{1}} ( z ) N_{\mathbf{U}_{2}} ( z ) .
\end{equation}
This formula may be generalized to matrices other than unitary, although caution must then be exercised. For instance, it may be applied to Hermitian matrices, even though their product is generically non-Hermitian. It does works for certain models such as in~\cite{BurdaJaroszLivanNowakSwiech2010,BurdaJaroszLivanNowakSwiech2011} or the below derivation, however, Sec.~\ref{sss:RotationalSymmetry} contains a counterexample. (Cf.~the end of App.~\ref{aaa:FreeLevyDistribution} for the free probability addition law in the Hermitian world, and the end of App.~\ref{aaa:NonHermitianGreenFunctions} in the non-Hermitian world.)

Now if the true covariance function of the returns is factorized (\ref{eq:Case2Definition}), they can be recast as \smash{$\mathbf{R} \equiv \mathbf{C}^{1 / 2} \mathbf{R}_{1} \mathbf{A}^{1 / 2}$}, where the new returns are IID of variance $1$. Combining cyclic shifts of terms with the multiplication law (\ref{eq:FreeProbabilityMultiplicationLaw}) allows then to relate the ETCE $\mathbf{c}$ for the original returns $\mathbf{R}$ to the ETCE \smash{$\mathbf{c}_{1}$} for the IID returns \smash{$\mathbf{R}_{1}$},
\begin{subequations}
\begin{align}
N_{\mathbf{c}} ( z ) &=\nonumber\\
\stackrel{\substack{\textrm{cycl.}\\\downarrow}}{=} &N_{\frac{1}{T} \mathbf{R}_{1} \mathbf{A} \mathbf{R}_{1}^{\dagger} \mathbf{C}} ( z ) =\nonumber\\
\stackrel{\substack{\textrm{(\ref{eq:FreeProbabilityMultiplicationLaw})}\\\downarrow}}{=} &\frac{z}{1 + z} N_{\frac{1}{T} \mathbf{R}_{1} \mathbf{A} \mathbf{R}_{1}^{\dagger}} ( z ) N_{\mathbf{C}} ( z ) =\nonumber\\
\stackrel{\substack{\textrm{cycl.}\\\downarrow}}{=} &\frac{z}{1 + z} N_{\frac{1}{T} \mathbf{R}_{1}^{\dagger} \mathbf{R}_{1} \mathbf{A}} ( r z ) N_{\mathbf{C}} ( z ) =\nonumber\\
\stackrel{\substack{\textrm{(\ref{eq:FreeProbabilityMultiplicationLaw})}\\\downarrow}}{=} &\frac{r z^{2}}{( 1 + z ) ( 1 + r z )} N_{\frac{1}{T} \mathbf{R}_{1}^{\dagger} \mathbf{R}_{1}} ( r z ) N_{\mathbf{A}} ( r z ) N_{\mathbf{C}} ( z ) =\label{eq:Case2ETCEEq07a}\\
\stackrel{\substack{\textrm{cycl.}\\\downarrow}}{=} &\frac{r z^{2}}{( 1 + z ) ( 1 + r z )} N_{\mathbf{c}_{1}} ( z ) N_{\mathbf{A}} ( r z ) N_{\mathbf{C}} ( z ) .\label{eq:Case2ETCEEq07b}
\end{align}
\end{subequations}
Finally, inserting \smash{$N_{\mathbf{c}_{1}} ( z ) = ( 1 + z ) ( 1 + r z ) / z$}, which readily follows from the MP holomorphic Green function (\ref{eq:TM1ETCEEq02}), reproduces the master equation (\ref{eq:Case2ETCEEq06}).

\emph{Generalization to free L\'{e}vy returns.} In the free L\'{e}vy instead of Gaussian world, formula (\ref{eq:Case2ETCEEq07b}) remains intact [actually, (\ref{eq:Case2ETCEEq07a}) will be used; also, the normalization is rather \smash{$T^{- 2 / \alpha}$}, cf.~(\ref{eq:WignerLevyRandomMatrixDefinition})], only a new expression for \smash{$N_{\mathbf{c}_{1}} ( z )$} needs to be worked out.

It can be accomplished by the ``projector trick''~\cite{BurdaJaroszJurkiewiczNowakPappZahed2010}, which is to define a rectangular ($N \times T$; assume here $N \leq T$) free L\'{e}vy random matrix as the projection with $\mathbf{P} \equiv \textrm{diag} ( \Id_{N} , \Zero_{T - N} )$ of a square ($T \times T$) free L\'{e}vy Hermitian matrix $\mathbf{L}$. This implies \smash{$T^{- 2 / \alpha} \mathbf{R}_{1}^{\dagger} \mathbf{R}_{1} = \mathbf{L} \mathbf{P} \mathbf{L}$}, in which changing cyclicly the order of terms and using the multiplication law (\ref{eq:FreeProbabilityMultiplicationLaw}) (again for a pair of Hermitian matrices; it works this time as well) yields \smash{$N_{T^{- 2 / \alpha} \mathbf{R}_{1}^{\dagger} \mathbf{R}_{1}} ( z ) = ( z / ( z + 1 ) ) N_{\mathbf{P}} ( z ) N_{\mathbf{L}^{2}} ( z )$} \smash{$= ( ( z + r ) / ( z + 1 ) ) N_{\mathbf{L}^{2}} ( z )$}.

Secondly, simple algebra applied to the definition (\ref{eq:NonHolomorphicMTransformDefinition}) yields \smash{$M_{\mathbf{L}^{2}} ( z^{2} ) = \frac{1}{2} ( M_{\mathbf{L}} ( z ) + M_{\mathbf{L}} ( - z ) )$}. Now, for $z$ in the upper half-plane, \smash{$M_{\mathbf{L}} ( z )$} follows from (\ref{eq:LevyStableDistributionCharacteristicFunction}), while \smash{$M_{\mathbf{L}} ( - z )$} from the same equation only with $\beta \to - \beta$.

Combining these results leads to the following set of two master equations for two complex unknowns, \smash{$M = M_{\mathbf{c}} ( z )$} and auxiliary $\delta$,
\begin{subequations}
\begin{align}
&\varphi \frac{( r M + 1 + \delta )^{\alpha}}{r M + \delta} = \frac{( r M + 1 - \delta )^{\alpha}}{r M - \delta} ,\label{eq:Case2ETCEEq08a}\\
&\left( r^{2} M^{2} \left( 1 - \frac{\delta^{2}}{( r M + 1 )^{2}} \right) N_{\mathbf{A}} ( r M ) N_{\mathbf{C}} ( M ) \right)^{\alpha} \cdot\nonumber\\
&\cdot \frac{1}{r^{2} M^{2} - \delta^{2}} \ee^{\ii \pi \alpha} = \frac{z^{\alpha}}{\gamma^{2}} ,\label{eq:Case2ETCEEq08b}
\end{align}
\end{subequations}
where for short,
\begin{equation}\label{eq:Case2ETCEEq09}
\varphi \equiv \left\{ \begin{array}{ll} \ee^{\ii \pi \alpha \beta} , & \textrm{for } \alpha \in ( 0 , 1 ) , \\ \ee^{\ii \pi ( \alpha - 2 ) \beta} , & \textrm{for } \alpha \in ( 1 , 2 ) . \end{array} \right.
\end{equation}

In particular, for a symmetric distribution ($\beta = 0$, i.e., $\varphi = 1$), Eq.~(\ref{eq:Case2ETCEEq08a}) is trivially satisfied with $\delta = 0$, upon which Eq.~(\ref{eq:Case2ETCEEq08b}) turns into
\begin{equation}\label{eq:Case2ETCEEq10}
\ee^{\ii \pi \alpha} ( r M )^{2 ( \alpha - 1 )} N_{\mathbf{A}} ( r M )^{\alpha} N_{\mathbf{C}} ( M )^{\alpha} = \frac{z}{\gamma^{2}} .
\end{equation}
For $\alpha = 2$ and $\gamma = 1$, it further reduces to the Gaussian result (\ref{eq:Case2ETCEEq06}).

%%%%%%%%%%%%%%%%%%%%%%%%%%%%%%%%%%%%%%%%%%%%%%%%%%%%%%%%%%%%%%%%%%%%%%

\subsubsection{Generalized time-lagged covariance estimator}
\label{sss:Case2TLCE}

For $\mathbf{b}$ (\ref{eq:Case1TLCEEq01}), one is able to follow the steps in the above planar diagrammatics derivation to a certain extent, recognizing similar structures, however, a reduction to a single equation like (\ref{eq:Case2ETCEEq06}) is yet to be accomplished.

Really, the only manageable simplification so far is to write the second set of DS equations [(\ref{eq:Case1TLCEEq05a})-(\ref{eq:Case1TLCEEq05h})] in a form analogous to [(\ref{eq:Case2ETCEEq01a}), (\ref{eq:Case2ETCEEq01b}), (\ref{eq:Case2ETCEEq02a}), (\ref{eq:Case2ETCEEq02b})],
\begin{subequations}
\begin{align}
\left( \begin{array}{c|c} \boldsymbol{\Sigma}^{N N} & \boldsymbol{\Sigma}^{N \overline{N}} \\ \hline \boldsymbol{\Sigma}^{\overline{N} N} & \boldsymbol{\Sigma}^{\overline{N} \overline{N}} \end{array} \right) &= \left( \begin{array}{c|c} \mathbf{C} \mathbf{F} \gamma_{1} & \mathbf{C} \eta_{1} \\ \hline \mathbf{F}^{\dagger} \mathbf{C} \mathbf{F} \eta_{2} & \mathbf{F}^{\dagger} \mathbf{C} \gamma_{2} \end{array} \right) ,\label{eq:Case2TLCEEq01a}\\
\left( \begin{array}{c|c} \boldsymbol{\Sigma}^{T T} & \boldsymbol{\Sigma}^{T \overline{T}} \\ \hline \boldsymbol{\Sigma}^{\overline{T} T} & \boldsymbol{\Sigma}^{\overline{T} \overline{T}} \end{array} \right) &= \left( \begin{array}{c|c} \mathbf{A} \mathbf{E} g_{1} & \mathbf{A} h_{1} \\ \hline \mathbf{E}^{\dagger} \mathbf{A} \mathbf{E} h_{2} & \mathbf{E}^{\dagger} \mathbf{A} g_{2} \end{array} \right) ,\label{eq:Case2TLCEEq01b}
\end{align}
\end{subequations}
where
\begin{subequations}
\begin{align}
\left( \begin{array}{c|c} \gamma_{1} & \eta_{1} \\ \hline \eta_{2} & \gamma_{2} \end{array} \right) &= \frac{1}{T} \bTr \left( \begin{array}{c|c} \mathbf{A} \mathbf{E} \mathbf{G}^{T T} & \mathbf{E}^{\dagger} \mathbf{A} \mathbf{E} \mathbf{G}^{T \overline{T}} \\ \hline \mathbf{A} \mathbf{G}^{\overline{T} T} & \mathbf{E}^{\dagger} \mathbf{A} \mathbf{G}^{\overline{T} \overline{T}} \end{array} \right) ,\label{eq:Case2TLCEEq02a}\\
\left( \begin{array}{c|c} g_{1} & h_{1} \\ \hline h_{2} & g_{2} \end{array} \right) &= \frac{1}{T} \bTr \left( \begin{array}{c|c} \mathbf{C} \mathbf{F} \mathbf{G}^{N N} & \mathbf{F}^{\dagger} \mathbf{C} \mathbf{F} \mathbf{G}^{N \overline{N}} \\ \hline \mathbf{C} \mathbf{G}^{\overline{N} N} & \mathbf{F}^{\dagger} \mathbf{C} \mathbf{G}^{\overline{N} \overline{N}} \end{array} \right) .\label{eq:Case2TLCEEq02b}
\end{align}
\end{subequations}
These along with the first set of DS equations [(\ref{eq:Case1TLCEEq07a}), (\ref{eq:Case1TLCEEq07b})] form a set of master equations in the considered case.

In order to compare them with their ETCE counterparts, define
\begin{equation}\label{eq:Case2TLCEEq03}
\mathbf{E}^{\textrm{up}} \equiv \left( \begin{array}{c|c} \mathbf{E} & \Zero \\ \hline \Zero & \Id_{T} \end{array} \right) , \quad \mathbf{E}^{\textrm{down}} \equiv \left( \begin{array}{c|c} \Id_{T} & \Zero \\ \hline \Zero & \mathbf{E}^{\dagger} \end{array} \right) ,
\end{equation}
and recall \smash{$\mathbf{A}^{\dupl} = \diag ( \mathbf{A} , \mathbf{A} )$} (\ref{eq:MatrixValuedGreenFunctionMatrixDefinition2}), plus analogously in the spatial sector. Then, [(\ref{eq:Case1TLCEEq07a}), (\ref{eq:Case1TLCEEq07b})] combined with [(\ref{eq:Case2TLCEEq01a}), (\ref{eq:Case2TLCEEq01b})] read
\begin{subequations}
\begin{align}
\mathbf{G}^{N} &\equiv \left( \begin{array}{c|c} \mathbf{G}^{N N} & \mathbf{G}^{N \overline{N}} \\ \hline \mathbf{G}^{\overline{N} N} & \mathbf{G}^{\overline{N} \overline{N}} \end{array} \right) = \frac{1}{\mathcal{Z} - \mathbf{F}^{\textrm{down}} \mathbf{C}^{\dupl} \Gamma \mathbf{F}^{\textrm{up}}} ,\label{eq:Case2TLCEEq04a}\\
\mathbf{G}^{T} &\equiv \left( \begin{array}{c|c} \mathbf{G}^{T T} & \mathbf{G}^{T \overline{T}} \\ \hline \mathbf{G}^{\overline{T} T} & \mathbf{G}^{\overline{T} \overline{T}} \end{array} \right) = \frac{1}{\mathcal{Z} - \mathbf{E}^{\textrm{down}} \mathbf{A}^{\dupl} \mathcal{G} \mathbf{E}^{\textrm{up}}} ,\label{eq:Case2TLCEEq04b}
\end{align}
\end{subequations}
while the definitions [(\ref{eq:Case2TLCEEq02a}), (\ref{eq:Case2TLCEEq02b})] become
\begin{subequations}
\begin{align}
\Gamma &\equiv \left( \begin{array}{c|c} \gamma_{1} & \eta_{1} \\ \hline \eta_{2} & \gamma_{2} \end{array} \right) = \frac{1}{T} \bTr \left( \mathbf{A}^{\dupl} \mathbf{E}^{\textrm{up}} \mathbf{G}^{T} \mathbf{E}^{\textrm{down}} \right) ,\label{eq:Case2TLCEEq05a}\\
\mathcal{G} &\equiv \left( \begin{array}{c|c} g_{1} & h_{1} \\ \hline h_{2} & g_{2} \end{array} \right) = \frac{1}{T} \bTr \left( \mathbf{C}^{\dupl} \mathbf{F}^{\textrm{up}} \mathbf{G}^{N} \mathbf{F}^{\textrm{down}} \right) .\label{eq:Case2TLCEEq05b}
\end{align}
\end{subequations}
One recognizes now similarities between [(\ref{eq:Case2TLCEEq01a}), (\ref{eq:Case2TLCEEq01b})] and [(\ref{eq:Case2ETCEEq01a}), (\ref{eq:Case2ETCEEq01b})], [(\ref{eq:Case2TLCEEq04a}), (\ref{eq:Case2TLCEEq04b})] and [(\ref{eq:Case2ETCEEq03a}), (\ref{eq:Case2ETCEEq03b})], [(\ref{eq:Case2TLCEEq05a}), (\ref{eq:Case2TLCEEq05b})] and [(\ref{eq:Case2ETCEEq02a}), (\ref{eq:Case2ETCEEq02b})].

A way to reduce these master equations to a simple single equation like (\ref{eq:Case2ETCEEq06}) for the ETCE has unfortunately not been found. The reason may essentially be traced back to the fact that the block trace (\ref{eq:BlockTraceDefinition})---unlike the usual trace---does not exhibit the cyclic property; this obscures a way to write an analog of Eq.~(\ref{eq:Case2ETCEEq05}), which relates the spectral information about the estimator to the constant matrices $\Gamma$, $\mathcal{G}$. In other words, just as (\ref{eq:Case2ETCEEq06}) utilizes the notion of the holomorphic $N$-transform (\ref{eq:HolomorphicNTransformDefinition}) (of the Hermitian matrices $\mathbf{C}$ and $\mathbf{A}$), one would expect the desired simplification to be possible only once a proper definition of the $N$-transform for non-Hermitian matrices has been found, which is not yet the case (cf.~\cite{BurdaJanikNowak2011} for recent progress in this direction).

%%%%%%%%%%%%%%%%%%%%%%%%%%%%%%%%%%%%%%%%%%%%%%%%%%%%%%%%%%%%%%%%%%%%%%

\subsubsection{Rotational symmetry}
\label{sss:RotationalSymmetry}

\emph{Rotational symmetry of the MSD.} There is an important property, often encountered throughout this paper, which the MSD of a non-Hermitian random matrix $\mathbf{X}$ (e.g., the TLCE) may possess---rotational symmetry around zero, i.e., dependence only on
\begin{equation}\label{eq:RadiusDefinition}
R \equiv | z | ,
\end{equation}
in which case it is sufficient to consider just its radial part,
\begin{equation}
\begin{split}\label{eq:RadialMSDDefinition}
\rho^{\textrm{rad.}}_{\mathbf{X}} ( R ) &\equiv 2 \pi R \left. \rho_{\mathbf{X}} ( z , \overline{z} ) \right|_{| z | = R} =\\
&= \frac{\dd}{\dd R} \left. M_{\mathbf{X}} ( z , \overline{z} ) \right|_{| z | = R} ,
\end{split}
\end{equation}
where the lower line originates from [(\ref{eq:NonHolomorphicMTransformDefinition}), (\ref{eq:MSDFromNonHolomorphicGreenFunction})].

\emph{$N$-transform conjecture.} In such a situation, a very helpful tool is the following hypothesis which has been put forth and extensively tested in~\cite{BurdaJaroszLivanNowakSwiech2010,BurdaJaroszLivanNowakSwiech2011,Jarosz2011-01,Jarosz2012-01}:

(i) For any non-Hermitian random matrix model $\mathbf{X}$ whose nonholomorphic $M$-transform (and therefore, the MSD) is rotationally symmetric around zero,
\begin{equation}\label{eq:RotationallySymmetricNonHolomorphicMTransformDefinition}
M_{\mathbf{X}} ( z , \overline{z} ) = \mathfrak{M}_{\mathbf{X}} \left( R^{2} \right) ,
\end{equation}
one may invert (\ref{eq:RotationallySymmetricNonHolomorphicMTransformDefinition}) functionally,
\begin{equation}\label{eq:RotationallySymmetricNonHolomorphicNTransformDefinition}
\mathfrak{M}_{\mathbf{X}} \left( \mathfrak{N}_{\mathbf{X}} ( z ) \right) = z , \quad \mathfrak{N}_{\mathbf{X}} \left( \mathfrak{M}_{\mathbf{X}} \left( R^{2} \right) \right) = R^{2} ,
\end{equation}
which is named the ``rotationally symmetric non-holomorphic $N$-transform.''

(ii) For the Hermitian random matrix \smash{$\mathbf{X}^{\dagger} \mathbf{X}$}, the holomorphic $N$-transform is the functional inverse of the holomorphic $M$-transform (\ref{eq:HolomorphicNTransformDefinition}).

(iii) The $N$-transform conjecture claims that these two quantities are related through
\begin{equation}\label{eq:NTransformConjecture}
N_{\mathbf{X}^{\dagger} \mathbf{X}} ( z ) = \frac{z + 1}{z} \mathfrak{N}_{\mathbf{X}} ( z ) .
\end{equation}

It is useful because often one of the models, $\mathbf{X}$ or \smash{$\mathbf{X}^{\dagger} \mathbf{X}$}, is more tractable than the other (concerning computation of the MSD), and therefore Eq.~(\ref{eq:NTransformConjecture}) gains access to that more complicated model.

\emph{Relationship between the TLCE and ETCE assuming the rotational symmetry.} A general procedure to discern whether the MSD of the TLCE for a given true covariance function has the rotational symmetry will not be attempted in this work. Potential implications of this property will nonetheless be analyzed and two specific examples outlined.

Application of the above hypothesis to the TLCE requires caution concerning the order of terms:

Firstly, introduce the random matrix \smash{$\check{\mathbf{c}} ( t ) \equiv \frac{1}{T} \mathbf{R}^{\dagger} \mathbf{R} \mathbf{D} ( t )$}, which differs from $\mathbf{c} ( t )$ only by the order of terms and dimension; consequently, their nonholomorphic $M$-transforms relate as \smash{$M_{\check{\mathbf{c}} ( t )} ( z , \overline{z} ) = r M_{\mathbf{c} ( t )} ( z , \overline{z} ) \equiv r M$}.

Secondly, assume the rotational symmetry around zero (to be verified a posteriori), and write the rotationally symmetric non-holomorphic $N$-transform (\ref{eq:RotationallySymmetricNonHolomorphicNTransformDefinition}), \smash{$\mathfrak{N}_{\check{\mathbf{c}} ( t )} ( M_{\check{\mathbf{c}} ( t )} ( z , \overline{z} ) ) = R^{2}$}.

Thirdly, consider the Hermitian random matrix \smash{$\check{\mathbf{c}} ( t )^{\dagger} \check{\mathbf{c}} ( t ) = \mathbf{D} ( t )^{\dagger} \check{\mathbf{c}}^{2} \mathbf{D} ( t )$}, where \smash{$\check{\mathbf{c}} \equiv \frac{1}{T} \mathbf{R}^{\dagger} \mathbf{R}$}. The $N$-transform conjecture (\ref{eq:NTransformConjecture}) reads \smash{$N_{\check{\mathbf{c}} ( t )^{\dagger} \check{\mathbf{c}} ( t )} ( z ) = ( ( z + 1 ) / z ) \mathfrak{N}_{\check{\mathbf{c}} ( t )} ( z )$}. On the other hand, changing the order of the constituents and using the unitarity of $\mathbf{D} ( t )$, one finds \smash{$M_{\check{\mathbf{c}} ( t )^{\dagger} \check{\mathbf{c}} ( t )} ( z ) = M_{\check{\mathbf{c}}^{2}} ( z )$}.

Fourthly, recall \smash{$M_{\check{\mathbf{c}}^{2}} ( z^{2} ) = \frac{1}{2} ( M_{\check{\mathbf{c}}} ( z ) + M_{\check{\mathbf{c}}} ( - z ) )$}.

Fifthly, changing again the order of terms, \smash{$M_{\check{\mathbf{c}}} ( z ) = r M_{\mathbf{c}} ( z )$}, where $\mathbf{c}$ is the ETCE (\ref{eq:cDefinition}).

Collecting all the above formulae leads to an equation which relates the $M$-transforms of the TLCE and ETCE,
\begin{equation}\label{eq:MTransformsOfTLCEAndETCEAssumingRotationalSymmetry}
M = \re \, M_{\mathbf{c}} \left( \pm \ii R \sqrt{- \left( 1 + \frac{1}{r M} \right)} \right) .
\end{equation}
[The expression under the square root is always nonnegative, cf.~the discussion below. The sign, reflecting the two possible square roots, is irrelevant.]

\emph{Single ring conjecture.} Another hypothesis for non-Hermitian random matrices with the rotational symmetry (\ref{eq:RotationallySymmetricNonHolomorphicMTransformDefinition}) has been proposed~\cite{Jarosz2011-01,Jarosz2012-01}---which may be regarded as an extension of the ``Feinberg-Zee single ring theorem''~\cite{FeinbergZee1997-02,FeinbergScalettarZee2001,Feinberg2006,GuionnetKrishnapurZeitouni2009}---that their mean spectral domain is either a ring or an annulus; the radii of the enclosing circles are denoted by \smash{$R_{\textrm{ext.}}$} and \smash{$R_{\textrm{int.}}$}. Assuming this is true, for the TLCE, it is a general caveat that $M = 0$ on the external circle and $M = - 1 / r$ on the internal one, where the latter exists only for $r > 1$. Indeed, on the borderline, the nonholomorphic and holomorphic $M$-transforms coincide. But the only holomorphic function compatible with the rotational symmetry (i.e., depending on $R$) is a constant, \smash{$M_{\mathbf{c} ( t )} ( z ) = M_{0}$}. Moreover, it is known from RMT that in the external outside (i.e., containing $z = \infty$), the holomorphic Green function must behave as \smash{$G_{\mathbf{c} ( t )} ( z ) = ( M_{0} + 1 ) / z \sim 1 / z$}, for $z \to \infty$---which implies \smash{$M_{0} = 0$}. On the other hand, in the internal outside (i.e., containing $z = 0$), the value of \smash{$M_{0}$} is related to the zero modes: The MSD (\ref{eq:MSDFromNonHolomorphicGreenFunction}) reads, \smash{$\frac{1}{\pi} \partial_{\overline{z}} G_{\mathbf{c} ( t )} ( z ) = ( M_{0} + 1 ) \delta^{( 2 )} ( z , \overline{z} )$}, where the representation of the complex Dirac delta (\ref{eq:ComplexDiracDeltaRepresentation}) has been used. Now, zero modes appear in the spectrum of the TLCE only if $N > T$, and their number is $( N - T )$, i.e., their density, \smash{$( 1 - 1 / r ) \delta^{( 2 )} ( z , \overline{z} )$}; indeed, the matrices \smash{$\mathbf{R} \mathbf{D} ( t ) \mathbf{R}^{\dagger}$} and \smash{$\mathbf{R}^{\dagger} \mathbf{R} \mathbf{D} ( t )$} have the same nonzero eigenvalues, while the larger one has additionally $| N - T |$ zero modes. This implies \smash{$M_{0} = - 1 / r$} in the internal outside, which exists only for $r > 1$. To summarize, $M$ is a smooth function of $R$ which monotonically grows from $\max ( - 1 / r , - 1 )$ at \smash{$R = R_{\textrm{int.}}$} to $0$ at \smash{$R = R_{\textrm{ext.}}$}.

\emph{Example 1: Trivial temporal covariances.} It will be proven (cf.~Sec.~\ref{sss:Case2Plus3TLCEA1}) that the MSD of the TLCE certainly exhibits the rotational symmetry in the case of the true covariance function factorized (\ref{eq:Case2Definition}) with arbitrary $\mathbf{C}$ and \smash{$\mathbf{A} = \Id_{T}$}. The master equation for the holomorphic $M$-transform of the ETCE (\ref{eq:Case2ETCEEq06}) becomes then \smash{$( r M_{\mathbf{c}} ( z ) + 1 ) N_{\mathbf{C}} ( M_{\mathbf{c}} ( z ) ) = z$}, therefore Eq.~(\ref{eq:MTransformsOfTLCEAndETCEAssumingRotationalSymmetry}) turns into a complex equation for two real unknowns, \smash{$M = M_{\mathbf{c} ( t )} ( z , \overline{z} )$} and auxiliary $m$,
\begin{equation}\label{eq:Case2A1TLCEMasterEq}
M + \ii m = M_{\mathbf{C}} \left( \pm R \sqrt{- \left( 1 + \frac{1}{r M} \right)} \frac{1}{r m - \ii ( 1 + r M )} \right) .
\end{equation}
Remark that if $\mathbf{C}$ has a finite number of distinct eigenvalues, this can be recast as a set of polynomial equations (cf.~Toy Models 1 and 2a below). For $\mathbf{C}$ with an infinite number of distinct eigenvalues, this is typically more complicated (cf.~Toy Model 2b).

The radii of the circles enclosing the mean spectral domain are obtained from (\ref{eq:Case2A1TLCEMasterEq}) by expanding it around, respectively, $M = 0$ or if $r > 1$ also $M = - 1 / r$, which both are consistent with assuming an expansion around $m = 0$. The external one is given by the first and second moments of $\mathbf{C}$ (\ref{eq:MomentsDefinition}), while the internal one by its holomorphic $M$- and $N$-transforms,
\begin{subequations}
\begin{align}
R_{\textrm{ext.}}^{2} &= r^{2} m_{\mathbf{C} , 1}^{2} + r m_{\mathbf{C} , 2} , \label{eq:Case2A1TLCERExt}\\
R_{\textrm{int.}}^{2} &= r N_{\mathbf{C}} ( - 1 / r )^{3} M_{\mathbf{C}}^{\prime} \left( N_{\mathbf{C}} ( - 1 / r ) \right) \Theta ( r - 1 ) .\label{eq:Case2A1TLCERInt}
\end{align}
\end{subequations}
[Remark that (\ref{eq:Case2A1TLCERInt}) may be recast in a shorter but perhaps less useful way, \smash{$R_{\textrm{int.}}^{2} = - 2 r / \partial_{z} N_{\mathbf{C}} ( z )^{- 2} |_{z = - 1 / r}$}.]

\emph{Example 1---a surprisingly wrong derivation using the multiplication law.} Before proceeding to a second example of a rotationally-symmetric TLCE, recall that the master equation (\ref{eq:Case2ETCEEq06}) has been rederived (cf.~the middle part of Sec.~\ref{sss:Case2ETCE}) using the free probability multiplication law (\ref{eq:FreeProbabilityMultiplicationLaw}) by relating (\ref{eq:Case2ETCEEq07b}) the ETCE for true covariances factorized into $\mathbf{C}$, $\mathbf{A}$ (\ref{eq:Case2Definition}) to the ETCE for IID returns. One may therefore ask whether the master equation (\ref{eq:Case2A1TLCEMasterEq}) could be obtained in a similar fashion by relating the TLCE with the spatial covariance matrix $\mathbf{C}$ to the TLCE \smash{$\mathbf{c}_{1} ( t )$} for IID returns (which is Toy Model 1 below). The derivation should start from \smash{$\mathbf{R} \equiv \mathbf{C}^{1 / 2} \mathbf{R}_{1}$} and proceed as
\begin{subequations}
\begin{align}
\mathfrak{N}_{\mathbf{c} ( t )} ( z ) &=\nonumber\\
\stackrel{\substack{\textrm{cycl.}\\\downarrow}}{=} &\mathfrak{N}_{\mathbf{c}_{1} ( t ) \mathbf{C}} ( z ) =\nonumber\\
\stackrel{\substack{\textrm{(\ref{eq:NTransformConjecture})}\\\downarrow}}{=} &\frac{z}{z + 1} N_{\mathbf{C} \mathbf{c}_{1} ( t )^{\dagger} \mathbf{c}_{1} ( t ) \mathbf{C}} ( z ) =\nonumber\\
\stackrel{\substack{\textrm{cycl.}\\\downarrow}}{=} &\frac{z}{z + 1} N_{\mathbf{C}^{2} \mathbf{c}_{1} ( t )^{\dagger} \mathbf{c}_{1} ( t )} ( z ) =\nonumber\\
\stackrel{\substack{\textrm{(\ref{eq:FreeProbabilityMultiplicationLaw})}\\\downarrow}}{=} &\frac{z^{2}}{( z + 1 )^{2}} N_{\mathbf{C}^{2}} ( z ) N_{\mathbf{c}_{1} ( t )^{\dagger} \mathbf{c}_{1} ( t )} ( z ) =\nonumber\\
\stackrel{\substack{\textrm{(\ref{eq:NTransformConjecture})}\\\downarrow}}{=} &\frac{z}{z + 1} N_{\mathbf{C}^{2}} ( z ) \mathfrak{N}_{\mathbf{c}_{1} ( t )} ( z ) \label{eq:Case2A1TLCEWrongEq01}\\
&\textrm{(generally incorrect!).}\nonumber
\end{align}
\end{subequations}
Now the rotationally-symmetric nonholomorphic $N$-transform of \smash{$\mathbf{c}_{1} ( t )$} is given by functionally inverting (\ref{eq:TM1TLCEEq02}), which leads to the following equation
\begin{equation}
\begin{split}\label{eq:Case2A1TLCEWrongEq02}
M = \, &M_{\mathbf{C}^{2}} \left( R^{2} \frac{1 + r + r M}{r M ( 1 + r + 2 r M )^{2}} \right) .\\
&\textrm{(generally incorrect!).}
\end{split}
\end{equation}
It is surprising to discover that it is different from Eq.~(\ref{eq:Case2A1TLCEMasterEq}). It seems that the only place the above derivation may have failed is the application of the multiplication law to a product of two Hermitian matrices, as explained in Sec.~\ref{sss:Case2ETCE}. It is even more surprising to find out that the right and wrong master equations produce solutions with only a very slight difference in the MSD; Figure~\ref{fig:TM2aTLCETwoTests} (b) confirms the agreement of Monte Carlo simulations with Eq.~(\ref{eq:Case2A1TLCEMasterEq}) and their disagreement, albeit very small, with Eq.~(\ref{eq:Case2A1TLCEWrongEq02}).

\emph{Example 1---generalization to free L\'{e}vy returns.} Another point is to recall that the master equation (\ref{eq:Case2ETCEEq06}) for the ETCE for factorized true covariances (\ref{eq:Case2Definition}) has been generalized (cf.~the last part of Sec.~\ref{sss:Case2ETCE}) to the situation of free L\'{e}vy returns [(\ref{eq:Case2ETCEEq08a}), (\ref{eq:Case2ETCEEq08b})]. These formulae [with \smash{$\mathbf{A} = \Id_{T}$}, i.e., \smash{$M_{\mathbf{A}} ( z ) = 1 / ( z - 1 )$}] can be combined with (\ref{eq:MTransformsOfTLCEAndETCEAssumingRotationalSymmetry}) to yield a set of two master equations for the nonholomorphic $M$-transform of the TLCE for arbitrary $\mathbf{C}$ in the free L\'{e}vy case,
\begin{subequations}
\begin{align}
&\varphi \frac{( r M + 1 + \delta + \ii r m )^{\alpha}}{r M + \delta + \ii r m} = \frac{( r M + 1 - \delta + \ii r m )^{\alpha}}{r M - \delta + \ii r m} ,\label{eq:Case2A1TLCELevyEq01a}\\
&\left( \frac{\left( ( r M + 1 + \ii r m )^{2} - \delta^{2} \right) ( M + \ii m )}{( r M + 1 + \ii r m ) \sqrt{- \left( 1 + \frac{1}{r M} \right)}} N_{\mathbf{C}} ( M + \ii m ) \right)^{\alpha} \cdot\nonumber\\
&\cdot \frac{1}{r^{2} ( M + \ii m )^{2} - \delta^{2}} \ee^{\ii \pi \frac{\alpha}{2}} r^{\alpha} = \frac{R^{\alpha}}{\gamma^{2}} ,\label{eq:Case2A1TLCELevyEq01b}
\end{align}
\end{subequations}
where real $m$ and complex $\delta$ are auxiliary unknowns.

In particular, for zero skewness, Eq.~(\ref{eq:Case2A1TLCELevyEq01a}) is trivially satisfied with $\delta = 0$, while Eq.~(\ref{eq:Case2A1TLCELevyEq01b}) becomes
\begin{equation}
\begin{split}\label{eq:Case2A1TLCELevyEq02}
&\left( \frac{( r M + 1 + \ii r m ) ( M + \ii m )}{\sqrt{- \left( 1 + \frac{1}{r M} \right)}} N_{\mathbf{C}} ( M + \ii m ) \right)^{\alpha} \cdot\\
&\cdot \frac{1}{( M + \ii m )^{2}} \ee^{\ii \pi \frac{\alpha}{2}} r^{\alpha - 2} = \frac{R^{\alpha}}{\gamma^{2}} .
\end{split}
\end{equation}
Further setting $\alpha = 2$ and $\gamma = 1$ brings it back to the Gaussian result (\ref{eq:Case2A1TLCEMasterEq}).

A derivation of the values of the external and internal radii of the borderline of the mean spectral domain will not be presented in full generality, just commented on for special values of the parameters in Sec.~\ref{sss:TM1TLCELevy}.

\emph{Example 2: Trivial spatial covariances and diagonal temporal covariances.} Another situation when the rotational symmetry is apparent from an inspection of Monte Carlo data is the factorized covariance function (\ref{eq:Case2Definition}) with \smash{$\mathbf{C} = \Id_{N}$} and arbitrary diagonal \smash{$\mathbf{A} = \textrm{diag} ( \sigma_{1}^{2} , \ldots , \sigma_{T}^{2} )$}. The master equations for the TLCE [(\ref{eq:Case2TLCEEq01a})-(\ref{eq:Case2TLCEEq02b}), (\ref{eq:Case1TLCEEq07a}), (\ref{eq:Case1TLCEEq07b})] happen to be very cumbersome [especially the matrix inversion in (\ref{eq:Case1TLCEEq07b})], therefore a proof will not be offered, rather the assumption will be made and numerically verified a posteriori. Since Eq.~(\ref{eq:Case2ETCEEq06}) becomes \smash{$r ( M_{\mathbf{c}} ( z ) + 1 ) N_{\mathbf{A}} ( r M_{\mathbf{c}} ( z ) ) = z$}, Eq.~(\ref{eq:MTransformsOfTLCEAndETCEAssumingRotationalSymmetry}) turns again into a complex equation for two real unknowns, $M$ and auxiliary $m$,
\begin{equation}\label{eq:Case2C1ADiagonalTLCEMasterEq}
M + \ii m = \frac{1}{r} M_{\mathbf{A}} \left( \pm R \sqrt{- \left( 1 + \frac{1}{r M} \right)} \frac{1}{r \left( m - \ii ( 1 + M ) \right)} \right) .
\end{equation}
Recall that this equation holds conjecturally for $\mathbf{A}$ diagonal, not arbitrary.

The radii of the enclosing circles are found in an analogous way as above to be
\begin{subequations}
\begin{align}
R_{\textrm{ext.}}^{2} &= r m_{\mathbf{A} , 1}^{2} + r^{2} m_{\mathbf{A} , 2} , \label{eq:Case2C1ADiagonalTLCERExt}\\
R_{\textrm{int.}}^{2} &= \frac{( r - 1 )^{3}}{m_{\mathbf{A} , - 1}^{2} + ( r - 1 ) m_{\mathbf{A} , - 2}} \Theta ( r - 1 ) ,\label{eq:Case2C1ADiagonalTLCERInt}
\end{align}
\end{subequations}
where the negative moments are defined analogously to (\ref{eq:MomentsDefinition}) but with negative powers, and remark a different placement of the $r$ factors in (\ref{eq:Case2C1ADiagonalTLCERExt}) as compared to (\ref{eq:Case2A1TLCERExt}).

One could also directly generalize Eq.~(\ref{eq:Case2C1ADiagonalTLCEMasterEq}) to the free L\'{e}vy case; the result is not needed in this article and will not be presented.

%%%%%%%%%%%%%%%%%%%%%%%%%%%%%%%%%%%%%%%%%%%%%%%%%%%%%%%%%%%%%%%%%%%%%%
%%%%%%%%%%%%%%%%%%%%%%%%%%%%%%%%%%%%%%%%%%%%%%%%%%%%%%%%%%%%%%%%%%%%%%

\subsection{Case 3: True covariances translationally symmetric in time}
\label{ss:Case3}

Another interesting special form of the true covariance function (\ref{eq:ArbitraryCovarianceFunctionDefinition1}) is when its dependence on time is translationally symmetric,
\begin{equation}\label{eq:Case3Definition}
\mathcal{C}_{i j , a b} = C_{i j} ( b - a ) .
\end{equation}
Note that the right-hand side may be treated as a time-dependent $N \times N$ matrix, $\mathbf{C} ( c )$; it obeys \smash{$\mathbf{C} ( c ) = \mathbf{C} ( - c )^{\dagger}$}.

%%%%%%%%%%%%%%%%%%%%%%%%%%%%%%%%%%%%%%%%%%%%%%%%%%%%%%%%%%%%%%%%%%%%%%

\subsubsection{Fourier transformation}
\label{sss:Case3Fourier}

Since the thermodynamic limit (\ref{eq:ThermodynamicLimit}) is assumed, it proves very convenient---especially with the symmetry (\ref{eq:Case3Definition}) present in the system---to encode all the $T \times T$ matrices ($T \to \infty$; assume that the temporal matrix indices range from $- \infty$ to $+ \infty$) by their Fourier transforms. Namely, instead of any such matrix \smash{$B_{a b}$}, use
\begin{equation}\label{eq:Case3FourierEq01}
\hat{B} ( p , q ) \equiv \sum_{a , b = - \infty}^{+ \infty} \ee^{2 \pi \ii ( a p - b q )} B_{a b} ,
\end{equation}
where \smash{$p , q \in [ - 1 / 2 , 1 / 2 ]$}, which conversely reads,
\begin{equation}\label{eq:Case3FourierEq02}
B_{a b} = \int_{- 1 / 2}^{1 / 2} \int_{- 1 / 2}^{1 / 2} \dd p \dd q \ee^{- 2 \pi \ii ( a p - b q )} \hat{B} ( p , q ) .
\end{equation}
Remark that in particular, the Kronecker delta \smash{$\delta_{a b}$} is mapped to the Dirac delta $\delta ( p - q )$, while the matrix multiplication \smash{$\mathbf{B} = \mathbf{B}_{1} \mathbf{B}_{2}$} translates into the integration \smash{$\hat{B} ( p , q ) = \int_{- 1 / 2}^{1 / 2} \dd r \hat{B}_{1} ( p , r ) \hat{B}_{2} ( r , q )$}.

If a matrix satisfies \smash{$B_{a b} = B ( a - b )$}, where $B$ is some function of time, then its Fourier transform is proportional to the Dirac delta, \smash{$\hat{B} ( p , q ) = \delta ( p - q ) \hat{B} ( p )$}, with
\begin{subequations}
\begin{align}
\hat{B} ( p ) &\equiv \sum_{c = - \infty}^{+ \infty} \ee^{2 \pi \ii c p} B ( c ) ,\label{eq:Case3FourierEq03a}\\
B ( c ) &= \int_{- 1 / 2}^{1 / 2} \dd p \ee^{- 2 \pi \ii c p} \hat{B} ( p ) .\label{eq:Case3FourierEq03b}
\end{align}
\end{subequations}
In particular, matrix multiplication is simply the multiplication of these reduced Fourier transforms, \smash{$\hat{B} ( p ) = \hat{B}_{1} ( p ) \hat{B}_{2} ( p )$}. This language of Fourier transforms for $T \times T$ matrices will henceforth always be used provided that the temporal translational symmetry (\ref{eq:Case3Definition}) holds.

Remark finally that in a situation when the entries \smash{$B_{a b}$} themselves depend on $T$, the above procedure may not work. This is the case e.g. for the EWMA estimators [(\ref{eq:RealWeightedcDefinition}), (\ref{eq:RealWeightedctDefinition}), (\ref{eq:EWMADefinition})] in the limit (\ref{eq:EWMALimit}), to which a different method needs to be applied, namely Eq.~(\ref{eq:Case2C1ADiagonalTLCEMasterEq}) based on the rotational symmetry of the MSD (cf.~Sec.~\ref{sss:TM1TLCEEWMA}).

%%%%%%%%%%%%%%%%%%%%%%%%%%%%%%%%%%%%%%%%%%%%%%%%%%%%%%%%%%%%%%%%%%%%%%

\subsubsection{Equal-time covariance estimator}
\label{sss:Case3ETCE}

The general master equations for the ETCE [(\ref{eq:Case1ETCEEq06a}), (\ref{eq:Case1ETCEEq06b}), (\ref{eq:Case1ETCEEq08a}), (\ref{eq:Case1ETCEEq08b})] will now be written in the situation (\ref{eq:Case3Definition}) in the Fourier language.

The second set of DS equations acquires the form,
\begin{subequations}
\begin{align}
\boldsymbol{\Sigma}^{N N} &= \int_{- 1 / 2}^{1 / 2} \dd p \hat{\mathbf{C}} ( p ) \hat{G}^{T T} ( p ) ,\label{eq:Case3ETCEEq01a}\\
\hat{\Sigma}^{T T} ( p ) &= \frac{1}{T} \Tr \left( \hat{\mathbf{C}} ( p ) \mathbf{G}^{N N} \right) ,\label{eq:Case3ETCEEq01b}
\end{align}
\end{subequations}
while in the first set of DS equations, the spatial sector is unchanged and the temporal sector is Fourier-transformed, becoming a scalar equation,
\begin{subequations}
\begin{align}
\mathbf{G}^{N N} &= \frac{1}{z \Id_{N} - \boldsymbol{\Sigma}^{N N}} ,\label{eq:Case3ETCEEq02a}\\
\hat{G}^{T T} ( p ) &= \frac{1}{z - \hat{\Sigma}^{T T} ( p )} .\label{eq:Case3ETCEEq02b}
\end{align}
\end{subequations}
Finding a solution to [(\ref{eq:Case3ETCEEq01a}), (\ref{eq:Case3ETCEEq01b}), (\ref{eq:Case3ETCEEq02a}), (\ref{eq:Case3ETCEEq02b})] may be attempted once $\mathbf{C} ( c )$ [and thus \smash{$\hat{\mathbf{C}} ( p )$} (\ref{eq:Case3FourierEq03a}); it is Hermitian for any $p$] is known.

%%%%%%%%%%%%%%%%%%%%%%%%%%%%%%%%%%%%%%%%%%%%%%%%%%%%%%%%%%%%%%%%%%%%%%

\subsubsection{Time-lagged covariance estimator}
\label{sss:Case3TLCE}

Consider now the TLCE [not the more general random matrix $\mathbf{b}$ (\ref{eq:Case1TLCEEq01}), since arbitrary $\mathbf{E}$, $\mathbf{F}$ may break the translational symmetry in time].

The second set of DS equations [(\ref{eq:Case1TLCEEq05a})-(\ref{eq:Case1TLCEEq05h})] can thus be rewritten as follows: In the spatial sector,
\begin{subequations}
\begin{align}
\boldsymbol{\Sigma}^{N N} &= \int_{- 1 / 2}^{1 / 2} \dd p \ee^{- 2 \pi \ii t p} \hat{\mathbf{C}} ( p ) \hat{G}^{T T} ( p ) ,\label{eq:Case3TLCEEq01a}\\
\boldsymbol{\Sigma}^{N \overline{N}} &= \int_{- 1 / 2}^{1 / 2} \dd p \hat{\mathbf{C}} ( p ) \hat{G}^{T \overline{T}} ( p ) ,\label{eq:Case3TLCEEq01b}\\
\boldsymbol{\Sigma}^{\overline{N} N} &= \int_{- 1 / 2}^{1 / 2} \dd p \hat{\mathbf{C}} ( p ) \hat{G}^{\overline{T} T} ( p ) ,\label{eq:Case3TLCEEq01c}\\
\boldsymbol{\Sigma}^{\overline{N} \overline{N}} &= \int_{- 1 / 2}^{1 / 2} \dd p \ee^{2 \pi \ii t p} \hat{\mathbf{C}} ( p ) \hat{G}^{\overline{T} \overline{T}} ( p ) ,\label{eq:Case3TLCEEq01d}
\end{align}
\end{subequations}
while in the temporal sector,
\begin{subequations}
\begin{align}
\hat{\Sigma}^{T T} ( p ) &= \frac{1}{T} \ee^{- 2 \pi \ii t p} \Tr \left( \hat{\mathbf{C}} ( p ) \mathbf{G}^{N N} \right) ,\label{eq:Case3TLCEEq02a}\\
\hat{\Sigma}^{T \overline{T}} ( p ) &= \frac{1}{T} \Tr \left( \hat{\mathbf{C}} ( p ) \mathbf{G}^{N \overline{N}} \right) ,\label{eq:Case3TLCEEq02b}\\
\hat{\Sigma}^{\overline{T} T} ( p ) &= \frac{1}{T} \Tr \left( \hat{\mathbf{C}} ( p ) \mathbf{G}^{\overline{N} N} \right) ,\label{eq:Case3TLCEEq02c}\\
\hat{\Sigma}^{\overline{T} \overline{T}} ( p ) &= \frac{1}{T} \ee^{2 \pi \ii t p} \Tr \left( \hat{\mathbf{C}} ( p ) \mathbf{G}^{\overline{N} \overline{N}} \right) .\label{eq:Case3TLCEEq02d}
\end{align}
\end{subequations}

In the first set of DS equations [(\ref{eq:Case1TLCEEq07a}), (\ref{eq:Case1TLCEEq07b})], the spatial sector is unchanged, while the temporal sector is Fourier-transformed, becoming a $2 \times 2$ matrix equation, featuring explicitly doable $2 \times 2$ matrix inversion,
\begin{subequations}
\begin{align}
&\left( \begin{array}{c|c} \mathbf{G}^{N N} & \mathbf{G}^{N \overline{N}} \\ \hline \mathbf{G}^{\overline{N} N} & \mathbf{G}^{\overline{N} \overline{N}} \end{array} \right) =\nonumber\\
&\qquad = \left( \begin{array}{c|c} z \Id_{N} - \mathbf{\Sigma}^{N N} & - \mathbf{\Sigma}^{N \overline{N}} \\ \hline - \mathbf{\Sigma}^{\overline{N} N} & \overline{z} \Id_{N} - \mathbf{\Sigma}^{\overline{N} \overline{N}} \end{array} \right)^{- 1} ,\label{eq:Case3TLCEEq03a}\\
&\left( \begin{array}{c|c} \hat{G}^{T T} ( p ) & \hat{G}^{T \overline{T}} ( p ) \\ \hline \hat{G}^{\overline{T} T} ( p ) & \hat{G}^{\overline{T} \overline{T}} ( p ) \end{array} \right) =\nonumber\\
&\qquad = \left( \begin{array}{c|c} z - \hat{\Sigma}^{T T} ( p ) & - \hat{\Sigma}^{T \overline{T}} ( p ) \\ \hline - \hat{\Sigma}^{\overline{T} T} ( p ) & \overline{z} - \hat{\Sigma}^{\overline{T} \overline{T}} ( p ) \end{array} \right)^{- 1} =\nonumber\\
&\qquad = \frac{1}{W^{T} ( p )} \left( \begin{array}{c|c} \overline{z} - \hat{\Sigma}^{\overline{T} \overline{T}} ( p ) & \hat{\Sigma}^{T \overline{T}} ( p ) \\ \hline \hat{\Sigma}^{\overline{T} T} ( p ) & z - \hat{\Sigma}^{T T} ( p ) \end{array} \right) ,\label{eq:Case3TLCEEq03b}
\end{align}
\end{subequations}
where for short,
\begin{equation}
\begin{split}\label{eq:Case3TLCEEq04}
W^{T} ( p ) &\equiv\\
&\equiv \left( z - \hat{\Sigma}^{T T} ( p ) \right) \left( \overline{z} - \hat{\Sigma}^{\overline{T} \overline{T}} ( p ) \right) -\\
&- \hat{\Sigma}^{T \overline{T}} ( p ) \hat{\Sigma}^{\overline{T} T} ( p ) .
\end{split}
\end{equation}

%%%%%%%%%%%%%%%%%%%%%%%%%%%%%%%%%%%%%%%%%%%%%%%%%%%%%%%%%%%%%%%%%%%%%%
%%%%%%%%%%%%%%%%%%%%%%%%%%%%%%%%%%%%%%%%%%%%%%%%%%%%%%%%%%%%%%%%%%%%%%

\subsection{Time-lagged covariance estimator for Case 2 + 3: True covariances factorized and translationally symmetric in time}
\label{ss:Case2Plus3TLCE}

Recall that in the case of factorized true covariances (\ref{eq:Case2Definition}), the master equations for the ETCE have been simplified to a single equation (\ref{eq:Case2ETCEEq06}) (cf.~Sec.~\ref{sss:Case2ETCE}), which has not been accomplished for the TLCE (cf.~Sec.~\ref{sss:Case2TLCE}). Some simplification becomes however possible if in addition the temporal translational symmetry (\ref{eq:Case3Definition}) is assumed,
\begin{equation}\label{eq:Case2Plus3Definition}
\mathcal{C}_{i j , a b} = C_{i j} A ( b - a ) .
\end{equation}

%%%%%%%%%%%%%%%%%%%%%%%%%%%%%%%%%%%%%%%%%%%%%%%%%%%%%%%%%%%%%%%%%%%%%%

\subsubsection{Arbitrary $\mathbf{C}$ and $A$}
\label{sss:Case2Plus3TLCEArbitraryCA}

Firstly, the temporal sector of the second set of DS equations [(\ref{eq:Case3TLCEEq02a})-(\ref{eq:Case3TLCEEq02d})] acquires the form
\begin{subequations}
\begin{align}
\hat{\Sigma}^{T T} ( p ) &= \ee^{- 2 \pi \ii t p} \hat{A} ( p ) t_{1} ,\label{eq:Case2Plus3TLCEEq01a}\\
\hat{\Sigma}^{T \overline{T}} ( p ) &=  \hat{A} ( p ) t_{2} ,\label{eq:Case2Plus3TLCEEq01b}\\
\hat{\Sigma}^{\overline{T} T} ( p ) &=  \hat{A} ( p ) t_{3} ,\label{eq:Case2Plus3TLCEEq01c}\\
\hat{\Sigma}^{\overline{T} \overline{T}} ( p ) &= \ee^{2 \pi \ii t p} \hat{A} ( p ) t_{4} ,\label{eq:Case2Plus3TLCEEq01d}
\end{align}
\end{subequations}
where the information about the model is now encoded in four complex parameters,
\begin{subequations}
\begin{align}
t_{1} &\equiv \frac{1}{T} \Tr \left( \mathbf{C} \mathbf{G}^{N N} \right) ,\label{eq:Case2Plus3TLCEEq02a}\\
t_{2} &\equiv \frac{1}{T} \Tr \left( \mathbf{C} \mathbf{G}^{N \overline{N}} \right) ,\label{eq:Case2Plus3TLCEEq02b}\\
t_{3} &\equiv \frac{1}{T} \Tr \left( \mathbf{C} \mathbf{G}^{\overline{N} N} \right) ,\label{eq:Case2Plus3TLCEEq02c}\\
t_{4} &\equiv \frac{1}{T} \Tr \left( \mathbf{C} \mathbf{G}^{\overline{N} \overline{N}} \right) .\label{eq:Case2Plus3TLCEEq02d}
\end{align}
\end{subequations}
Notice that the definition of the Green function matrix [(\ref{eq:MatrixValuedGreenFunctionMatrixDefinition1}), (\ref{eq:MatrixValuedGreenFunctionMatrixDefinition2})] implies \smash{$t_{4} = \overline{t_{1}}$}; it will moreover be shown [cf.~Eq.~(\ref{eq:Case2Plus3TLCEEq16})] that
\begin{equation}\label{eq:Case2Plus3TLCEEq02e}
h \equiv - t_{2} t_{3} \in \mathbb{R} , \quad h \geq 0 ,
\end{equation}
where $h = 0$ only on the borderline of the mean spectral domain (\ref{eq:Case2Plus3TLCEEq17}). Hence, there really is one complex and one real unknown.

Secondly, substitute [(\ref{eq:Case2Plus3TLCEEq01a})-(\ref{eq:Case2Plus3TLCEEq01d})] into the temporal sector of the first set of DS equations (\ref{eq:Case3TLCEEq03b}),
\begin{equation}
\begin{split}\label{eq:Case2Plus3TLCEEq03}
&\left( \begin{array}{c|c} \hat{G}^{T T} ( p ) & \hat{G}^{T \overline{T}} ( p ) \\ \hline \hat{G}^{\overline{T} T} ( p ) & \hat{G}^{\overline{T} \overline{T}} ( p ) \end{array} \right) =\\
&= \frac{1}{W^{T} ( p )} \left( \begin{array}{c|c} \overline{z} - \ee^{2 \pi \ii t p}  \hat{A} ( p ) t_{4} & \hat{A} ( p ) t_{2} \\ \hline \hat{A} ( p ) t_{3} & z - \ee^{- 2 \pi \ii t p}  \hat{A} ( p ) t_{1} \end{array} \right) ,
\end{split}
\end{equation}
where (\ref{eq:Case3TLCEEq04}),
\begin{equation}
\begin{split}\label{eq:Case2Plus3TLCEEq04}
W^{T} ( p ) &=\\
&= | z |^{2} - \hat{A} ( p ) \left( \ee^{- 2 \pi \ii t p} \overline{z} t_{1} + \ee^{2 \pi \ii t p} z t_{4} \right) +\\
&+ \hat{A} ( p )^{2} \left( t_{1} t_{4} - t_{2} t_{3} \right) .
\end{split}
\end{equation}

Thirdly, knowing (the Fourier transforms of) the temporal blocks of the Green function matrix [(\ref{eq:Case2Plus3TLCEEq03}), (\ref{eq:Case2Plus3TLCEEq04})], one may compute the spatial blocks of the self-energy matrix [(\ref{eq:Case3TLCEEq01a})-(\ref{eq:Case3TLCEEq01d})],
\begin{subequations}
\begin{align}
\boldsymbol{\Sigma}^{N N} &= \mathbf{C} \left( \overline{z} I_{1 -} - t_{4} I_{2} \right) ,\label{eq:Case2Plus3TLCEEq05a}\\
\boldsymbol{\Sigma}^{N \overline{N}} &= \mathbf{C} t_{2} I_{2} ,\label{eq:Case2Plus3TLCEEq05b}\\
\boldsymbol{\Sigma}^{\overline{N} N} &= \mathbf{C} t_{3} I_{2} ,\label{eq:Case2Plus3TLCEEq05c}\\
\boldsymbol{\Sigma}^{\overline{N} \overline{N}} &= \mathbf{C} \left( z I_{1 +} - t_{1} I_{2} \right) ,\label{eq:Case2Plus3TLCEEq05d}
\end{align}
\end{subequations}
where for short,
\begin{subequations}
\begin{align}
I_{1 \pm} &\equiv \int_{- 1 / 2}^{1 / 2} \dd p \frac{1}{W^{T} ( p )} \hat{A} ( p ) \ee^{\pm 2 \pi \ii t p} ,\label{eq:Case2Plus3TLCEEq06a}\\
I_{2} &\equiv \int_{- 1 / 2}^{1 / 2} \dd p \frac{1}{W^{T} ( p )} \hat{A} ( p )^{2} .\label{eq:Case2Plus3TLCEEq06b}
\end{align}
\end{subequations}
(Note, \smash{$I_{1 +} = \overline{I_{1 -}}$}, while \smash{$I_{2}$} is real.) Remark that the matrix structure of the spatial blocks of the self-energy matrix is discovered---they are all proportional to $\mathbf{C}$.

Fourthly, thanks to this proportionality, calculating the spatial blocks of the Green function matrix (\ref{eq:Case3TLCEEq03a}) only requires inverting a $2 \times 2$ matrix,
\begin{subequations}
\begin{align}
\mathbf{G}^{N N} &= \frac{\overline{z} \Id_{N} - \mathbf{C} \left( z I_{1 +} - t_{1} I_{2} \right)}{\mathbf{W}^{N}} ,\label{eq:Case2Plus3TLCEEq07a}\\
\mathbf{G}^{N \overline{N}} &= \frac{\mathbf{C} t_{2} I_{2}}{\mathbf{W}^{N}} ,\label{eq:Case2Plus3TLCEEq07b}\\
\mathbf{G}^{\overline{N} N} &= \frac{\mathbf{C} t_{3} I_{2}}{\mathbf{W}^{N}} ,\label{eq:Case2Plus3TLCEEq07c}\\
\mathbf{G}^{\overline{N} \overline{N}} &= \frac{z \Id_{N} - \mathbf{C} \left( \overline{z} I_{1 -} - t_{4} I_{2} \right)}{\mathbf{W}^{N}} ,\label{eq:Case2Plus3TLCEEq07d}
\end{align}
\end{subequations}
where for short,
\begin{equation}
\begin{split}\label{eq:Case2Plus3TLCEEq08}
\mathbf{W}^{N} &\equiv\\
&\equiv | z |^{2} + \mathbf{C} \left( I_{2} \left( z t_{1} + \overline{z} t_{4} \right) - z^{2} I_{1 +} - \overline{z}^{2} I_{1 -} \right) +\\
&+ \mathbf{C}^{2} \Big( I_{2}^{2} \left( t_{1} t_{4} - t_{2} t_{3} \right) - I_{2} \left( z t_{4} I_{1 +} + \overline{z} t_{1} I_{1 -} \right) +\\
&+ | z |^{2} I_{1 +} I_{1 -} \Big) .
\end{split}
\end{equation}
(Here all the matrices commute, so they may be treated as numbers.)

Finally, in order to obtain a set of equations for the basic unknowns, \smash{$t_{1 , 2 , 3 , 4}$}, the results [(\ref{eq:Case2Plus3TLCEEq07a})-(\ref{eq:Case2Plus3TLCEEq07d})] should be plugged back into the definitions [(\ref{eq:Case2Plus3TLCEEq02a})-(\ref{eq:Case2Plus3TLCEEq02d})], which yields
\begin{subequations}
\begin{align}
t_{1} &= \overline{z} J_{1} - \left( z I_{1 +} - t_{1} I_{2} \right) J_{2} ,\label{eq:Case2Plus3TLCEEq09a}\\
t_{2} &= t_{2} I_{2} J_{2} ,\label{eq:Case2Plus3TLCEEq09b}\\
t_{3} &= t_{3} I_{2} J_{2} ,\label{eq:Case2Plus3TLCEEq09c}\\
t_{4} &= z J_{1} - \left( \overline{z} I_{1 -} - t_{4} I_{2} \right) J_{2} ,\label{eq:Case2Plus3TLCEEq09d}
\end{align}
\end{subequations}
where for short,
\begin{subequations}
\begin{align}
J_{1} &\equiv \frac{1}{T} \Tr \frac{\mathbf{C}}{\mathbf{W}^{N}} ,\label{eq:Case2Plus3TLCEEq10a}\\
J_{2} &\equiv \frac{1}{T} \Tr \frac{\mathbf{C}^{2}}{\mathbf{W}^{N}} .\label{eq:Case2Plus3TLCEEq10b}
\end{align}
\end{subequations}
Note that \smash{$I_{1 \pm}$}, \smash{$I_{2}$}, \smash{$J_{1}$}, \smash{$J_{2}$} all depend---generically in a very complicated way---on the unknowns \smash{$t_{1 , 2 , 3 , 4}$}.

\emph{Nonholomorphic solution.} As a general caveat (cf.~App.~\ref{aa:NonHermitianRandomMatrices}), the master equations [(\ref{eq:Case2Plus3TLCEEq09a})-(\ref{eq:Case2Plus3TLCEEq09d})] will have a nonholomorphic solution---valid inside the mean spectral domain, \smash{$z \in \mathcal{D}_{\mathbf{c} ( t )^{\lin}}$}, and describing the MSD there---and a holomorphic one---for the outside of the domain, whose matching with the nonholomorphic solution provides the equation of the borderline of the domain.

In the nonholomorphic sector, there must necessarily be \smash{$t_{2 , 3} \neq 0$} [cf.~the discussion around Eqs.~(\ref{eq:Case2Plus3TLCEEq16}), (\ref{eq:Case2Plus3TLCEEq17})]. Simplifying [(\ref{eq:Case2Plus3TLCEEq09a})-(\ref{eq:Case2Plus3TLCEEq09d})] accordingly yields
\begin{subequations}
\begin{align}
J_{1} &= \frac{z I_{1 +}}{\overline{z} I_{2}} ,\label{eq:Case2Plus3TLCEEq11a}\\
J_{2} &= \frac{1}{I_{2}} ,\label{eq:Case2Plus3TLCEEq11b}\\
z^{2} I_{1 +} &= \overline{z}^{2} I_{1 -} \in \mathbb{R} .\label{eq:Case2Plus3TLCEEq11c}
\end{align}
\end{subequations}

Once \smash{$t_{1 , 2 , 3 , 4}$} are found from [(\ref{eq:Case2Plus3TLCEEq11a})-(\ref{eq:Case2Plus3TLCEEq11c})]---which, as mentioned, seems to be a formidable task---the nonholomorphic Green function of the linearized estimator \smash{$\mathbf{c} ( t )^{\lin}$} (\ref{eq:Case1TLCEEq02}) is given by an analog of (\ref{eq:Case1ETCEEq09}), with the respective traces calculated from (\ref{eq:Case2Plus3TLCEEq07a}) and (\ref{eq:Case2Plus3TLCEEq03}),
\begin{subequations}
\begin{align}
\frac{1}{T} \Tr \mathbf{G}^{N N} &= \overline{z} J_{0} - \left( z I_{1 +} - t_{1} I_{2} \right) J_{1} ,\label{eq:Case2Plus3TLCEEq12a}\\
\frac{1}{T} \Tr \mathbf{G}^{T T} &= \overline{z} I_{0} - t_{4} I_{1 +} ,\label{eq:Case2Plus3TLCEEq12b}
\end{align}
\end{subequations}
where for short,
\begin{subequations}
\begin{align}
J_{0} &\equiv \frac{1}{T} \Tr \frac{1}{\mathbf{W}^{N}} ,\label{eq:Case2Plus3TLCEEq13a}\\
I_{0} &\equiv \int_{- 1 / 2}^{1 / 2} \dd p \frac{1}{W^{T} ( p )} .\label{eq:Case2Plus3TLCEEq13b}
\end{align}
\end{subequations}
These quantities can be expressed through \smash{$J_{1}$}, \smash{$J_{2}$} and \smash{$I_{1 \pm}$}, \smash{$I_{2}$}, respectively, by using \smash{$\frac{1}{T} \Tr \Id_{N} = r$} and \smash{$\int_{- 1 / 2}^{1 / 2} \dd p = 1$},
\begin{subequations}
\begin{align}
| z |^{2} J_{0} &=\nonumber\\
&= r - \left( I_{2} \left( z t_{1} + \overline{z} t_{4} \right) - z^{2} I_{1 +} - \overline{z}^{2} I_{1 -} \right) J_{1} -\nonumber\\
&- \Big( I_{2}^{2} \left( t_{1} t_{4} - t_{2} t_{3} \right) - I_{2} \left( z t_{4} I_{1 +} + \overline{z} t_{1} I_{1 -} \right) +\nonumber\\
&+ | z |^{2} I_{1 +} I_{1 -} \Big) J_{2} ,\label{eq:Case2Plus3TLCEEq14a}\\
| z |^{2} I_{0} &= 1 + \overline{z} t_{1} I_{1 -} + z t_{4} I_{1 +} - \left( t_{1} t_{4} - t_{2} t_{3} \right) I_{2} .\label{eq:Case2Plus3TLCEEq14b}
\end{align}
\end{subequations}
Inserting [(\ref{eq:Case2Plus3TLCEEq14a}), (\ref{eq:Case2Plus3TLCEEq14b})] into [(\ref{eq:Case2Plus3TLCEEq12a}), (\ref{eq:Case2Plus3TLCEEq12b})], and then into the analog of (\ref{eq:Case1ETCEEq09}), using also (\ref{eq:NonHolomorphicMTransformDefinition}), and finally into the analog of (\ref{eq:Case1ETCEEq03})---we express the desired nonholomorphic $M$-transform of the TLCE (in the argument \smash{$z^{2}$}) through \smash{$t_{1 , 2 , 3 , 4}$},
\begin{equation}\label{eq:Case2Plus3TLCEEq15}
M_{\mathbf{c} ( t )} \left( z^{2} , \overline{z}^{2} \right) = \frac{1}{r} \Big( \overline{z} t_{1} I_{1 -} - \left( t_{1} t_{4} - t_{2} t_{3} \right) I_{2} \Big) .
\end{equation}
Eqs.~[(\ref{eq:Case2Plus3TLCEEq11a})-(\ref{eq:Case2Plus3TLCEEq11c}), (\ref{eq:Case2Plus3TLCEEq06a}), (\ref{eq:Case2Plus3TLCEEq06b}), (\ref{eq:Case2Plus3TLCEEq04}), (\ref{eq:Case2Plus3TLCEEq10a}), (\ref{eq:Case2Plus3TLCEEq10b}), (\ref{eq:Case2Plus3TLCEEq08}), (\ref{eq:Case2Plus3TLCEEq15})] thus conclude the derivation of the master equations under the considered circumstances (\ref{eq:Case2Plus3Definition}).

\emph{Holomorphic solution.} One may also evaluate the order parameter [(\ref{eq:BDefinition}), (\ref{eq:Case1TLCEEq08})] inside the mean spectral domain by using the identities (\ref{eq:Case2Plus3TLCEEq03}) (its off-diagonal equations) and [(\ref{eq:Case2Plus3TLCEEq07b}), (\ref{eq:Case2Plus3TLCEEq07c})],
\begin{equation}\label{eq:Case2Plus3TLCEEq16}
\mathcal{B}_{\mathbf{c} ( t )^{\lin}} ( z , \overline{z} ) = h \left( \frac{I_{2} J_{1} + I_{1}}{1 + r} \right)^{2} ,
\end{equation}
where for short, \smash{$I_{1} \equiv \int_{- 1 / 2}^{1 / 2} \dd p \hat{A} ( p ) / W^{T} ( p )$}. Since the left-hand side is real and nonnegative, and so is the bracket on the right-hand side---so must $h$ be (\ref{eq:Case2Plus3TLCEEq02e}). Therefore, the equation of the borderline (\ref{eq:BorderlineEquation}) reads
\begin{equation}\label{eq:Case2Plus3TLCEEq17}
h = 0 .
\end{equation}

\emph{Special cases.} In the remaining part of this Section, two subcases of (\ref{eq:Case2Plus3Definition}) will be investigated: either \smash{$\mathbf{C} = \Id_{N}$} (Sec.~\ref{sss:Case2Plus3TLCEC1}) or \smash{$A ( b - a ) = \delta_{a b}$} (Sec.~\ref{sss:Case2Plus3TLCEA1}).

%%%%%%%%%%%%%%%%%%%%%%%%%%%%%%%%%%%%%%%%%%%%%%%%%%%%%%%%%%%%%%%%%%%%%%

\subsubsection{Trivial $\mathbf{C}$ and arbitrary $A$}
\label{sss:Case2Plus3TLCEC1}

As a first subcase, suppose
\begin{equation}\label{eq:Case2Plus3TLCEC1Eq01}
\mathbf{C} = \Id_{N} , \quad A ( b - a ) = \textrm{arbitrary} .
\end{equation}

Then, the spatial traces [(\ref{eq:Case2Plus3TLCEEq10a}), (\ref{eq:Case2Plus3TLCEEq10b})] greatly simplify, and so do the master equations [(\ref{eq:Case2Plus3TLCEEq11a})-(\ref{eq:Case2Plus3TLCEEq11c})], becoming \smash{$I_{1 +} = \overline{z} / z$}, \smash{$I_{1 -} = z / \overline{z}$} and \smash{$I_{2} = r / ( t_{1} t_{4} - t_{2} t_{3} )$}. Moreover, (\ref{eq:Case2Plus3TLCEEq15}) turns into (here it proves simpler to use the nonholomorphic Green function instead of the $M$-transform) \smash{$G_{\mathbf{c} ( t )} ( z^{2} , \overline{z}^{2} ) = t_{1} / ( r z )$}. Collecting all these results (and changing the argument \smash{$z^{2}$} to $z$) yields the final set of master equations,
\begin{subequations}
\begin{align}
\int_{- 1 / 2}^{1 / 2} \dd p \frac{1}{w ( p )} &= \frac{r}{| G |^{2} + \tilde{h}} ,\label{eq:Case2Plus3TLCEC1Eq02a}\\
\int_{- 1 / 2}^{1 / 2} \dd p \frac{1}{w ( p )} \frac{1}{r^{2} \hat{A} ( p ) \ee^{2 \pi \ii t p}} &= z ,\label{eq:Case2Plus3TLCEC1Eq02b}
\end{align}
\end{subequations}
where for short,
\begin{equation}\label{eq:Case2Plus3TLCEC1Eq03}
w ( p ) \equiv \left| G - \frac{\ee^{2 \pi \ii t p}}{r \hat{A} ( p )} \right|^{2} + \tilde{h} ,
\end{equation}
where the unknowns are: complex \smash{$G \equiv G_{\mathbf{c} ( t )} ( z , \overline{z} )$} and real and nonnegative \smash{$\tilde{h} \equiv \tilde{h}_{\mathbf{c} ( t )} ( z , \overline{z} )$}, with \smash{$\tilde{h}_{\mathbf{c} ( t )} ( z^{2} , \overline{z}^{2} ) \equiv h / ( r^{2} | z |^{2} )$}. Notice that there generically is no rotational symmetry here.

%%%%%%%%%%%%%%%%%%%%%%%%%%%%%%%%%%%%%%%%%%%%%%%%%%%%%%%%%%%%%%%%%%%%%%

\subsubsection{Arbitrary $\mathbf{C}$ and trivial $A$}
\label{sss:Case2Plus3TLCEA1}

As a second subcase, suppose on the contrary,
\begin{equation}\label{eq:Case2Plus3TLCEA1Eq01}
\mathbf{C} = \textrm{arbitrary} , \quad A ( b - a ) = \delta_{a b} .
\end{equation}
Recall that the master equation in this situation has already been derived (\ref{eq:Case2A1TLCEMasterEq}) using the $N$-transform conjecture; now it will be rigorously proven.

\emph{Nonholomorphic master equations.} Immediately, \smash{$\hat{A} ( p ) = 1$}, and the integrals [(\ref{eq:Case2Plus3TLCEEq06a}), (\ref{eq:Case2Plus3TLCEEq06b})] can be performed explicitly: Introducing a new variable
\begin{equation}\label{eq:Case2Plus3TLCEA1Eq02}
v \equiv \ee^{- 2 \pi \ii t p} , \, \textrm{i.e.,} \, \int_{- 1 / 2}^{1 / 2} \dd p ( \ldots ) = \frac{1}{2 \pi \ii} \ointctrclockwise_{C ( 0 , 1 )} \dd v \frac{1}{v} ( \ldots )
\end{equation}
[notice that the dependence on $t$ is lost---this subcase is independent of the value of the time lag, of course as long as $t \ll T$, cf.~Fig.~\ref{fig:TM2aTLCETwoTests} (a)], and applying the method of residues---the integration requires finding roots of a quadratic polynomial in $v$,
\begin{equation}\label{eq:Case2Plus3TLCEA1Eq03}
V ( v ) \equiv - \overline{z} t_{1} v^{2} + \left( | z |^{2} + t_{1} t_{4} + h \right) v - z t_{4} ,
\end{equation}
which leads to
\begin{subequations}
\begin{align}
I_{1 +} &= \frac{1}{2 \pi \ii} \ointctrclockwise_{C ( 0 , 1 )} \dd v \frac{1}{v V ( v )} =\nonumber\\
&= \frac{1}{2 z t_{4}} \left( \frac{| z |^{2} + t_{1} t_{4} + h}{s} - 1 \right) ,\label{eq:Case2Plus3TLCEA1Eq04a}\\
I_{1 -} &= \frac{1}{2 \pi \ii} \ointctrclockwise_{C ( 0 , 1 )} \dd v \frac{v}{V ( v )} =\nonumber\\
&= \frac{1}{2 \overline{z} t_{1}} \left( \frac{| z |^{2} + t_{1} t_{4} + h}{s} - 1 \right) ,\label{eq:Case2Plus3TLCEA1Eq04b}\\
I_{2} &= \frac{1}{2 \pi \ii} \ointctrclockwise_{C ( 0 , 1 )} \dd v \frac{1}{V ( v )} = \frac{1}{s} ,\label{eq:Case2Plus3TLCEA1Eq04c}
\end{align}
\end{subequations}
where for short,
\begin{equation}\label{eq:Case2Plus3TLCEA1Eq05}
s \equiv \sqrt{\left( | z |^{2} + t_{1} t_{4} + h \right)^{2} - 4 | z |^{2} t_{1} t_{4}} .
\end{equation}
[The two roots of $V ( v )$ (\ref{eq:Case2Plus3TLCEA1Eq03}) obey \smash{$| v_{1} | | v_{2} | = 1$}, hence, one of them lies inside the centered unit circle $C ( 0 , 1 )$, and one outside it; only the interior one contributes to the integrals.]

Consequently, Eq.~(\ref{eq:Case2Plus3TLCEEq11c}) implies \smash{$t_{1} = f / z$}, \smash{$t_{4} = f / \overline{z}$}, with $f \in \mathbb{R}$. The other two master equations [(\ref{eq:Case2Plus3TLCEEq11a}), (\ref{eq:Case2Plus3TLCEEq11b})], turn finally into two real equations for two real unknowns $f$, $h$,
\begin{subequations}
\begin{align}
\frac{1}{T} \Tr \frac{\mathbf{C}}{\mathbf{W}^{N}} &= \frac{1}{2 f} \left( R + \frac{f^{2}}{R} + h - s \right) ,\label{eq:Case2Plus3TLCEA1Eq06a}\\
\frac{1}{T} \Tr \frac{\mathbf{C}^{2}}{\mathbf{W}^{N}} &= s ,\label{eq:Case2Plus3TLCEA1Eq06b}
\end{align}
\end{subequations}
where substituting the results [(\ref{eq:Case2Plus3TLCEA1Eq04a})-(\ref{eq:Case2Plus3TLCEA1Eq04c})] into (\ref{eq:Case2Plus3TLCEEq08}) yields
\begin{equation}\label{eq:Case2Plus3TLCEA1Eq07}
\mathbf{W}^{N} = R + \mathbf{C} ( \mathbf{C} + 2 f ) \frac{1}{2 f^{2}} \left( R + \frac{f^{2} - R^{2} - h R}{s} \right) ,
\end{equation}
and recall that in this notation Eq.~(\ref{eq:Case2Plus3TLCEA1Eq05}) reads
\begin{equation}\label{eq:Case2Plus3TLCEA1Eq08}
s \equiv \sqrt{\left( R + \frac{f^{2}}{R} + h \right)^{2} - 4 f^{2}} .
\end{equation}

Once these equations [(\ref{eq:Case2Plus3TLCEA1Eq06a})-(\ref{eq:Case2Plus3TLCEA1Eq08})] are solved (after specifying the true spatial covariance matrix $\mathbf{C}$), then the nonholomorphic $M$-transform (\ref{eq:Case2Plus3TLCEEq15}) is expressed through $f$, $h$ as
\begin{equation}\label{eq:Case2Plus3TLCEA1Eq09}
M \equiv M_{\mathbf{c} ( t )} ( z , \overline{z} ) = \frac{1}{2 r} \left( \frac{R - \frac{f^{2}}{R} - h}{s} - 1 \right) .
\end{equation}

In [(\ref{eq:Case2Plus3TLCEA1Eq06a})-(\ref{eq:Case2Plus3TLCEA1Eq09})] the argument has already been changed from \smash{$z^{2}$} to $z$. Moreover, all these identities depend only on the radius $R = | z |$, so the rotational symmetry has been demonstrated. Conversely, one may say that deviations from rotational symmetry in the MSD of the TLCE point toward nontrivial true temporal covariances.

\emph{Nonholomorphic master equations---a simpler form.} The above master equations can be considerably simplified. One should start from rewriting \smash{$J_{1}$} and \smash{$J_{2}$} [i.e., the left-hand sides of (\ref{eq:Case2Plus3TLCEA1Eq06a}), (\ref{eq:Case2Plus3TLCEA1Eq06b}) with (\ref{eq:Case2Plus3TLCEA1Eq07})] in the language of the holomorphic $M$-transform of $\mathbf{C}$, which is done by factorizing \smash{$\mathbf{W}^{N} = \varpi ( \mathbf{C} - \zeta ) ( \mathbf{C} - \overline{\zeta} )$},
\begin{subequations}
\begin{align}
J_{1} &= - r \frac{M_{\mathbf{C}} ( \zeta ) - M_{\mathbf{C}} ( \overline{\zeta} )}{\varpi \left( \zeta - \overline{\zeta} \right)} ,\label{eq:Case2Plus3TLCEA1Eq12a}\\
J_{2} &= - r \frac{\zeta M_{\mathbf{C}} ( \zeta ) - \overline{\zeta} M_{\mathbf{C}} ( \overline{\zeta} )}{\varpi \left( \zeta - \overline{\zeta} \right)} .\label{eq:Case2Plus3TLCEA1Eq12b}
\end{align}
\end{subequations}
This leads to
\begin{subequations}
\begin{align}
M + \ii m &= M_{\mathbf{C}} \left( - f + \ii \frac{f^{2} - ( h + s ) R - R^{2}}{2 \sqrt{h R}} \right) ,\label{eq:Case2Plus3TLCEA1Eq13a}\\
m &\equiv \sqrt{h R} \frac{f^{2} + ( h - s ) R + R^{2}}{2 r R f s} ,\label{eq:Case2Plus3TLCEA1Eq13b}
\end{align}
\end{subequations}
which is supplemented by [(\ref{eq:Case2Plus3TLCEA1Eq08}), (\ref{eq:Case2Plus3TLCEA1Eq09})]. Using these last three equations, one expresses $f$, $h$, $s$ through $M$, $m$, and substitutes them into (\ref{eq:Case2Plus3TLCEA1Eq13a}), which finally yields a concise complex equation (\ref{eq:Case2A1TLCEMasterEq}) for two real unknowns, $M$ and auxiliary $m$.

\emph{Borderline of the mean spectral domain.} It will now be proven that $\mathcal{D}$ is a ring for $r \leq 1$ and an annulus for $r > 1$, with the radii [(\ref{eq:Case2A1TLCERExt}), (\ref{eq:Case2A1TLCERInt})]. To find the borderline, the original form of the master equations is more convenient, in which one should set $h = 0$ (\ref{eq:Case2Plus3TLCEEq17}). To begin with, Eq.~(\ref{eq:Case2Plus3TLCEA1Eq08}) becomes \smash{$s = | R - f^{2} / R |$}, which inserted into Eq.~(\ref{eq:Case2Plus3TLCEA1Eq09}) confirms the two possible values that the nonholomorphic $M$-transform of the TLCE may acquire on the borderline, $M = 0$ or $M = - 1 / r$. Furthermore, manipulating Eqs.~[(\ref{eq:Case2Plus3TLCEA1Eq06a}), (\ref{eq:Case2Plus3TLCEA1Eq06b}), (\ref{eq:Case2Plus3TLCEA1Eq07})] with $h = 0$ leads to two solutions:

(i) External circle: If $| f | < R$ (which corresponds to $M = 0$), the master equations become
\begin{subequations}
\begin{align}
f &= r m_{\mathbf{C} , 1} ,\label{eq:Case2Plus3TLCEA1Eq14a}\\
R^{2} - f^{2} &= r m_{\mathbf{C} , 2} ,\label{eq:Case2Plus3TLCEA1Eq14b}
\end{align}
\end{subequations}
which reproduces (\ref{eq:Case2A1TLCERExt}).

(ii) Internal circle: If $| f | > R$ (which corresponds to $M = - 1 / r$; there must be $r > 1$), the master equations become
\begin{subequations}
\begin{align}
\frac{R^{2}}{f} &= r \frac{1}{N} \Tr \frac{\mathbf{C}}{\left( \frac{\mathbf{C}}{f} + 1 \right)^{2}} ,\label{eq:Case2Plus3TLCEA1Eq15a}\\
f^{2} - R^{2} &= r \frac{1}{N} \Tr \frac{\mathbf{C}^{2}}{\left( \frac{\mathbf{C}}{f} + 1 \right)^{2}} ,\label{eq:Case2Plus3TLCEA1Eq15b}
\end{align}
\end{subequations}
which, using the definitions [(\ref{eq:HolomorphicGreenFunctionDefinition}), (\ref{eq:HolomorphicMTransformDefinition}), (\ref{eq:HolomorphicNTransformDefinition})], reproduces (\ref{eq:Case2A1TLCERInt}) [with \smash{$f = - N_{\mathbf{C}} ( - 1 / r )$}].

%%%%%%%%%%%%%%%%%%%%%%%%%%%%%%%%%%%%%%%%%%%%%%%%%%%%%%%%%%%%%%%%%%%%%%
%%%%%%%%%%%%%%%%%%%%%%%%%%%%%%%%%%%%%%%%%%%%%%%%%%%%%%%%%%%%%%%%%%%%%%
%%%%%%%%%%%%%%%%%%%%%%%%%%%%%%%%%%%%%%%%%%%%%%%%%%%%%%%%%%%%%%%%%%%%%%

\section{Toy Models}
\label{s:TM}

%%%%%%%%%%%%%%%%%%%%%%%%%%%%%%%%%%%%%%%%%%%%%%%%%%%%%%%%%%%%%%%%%%%%%%
%%%%%%%%%%%%%%%%%%%%%%%%%%%%%%%%%%%%%%%%%%%%%%%%%%%%%%%%%%%%%%%%%%%%%%

\subsection{Outline}
\label{ss:TMOutline}

\begin{table*}
\begin{tabular}[t]{|p{0.0692\textwidth}|p{0.0692\textwidth}||p{0.0692\textwidth}|p{0.0692\textwidth}|p{0.0692\textwidth}|p{0.0692\textwidth}|p{0.0692\textwidth}|p{0.0692\textwidth}|p{0.0692\textwidth}|p{0.0692\textwidth}|p{0.0692\textwidth}|p{0.0692\textwidth}|p{0.0692\textwidth}|}
\hline
\multicolumn{2}{|l||}{} & \multicolumn{5}{l|}{TM 1} & \multicolumn{2}{l|}{TM 2} & TM 3 & \multicolumn{3}{l|}{TM 4 (comments)}\\
\cline{3-9}\cline{11-13}
\multicolumn{2}{|l||}{} & Gaussian & EWMA & \multicolumn{2}{l|}{Student} & free L\'{e}vy & TM 2a & TM 2b & & TM 4a & TM 4b & TM 4c\\
\cline{5-6}
\multicolumn{2}{|l||}{} & & & Ver. 1 & Ver. 2 & & & & & & &\\
\hline\hline
ETCE & Sec. & \ref{aa:TM1ETCE} & --- & --- & --- & --- & \ref{aa:TM2aETCE} & --- & \ref{aa:TM3ETCE} & \ref{sss:TM4a} & \ref{sss:TM4b} & \ref{sss:TM4c}\\
\cline{2-13}
& MSD & (\ref{eq:TM1ETCEEq03}) & --- & --- & --- & --- & (\ref{eq:TM2ETCEEq01}) & --- & (\ref{eq:TM3ETCEEq02}) & [(\ref{eq:TM4aETCEEq01a}), (\ref{eq:TM4aETCEEq01b})] & (\ref{eq:TM4bETCEEq02}) & (\ref{eq:TM4cETCEEq02})\\
\cline{2-13}
& Fig. & --- & --- & --- & --- & --- & \ref{fig:TM2aETCE} & --- & \ref{fig:TM3ETCE} & \ref{fig:TM4aETCE} & \ref{fig:TM4bETCE} & \ref{fig:TM4cETCE}\\
\hline
TLCE & Sec. & [\ref{sss:TM1TLCEBorderline}, \ref{sss:TM1TLCEMSD}, \ref{sss:TM1TLCEErfc}] & \ref{sss:TM1TLCEEWMA} & \multicolumn{2}{l|}{\ref{sss:TM1TLCEStudent}} & \ref{sss:TM1TLCELevy} & \ref{sss:TM2a} & \ref{sss:TM2b} & \ref{ss:TM3} & \ref{sss:TM4a} & \ref{sss:TM4b} & \ref{sss:TM4c}\\
\cline{2-13}
& MSD & [(\ref{eq:TM1TLCEEq04}), (\ref{eq:TM1TLCEEq03b})-(\ref{eq:TM1TLCEEq03d})] & [(\ref{eq:Case2C1ADiagonalTLCEMasterEq}), (\ref{eq:TM1TLCEEWMAEq01})] & (\ref{eq:TM1TLCEStudent1Eq01}) & [(\ref{eq:Case2C1ADiagonalTLCEMasterEq}), (\ref{eq:TM1TLCEStudent2Eq01})] & [(\ref{eq:TM1TLCELevyEq01a}), (\ref{eq:TM1TLCELevyEq01b})] or (\ref{eq:TM1TLCELevyEq02}) & [(\ref{eq:Case2A1TLCEMasterEq}), (\ref{eq:TM2aEq02})] & [(\ref{eq:Case2A1TLCEMasterEq}), (\ref{eq:TM2bEq01})] & [(\ref{eq:TM3TLCEEq01a})-(\ref{eq:TM3TLCEEq03})] & [(\ref{eq:Case2Plus3TLCEEq11a})-(\ref{eq:Case2Plus3TLCEEq11c}), (\ref{eq:Case2Plus3TLCEEq06a}), (\ref{eq:Case2Plus3TLCEEq06b}), (\ref{eq:Case2Plus3TLCEEq04}), (\ref{eq:Case2Plus3TLCEEq10a}), (\ref{eq:Case2Plus3TLCEEq10b}), (\ref{eq:Case2Plus3TLCEEq08}), (\ref{eq:TM4aDefinitionEq01})] & [(\ref{eq:Case3TLCEEq01a})-(\ref{eq:Case3TLCEEq04}), (\ref{eq:TM4bDefinitionEq02})] & [(\ref{eq:Case3TLCEEq01a})-(\ref{eq:Case3TLCEEq04}), (\ref{eq:TM4cDefinitionEq02a})-(\ref{eq:TM4cDefinitionEq02d})]\\
\cline{2-10}
& $\partial \mathcal{D}$ & [(\ref{eq:TM1TLCEEq01a}), (\ref{eq:TM1TLCEEq01b})] & [(\ref{eq:TM1TLCEEWMAEq02a}), (\ref{eq:TM1TLCEEWMAEq02b})] & $R \geq 0$ & [(\ref{eq:TM1TLCEStudent2Eq02a}), (\ref{eq:TM1TLCEStudent2Eq02b})] & $R \geq 0$ & [(\ref{eq:TM2aTLCEEq01})-(\ref{eq:TM2aTLCEEq02b})] & [(\ref{eq:TM2bTLCEEq01})-(\ref{eq:TM2bTLCEEq04})] & [(\ref{eq:TM3TLCEEq05a})-(\ref{eq:TM3TLCEEq06b})], [(\ref{eq:TM3TLCEEq09})-(\ref{eq:TM3TLCEEq12b})] ($t = 1$) & & &\\
\cline{2-13}
& Fig. & \ref{fig:TM1TLCE} & \ref{fig:TM1TLCEEWMA} & \ref{fig:TM1TLCEStudent} (a) & \ref{fig:TM1TLCEStudent} (b) & \ref{fig:TM1TLCELevy} & \ref{fig:TM2aTLCE} & \ref{fig:TM2bTLCE} & [\ref{fig:TM3TLCEEIGt1}, \ref{fig:TM3TLCEEIGt5}, \ref{fig:TM3TLCEEIGt15}, \ref{fig:TM3TLCEMSD}] & \ref{fig:TM4aTLCE} & \ref{fig:TM4bTLCE} & \ref{fig:TM4cTLCE}\\
\hline
\end{tabular}
\caption{Summary of the main results obtained in Sec.~\ref{s:TM} and App.~\ref{a:TM12a3ETCE} for the ETCE and TLCE: the pertinent (i) Section, (ii) Figures, (iii) formula for the MSD or master equations from which it derives, (iv) equation of the borderline of the mean spectral domain (for the TLCE).}
\label{tab:TM}
\end{table*}

%%%%%%%%%%%%%%%%%%%%%%%%%%%%%%%%%%%%%%%%%%%%%%%%%%%%%%%%%%%%%%%%%%%%%%

\subsubsection{Toy Models 1, 2a, 2b, 3}
\label{sss:TM12a2b3Outline}

In this Section, the general master equations developed in Sec.~\ref{s:MasterEquations} (and summarized in Tab.~\ref{tab:MasterEquations})---which are a gateway to the MSD [(\ref{eq:HermitianMSDDefinition}), (\ref{eq:NonHermitianMSDDefinition})] of the ETCE (\ref{eq:cDefinition}) and TLCE (\ref{eq:ctDefinition}) for arbitrary complex Gaussian assets---are applied to three specific models of the true covariance function (\ref{eq:ArbitraryCovarianceFunctionDefinition1}) which are designed to approximate certain aspects of the real-world financial markets and other complex multivariate systems, according to the introduction in Sec.~\ref{ss:ModelsOfSpatialAndTemporalCorrelations}:

(i) Toy Model 1 (Sec.~\ref{ss:TM1}): The assets are IID. This is the simplest case; nevertheless, it is useful as the ``null hypothesis'' (cf.~Sec.~\ref{sss:MeasurementNoise}), i.e., any discrepancy between the spectra derived from experimental data and from this model signifies genuine correlations present in the system.

The MSD of the TLCE has first been calculated in~\cite{BielyThurner2006,ThurnerBiely2007}, but their formula is demonstrated to be wrong (Sec.~\ref{sss:TM1TLCEBielyThurner}), and the right one is presented (Secs.~[\ref{sss:TM1TLCEBorderline}, \ref{sss:TM1TLCEMSD}, \ref{sss:TM1TLCEErfc}]); it is noteworthy that the only analytical result to date concerning the spectrum of the TLCE was incorrect---it reveals how involved the subject is.

This result is generalized to the EWMA estimator for Gaussian assets (Sec.~\ref{sss:TM1TLCEEWMA}), as well as the standard estimator for two models of non-Gaussian behavior of the returns: Student (Sec.~\ref{sss:TM1TLCEStudent}; it comes in two versions: with a common random volatility and IID temporal random volatilities) and free L\'{e}vy (Sec.~\ref{sss:TM1TLCELevy}) distributions.

(ii) Toy Model 2 (Sec.~\ref{ss:TM2}): The assets have arbitrary variances but no temporal correlations; they come in two versions (cf.~Sec.~\ref{sss:IndustrialSectors}): either a finite number of variance sectors (Toy Model 2a, Sec.~\ref{sss:TM2a}), or variances distributed according to a power law with a lower cutoff (Toy Model 2b, Sec.~\ref{sss:TM2b}). This and the following models signify another than the null hypothesis approach which is to build more and more realistic toy models, comparing them to data, thereby verifying their validity and assessing their parameters.

(iii) Toy Model 3 (Sec.~\ref{ss:TM3}): The assets are independent from each other but exhibit exponentially-decaying autocorrelations (identical for all the variables), which is a simplest version of the SVAR($1$) model (cf.~Sec.~\ref{sss:VAR}).

It is enlightening to compare the results for the TLCE and ETCE, hence the MSD of the latter is rederived and illustrated in App.~\ref{a:TM12a3ETCE} for Gaussian assets and Toy Models 1 (App.~\ref{aa:TM1ETCE}), 2a (App.~\ref{aa:TM2aETCE}) and 3 (App.~\ref{aa:TM3ETCE}).

%%%%%%%%%%%%%%%%%%%%%%%%%%%%%%%%%%%%%%%%%%%%%%%%%%%%%%%%%%%%%%%%%%%%%%

\subsubsection{Toy Models 4a, 4b, 4c}
\label{sss:TM4a4b4cOutline}

Two reflections naturally arise after analyzing the above models---firstly, they are still quite far away from financial reality; secondly, the analytical derivations of the MSD are already increasingly complicated. Consequently, it is a valid research program to both construct more realistic toy models and to enhance the techniques presented in this paper (especially on the numerics side) to be capable of handling such models. Even though the second part is beyond the scope of this article, three more models are announced and initially analyzed (i.e., the pertinent master equations are explicitly written down or commented on, and the Monte Carlo simulations are presented; Sec.~\ref{ss:TM4})---in contrast to the above examples, they are mixes of both the spatial and temporal correlations from Toy Models 2a and 3, being more advanced examples of the SVAR($1$) class:

(i) Toy Model 4a (Sec.~\ref{sss:TM4a}): There is a number of variance sectors and exponentially-decaying autocorrelations identical for all the variables.

(ii) Toy Model 4b (Sec.~\ref{sss:TM4b}): There is a number of variance sectors and exponentially-decaying autocorrelations different for each sector.

(iii) Toy Model 4c (Sec.~\ref{sss:TM4c}): A SVAR($1$) model with a simple nondiagonal covariance function originating from the market mode.

For these examples also the ETCE is analyzed because for models 4b and 4c the true spatial and temporal correlations are not factorized, unlike in the remaining cases, which causes the need to use the new master equations developed in Sec.~\ref{ss:Case1}.

The main results obtained for the above six models are summarized in Tab.~\ref{tab:TM}.

%%%%%%%%%%%%%%%%%%%%%%%%%%%%%%%%%%%%%%%%%%%%%%%%%%%%%%%%%%%%%%%%%%%%%%
%%%%%%%%%%%%%%%%%%%%%%%%%%%%%%%%%%%%%%%%%%%%%%%%%%%%%%%%%%%%%%%%%%%%%%

\subsection{Toy Model 1 (null hypothesis): Uncorrelated Gaussian/Student/free L\'{e}vy assets and standard/EWMA estimator}
\label{ss:TM1}

\begin{figure*}[t]
\includegraphics[width=\columnwidth]{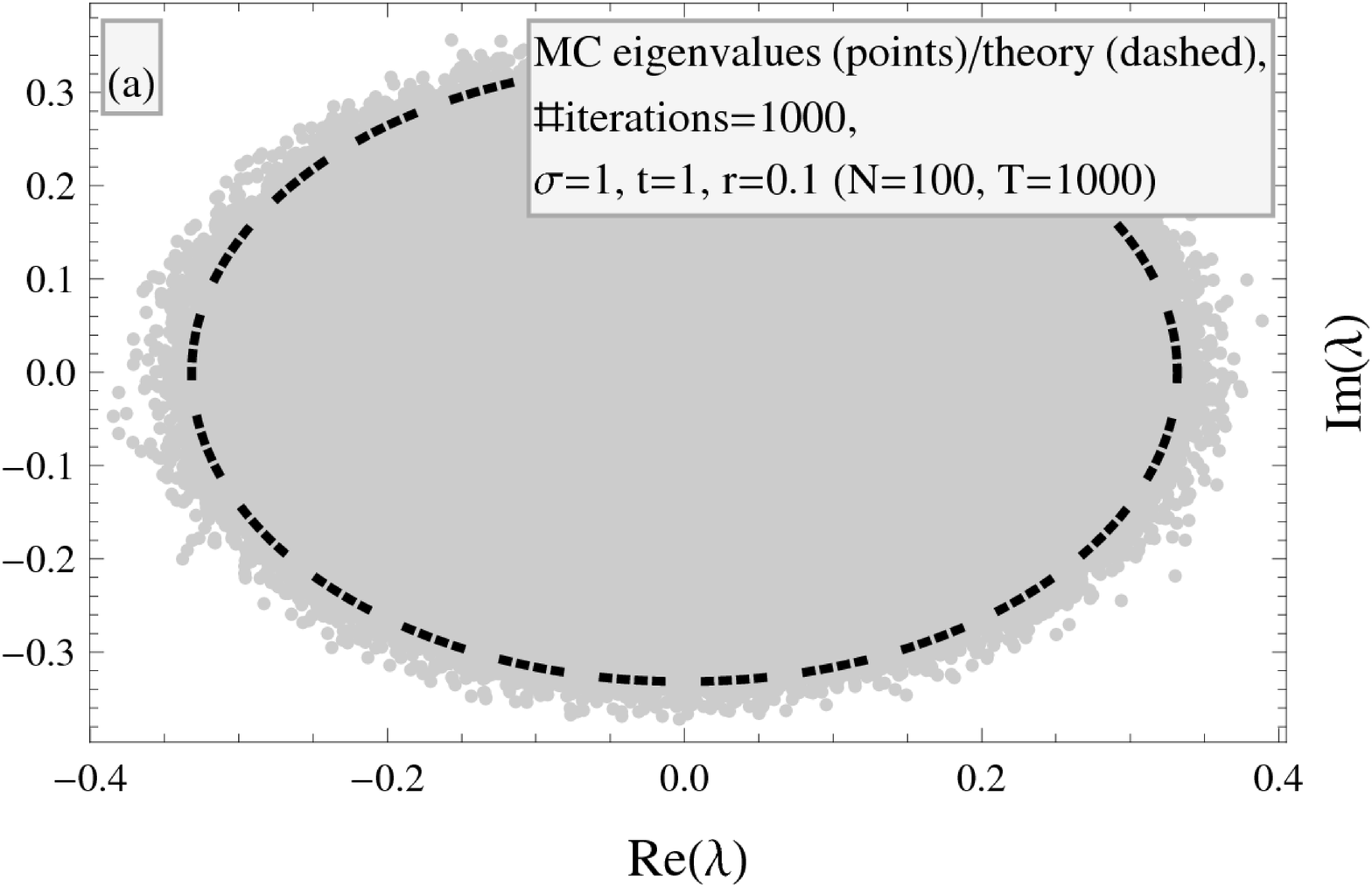}
\includegraphics[width=\columnwidth]{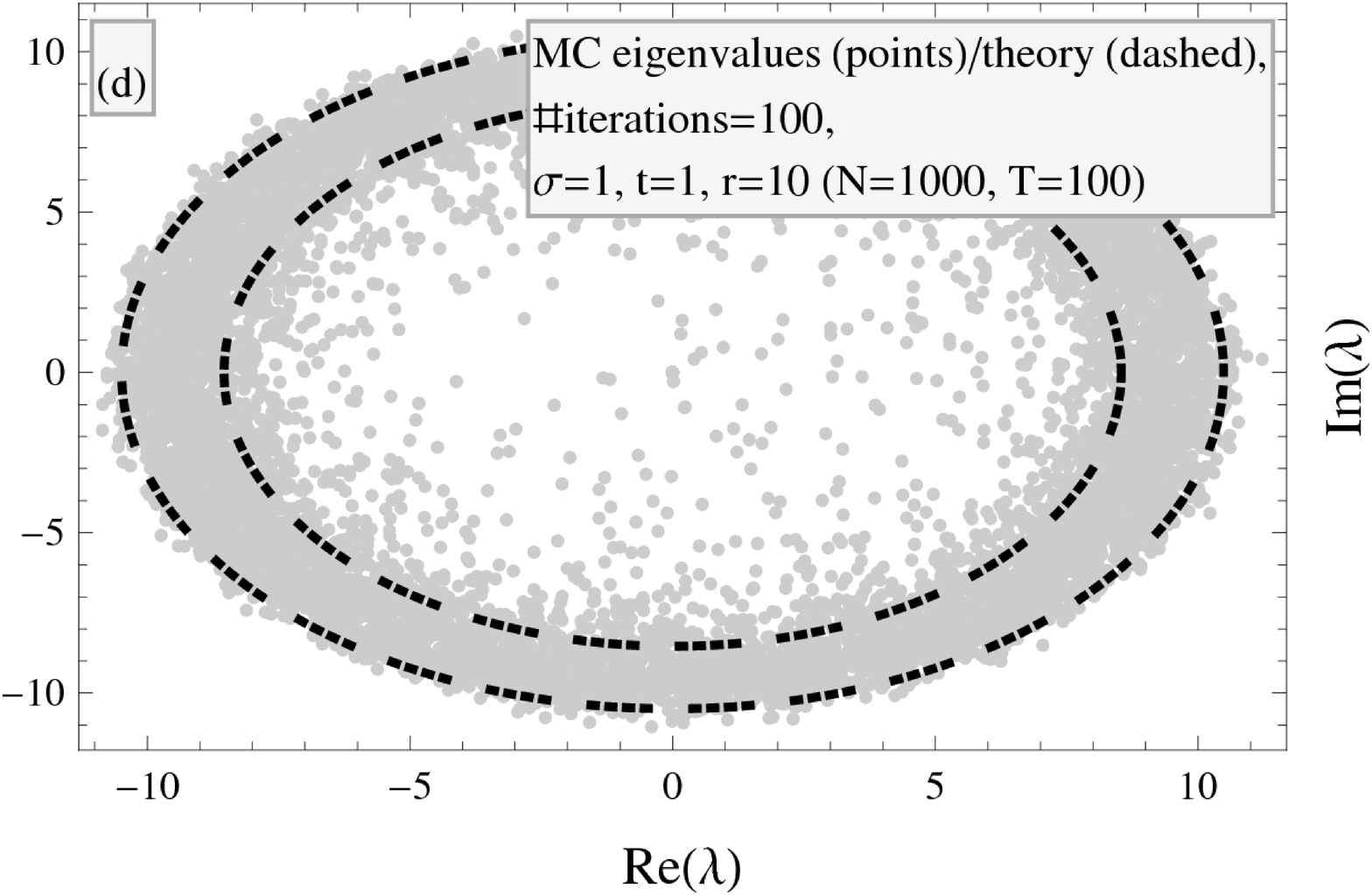}
\includegraphics[width=\columnwidth]{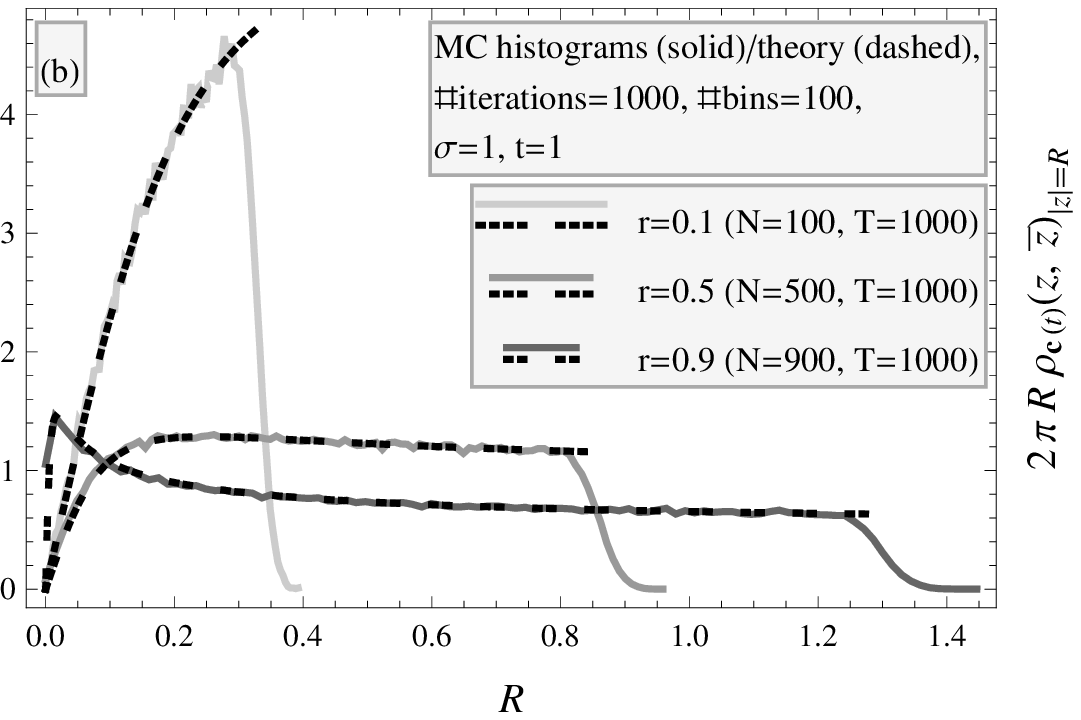}
\includegraphics[width=\columnwidth]{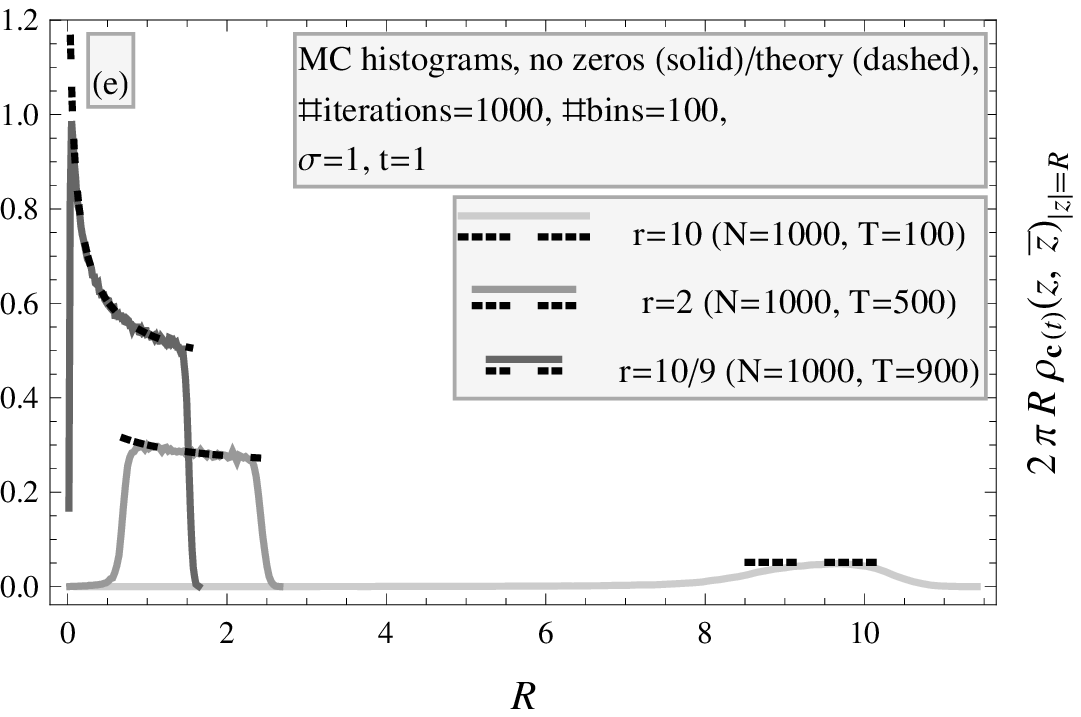}
\includegraphics[width=\columnwidth]{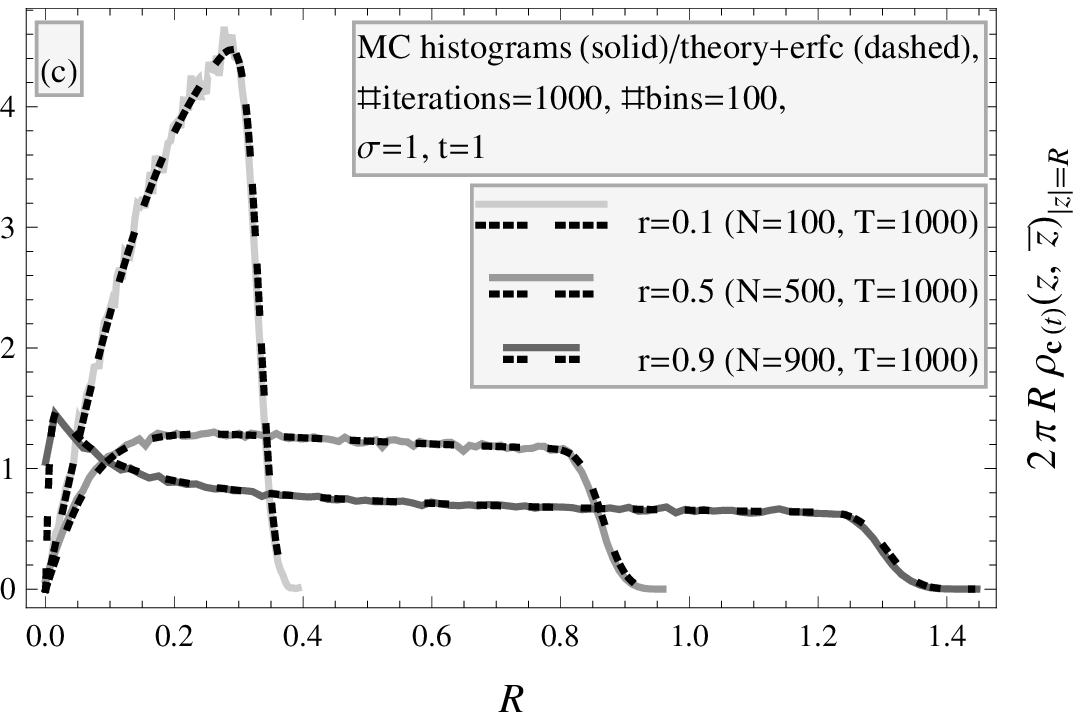}
\includegraphics[width=\columnwidth]{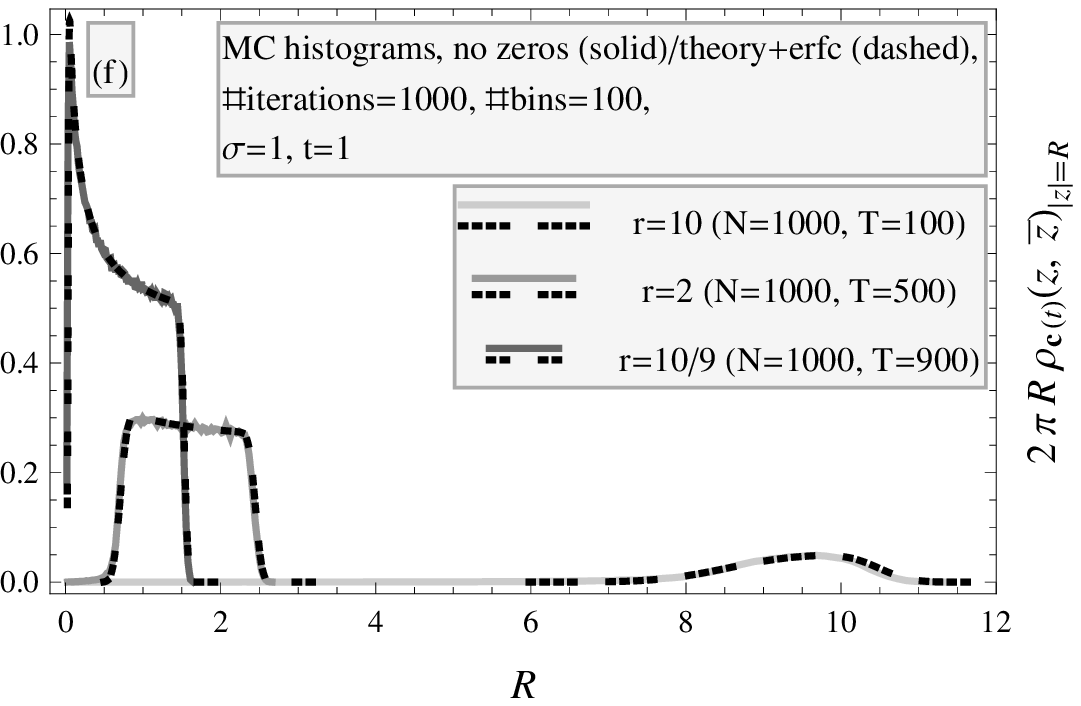}
\caption{Monte Carlo eigenvalues versus theory for the TLCE for Toy Model 1.}
\label{fig:TM1TLCE}
\end{figure*}

\begin{figure*}[t]
\includegraphics[width=\columnwidth]{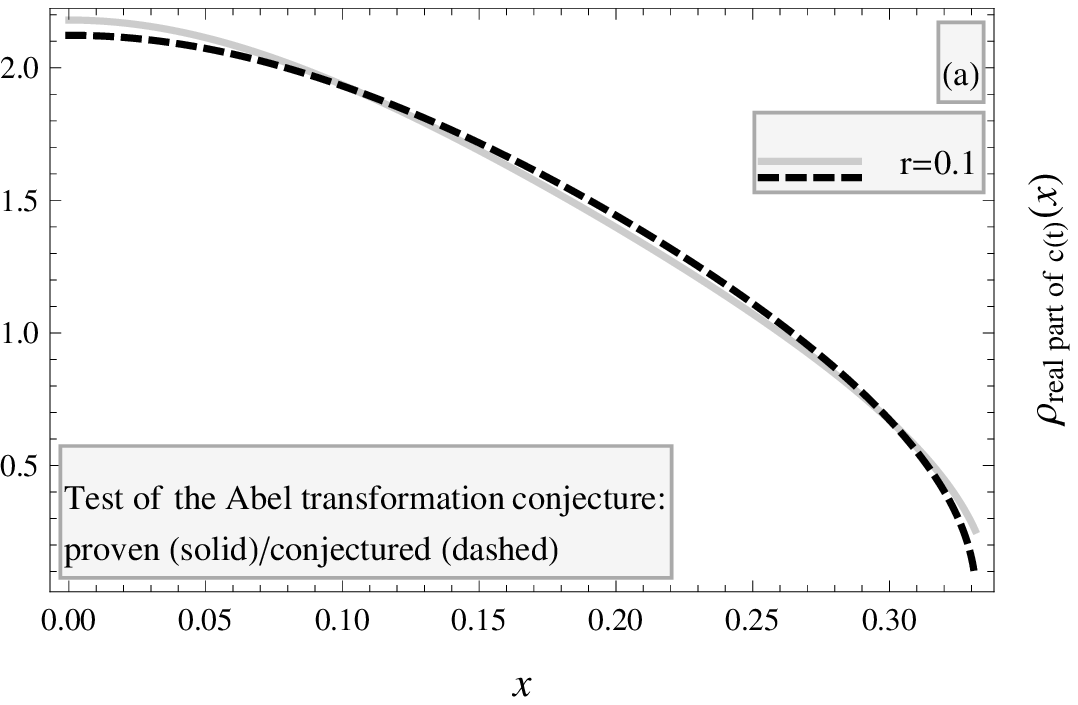}
\includegraphics[width=\columnwidth]{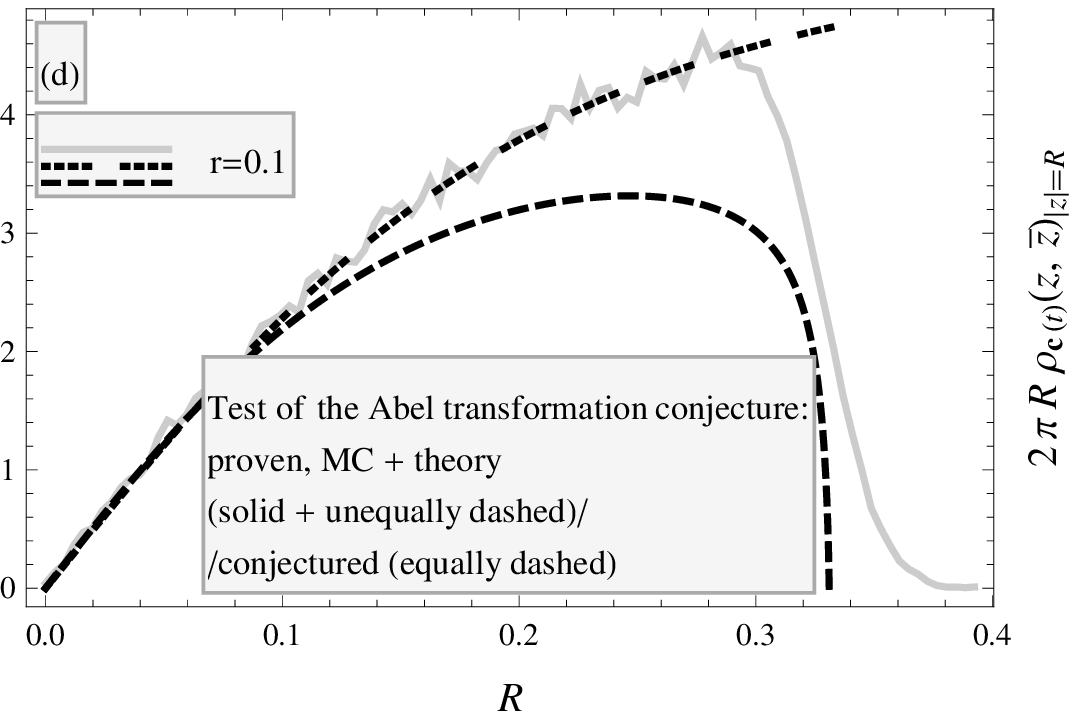}
\includegraphics[width=\columnwidth]{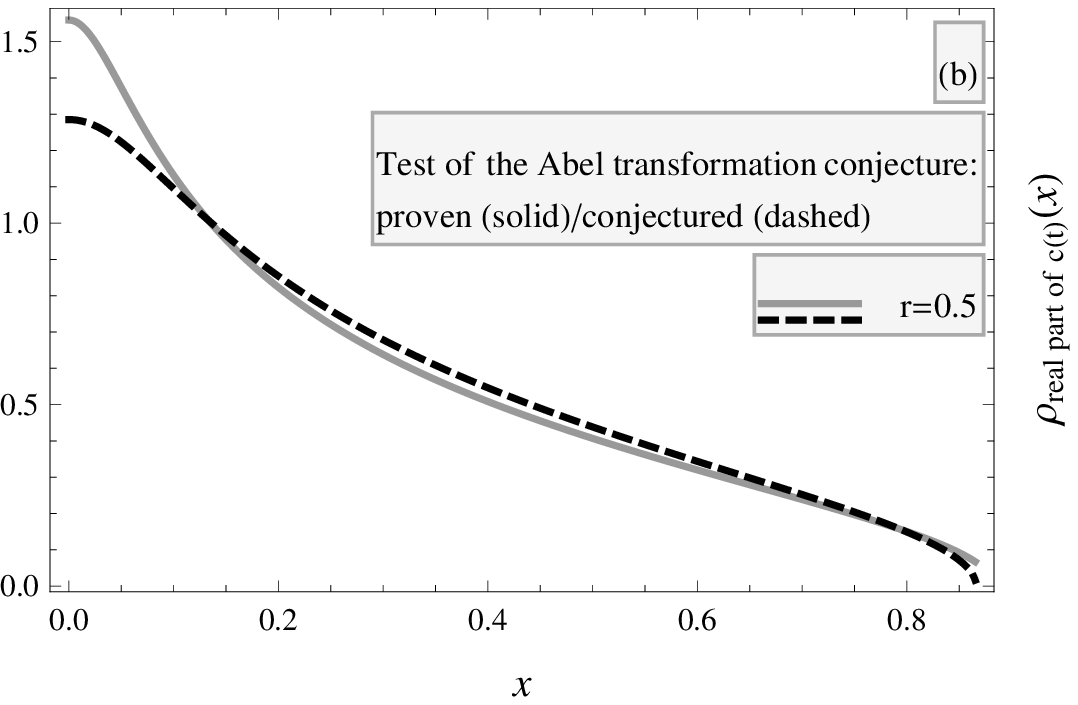}
\includegraphics[width=\columnwidth]{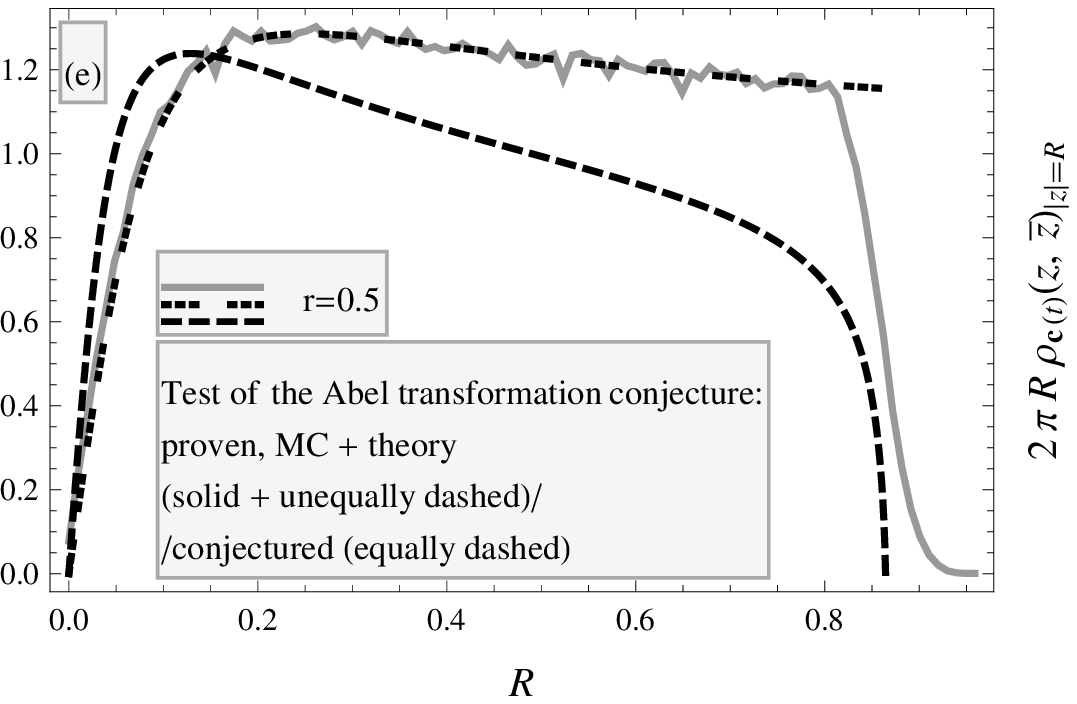}
\includegraphics[width=\columnwidth]{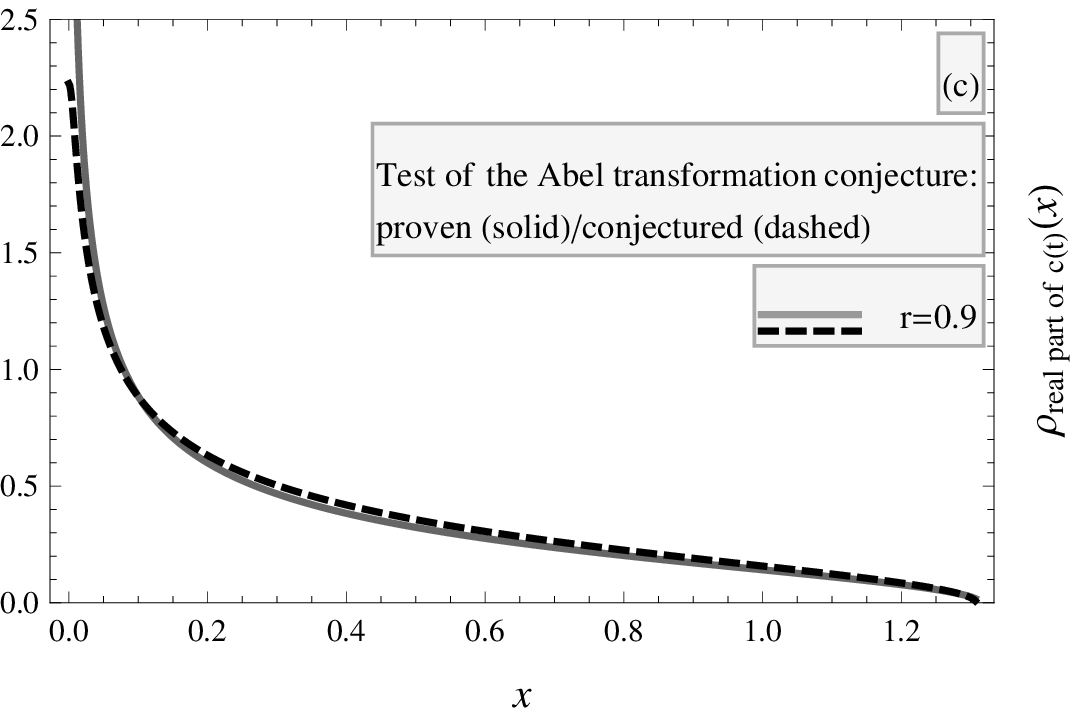}
\includegraphics[width=\columnwidth]{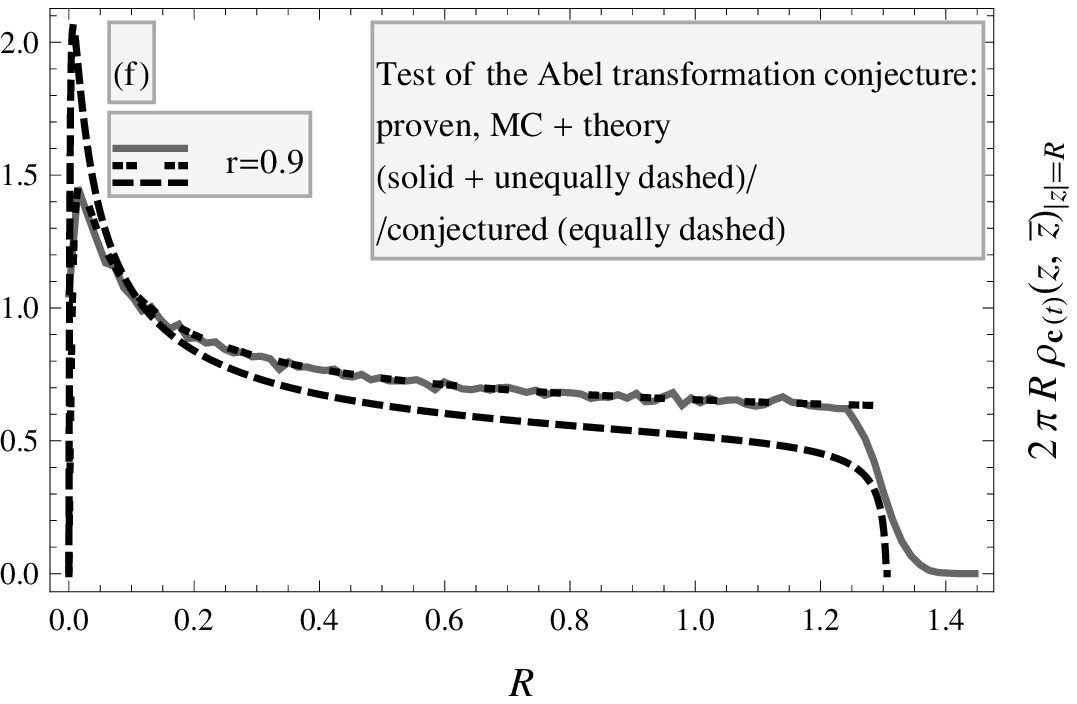}
\caption{Demonstration that the result for the MSD for the TLCE for Toy Model 1 based on the Abel transformation is incorrect.}
\label{fig:TM1TLCEBielyThurner}
\end{figure*}

\begin{figure*}[t]
\includegraphics[width=\columnwidth]{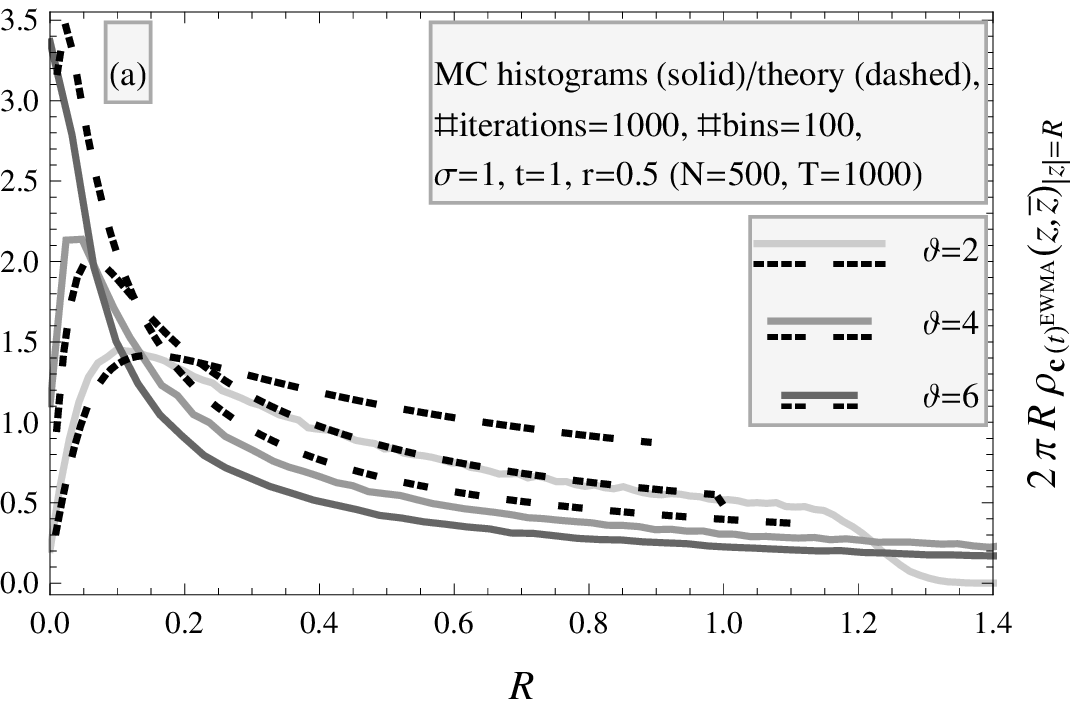}
\includegraphics[width=\columnwidth]{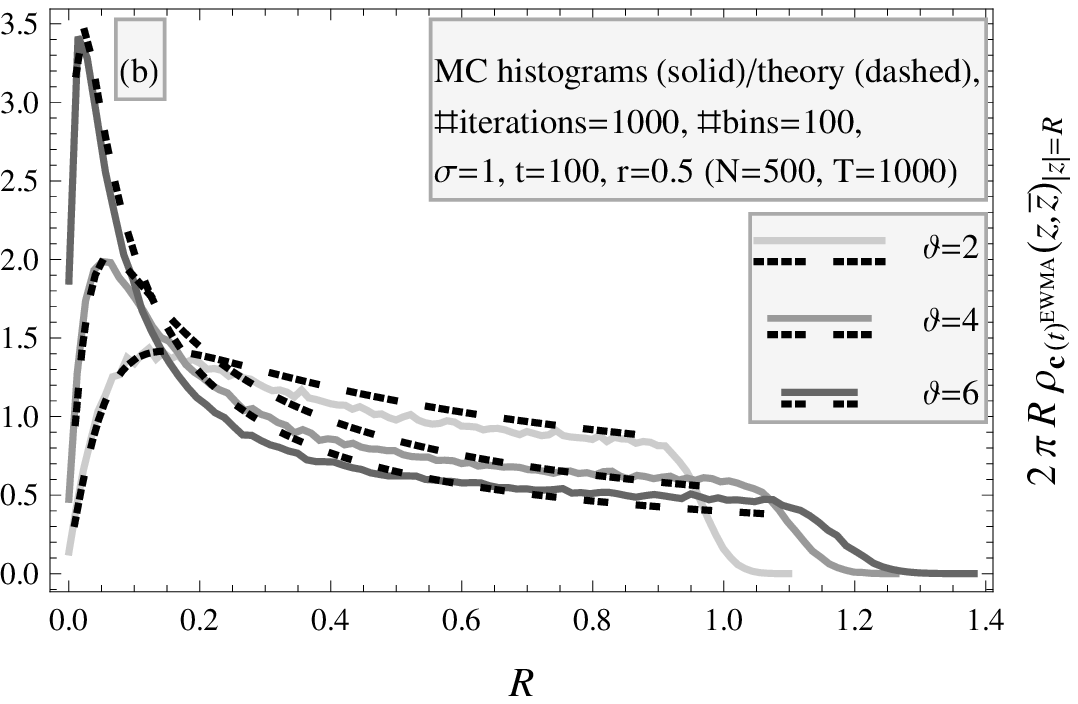}
\caption{Monte Carlo eigenvalues versus theory for the EWMA-TLCE for Toy Model 1.}
\label{fig:TM1TLCEEWMA}
\end{figure*}

\begin{figure*}[t]
\includegraphics[width=\columnwidth]{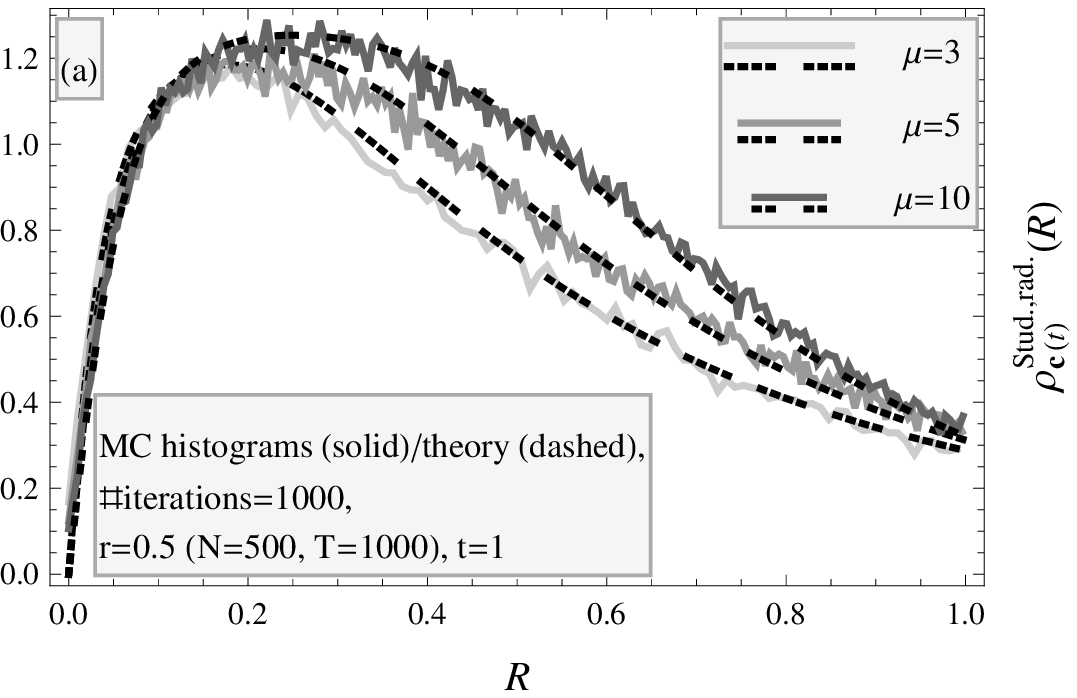}
\includegraphics[width=\columnwidth]{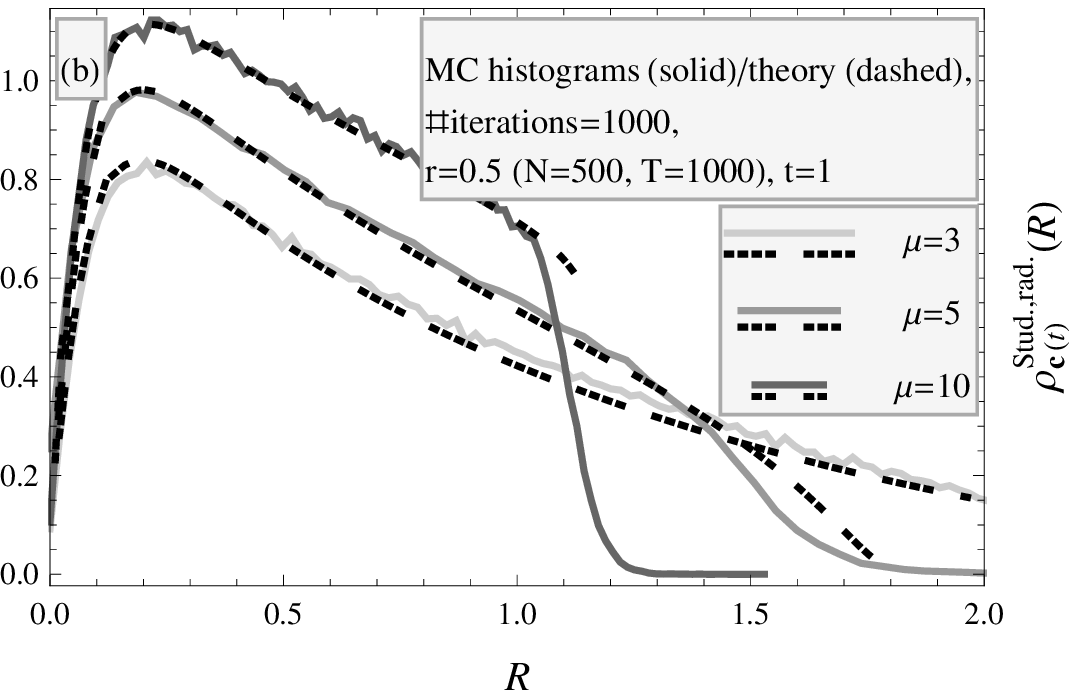}
\caption{Monte Carlo eigenvalues versus theory for the TLCE for Toy Model 1 and two versions of the Student t-distribution of the assets.}
\label{fig:TM1TLCEStudent}
\end{figure*}

\begin{figure*}[t]
\includegraphics[width=\columnwidth]{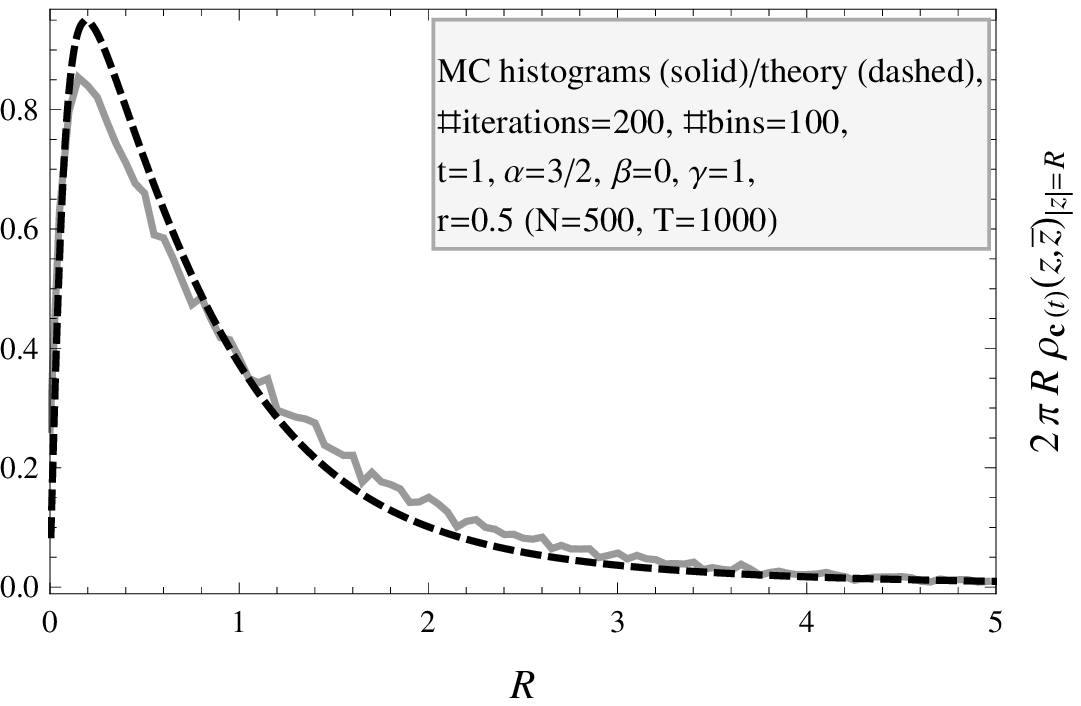}
\includegraphics[width=\columnwidth]{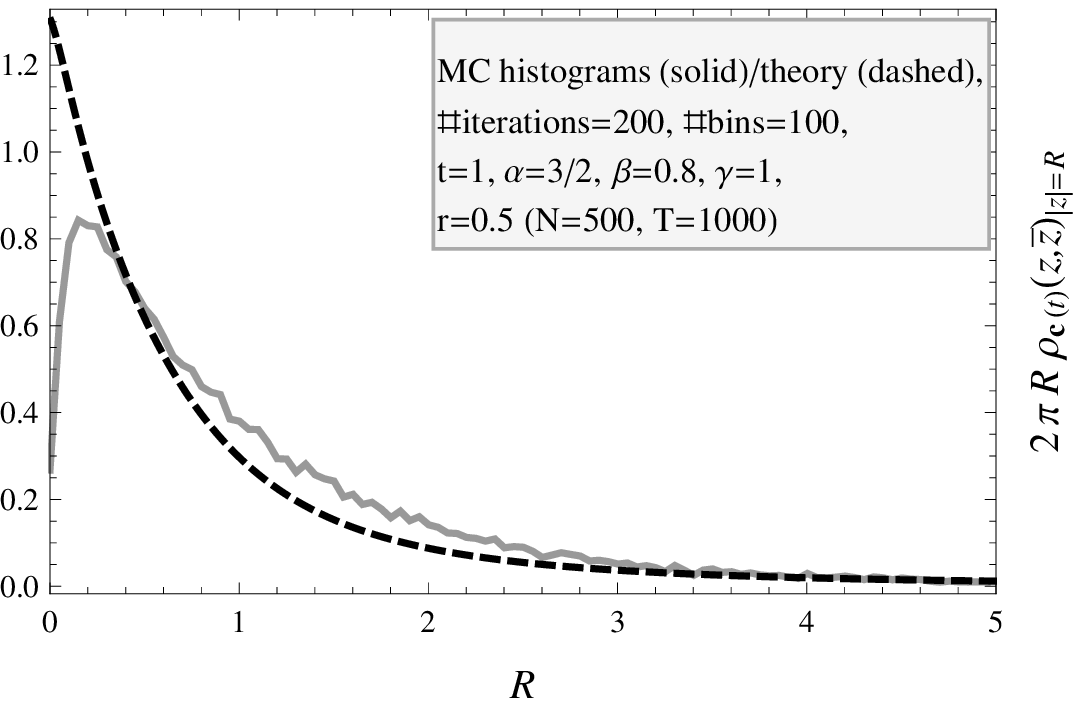}
\caption{Monte Carlo eigenvalues versus theory for the TLCE for Toy Model 1 and the free L\'{e}vy distribution of the assets.}
\label{fig:TM1TLCELevy}
\end{figure*}

%%%%%%%%%%%%%%%%%%%%%%%%%%%%%%%%%%%%%%%%%%%%%%%%%%%%%%%%%%%%%%%%%%%%%%

\subsubsection{Definition}
\label{sss:TM1Definition}

The simplest instance of the true covariance function (\ref{eq:ArbitraryCovarianceFunctionDefinition1}), serving also as the null hypothesis, is the factorized version (\ref{eq:Case2Definition}) with
\begin{equation}\label{eq:TM1DefinitionEq01}
\mathbf{C} = \sigma^{2} \Id_{N} , \quad \mathbf{A} = \Id_{T} ,
\end{equation}
i.e., the variables \smash{$R_{i a}$} are IID Gaussian random numbers of zero mean and some variance \smash{$\sigma^{2}$}.

%%%%%%%%%%%%%%%%%%%%%%%%%%%%%%%%%%%%%%%%%%%%%%%%%%%%%%%%%%%%%%%%%%%%%%

\subsubsection{Borderline of the mean spectral domain}
\label{sss:TM1TLCEBorderline}

For the TLCE, the considered structure of the true covariance function falls under the discussion in Secs.~\ref{sss:Case2Plus3TLCEA1} and~\ref{sss:RotationalSymmetry}. The mean spectral domain is either a disk (for $r \leq 1$) or an annulus (for $r > 1$), with the radii of the enclosing circles [(\ref{eq:Case2A1TLCERExt}), (\ref{eq:Case2A1TLCERInt})],
\begin{subequations}
\begin{align}
R_{\textrm{ext.}}^{2} &= \sigma^{4} r ( 1 + r ) ,\label{eq:TM1TLCEEq01a}\\
R_{\textrm{int.}}^{2} &= \sigma^{4} \frac{( r - 1 )^{3}}{r} \Theta ( r - 1 ) .\label{eq:TM1TLCEEq01b}
\end{align}
\end{subequations}
(One easily checks that \smash{$R_{\textrm{int.}} < R_{\textrm{ext.}}$} for all $r$.)

%%%%%%%%%%%%%%%%%%%%%%%%%%%%%%%%%%%%%%%%%%%%%%%%%%%%%%%%%%%%%%%%%%%%%%

\subsubsection{Mean spectral density}
\label{sss:TM1TLCEMSD}

The relevant master equation is (\ref{eq:Case2A1TLCEMasterEq}) with \smash{$M_{\mathbf{C}} ( z ) = \sigma^{2} / ( z - \sigma^{2} )$}, and elimination of $m$ leads to a cubic polynomial equation for the nonholomorphic $M$-transform, \smash{$M \equiv M_{\mathbf{c} ( t )} ( z , \overline{z} )$},
\begin{equation}
\begin{split}\label{eq:TM1TLCEEq02}
&4 r^{3} M^{3} + 4 r^{2} ( 1 + 2 r ) M^{2} +\\
&+ r \left( ( 1 + r ) ( 1 + 5 r ) - R_{\sigma}^{2} \right) M +\\
&+ ( 1 + r ) \left( r (1 + r) - R_{\sigma}^{2} \right) = 0 ,
\end{split}
\end{equation}
where for short, \smash{$R_{\sigma} \equiv R / \sigma^{2}$}, and \smash{$R \in [ R_{\textrm{int.}} , R_{\textrm{ext.}} ]$}. The rotational symmetry around zero is evident. Once $M$ is found from (\ref{eq:TM1TLCEEq02}), the (rescaled) radial MSD \smash{$\rho^{\textrm{rad.}}_{\sigma} \equiv \sigma^{2} \rho^{\textrm{rad.}}_{\mathbf{c} ( t )} ( R ) = \dd M / \dd R_{\sigma}$} (\ref{eq:RadialMSDDefinition}). Actually, this relationship allows to write a depressed cubic equation directly for the MSD,
\begin{subequations}
\begin{align}
0 &= a \left( \rho^{\textrm{rad.}}_{\sigma} \right)^{3} + b \rho^{\textrm{rad.}}_{\sigma} + c ,\label{eq:TM1TLCEEq03a}\\
a &\equiv 4 r^{3} \Big( R_{\sigma}^{4} - \left( 11 + 14 r + 2 r^{2} \right) R_{\sigma}^{2} - \nonumber\\
&- \left( 1 - r^{2} \right) ( 1 - r )^{2} \Big) ,\label{eq:TM1TLCEEq03b}\\
b &\equiv r \Big( - R_{\sigma}^{4} + 2 ( 1 + r ) ( 5 + r ) R_{\sigma}^{2} - \nonumber\\
&- \left( 1 - r^{2} \right)^{2} \Big) ,\label{eq:TM1TLCEEq03c}\\
c &\equiv 2 ( 1 + r )^{2} R_{\sigma} .\label{eq:TM1TLCEEq03d}
\end{align}
\end{subequations}
One may check that the discriminant of (\ref{eq:TM1TLCEEq03a}) is always negative, i.e., it has one real root which is the desired MSD,
\begin{equation}\label{eq:TM1TLCEEq04}
\rho^{\textrm{rad.}}_{\sigma} = \tilde{a} - \frac{\tilde{b}}{3 \tilde{a}} , \quad \tilde{a} \equiv \left(  - \frac{\tilde{c}}{2} + \sqrt{\frac{\tilde{b}^{3}}{27} + \frac{\tilde{c}^{2}}{4}} \right)^{1 / 3} ,
\end{equation}
where for short, \smash{$\tilde{b} \equiv b / a$} and \smash{$\tilde{c} \equiv c / a$}, and two redundant complex conjugate roots. Recall also that this case is independent of the time lag $t$ [cf.~(\ref{eq:Case2Plus3TLCEA1Eq02})]. This is the time-lagged counterpart of the MP distribution (\ref{eq:TM1ETCEEq03}) (cf.~App.~\ref{aa:TM1ETCE}).

Figures~\ref{fig:TM1TLCE} [(a), (d)] depict, respectively for $r = 0.1$ or $r = 10$, the numerically-generated eigenvalues of the TLCE and the theoretical borderline(s) [(\ref{eq:TM1TLCEEq01a})-(\ref{eq:TM1TLCEEq01b})], which are seen to accurately enclose the mean spectrum, with an exception of a small number of eigenvalues leaking out (this being a finite-size effect). In Figs.~\ref{fig:TM1TLCE} [(b), (e)], formula (\ref{eq:TM1TLCEEq04}) is shown to precisely reproduce the histograms of (the absolute values of) these numerical eigenvalues, respectively for $r = 0.1 , 0.5 , 0.9$ or $r = 10 , 2 , 10 / 9$, again except the vicinity of the borderline(s), to which a finite-size reasoning needs to be applied.

%%%%%%%%%%%%%%%%%%%%%%%%%%%%%%%%%%%%%%%%%%%%%%%%%%%%%%%%%%%%%%%%%%%%%%

\subsubsection{Universal erfc scaling}
\label{sss:TM1TLCEErfc}

As explained in App.~\ref{aaa:DSAndGUE}, the method of planar diagrammatic expansion employed in this paper works for infinite matrix dimensions (\ref{eq:ThermodynamicLimit}). However, for any non-Hermitian random matrix model $\mathbf{X}$ (of dimensions $N \times N$) whose MSD exhibits rotational symmetry around zero (cf.~Sec.~\ref{sss:RotationalSymmetry}) there has been put forward the ``erfc conjecture''~\cite{BurdaJaroszLivanNowakSwiech2010,BurdaJaroszLivanNowakSwiech2011,Jarosz2011-01,Jarosz2012-01} which claims that the universal finite-size behavior of the MSD close to the borderline of the domain (i.e., as it seems, always a ring or an annulus) is obtained simply by multiplying the density with the following form-factor,
\begin{equation}\label{eq:ErfcFormFactorDefinition}
f_{N , q_{\textrm{b}} , R_{\textrm{b}} , s_{\textrm{b}}} ( R ) \equiv \frac{1}{2} \erfc \left( q_{\textrm{b}} s_{\textrm{b}} \left( R - R_{\textrm{b}} \right) \sqrt{N} \right) ,
\end{equation}
for each circle (one or two) \smash{$R = R_{\textrm{b}}$} which constitutes the borderline. Moreover, the sign \smash{$s_{\textrm{b}}$} is $+ 1$ for the external borderline and $- 1$ for the internal one; \smash{$q_{\textrm{b}}$} is a parameter (one for each circle) depending on a particular model, whose derivation will not be attempted (this requires genuinely finite-size techniques), but its value adjusted by fitting to Monte Carlo data; finally, \smash{$\erfc ( x ) \equiv \frac{2}{\sqrt{\pi}} \int_{x}^{\infty} \dd y \exp ( - y^{2} )$} is the complementary error function.

This universal form-factor has been first calculated for the Ginibre unitary ensemble (\ref{eq:GinUEMeasureDefinition})~\cite{ForresterHonner1999,Kanzieper2005} (cf.~also~\cite{KhoruzhenkoSommers2009}), then for a product of two rectangular Gaussian random matrices with IID entries~\cite{KanzieperSingh2010}; in both cases, the mean spectral domain is a disk, and the parameter $q$ has been analytically determined. Inspired by this performance, the result has been conjectured and tested numerically for a product of an arbitrary number of rectangular Gaussian matrices~\cite{BurdaJaroszLivanNowakSwiech2010,BurdaJaroszLivanNowakSwiech2011} (one borderline), as well as a weighted sum of unitary random matrices from the Dyson circular unitary ensemble (CUE)~\cite{Jarosz2011-01} (one or two borderlines), or an arbitrary product of these last two models~\cite{Jarosz2012-01} (one or two borderlines).

In Fig.~\ref{fig:TM1TLCE} [(c), (f)], numerical tests of this hypothesis are presented, with excellent agreement; in particular, the steep decline of the MSD close to the borderline(s) is reproduced.

%%%%%%%%%%%%%%%%%%%%%%%%%%%%%%%%%%%%%%%%%%%%%%%%%%%%%%%%%%%%%%%%%%%%%%

\subsubsection{Previous solution based on the Abel transformation is incorrect}
\label{sss:TM1TLCEBielyThurner}

The MSD of the TLCE in the situation (\ref{eq:TM1DefinitionEq01}) has first been evaluated in~\cite{BielyThurner2006,ThurnerBiely2007}. It will now be demonstrated that their solution is incorrect.

\emph{Abel transformation.} The main idea of~\cite{BielyThurner2006,ThurnerBiely2007} is conceptually similar to the $N$-transform conjecture (cf.~Sec.~\ref{sss:RotationalSymmetry})---which recall constitutes one way to derive the above formula (\ref{eq:TM1TLCEEq04})---in which one relates the rotationally symmetric MSD of a non-Hermitian random matrix $\mathbf{X}$ to its ``absolute value squared'' \smash{$\mathbf{X}^{\dagger} \mathbf{X}$}. On the other hand, the authors of~\cite{BielyThurner2006,ThurnerBiely2007} conjecture that under the same assumption of the rotational symmetry, one may relate the spectra of $\mathbf{X}$ and its ``real part'' \smash{$( \mathbf{X} + \mathbf{X}^{\dagger} ) / 2$} through the so-called ``Abel transformation,'' which in both ways reads
\begin{subequations}
\begin{align}
\rho_{( \mathbf{X} + \mathbf{X}^{\dagger} ) / 2} \left( \sqrt{2} x \right) =& \frac{1}{\pi} \int_{x}^{\infty} \dd R \frac{\rho^{\textrm{rad.}}_{\mathbf{X}} ( R )}{\sqrt{R^{2} - x^{2}}} ,\label{eq:TM1TLCEBielyThurnerEq01a}\\
\rho^{\textrm{rad.}}_{\mathbf{X}} ( R ) =& - 2 R \int_{R}^{\infty} \dd x \frac{\frac{\dd}{\dd x} \rho_{( \mathbf{X} + \mathbf{X}^{\dagger} ) / 2} \left( \sqrt{2} x \right)}{\sqrt{x^{2} - R^{2}}}\label{eq:TM1TLCEBielyThurnerEq01b}\\
&\textrm{(generally incorrect!).}\nonumber
\end{align}
\end{subequations}

They then reinforce this hypothesis by verifying that [(\ref{eq:TM1TLCEBielyThurnerEq01a})-(\ref{eq:TM1TLCEBielyThurnerEq01b})] work for $\mathbf{X}$ being the Ginibre unitary ensemble with some variance \smash{$\sigma^{2}$} (\ref{eq:GinUEMeasureDefinition}), in which case the real part is the Gaussian unitary ensemble (\ref{eq:GUEMeasureDefinition}) with variance \smash{$\sigma^{2} / 2$} (this may be easily checked by computing the pertinent propagator). Plugging both MSD [(\ref{eq:MSDForGinUE}), (\ref{eq:MSDForGUE})] into the above formulae, one proves that they are indeed related by the Abel transformation.

\emph{Arguments against [(\ref{eq:TM1TLCEBielyThurnerEq01a})-(\ref{eq:TM1TLCEBielyThurnerEq01b})].} This coincidence is however accidental. The authors of~\cite{BurdaJanikWaclaw2010} have first falsified the Abel transformation conjecture by constructing a counterexample, \smash{$\mathbf{X} = \mathbf{H}_{1} \mathbf{H}_{2}$}, where \smash{$\mathbf{H}_{1 , 2}$} are two independent GUE random matrices.

This conjecture does not work for Toy Model 1 either. To show it, notice two facts: Firstly, formula (\ref{eq:TM1TLCEEq04}) is certainly correct, as it has been obtained not only through the $N$-transform conjecture but also the diagrammatic expansion method (which is a solid proof), plus it perfectly agrees in the bulk with the Monte Carlo data (cf.~Fig.~\ref{fig:TM1TLCE}). Secondly, the MSD of the real part of the TLCE is known (cf.~e.g.~\cite{BurdaJaroszJurkiewiczNowakPappZahed2010} for derivation)---the holomorphic $M$-transform \smash{$M \equiv M_{( \mathbf{c} ( t ) + \mathbf{c} ( t )^{\dagger} ) / 2} ( z )$} obeys a quartic polynomial equation,
\begin{equation}
\begin{split}\label{eq:TM1TLCEBielyThurnerEq02}
&r^{2} M^{4} + 2 r ( 1 + r ) M^{3} + \left( 1 + 4 r + r^{2} - z^{2} \right) M^{2} +\\
&+ 2 \left( 1 + r - \frac{z^{2}}{r} \right) M + 1 = 0 .
\end{split}
\end{equation}
Substituting these results into [(\ref{eq:TM1TLCEBielyThurnerEq01a})-(\ref{eq:TM1TLCEBielyThurnerEq01b})], one finds a discrepancy between the left- and right-hand sides of both equations, as demonstrated in Fig.~\ref{fig:TM1TLCEBielyThurner}.

To summarize, it is fortunate that the MSD of this model is given by a solution (\ref{eq:TM1TLCEEq04}) to a simple depressed cubic equation, and not by a much more complicated combination of the quartic equation (\ref{eq:TM1TLCEBielyThurnerEq02}) and the inverse Abel transformation (\ref{eq:TM1TLCEBielyThurnerEq01b}).

%%%%%%%%%%%%%%%%%%%%%%%%%%%%%%%%%%%%%%%%%%%%%%%%%%%%%%%%%%%%%%%%%%%%%%

\subsubsection{Generalization to the EWMA estimator}
\label{sss:TM1TLCEEWMA}

A first extension to be considered of the TLCE in the situation (\ref{eq:TM1DefinitionEq01}) (let for simplicity $\sigma = 1$) is its EWMA version [(\ref{eq:RealWeightedctDefinition}), (\ref{eq:EWMADefinition})] in the limit (\ref{eq:EWMALimit}). As noted in Sec.~\ref{sss:MeasurementNoise}, this limit requires $t \sim T$, which however is incompatible with the method used---therefore, the results will not reproduce Monte Carlo simulations precisely.

The master equation is the conjectured (\ref{eq:Case2C1ADiagonalTLCEMasterEq}) with diagonal $\mathbf{A} = \mathbf{W}$, since as explained in Sec.~\ref{sss:RotationalSymmetry}, the exact master equations seem hard to solve. The limit (\ref{eq:EWMALimit}) allows to pass in it from a discrete temporal description (summation over $a$) to a continuous one (integration over $a / T \in [ 0 , 1 ]$) since \smash{$w_{a}^{2} \to ( \vartheta / ( \ee^{\vartheta} - 1 ) ) \ee^{\vartheta ( a / T )}$}, which yields an explicit expression for the holomorphic $M$-transform of $\mathbf{A}$,
\begin{equation}
\begin{split}\label{eq:TM1TLCEEWMAEq01}
M_{\mathbf{A}} ( z ) &=\\
&= \frac{1}{\vartheta} \Big( \log \left( \left( 1 - \ee^{\vartheta} \right) z + \vartheta \right) -\\
&- \log \left( \left( 1 - \ee^{\vartheta} \right) z + \vartheta \ee^{\vartheta} \right) \Big) .
\end{split}
\end{equation}
It is straightforward to numerically construct the MSD from this master equation.

The borderline is a ring ($r \leq 1$) or an annulus ($r > 1$) with the radii [(\ref{eq:Case2C1ADiagonalTLCERExt}), (\ref{eq:Case2C1ADiagonalTLCERInt})],
\begin{subequations}
\begin{align}
R_{\textrm{ext.}}^{2} &= r \left( 1 + r \frac{\vartheta}{2} \coth \left( \frac{\vartheta}{2} \right) \right) ,\label{eq:TM1TLCEEWMAEq02a}\\
R_{\textrm{int.}}^{2} &= \frac{( r - 1 )^{3} \left( \frac{\vartheta}{2} \right)^{4} \csch^{4} \left( \frac{\vartheta}{2} \right)}{( r - 1 ) \frac{\vartheta}{2} \coth \left( \frac{\vartheta}{2} \right) + 1} \Theta ( r - 1 ) .\label{eq:TM1TLCEEWMAEq02b}
\end{align}
\end{subequations}
They tend to the standard values [(\ref{eq:TM1TLCEEq01a}), (\ref{eq:TM1TLCEEq01b})] when $\vartheta \to 0$.

Figures~\ref{fig:TM1TLCEEWMA} [(a), (b)] present the MSD derived from [(\ref{eq:Case2C1ADiagonalTLCEMasterEq}), (\ref{eq:TM1TLCEEWMAEq01})] and compare it with Monte Carlo simulations, for $r = 0.5$, three values of $\vartheta = 2 , 4 , 6$ and, respectively, $t = 1$ or $t = 100$. The concord is satisfactory for $t = 100$, since this value is large enough to be comparable with \smash{$\tau_{\textrm{EWMA}}$} and small enough for the finite-size effects not to be manifested. On the other hand, for $t = 1$ there are anticipated discrepancies, especially close to the border of the domain. They may be checked to decrease with decreasing $\vartheta$ (if $\vartheta$ is less than about $1$, one may quite safely use small $t$). Moreover, one may verify that for \smash{$\tau_{\textrm{EWMA}} \ll T$} and $t \ll T$ the agreement between theory and numerics is excellent (not shown here). One may also infer from these plots that for decreasing $\vartheta$ they approach the MSD of the standard TLCE, while larger $\vartheta$ means a shorter effective length of the time series, hence a more ``smeared'' density.

%%%%%%%%%%%%%%%%%%%%%%%%%%%%%%%%%%%%%%%%%%%%%%%%%%%%%%%%%%%%%%%%%%%%%%

\subsubsection{Generalization to the Student t-distribution}
\label{sss:TM1TLCEStudent}

As explained in App.~\ref{aaa:StudenttDistribution}, a simplest model of a non-Gaussian behavior of financial assets assumes that they are uncorrelated and distributed according to the Student t PDF (\ref{eq:MonovariateStudenttDistributionDefinition}). Moreover, it is known (cf.~App.~\ref{aaa:MonovariateRandomVolatilityModels}) that a Student random variable may be thought of as a Gaussian random variable whose volatility $\sigma$ is itself a random variable such that \smash{$1 / \sigma^{2}$} has the gamma distribution (\ref{eq:SquareRootInverseGammaDistributionDefinition}). This prescription will now be applied to transform the MSD of the TLCE to the Student case from two versions of the Gaussian model, both with independent assets and either (i) a common random volatility $\sigma$ for all the assets~\cite{BurdaGorlichWaclaw2006,BertuolaBohigasPato2004}; or (ii) IID random volatilities \smash{$\sigma_{a}$} at distinct time moments, but identical for all assets $i$~\cite{BiroliBouchaudPotters2007-1}.

\emph{Version 1: Common random volatility.} Assume first that the \smash{$R_{i a}$} are IID Gaussian with volatility $\sigma$, as above (\ref{eq:TM1DefinitionEq01}). Multiplying the MSD (\ref{eq:TM1TLCEEq04}) with the weight function (\ref{eq:SquareRootInverseGammaDistributionDefinition}) and integrating it over $\sigma$ (and changing the integration variable to \smash{$\xi \equiv R_{\sigma}$}) leads to the following radial MSD of the TLCE,
\begin{equation}
\begin{split}\label{eq:TM1TLCEStudent1Eq01}
\rho^{\textrm{Stud.,rad.}}_{\mathbf{c} ( t )} ( R ) &= \frac{\theta^{\mu}}{2^{\mu / 2} \Gamma \left( \frac{\mu}{2} \right)} \frac{1}{R^{1 + \mu / 2}} \cdot\\
&\cdot \int_{R_{\sigma = 1 , \textrm{int.}}}^{R_{\sigma = 1 , \textrm{ext.}}} \dd \xi \rho^{\textrm{rad.}}_{\sigma = 1} ( \xi ) \xi^{\mu / 2} \ee^{- \frac{\theta^{2} \xi}{2 R}} ,
\end{split}
\end{equation}
for $R \geq 0$ (i.e., it is unbounded for all values of the parameters), where the limits of integration are [(\ref{eq:TM1TLCEEq01a}), (\ref{eq:TM1TLCEEq01b})].

This function is plotted and tested against Monte Carlo simulations in Fig.~\ref{fig:TM1TLCEStudent} (a), for $r = 0.5$, \smash{$\theta = \sqrt{\mu}$} and three values of $\mu = 3 , 5 , 10$, with excellent concord in the whole domain---since there are no boundaries, the erfc form-factor is unnecessary. Even though the spectrum spreads to infinity, it decays fast and is effectively more restrained to the region around zero than for Gaussian assets (formally corresponding to $\mu \to \infty$), the more so the smaller $\mu$ is.

\emph{Version 2: IID temporal random volatilities.} A more realistic approach would however be to choose the \smash{$R_{i a}$} to be independent Gaussian with arbitrary volatilities \smash{$\sigma_{a}$}, dependent on $a$ but not $i$ (\ref{eq:MonovariateRandomVolatilityModelDefinition}), and eventually to assume these to be IID random variables distributed according to (\ref{eq:SquareRootInverseGammaDistributionDefinition}). In this case, analogously as for the EWMA estimator (cf.~Sec.~\ref{sss:TM1TLCEEWMA}), the master equation is the conjectured (\ref{eq:Case2C1ADiagonalTLCEMasterEq}) with
\begin{equation}\label{eq:TM1TLCEStudent2Eq01}
M_{\mathbf{A}} ( z ) = \int_{0}^{\infty} \dd \sigma \frac{1}{2^{\mu / 2 - 1} \Gamma \left( \frac{\mu}{2} \right)} \frac{\theta^{\mu}}{\sigma^{\mu + 1}} \ee^{- \frac{\theta^{2}}{2 \sigma^{2}}} \frac{1}{\frac{z}{\sigma^{2}} - 1} ,
\end{equation}
and it can be straightforwardly solved numerically---with proper care, as (\ref{eq:TM1TLCEStudent2Eq01}) cannot be expressed by elementary functions---to produce the MSD of the TLCE.

The borderline is a ring ($r \leq 1$) or an annulus ($r > 1$), extending to infinity for $\mu \leq 4$, with the radii [(\ref{eq:Case2C1ADiagonalTLCERExt}), (\ref{eq:Case2C1ADiagonalTLCERInt})],
\begin{subequations}
\begin{align}
R_{\textrm{ext.}}^{2} &= \left\{ \begin{array}{ll} \infty , & \textrm{for } \mu \leq 4 , \\ \frac{r \theta^{4} \left( ( r + 1 ) \mu - 2 ( r + 2 ) \right)}{( \mu - 2 )^{2} ( \mu - 4 )} , & \textrm{for } \mu > 4 , \end{array} \right.\label{eq:TM1TLCEStudent2Eq02a}\\
R_{\textrm{int.}}^{2} &= \frac{( r - 1 )^{3} \theta^{4}}{\mu \left( r \mu + 2 ( r - 1 ) \right)} .\label{eq:TM1TLCEStudent2Eq02b}
\end{align}
\end{subequations}
They tend to the standard values [(\ref{eq:TM1TLCEEq01a}), (\ref{eq:TM1TLCEEq01b})] in the Gaussian limit, \smash{$\theta = \sqrt{\mu}$}, $\mu \to \infty$.

The solutions have been found and checked with Monte Carlo data in Fig.~\ref{fig:TM1TLCEStudent} (b), for $r = 0.5$, \smash{$\theta = \sqrt{\mu}$} and three values of $\mu = 3 , 5 , 10$, with perfect agreement, except of a small area close to the border (for $\mu = 5 , 10$, i.e., when the MSD is bounded), to amend which the erfc form-factor should be used (not shown here). Note by comparing Figs.~(a) and (b) that although the qualitative behavior of the MSD is analogous for Versions 1 and 2, its precise form is quite different.

%%%%%%%%%%%%%%%%%%%%%%%%%%%%%%%%%%%%%%%%%%%%%%%%%%%%%%%%%%%%%%%%%%%%%%

\subsubsection{Generalization to the free L\'{e}vy distribution}
\label{sss:TM1TLCELevy}

Another non-Gaussian model considered in this paper is the free L\'{e}vy law (cf.~App.~\ref{aaa:FreeLevyDistribution}). The master equations for the TLCE in the situation of the true covariance function factorized (\ref{eq:Case2Definition}) into arbitrary $\mathbf{C}$ and trivial \smash{$\mathbf{A} = \Id_{T}$} has been developed in Sec.~\ref{sss:RotationalSymmetry} to be [(\ref{eq:Case2A1TLCELevyEq01a}), (\ref{eq:Case2A1TLCELevyEq01b})] or (\ref{eq:Case2A1TLCELevyEq02}) for zero skewness. Setting in them \smash{$\mathbf{C} = \Id_{N}$} (i.e., \smash{$N_{\mathbf{C}} ( z ) = 1 + 1 / z$}; let $\sigma = 1$ since the variance is generalized by $\gamma$) simplifies them to
\begin{subequations}
\begin{align}
&\varphi \frac{( r M + 1 + \delta + \ii r m )^{\alpha}}{r M + \delta + \ii r m} = \frac{( r M + 1 - \delta + \ii r m )^{\alpha}}{r M - \delta + \ii r m} ,\label{eq:TM1TLCELevyEq01a}\\
&\left( \frac{\left( ( r M + 1 + \ii r m )^{2} - \delta^{2} \right) ( M + 1 + \ii m )}{( r M + 1 + \ii r m ) \sqrt{- \left( 1 + \frac{1}{r M} \right)}} \right)^{\alpha} \cdot\nonumber\\
&\cdot \frac{1}{r^{2} ( M + \ii m )^{2} - \delta^{2}} \ee^{\ii \pi \frac{\alpha}{2}} r^{\alpha} = \frac{R^{\alpha}}{\gamma^{2}} ,\label{eq:TM1TLCELevyEq01b}
\end{align}
\end{subequations}
with real $m$ and complex $\delta$ auxiliary unknowns, or in the case of zero skewness,
\begin{equation}
\begin{split}\label{eq:TM1TLCELevyEq02}
&\left( \frac{( r M + 1 + \ii r m ) ( M + 1 + \ii m )}{\sqrt{- \left( 1 + \frac{1}{r M} \right)}} \right)^{\alpha} \cdot\\
&\cdot \frac{1}{( M + \ii m )^{2}} \ee^{\ii \pi \frac{\alpha}{2}} r^{\alpha - 2} = \frac{R^{\alpha}}{\gamma^{2}} .
\end{split}
\end{equation}

Figure~\ref{fig:TM1TLCELevy} shows Monte Carlo histograms and the MSD following from numerical solutions to these equations for $r = 0.5$ and the L\'{e}vy parameters $\alpha = 3 / 2$ (cf.~the end of App.~\ref{aaa:FreeLevyDistribution}), $\gamma = 1$ and respectively $\beta = 0$ (a) or $\beta = 0.8$ (b). The numerical eigenvalues have been obtained from Wigner-L\'{e}vy random matrices, as being more relevant for finances, hence some expected discrepancies are observed (yet not big, especially for no skewness). It is also notable that numerical histograms are nearly identical for zero and nonzero skewness, while the theoretical results differ, especially close to zero. Moreover, one may argue (by expanding $M$ and $m$ around zero, cf.~Sec.~\ref{sss:RotationalSymmetry}) that the domain in these cases extends to infinity, so only a part of it is shown.

%%%%%%%%%%%%%%%%%%%%%%%%%%%%%%%%%%%%%%%%%%%%%%%%%%%%%%%%%%%%%%%%%%%%%%
%%%%%%%%%%%%%%%%%%%%%%%%%%%%%%%%%%%%%%%%%%%%%%%%%%%%%%%%%%%%%%%%%%%%%%

\subsection{Toy Model 2: Gaussian assets with distinct variances}
\label{ss:TM2}

\begin{figure*}[t]
\includegraphics[width=\columnwidth]{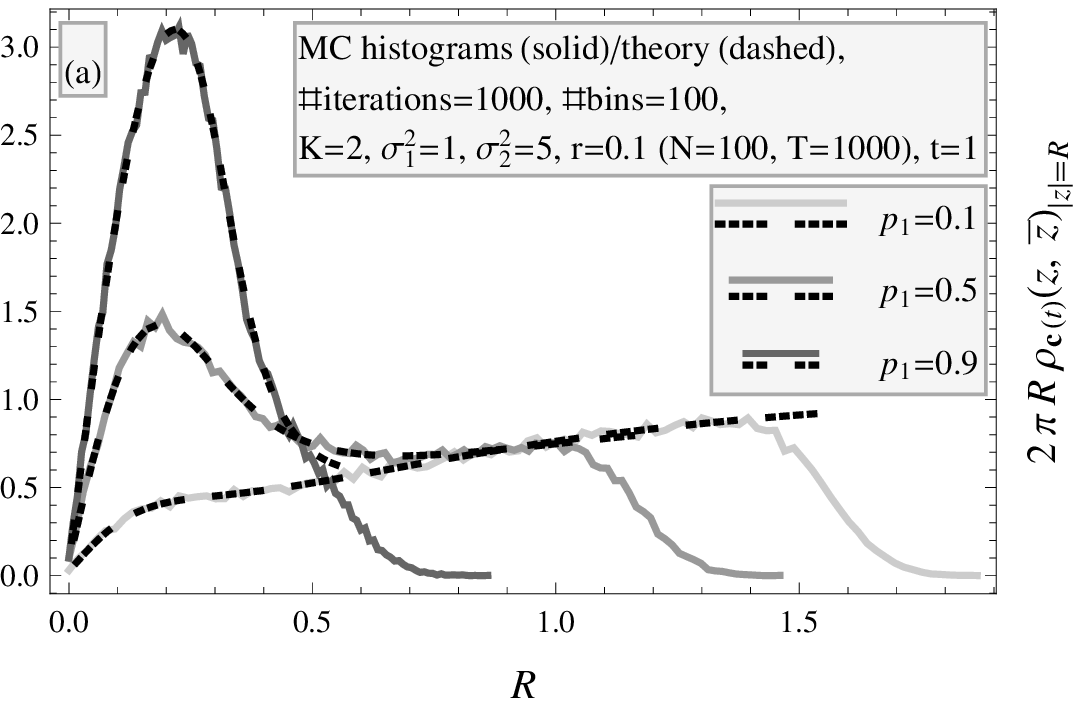}
\includegraphics[width=\columnwidth]{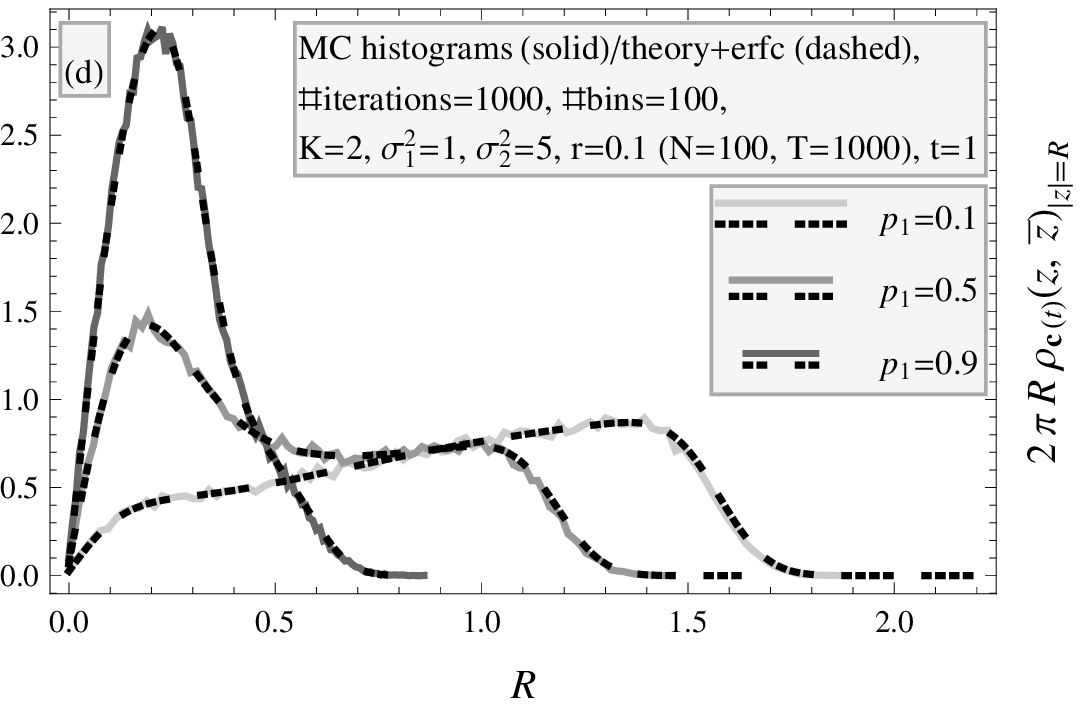}
\includegraphics[width=\columnwidth]{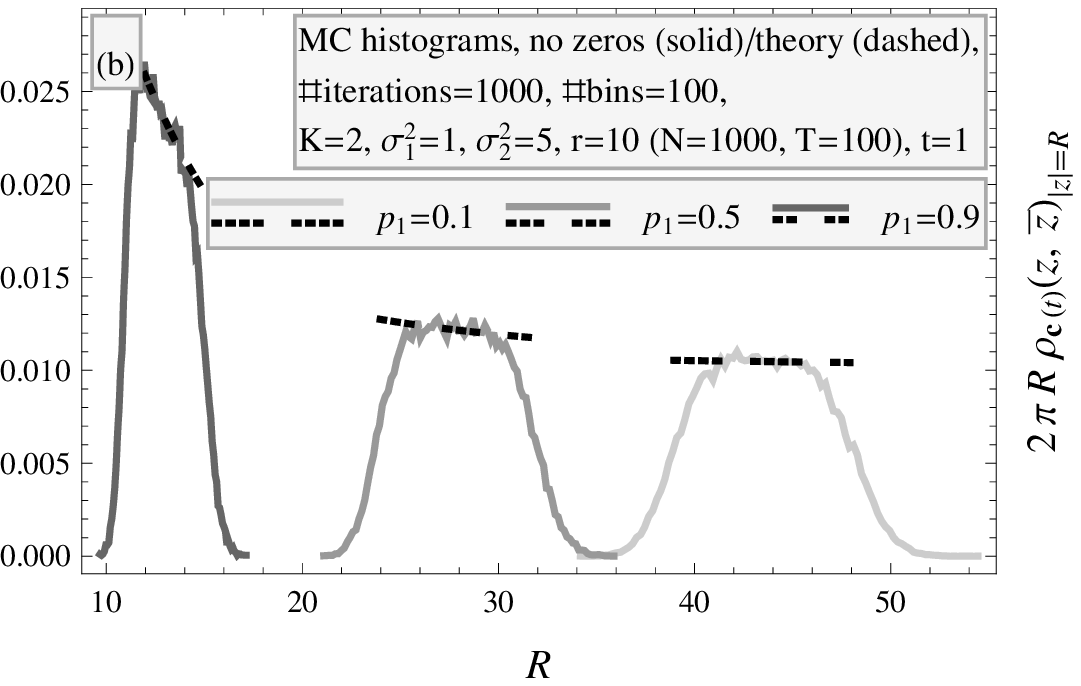}
\includegraphics[width=\columnwidth]{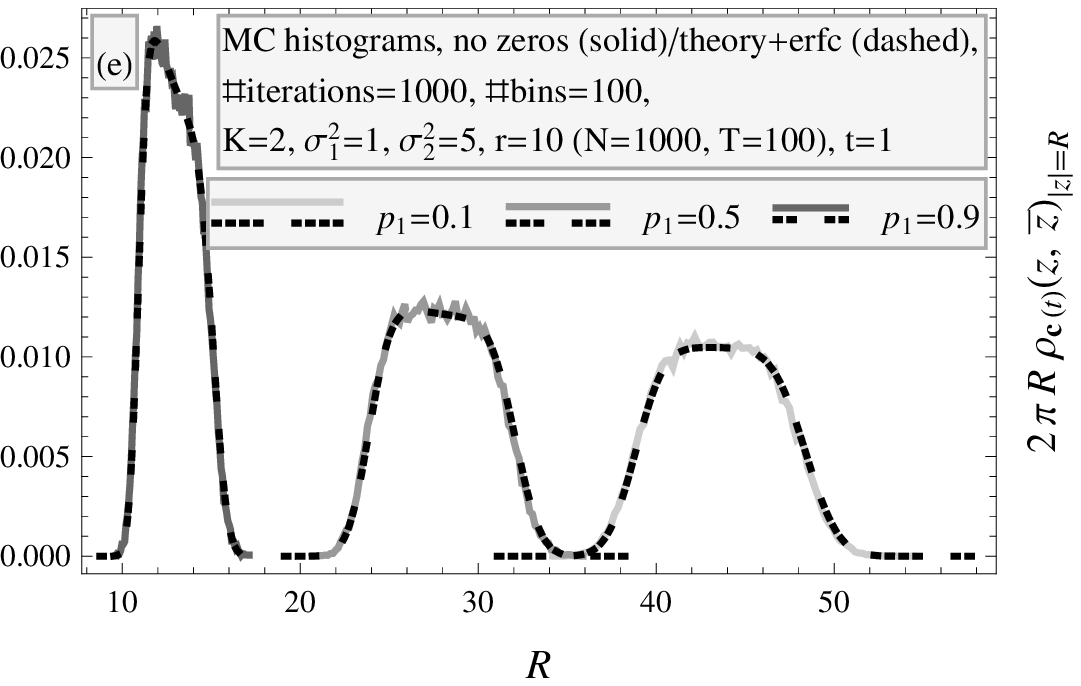}
\includegraphics[width=\columnwidth]{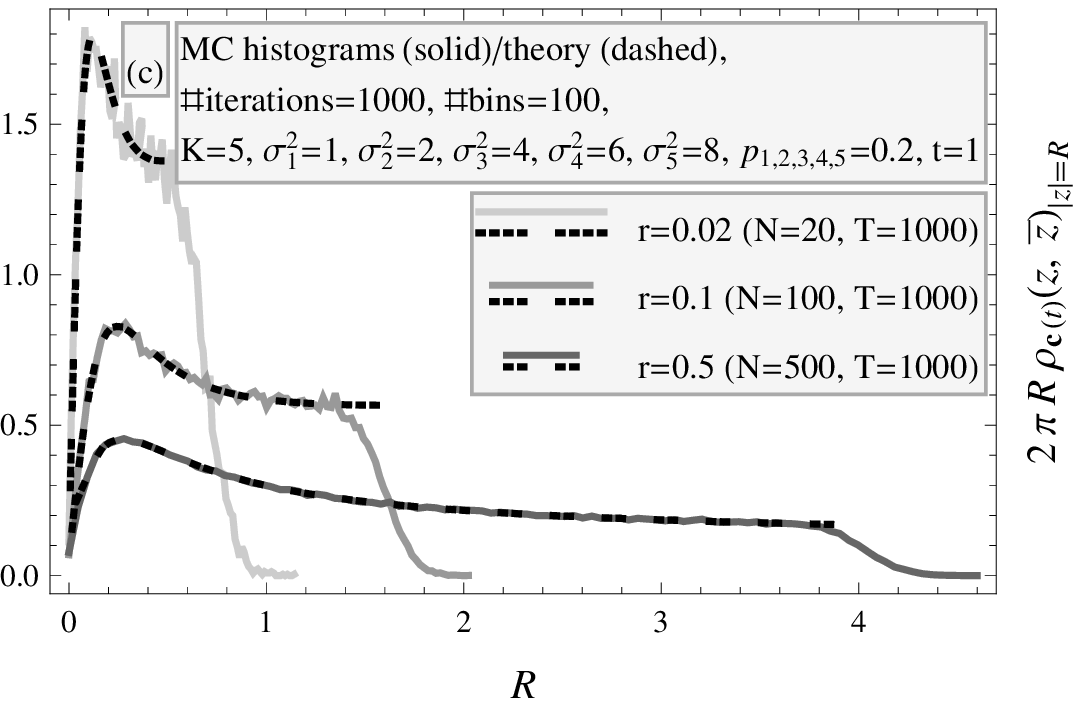}
\includegraphics[width=\columnwidth]{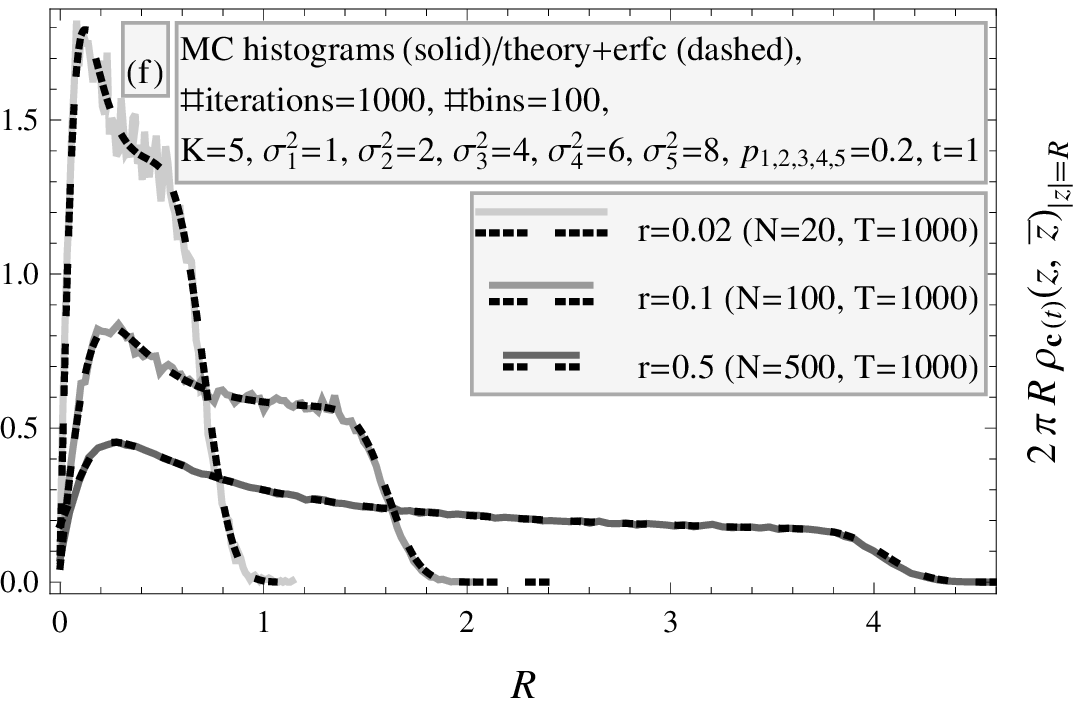}
\caption{Monte Carlo eigenvalues versus theory for the TLCE for Toy Model 2a.}
\label{fig:TM2aTLCE}
\end{figure*}

\begin{figure*}[t]
\includegraphics[width=\columnwidth]{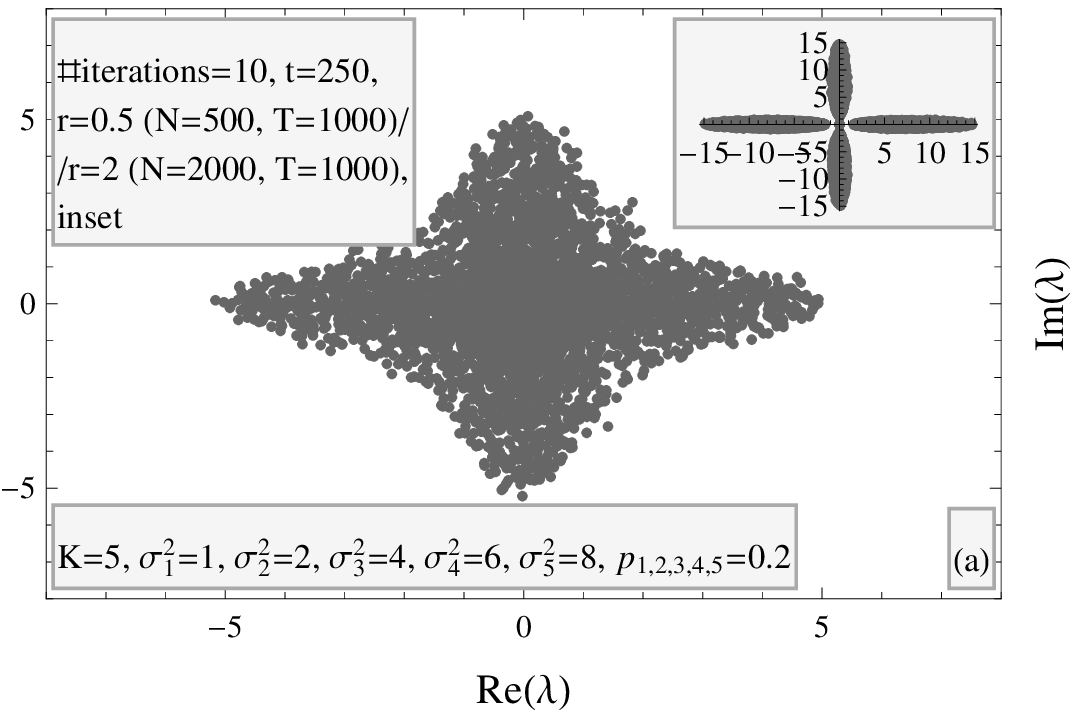}
\includegraphics[width=\columnwidth]{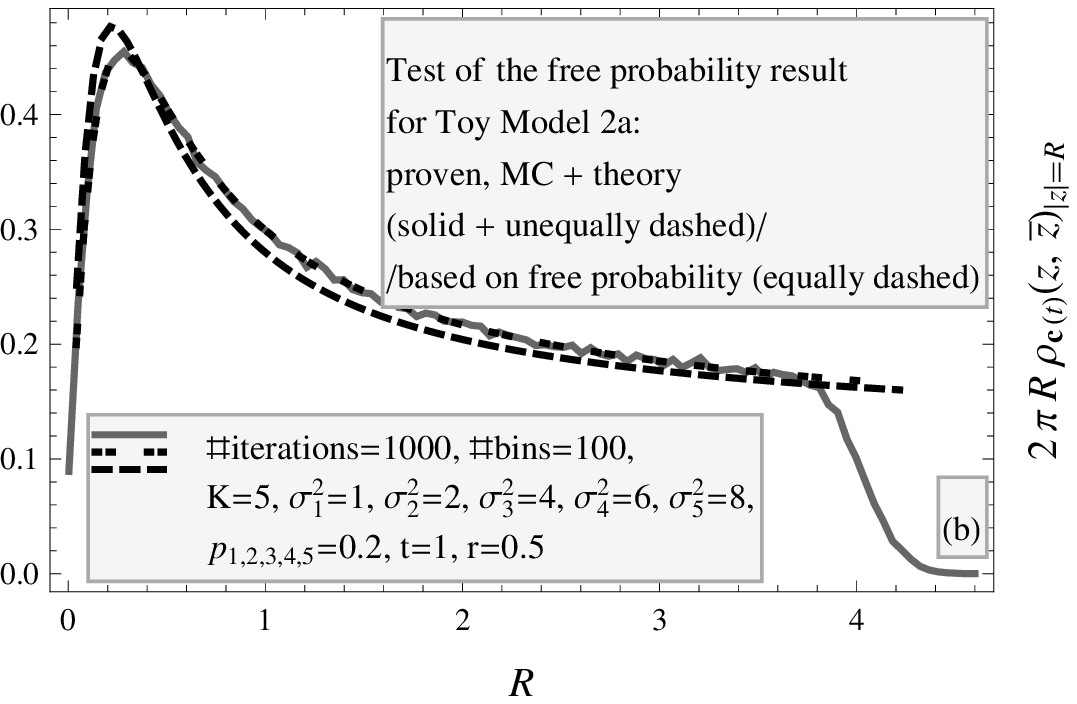}
\caption{Two tests for the Monte Carlo eigenvalues and theory for the TLCE for Toy Model 2a: the spectrum is not rotationally-symmetric for large $t$ (a); the derivation of the master equation based on free probability is surprisingly incorrect (b).}
\label{fig:TM2aTLCETwoTests}
\end{figure*}

In this Section, effects of the presence of nontrivial true spatial covariances will be analyzed; since there are no temporal correlations here, it is enough to consider a diagonal prior $\mathbf{C}$.

Only the standard TLCE and Gaussian assets will be investigated below, albeit generalizations to the EWMA version of the TLCE or to Student or free L\'{e}vy assets---as for Toy Model 1 (cf.~Secs.~\ref{sss:TM1TLCEEWMA}, \ref{sss:TM1TLCEStudent}, \ref{sss:TM1TLCELevy})---are within reach of the methods described in Sec.~\ref{sss:RotationalSymmetry}.

%%%%%%%%%%%%%%%%%%%%%%%%%%%%%%%%%%%%%%%%%%%%%%%%%%%%%%%%%%%%%%%%%%%%%%

\subsubsection{Toy Model 2a: Variance sectors}
\label{sss:TM2a}

\emph{Definition.} To begin with, in the fashion of the factor models (cf.~the first part of Sec.~\ref{sss:IndustrialSectors}), consider a finite number $K$ of distinct variances,
\begin{equation}\label{eq:TM2aEq01}
\mathbf{C} = \diag ( \underbrace{\sigma_{1}^{2}}_{N_{1} \textrm{ times}} , \underbrace{\sigma_{2}^{2}}_{N_{2} \textrm{ times}} , \ldots , \underbrace{\sigma_{K}^{2}}_{N_{K} \textrm{ times}} ) , \quad \mathbf{A} = \Id_{T} ,
\end{equation}
where denote \smash{$p_{k} \equiv N_{k} / N$}, for $k = 1 , 2 , \ldots , K$, which obey \smash{$\sum_{k = 1}^{K} p_{k} = 1$}, and which are also assumed finite in the thermodynamic limit. The holomorphic $M$-transform of $\mathbf{C}$, which is the basic ingredient of the master equation, thus reads
\begin{equation}\label{eq:TM2aEq02}
M_{\mathbf{C}} ( z ) = \sum_{k = 1}^{K} \frac{p_{k}}{\frac{z}{\sigma_{k}^{2}} - 1} .
\end{equation}

\emph{Borderline of the mean spectral domain.} As explained in Secs.~\ref{sss:RotationalSymmetry} and~\ref{sss:Case2Plus3TLCEA1}, the MSD is rotationally symmetric around zero, and its domain is either a disk ($r \leq 1$) or an annulus ($r > 1$). The external radius (\ref{eq:Case2A1TLCERExt}) reads
\begin{equation}\label{eq:TM2aTLCEEq01}
R_{\textrm{ext.}}^{2} = r^{2} \left( \sum_{k = 1}^{K} p_{k} \sigma_{k}^{2} \right)^{2} + r \sum_{k = 1}^{K} p_{k} \sigma_{k}^{4} .
\end{equation}

The internal radius (\ref{eq:Case2A1TLCERInt}) is a function of a solution $f$ of a polynomial equation of order $K$,
\begin{subequations}
\begin{align}
R_{\textrm{int.}}^{2} &= r f^{3} \sum_{k = 1}^{K} \frac{p_{k} \sigma_{k}^{2}}{\left( f + \sigma_{k}^{2} \right)^{2}} ,\label{eq:TM2aTLCEEq02a}\\
\sum_{k = 1}^{K} \frac{p_{k} \sigma_{k}^{2}}{f + \sigma_{k}^{2}} &= \frac{1}{r} .\label{eq:TM2aTLCEEq02b}
\end{align}
\end{subequations}
This equation has no positive solutions ($f > 0$, required for positive \smash{$R_{\textrm{int.}}^{2}$}) for $r \leq 1$, and exactly one positive solution for $r > 1$. Indeed, the function \smash{$\mathcal{F} ( f ) \equiv \sum_{k = 1}^{K} \frac{p_{k} \sigma_{k}^{2}}{f + \sigma_{k}^{2}} - 1 / r$} has the first derivative always negative, and singularities (vertical asymptotes) at the points \smash{$- \sigma_{k}^{2}$}; in other words, $\mathcal{F} ( f )$ is piecewise decreasing in the intervals \smash{$( - \infty , - \sigma_{K}^{2} ]$}, \smash{$[ - \sigma_{K}^{2} , - \sigma_{K - 1}^{2} ]$}, \ldots, \smash{$[ - \sigma_{1}^{2} , + \infty )$}. A first implication is that all its zeros are real. Moreover, $\mathcal{F} ( 0 ) = 1 - 1 / r$, which is negative or zero for $r \leq 1$ and positive for $r > 1$; and $\mathcal{F} ( + \infty ) = - 1 / r$, which is always negative. This completes the proof.

\emph{Mean spectral density.} The master equation is (\ref{eq:Case2A1TLCEMasterEq}) with (\ref{eq:TM2aEq02}), and it is straightforward to solve it numerically for any $K$.

Figures~\ref{fig:TM2aTLCE} [(a), (b), (c)] show comparisons of the numerically solved master equation with Monte Carlo data, all perfectly supporting the theoretical results. Figures~(a) and (b) differ by having $r = 0.1$ and $r = 10$, respectively, but both assume $K = 2$ with \smash{$\sigma_{1}^{2} = 1$}, \smash{$\sigma_{2}^{2} = 5$} and three values of \smash{$p_{1} = 0.1 , 0.5 , 0.9$}. In Fig.~(a), one may recognize two ``lumps,'' corresponding to the two variances smeared with the measurement noise, of sizes proportional to the relative multiplicities. However, this is rather an exception due to a small value of $r$ and a considerable difference between the variances; in general, unfortunately, any structure originating from separate variances will be lost under the noise, cf.~Fig.~(c), where three values of $r = 0.02 , 0.1 , 0.5$ are considered for $K = 5$ with \smash{$\sigma_{1}^{2} = 1$}, \smash{$\sigma_{2}^{2} = 2$}, \smash{$\sigma_{3}^{2} = 4$}, \smash{$\sigma_{4}^{2} = 6$}, \smash{$\sigma_{5}^{2} = 8$} and equal relative multiplicities $0.2$. In other words, these plots reveal that the MSD of the TLCE is far less sensitive to true spatial covariances than it is the case for the ETCE (cf.~App.~\ref{aa:TM2aETCE} and~Fig.~\ref{fig:TM2aETCE})---it is thus the ETCE which should be the main probe of true spatial covariances.

\emph{Universal erfc scaling.} Since the MSD is rotationally symmetric around zero, the erfc hypothesis (\ref{eq:ErfcFormFactorDefinition}) can be applied, finding accurate numerical confirmation depicted in Figs.~\ref{fig:TM2aTLCE} [(d), (e), (f)].

\emph{Miscellaneous.} Two facts announced above can now be verified numerically. First, it has been said that $t$ needs to be much smaller than $T$ for the planar diagrammatic method used in this paper to work (cf.~Sec~\ref{sss:MeasurementNoise}). Indeed, Fig.~\ref{fig:TM2aTLCETwoTests} (a) shows the Monte Carlo eigenvalues of the TLCE for Toy Model 2a, with $K = 5$, \smash{$\sigma_{1}^{2} = 1$}, \smash{$\sigma_{2}^{2} = 2$}, \smash{$\sigma_{3}^{2} = 4$}, \smash{$\sigma_{4}^{2} = 6$}, \smash{$\sigma_{5}^{2} = 8$}, equal relative multiplicities $0.2$, as well as $r = 0.5$ or $r = 2$ (in the inset), and $t = 250$, i.e., of magnitude comparable with $T = 1000$. In both cases, the rotational symmetry of the MSD is lost, even though it has been proven for $t \ll T$ in Sec.~\ref{sss:Case2Plus3TLCEA1}. Actually, one may infer from these (and other not shown) plots a conjecture that the mean spectral domain for $t \sim T$ has rather a $( T / t )$-fold symmetry around zero, polygon-shaped for $r < 1$ and star-shaped for $r > 1$.

Second, in Sec.~\ref{sss:RotationalSymmetry}, an alternative derivation of the master equation (\ref{eq:Case2A1TLCEMasterEq}) has been presented---based on the free probability multiplication law (\ref{eq:FreeProbabilityMultiplicationLaw}) applied to a product of Hermitian matrices, which a priori may not work, and inspired by an analogous derivation for the ETCE (cf.~Sec.~\ref{sss:Case2ETCE})---which resulted in Eq.~(\ref{eq:Case2A1TLCEWrongEq02}). Figure~\ref{fig:TM2aTLCETwoTests} (b) compares, for the same values of the parameters as in Fig.~(a) except for $t = 1$, a Monte Carlo histogram with numerical solutions of these two equations, demonstrating that Eq.~(\ref{eq:Case2A1TLCEMasterEq}) works perfectly, while Eq.~(\ref{eq:Case2A1TLCEWrongEq02}) leads to some discrepancy, albeit very small (it seems to be smaller for decreasing $r$). It deserves a deeper explanation why the multiplication law produces a wrong result and why these two quite different equations lead to very close solutions.

%%%%%%%%%%%%%%%%%%%%%%%%%%%%%%%%%%%%%%%%%%%%%%%%%%%%%%%%%%%%%%%%%%%%%%

\subsubsection{Toy Model 2b: Power-law-distributed variances}
\label{sss:TM2b}

\begin{figure}[t]
\includegraphics[width=\columnwidth]{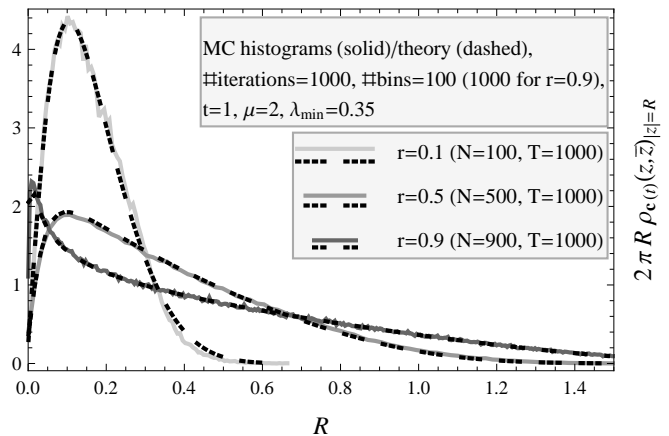}
\caption{Monte Carlo eigenvalues versus theory for the TLCE for Toy Model 2b.}
\label{fig:TM2bTLCE}
\end{figure}

\emph{Definition.} Suppose now that the variances are distributed according to the power law (\ref{eq:PowerLawCDefinition}) with $\mu = 2$ (cf.~the second part of Sec.~\ref{sss:IndustrialSectors}). The holomorphic $M$-transform of $\mathbf{C}$ is straightforward to calculate,
\begin{equation}
\begin{split}\label{eq:TM2bEq01}
M_{\mathbf{C}} ( z ) &= \frac{1}{\left( 1 - 2 \lambda_{\min} + z \right)^{3}} \cdot \\
&\cdot \bigg( \left( 1 - 2 \lambda_{\min} + z \right) \left( - \left( 1 - 2 \lambda_{\min} \right)^{2} + z \right) +\\
&+ 2 \left( 1 - \lambda_{\min} \right)^{2} z \cdot \\
&\cdot \Big( \log \left( z - \lambda_{\min} \right) - \log \left( 1 - \lambda_{\min} \right) - \pi \ii \Big) \bigg) .
\end{split}
\end{equation}

\emph{Borderline of the mean spectral domain.} As usual with the rotational symmetry, the domain is a disk for $r \leq 1$ or an annulus for $r > 1$. The external radius (\ref{eq:Case2A1TLCERExt}) for an arbitrary value of $\mu$ reads
\begin{equation}\label{eq:TM2bTLCEEq01}
R_{\textrm{ext.}}^{2} = \left\{ \begin{array}{ll} \infty , & \textrm{for } \mu \leq 2 , \\ r \left( r + 1 + \left( 1 - \lambda_{\min} \right)^{2} \frac{\mu}{\mu - 2} \right) , & \textrm{for } \mu > 2 , \end{array} \right.
\end{equation}
i.e., for $\mu \leq 2$ (which encompasses the model considered here) the eigenvalues of the TLCE spread to complex infinity [cf.~(\ref{eq:TM1TLCEStudent2Eq02a})].

For the internal radius (\ref{eq:Case2A1TLCERInt}) ($\mu = 2$ and $r > 1$ assumed), the following expression is obtained,
\begin{equation}\label{eq:TM2bTLCEEq02}
R_{\textrm{int.}}^{2} = \frac{2 f^{2} \left( f - f_{+} \right) \left( f - f_{-} \right)}{\left( f - f_{0} \right) \left( f + \lambda_{\min} \right)} ,
\end{equation}
where for short,
\begin{subequations}
\begin{align}
f_{\pm} &\equiv \frac{1}{4} \Big( r - 1 \pm \nonumber\\
&\pm \sqrt{( r - 1 ) \left( r - 1 + 8 \lambda_{\min} \left( 1 - 2 \lambda_{\min} \right) \right)} \Big) ,\label{eq:TM2bTLCEEq03a}\\
f_{0} &\equiv 1 - 2 \lambda_{\min} ,\label{eq:TM2bTLCEEq03b}
\end{align}
\end{subequations}
and where real \smash{$f > - \lambda_{\min}$} is a solution to the nonlinear equation
\begin{equation}
\begin{split}\label{eq:TM2bTLCEEq04}
0 = \mathcal{F} ( f ) &\equiv \\
&\equiv 2 \left( 1 - \lambda_{\min} \right)^{2} \log \left( \frac{f + \lambda_{\min}}{1 - \lambda_{\min}} \right) + \\
&\frac{f - f_{0}}{r f} \cdot \\
&\cdot \Big( - ( r - 1 ) \left( 1 - 2 \lambda_{\min} \right)^{2} - \\
&- \left( 2 + r - 4 \lambda_{\min} \right) f + f^{2} \Big) .
\end{split}
\end{equation}
Observe now the following properties of $\mathcal{F} ( f )$: (i) It decreases inside the interval \smash{$[ f_{-} , f_{+} ]$} and increases outside of it (notice that \smash{$f_{\pm}$}, \smash{$f_{0}$} are real and \smash{$- \lambda_{\min} < f_{-} < 0$}, \smash{$f_{+} > 0$} \smash{$f_{0} > 0$}). (ii) It has infinite limits, \smash{$\lim_{f \to - \lambda_{\min}^{+}} \mathcal{F} ( f ) = - \infty$}, \smash{$\lim_{f \to 0^{-}} \mathcal{F} ( f ) = - \infty$}, \smash{$\lim_{f \to 0^{+}} \mathcal{F} ( f ) = + \infty$}, \smash{$\lim_{f \to + \infty} \mathcal{F} ( f ) = + \infty$}. (iii) \smash{$\mathcal{F} ( f_{0} ) = 0$}. All this implies that $\mathcal{F} ( f )$ increases from $- \infty$ at \smash{$f = - \lambda_{\min}$} (a vertical asymptote) to some value (a maximum) at \smash{$f = f_{-}$}, then decreases to $- \infty$ at $f = 0$ (a vertical asymptote); on the other side of this asymptote, it decreases from $+ \infty$ to some value (a minimum) at \smash{$f = f_{+}$}, then increases to $+ \infty$ as $f \to + \infty$. Since \smash{$f_{0} > 0$} is a root of this function (but uninteresting as it yields an infinite \smash{$R_{\textrm{int.}}^{2}$}), it means there is exactly one other root \smash{$f_{1} > 0$}; it leads to a positive \smash{$R_{\textrm{int.}}^{2}$}, and therefore is the right solution. Moreover, there would be two roots in the interval \smash{$( - \lambda_{\min} , 0 )$} if and only if there were \smash{$\mathcal{F} ( f_{-} ) > 0$}; however, this may be verified to never be true. To summarize, Eq.~(\ref{eq:TM2bTLCEEq04}) has always precisely one real and positive solution \smash{$f_{1}$} besides \smash{$f_{0}$} which yields a real and positive \smash{$R_{\textrm{int.}}^{2}$} (\ref{eq:TM2bTLCEEq02}).

\emph{Mean spectral density.} The master equation is (\ref{eq:Case2A1TLCEMasterEq}) with (\ref{eq:TM2bEq01}), and with proper care can be solved numerically.

Figure~\ref{fig:TM2bTLCE} compares its only solution yielding a positive-definite and normalized MSD to Monte Carlo simulations for \smash{$\lambda_{\min} = 0.35$} (cf.~the end of Sec.~\ref{sss:IndustrialSectors}) and three values of $r = 0.1 , 0.5 , 0.9$. It is notable that the agreement is perfect in the whole domain---there is no upper edge, thus no need to use the erfc form-factor.

%%%%%%%%%%%%%%%%%%%%%%%%%%%%%%%%%%%%%%%%%%%%%%%%%%%%%%%%%%%%%%%%%%%%%%
%%%%%%%%%%%%%%%%%%%%%%%%%%%%%%%%%%%%%%%%%%%%%%%%%%%%%%%%%%%%%%%%%%%%%%

\subsection{Toy Model 3: Gaussian assets with temporal covariances}
\label{ss:TM3}

\begin{figure*}[t]
\includegraphics[width=\columnwidth]{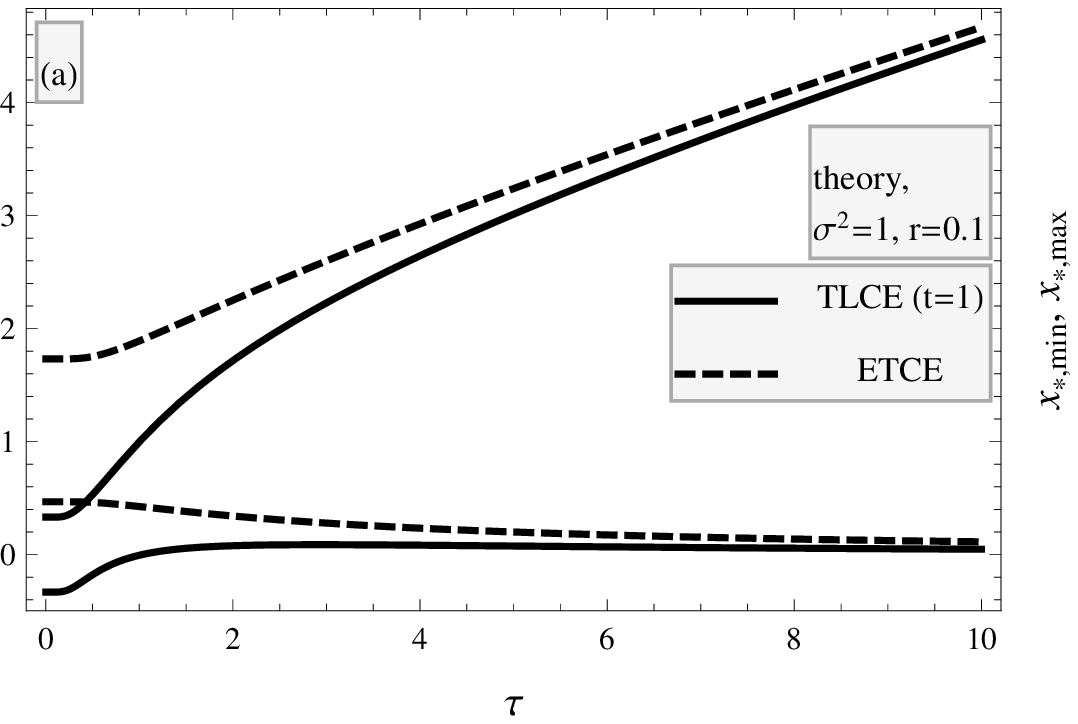}
\includegraphics[width=\columnwidth]{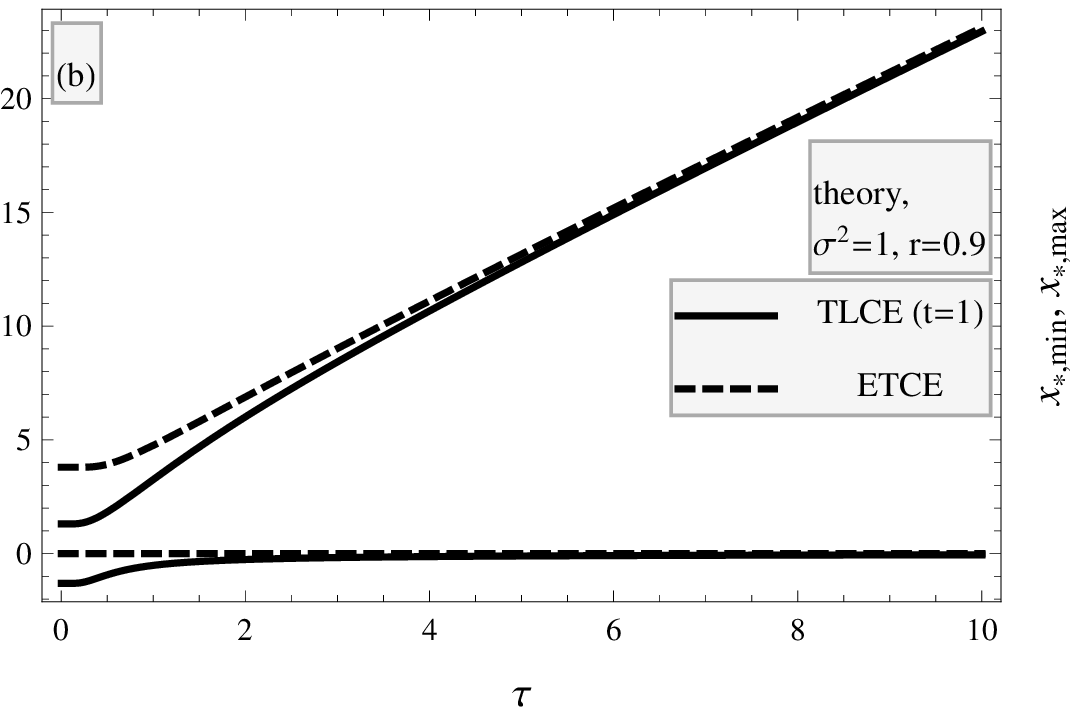}
\caption{The theoretical edges of the MSD of the ETCE versus the crossings of the external borderline of the mean spectral domain of the TLCE with the abscissa axis for Toy Model 3.}
\label{fig:TM3TLCEEIGEdges}
\end{figure*}

\begin{figure}[ht]
\includegraphics[width=\columnwidth]{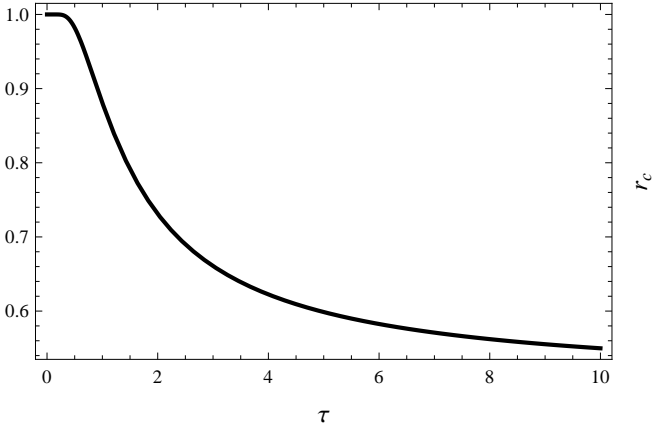}
\caption{The critical rectangularity ratio (in the sector $r < 1$) above which the internal borderline arises for the TLCE for Toy Model 3.}
\label{fig:TM3TLCEEIGrc}
\end{figure}

\begin{figure*}[t]
\includegraphics[width=\columnwidth]{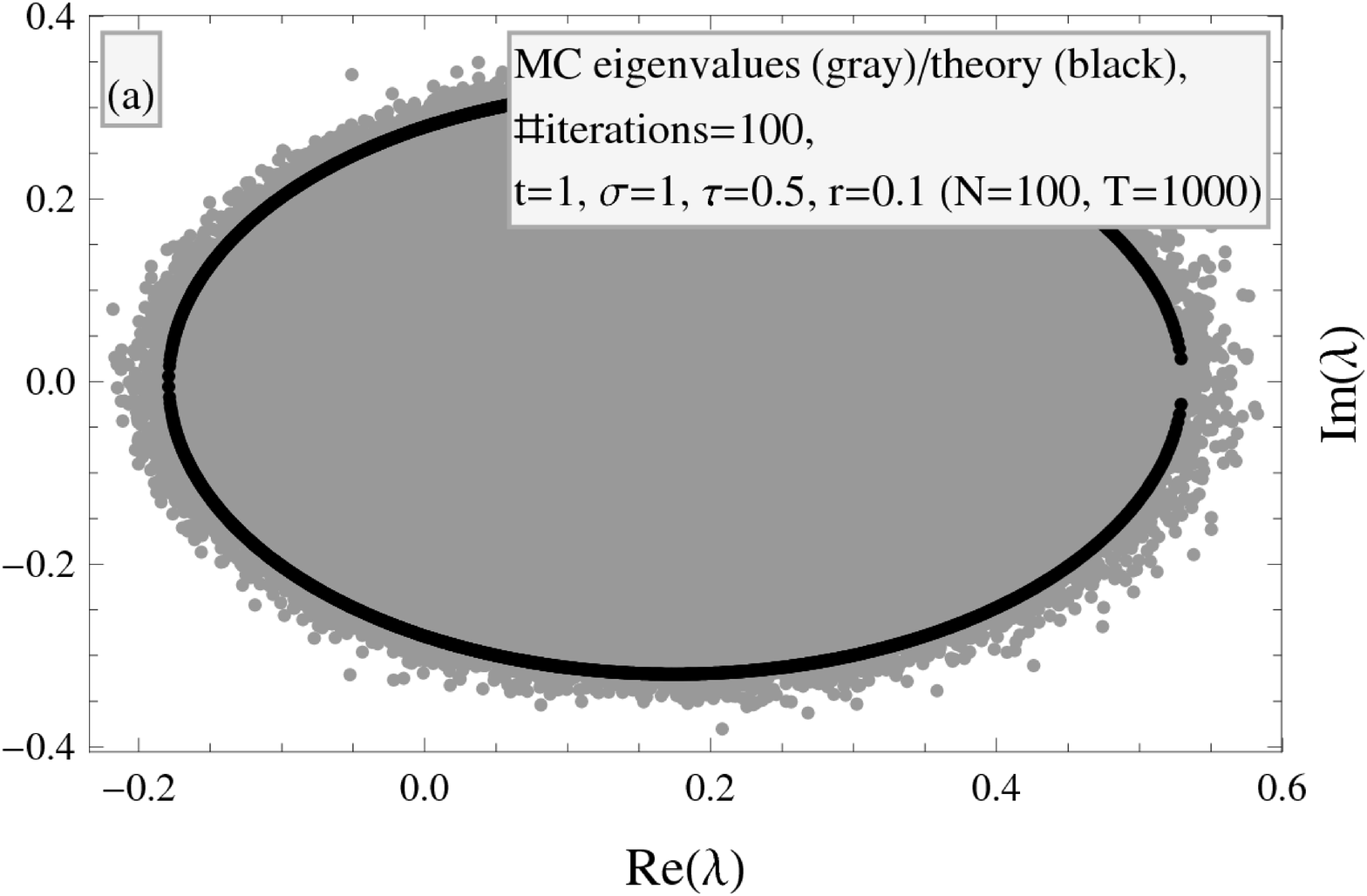}
\includegraphics[width=\columnwidth]{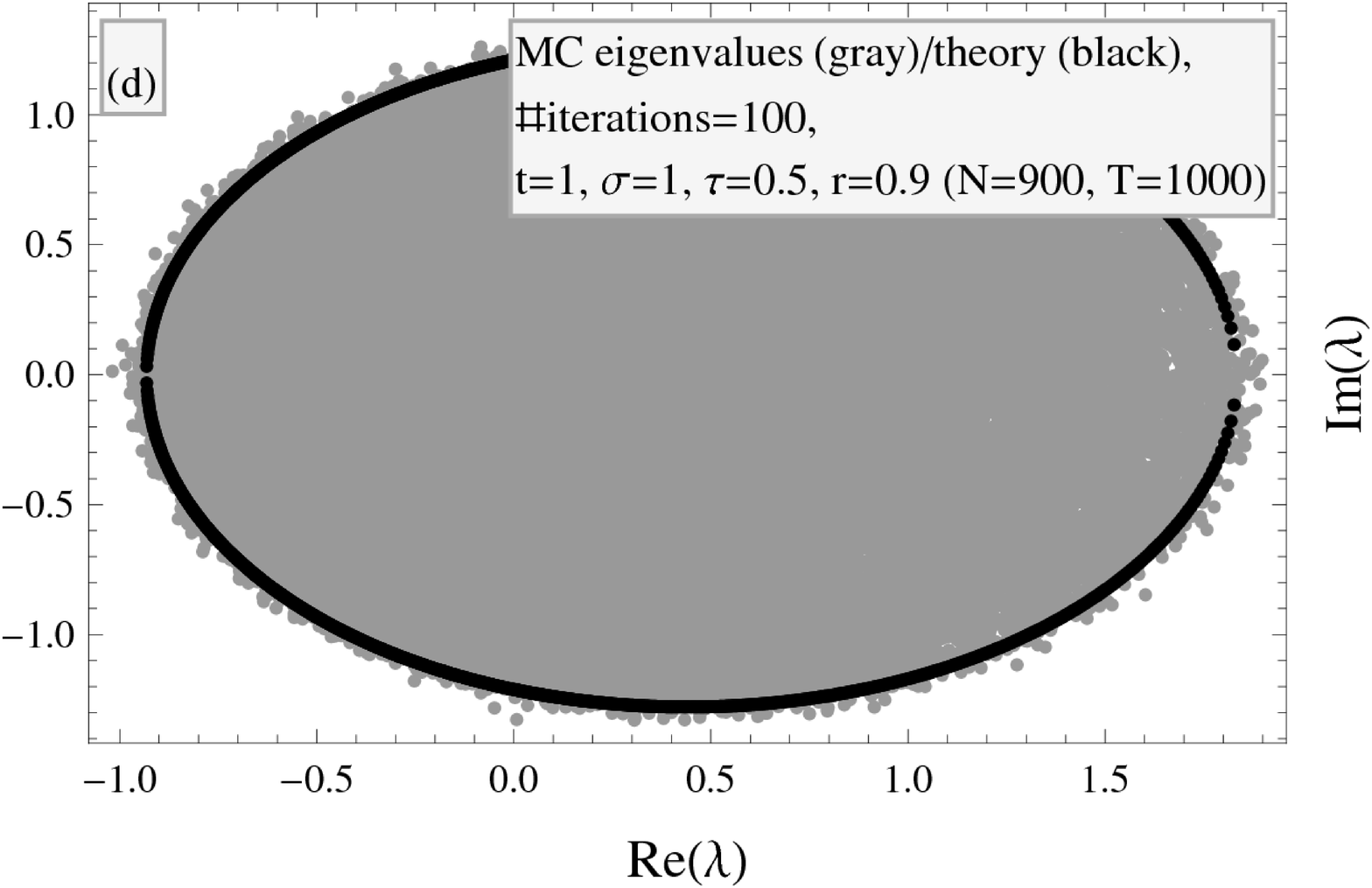}
\includegraphics[width=\columnwidth]{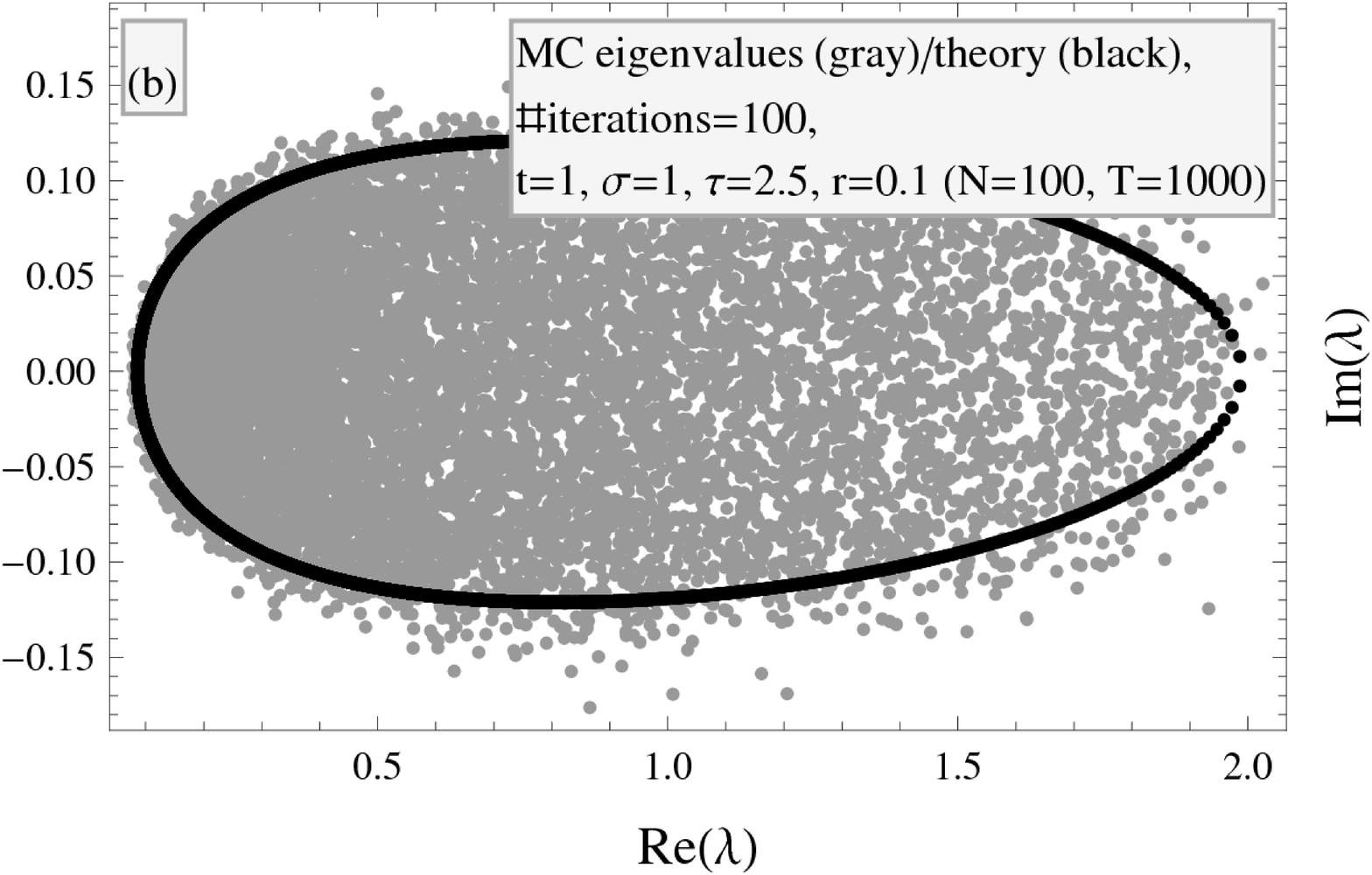}
\includegraphics[width=\columnwidth]{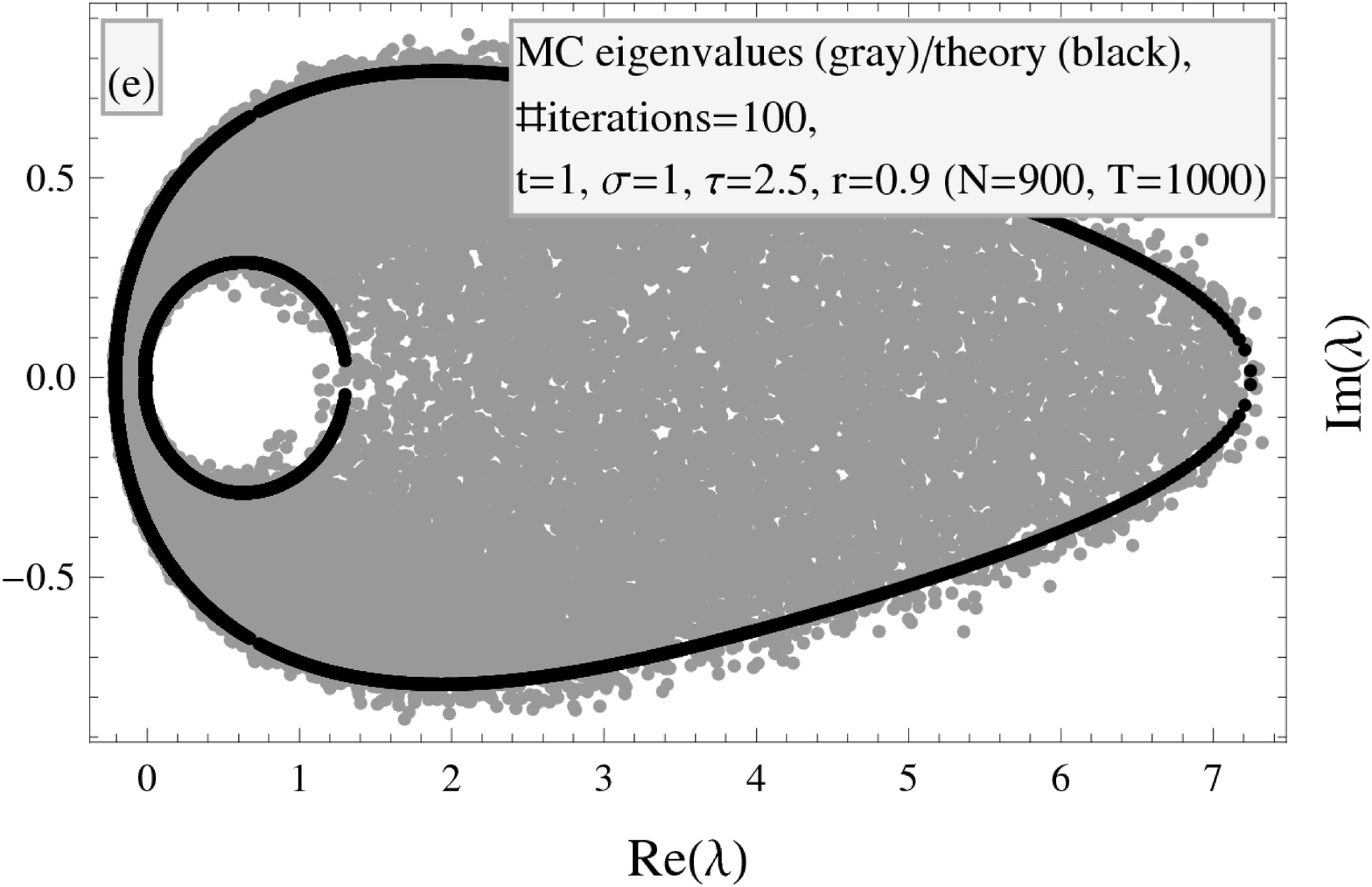}
\includegraphics[width=\columnwidth]{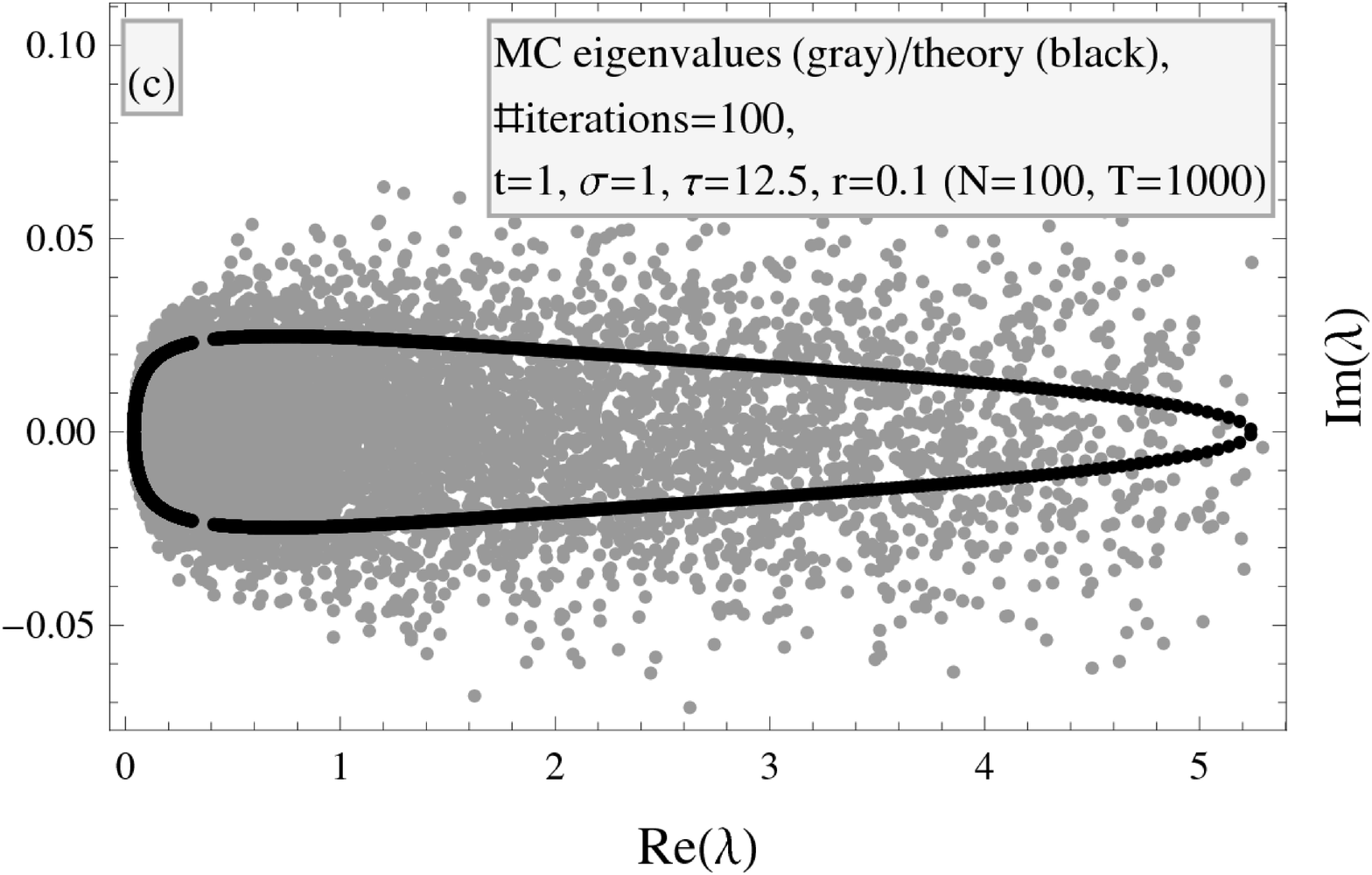}
\includegraphics[width=\columnwidth]{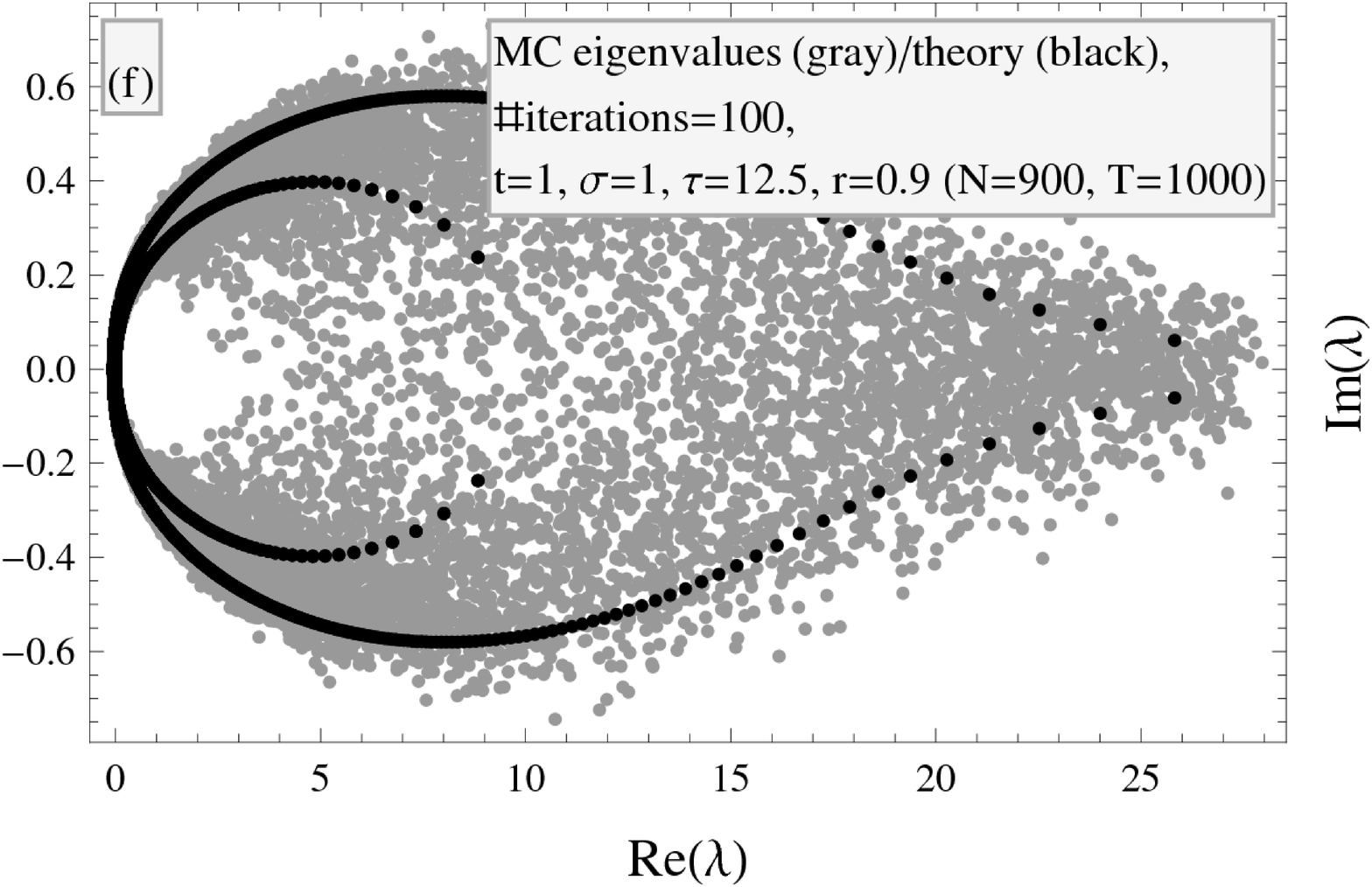}
\caption{Monte Carlo eigenvalues versus theory for the TLCE ($t = 1$) for Toy Model 3.}
\label{fig:TM3TLCEEIGt1}
\end{figure*}

\begin{figure*}[t]
\includegraphics[width=\columnwidth]{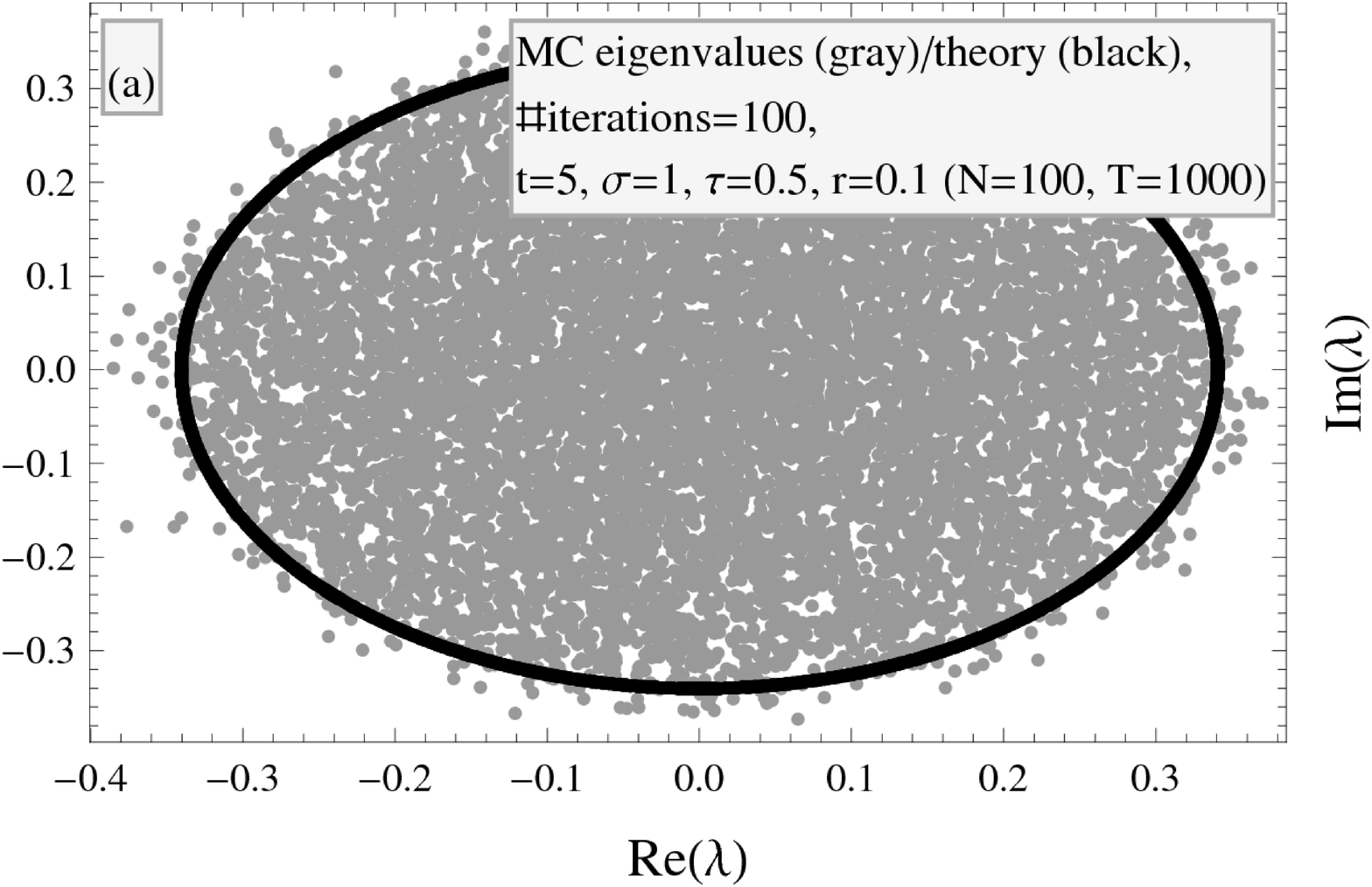}
\includegraphics[width=\columnwidth]{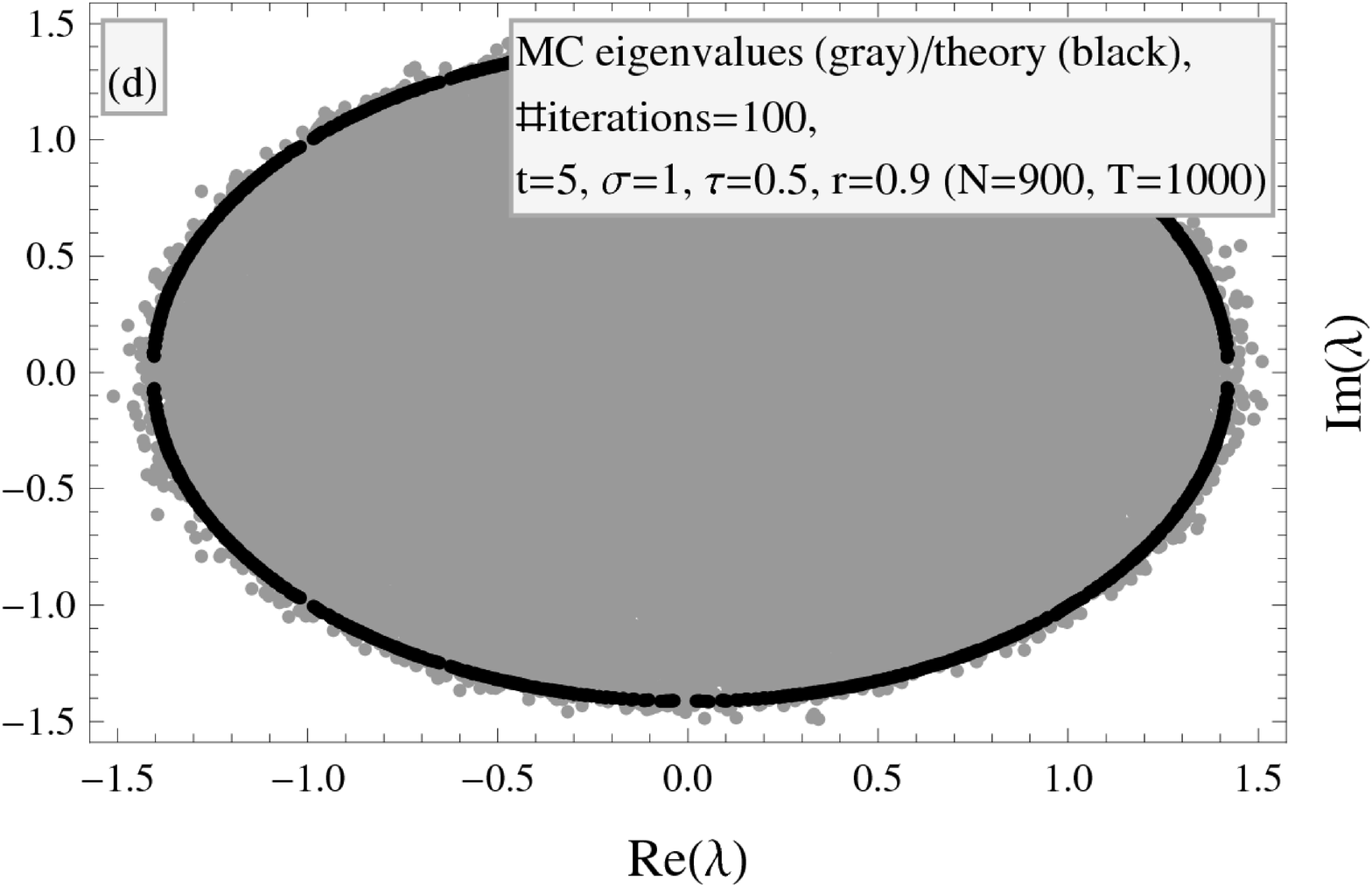}
\includegraphics[width=\columnwidth]{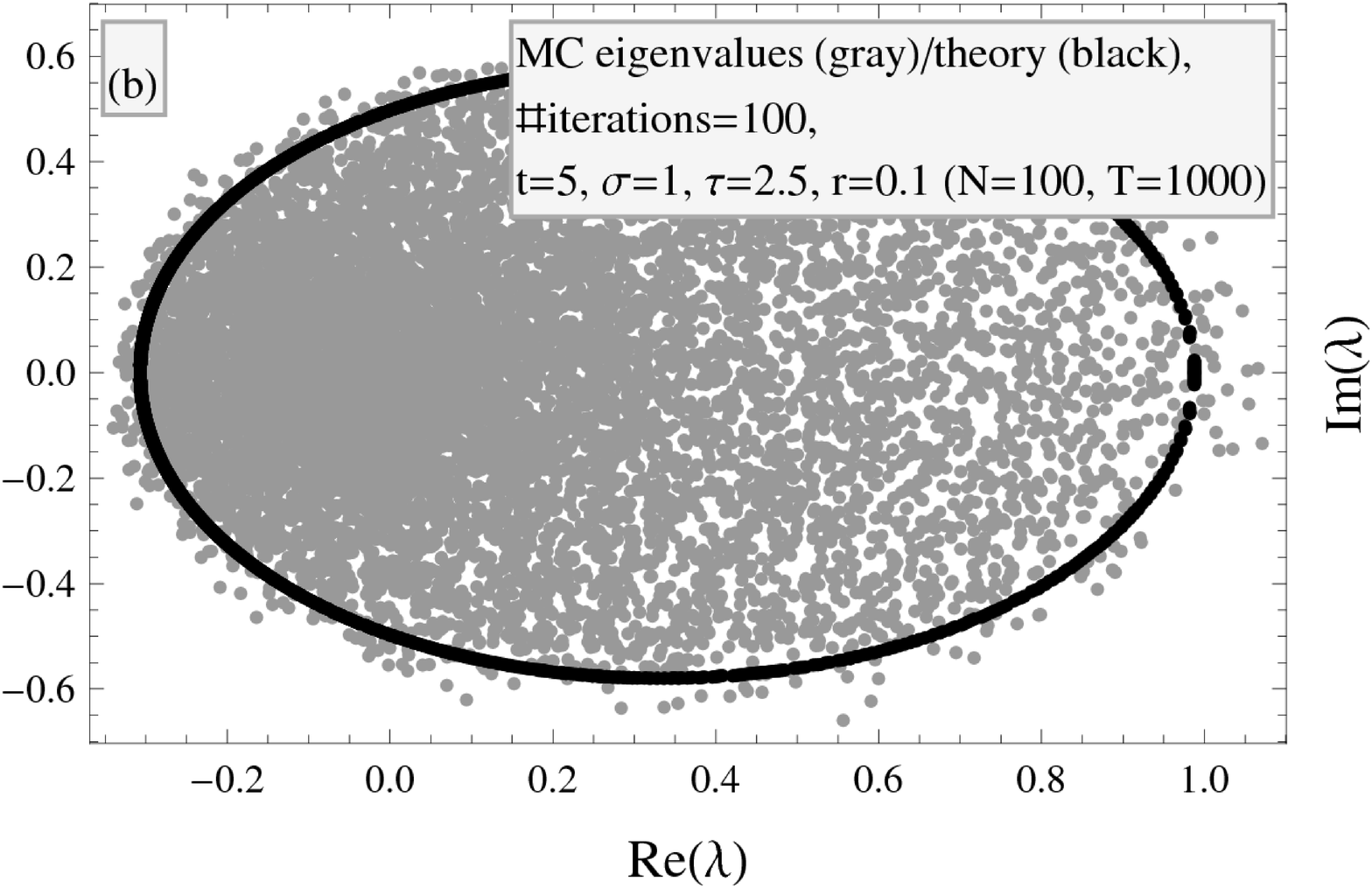}
\includegraphics[width=\columnwidth]{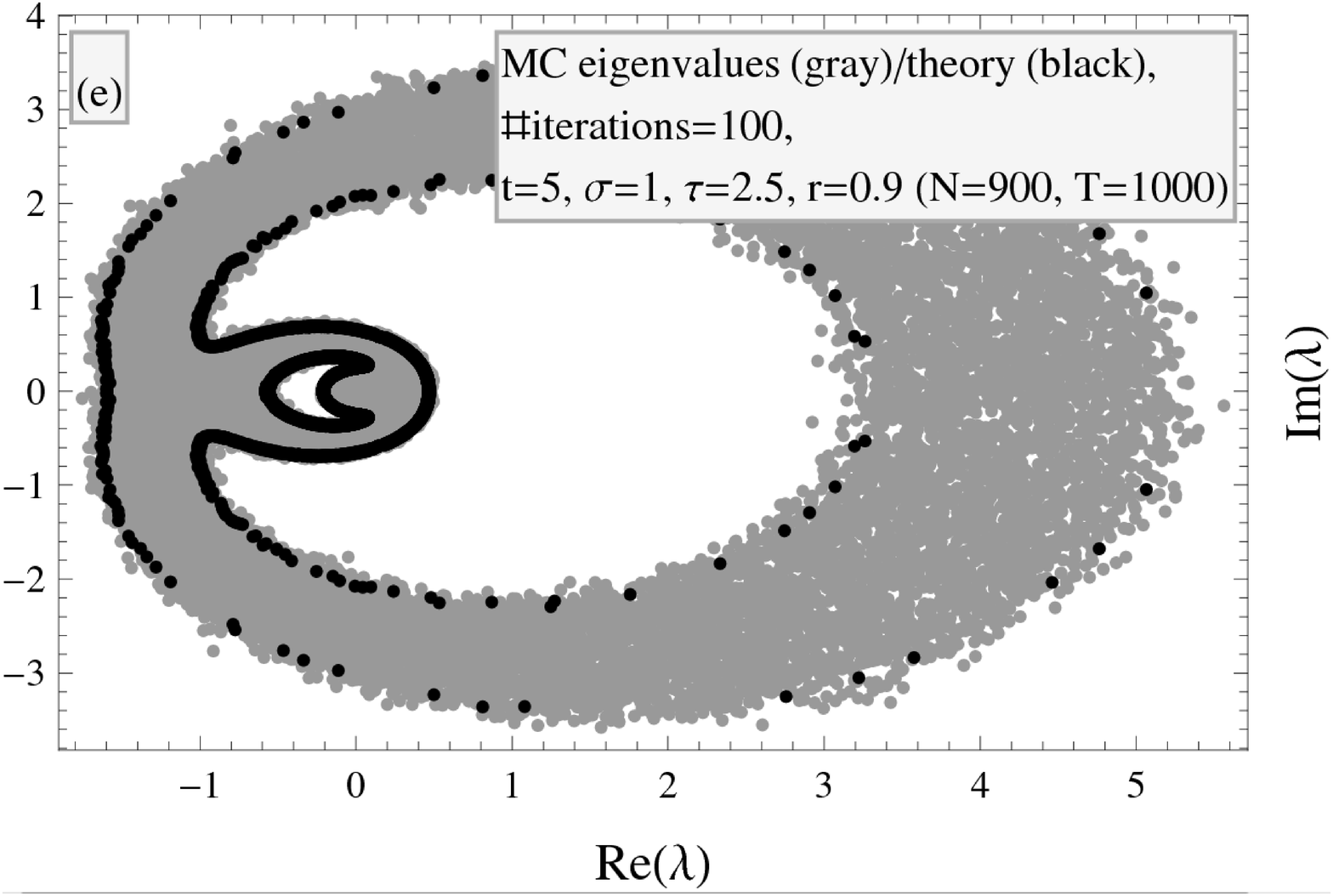}
\includegraphics[width=\columnwidth]{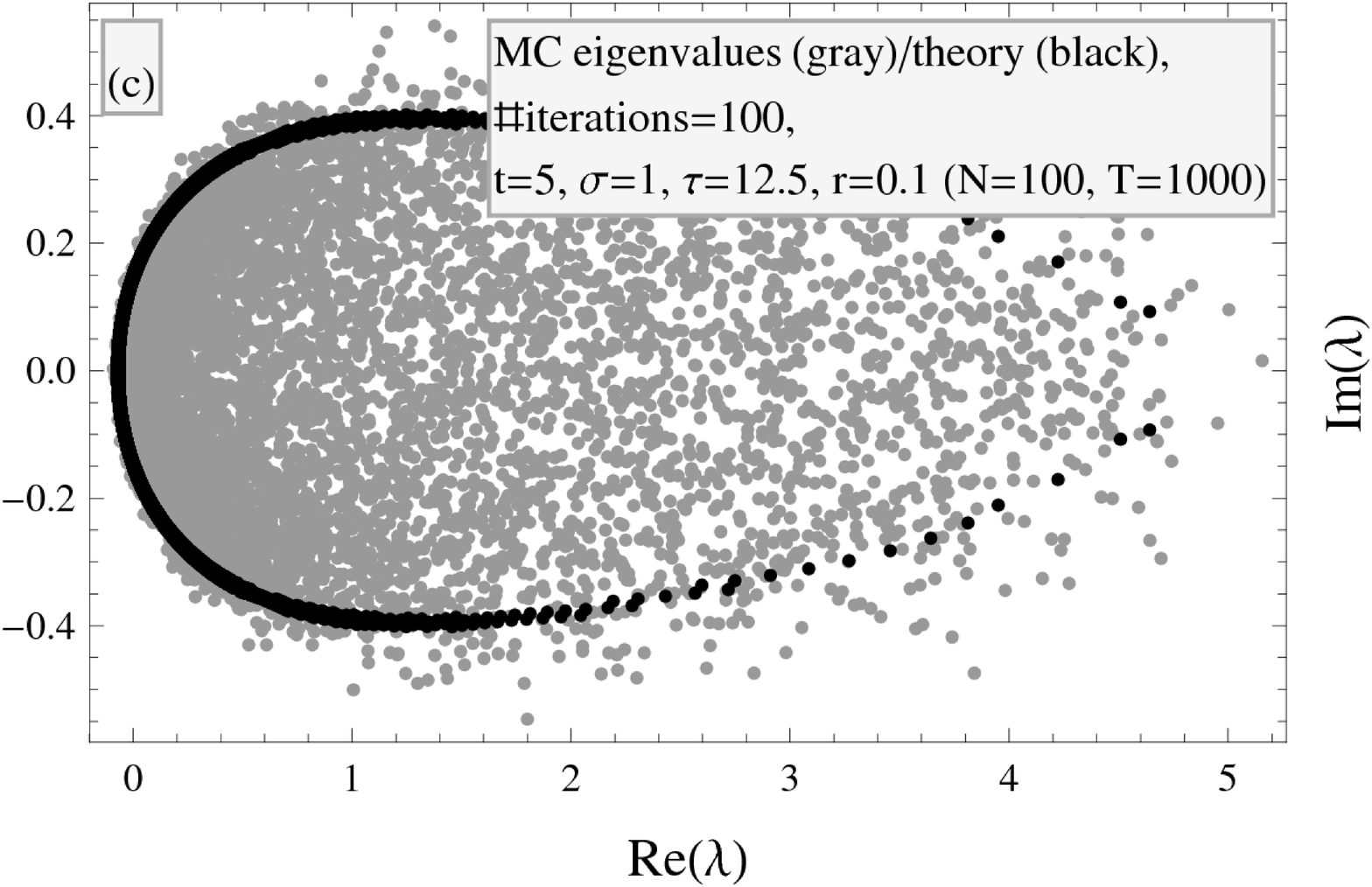}
\includegraphics[width=\columnwidth]{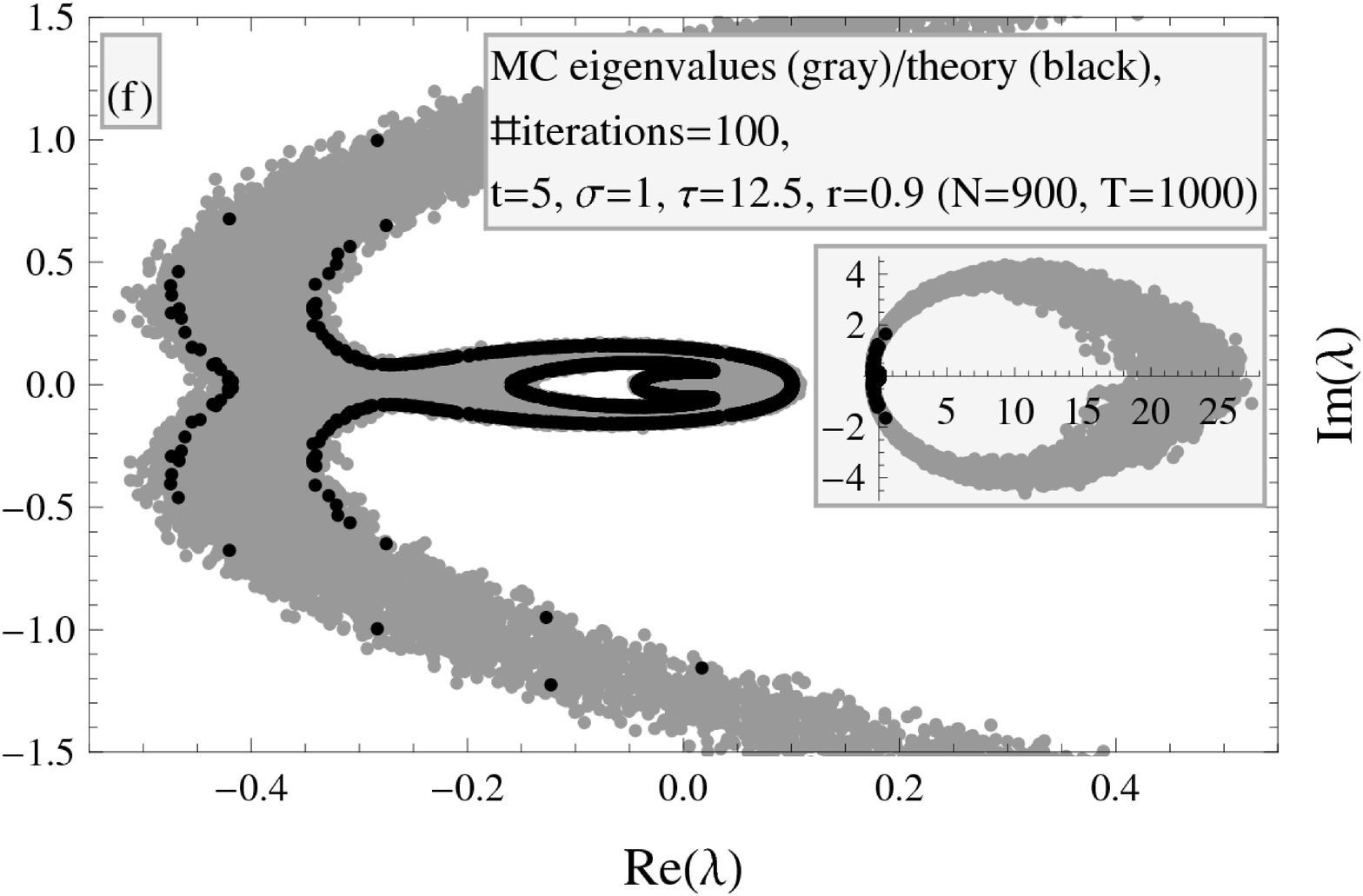}
\caption{Monte Carlo eigenvalues versus theory for the TLCE ($t = 5$) for Toy Model 3. In (f), the inset presents the complete domain, while the main figure magnifies an area around zero.}
\label{fig:TM3TLCEEIGt5}
\end{figure*}

\begin{figure*}[t]
\includegraphics[width=\columnwidth]{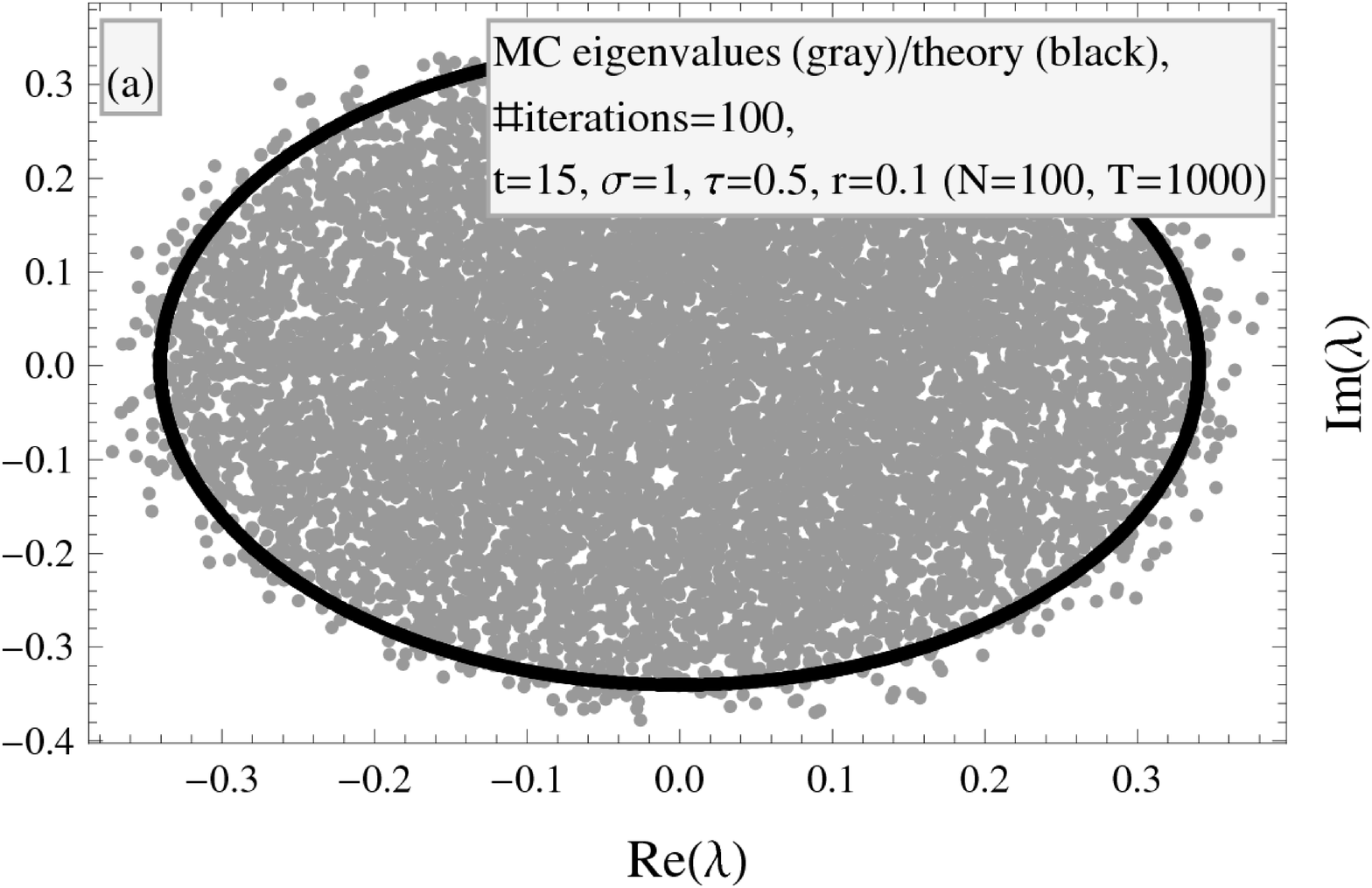}
\includegraphics[width=\columnwidth]{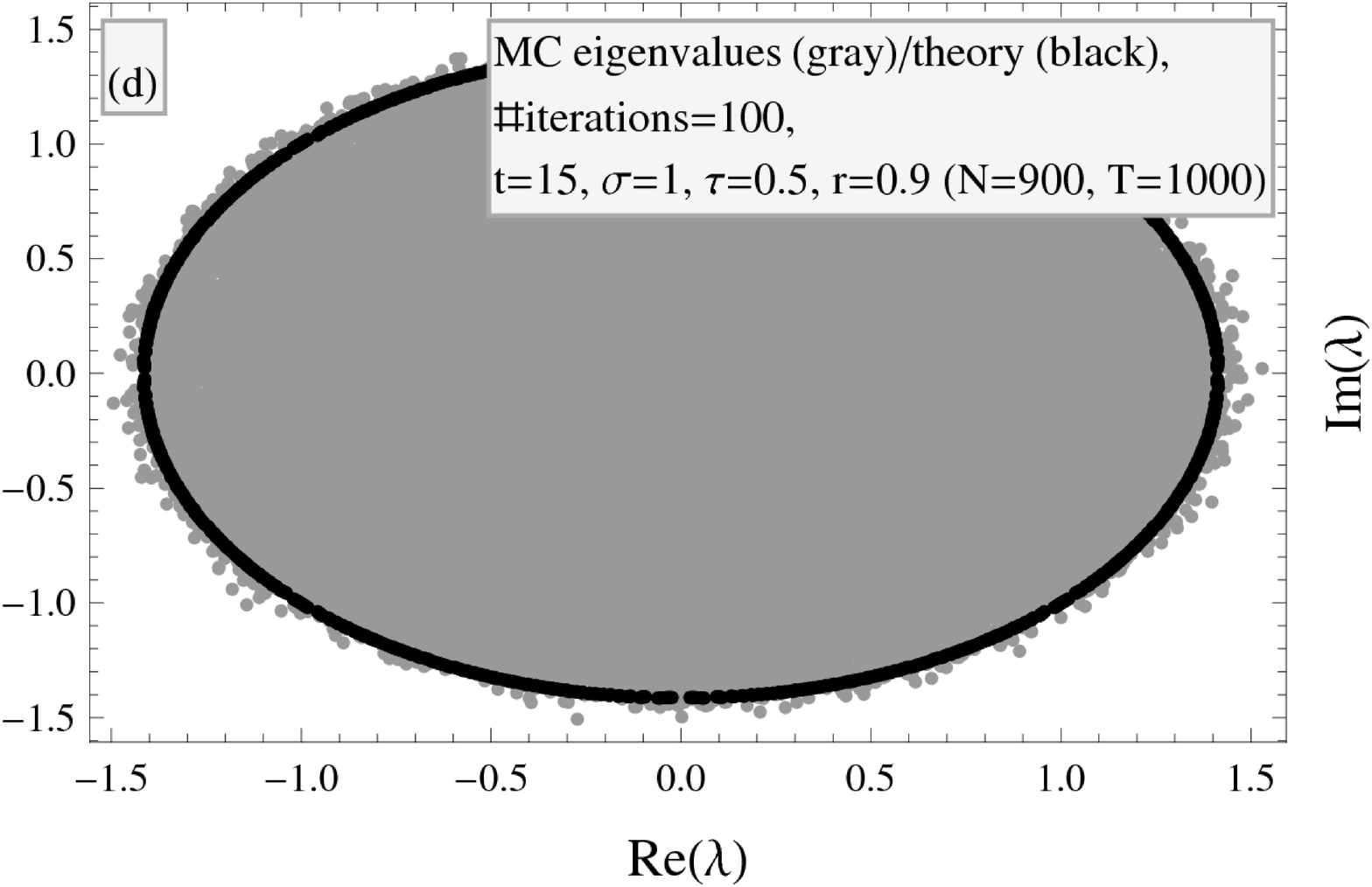}
\includegraphics[width=\columnwidth]{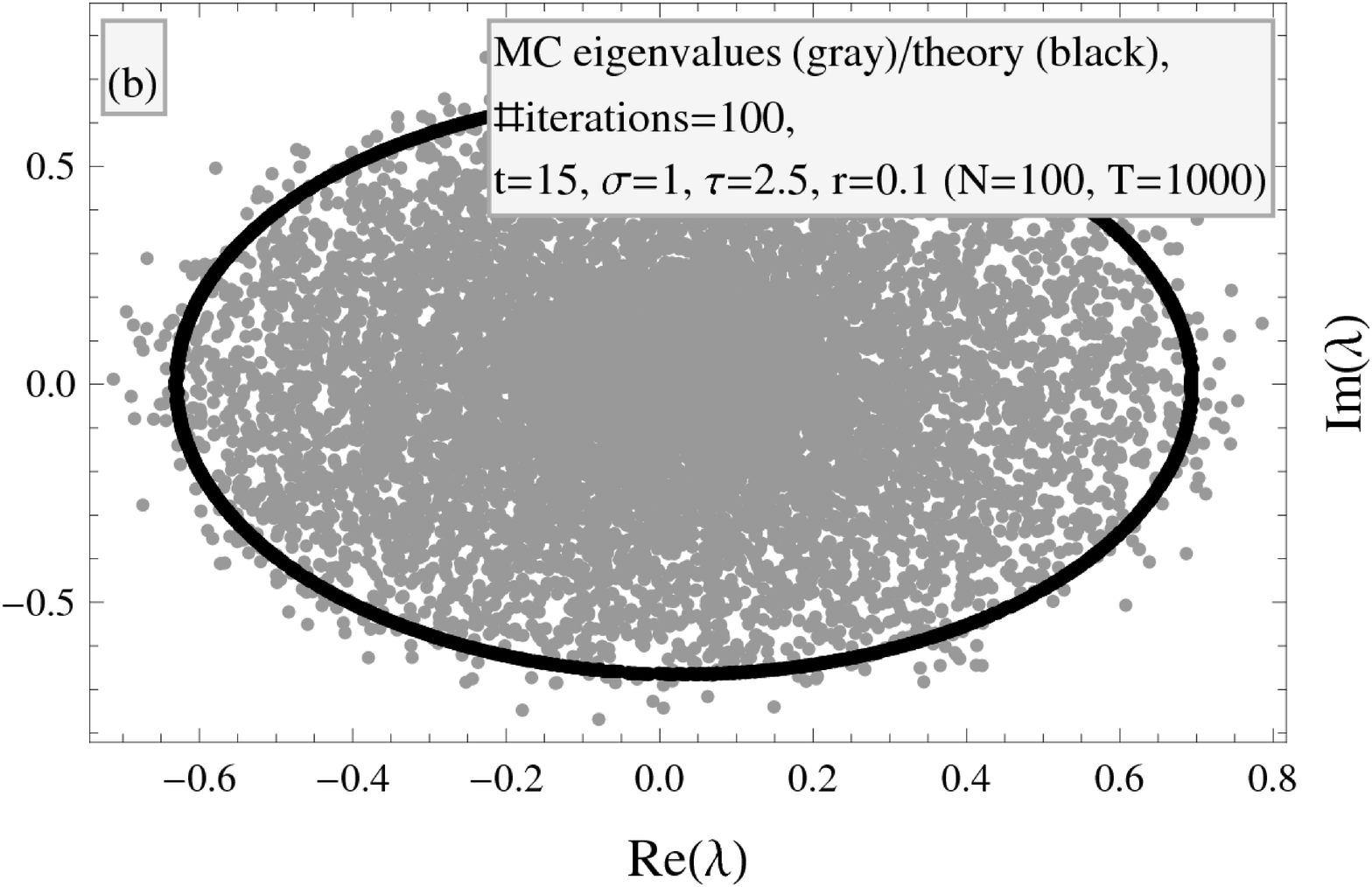}
\includegraphics[width=\columnwidth]{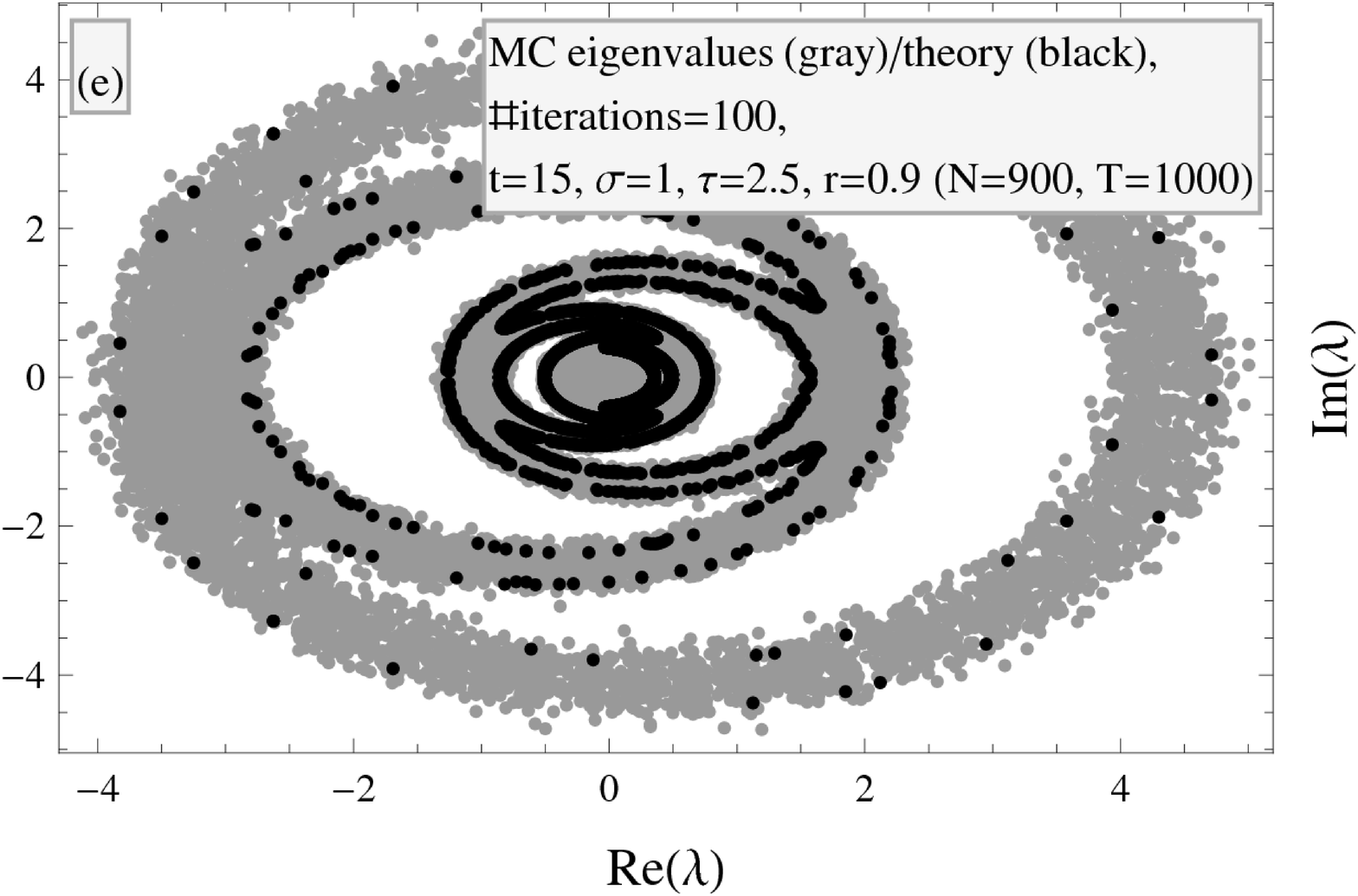}
\includegraphics[width=\columnwidth]{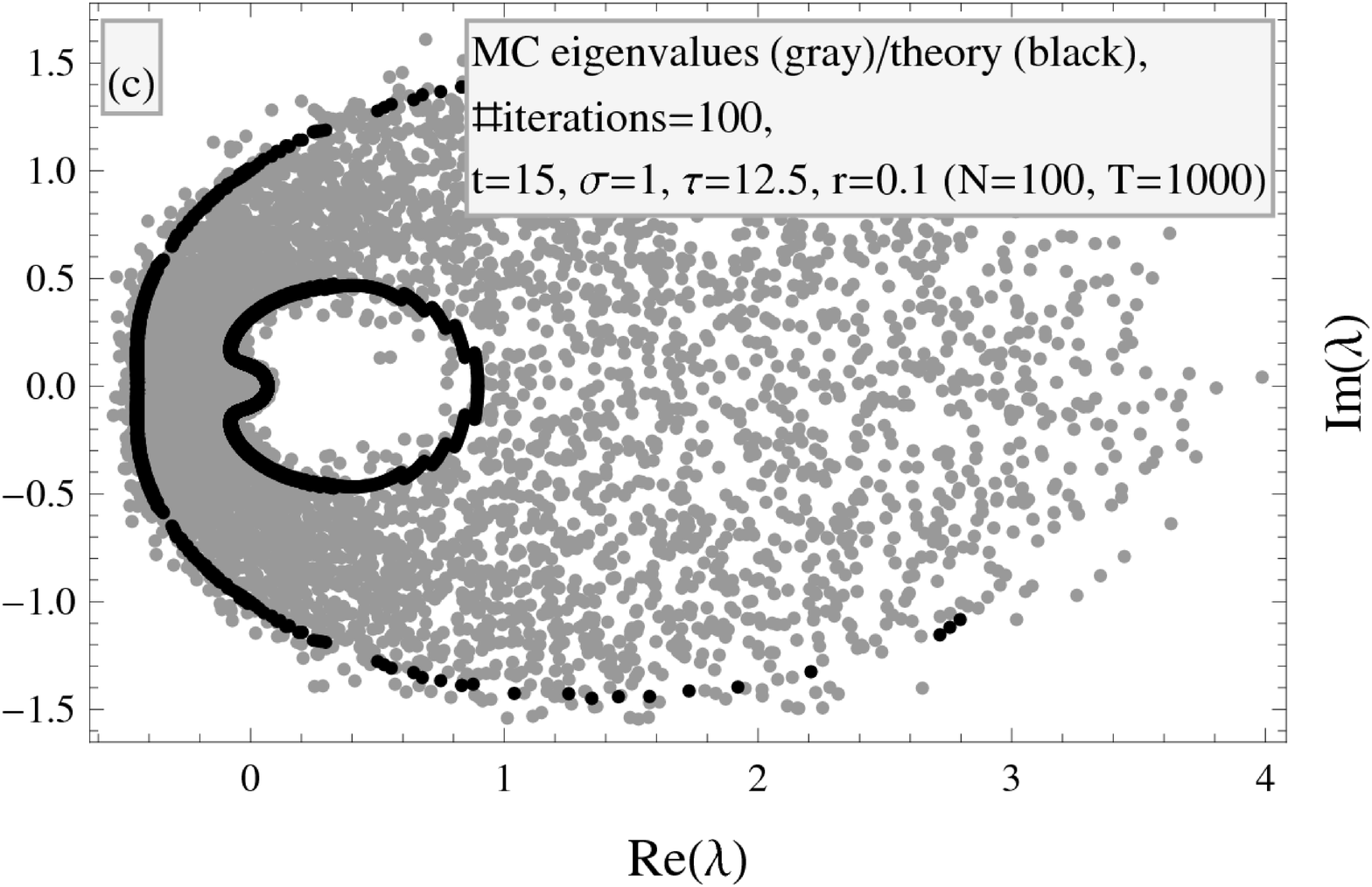}
\includegraphics[width=\columnwidth]{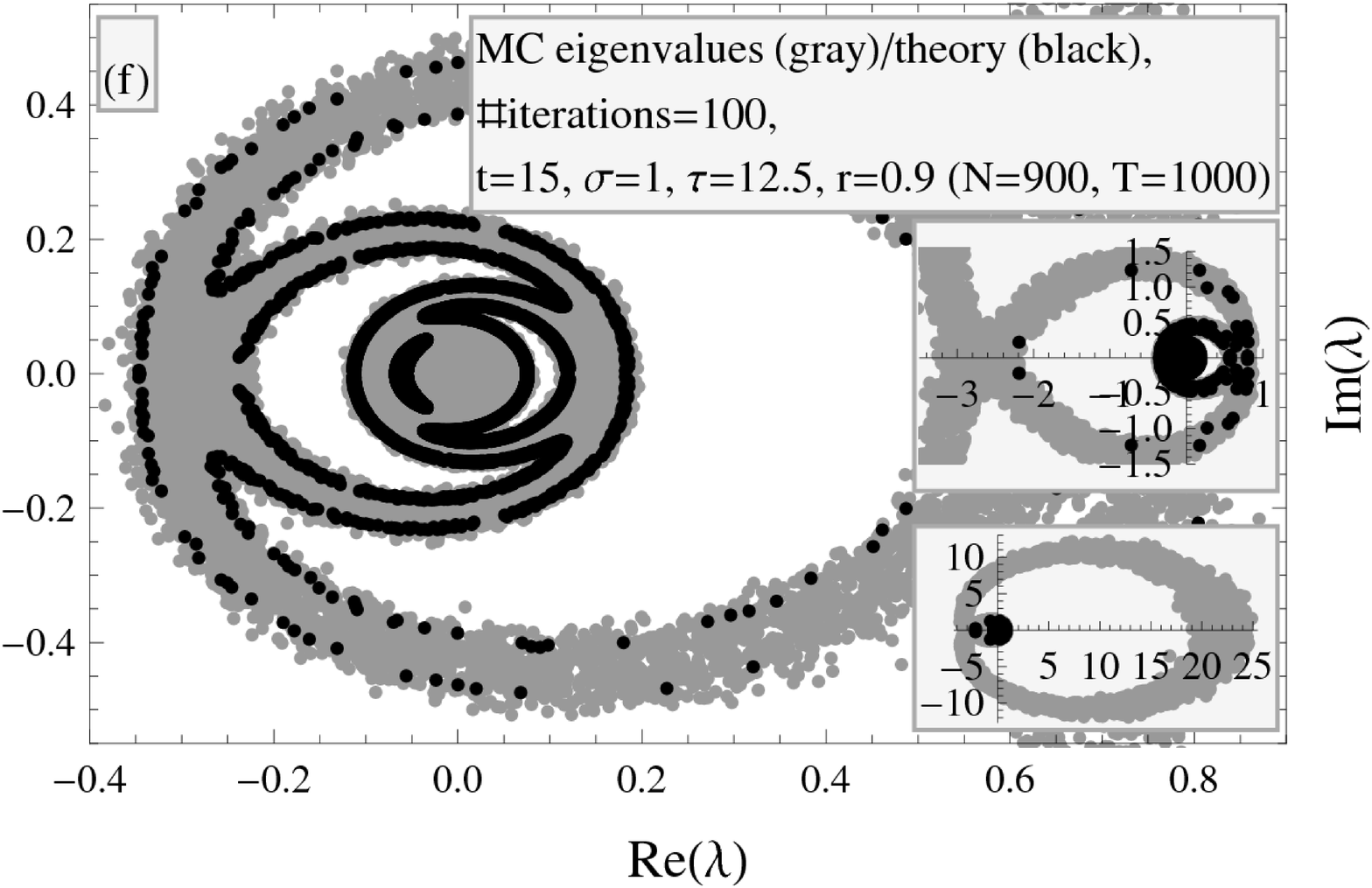}
\caption{Monte Carlo eigenvalues versus theory for the TLCE ($t = 15$) for Toy Model 3. In (f), the lower inset presents the complete domain, while the upper inset and main figure magnify an area around zero.}
\label{fig:TM3TLCEEIGt15}
\end{figure*}

\begin{figure*}[t]
\includegraphics[width=\columnwidth]{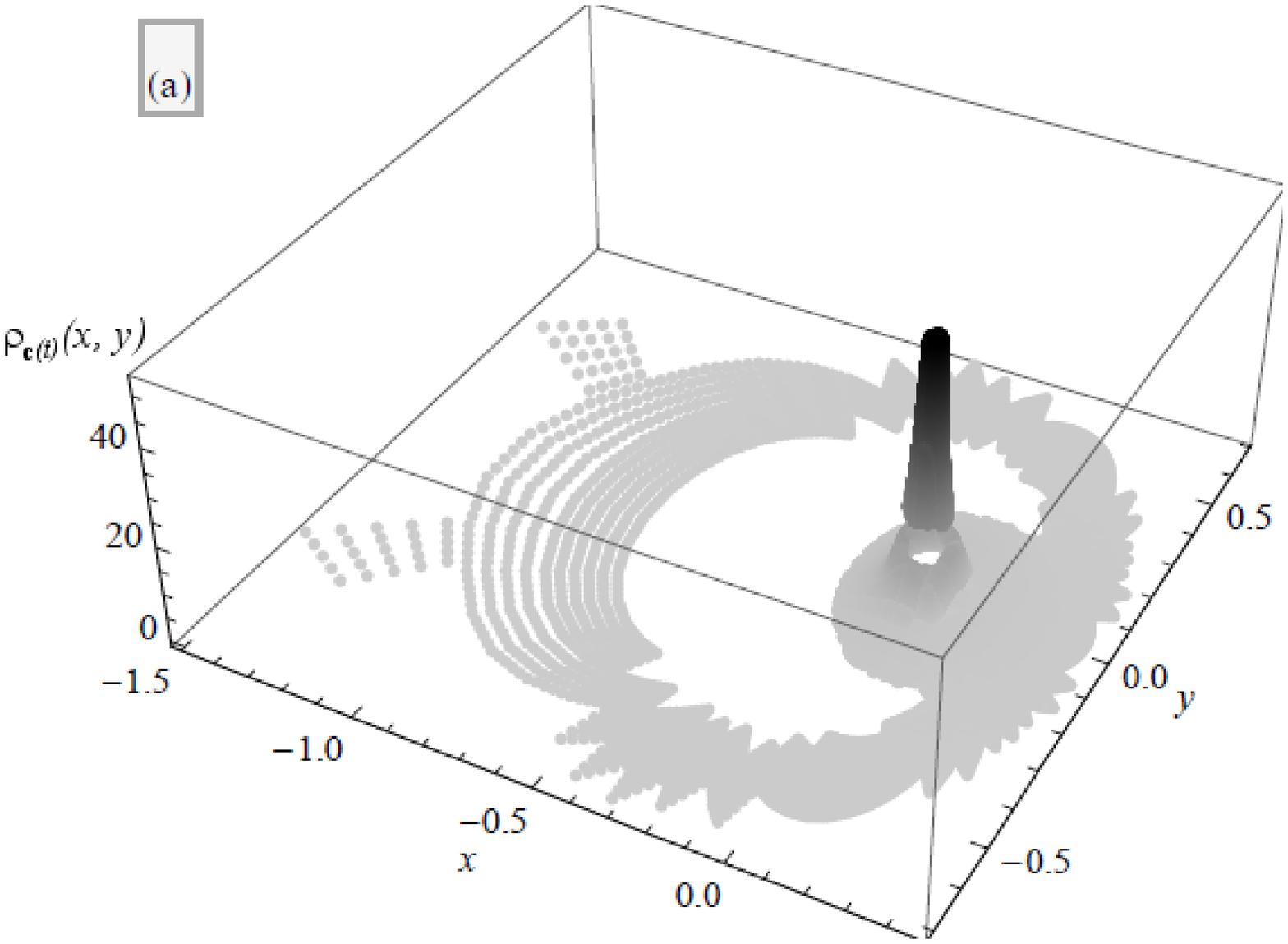}
\includegraphics[width=\columnwidth]{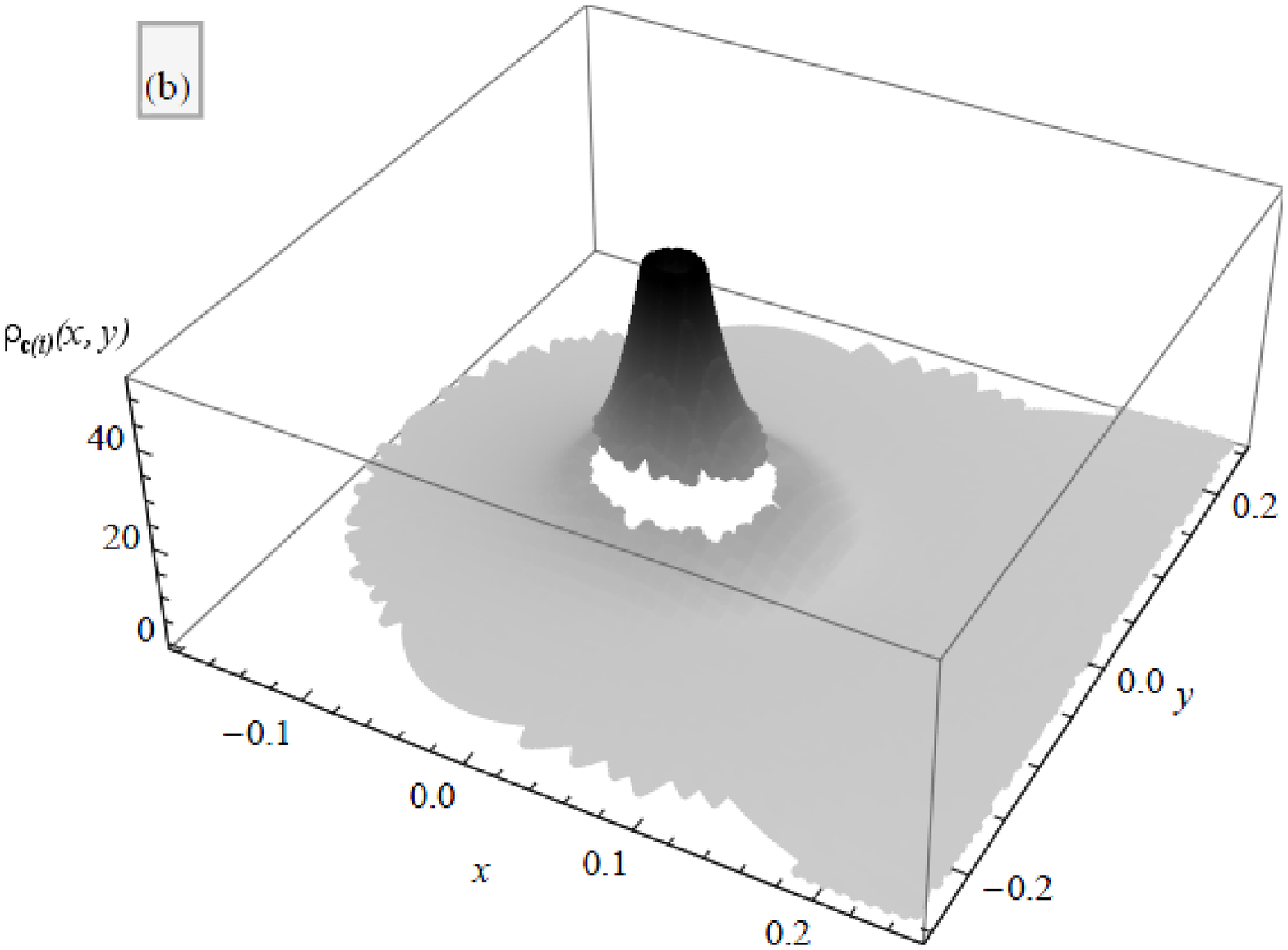}
\caption{The MSD of the TLCE for Toy Model 3 obtained from the master equations using the numerical algorithm described in App.~\ref{a:NumericalEvaluationOfTheMSDOfTheTLCEForToyModel3}. In~(b), the magnification of the area $[ - 0.25 , 0.25 ] \times [ - 0.25 , 0.25 ]$.}
\label{fig:TM3TLCEMSD}
\end{figure*}

%%%%%%%%%%%%%%%%%%%%%%%%%%%%%%%%%%%%%%%%%%%%%%%%%%%%%%%%%%%%%%%%%%%%%%

\subsubsection{Definition}
\label{sss:TM3Definition}

In the last example, there were nontrivial true spatial covariances and no temporal correlation, while now an opposite situation will be considered---trivial true spatial covariances and a simple but fairly realistic model of temporal correlations, namely (cf.~Sec.~\ref{sss:VAR}), the SVAR(1) with $\mathbf{B} = \beta \Id_{N}$, $\beta \in ( 0 , 1 )$, and idiosyncratic variances equal to each other. After a redefinition of parameters [into a dimensionless autocorrelation time, \smash{$\tau \equiv - 1 / \log \beta$}, and variance \smash{$\sigma^{2} \equiv ( \sigma^{\textrm{id.}} )^{2} / ( 1 - \beta^{2} )$}], the covariance function (\ref{eq:SVAR1BDiagonalCovarianceFunction}) is of the form (\ref{eq:Case2Plus3TLCEC1Eq01}) and reads
\begin{equation}\label{eq:TM3DefinitionEq01}
\mathbf{C} = \Id_{N} , \quad A ( b - a ) = \sigma^{2} \ee^{- \frac{| b - a |}{\tau}} .
\end{equation}

It is convenient to use the language of Fourier transforms (cf.~Sec.~\ref{sss:Case3Fourier}), where (\ref{eq:Case3FourierEq03a}),
\begin{equation}\label{eq:TM3DefinitionEq02}
\frac{1}{\hat{A} ( u )} = \frac{1}{A_{1}} \left( A_{2} - u - \frac{1}{u} \right) ,
\end{equation}
where the two parameters of the model have been again changed to \smash{$A_{1} \equiv 2 \sigma^{2} \sinh ( 1 / \tau )$} and \smash{$A_{2} \equiv 2 \cosh ( 1 / \tau )$}, and the argument $p$ to a variable on the centered unit circle [cf.~(\ref{eq:Case2Plus3TLCEA1Eq02}); \smash{$v = u^{t}$}],
\begin{equation}\label{eq:TM3DefinitionEq03}
u \equiv \ee^{- 2 \pi \ii p} , \, \textrm{i.e.,} \, \int_{- 1 / 2}^{1 / 2} \dd p ( \ldots ) = \frac{1}{2 \pi \ii} \ointctrclockwise_{C ( 0 , 1 )} \dd u \frac{1}{u} ( \ldots ) ,
\end{equation}
in which it is most straightforward to apply the integration method of residues.

%%%%%%%%%%%%%%%%%%%%%%%%%%%%%%%%%%%%%%%%%%%%%%%%%%%%%%%%%%%%%%%%%%%%%%

\subsubsection{Master equations}
\label{sss:TM3TLCEMasterEquations}

The discussion from Sec.~\ref{sss:Case2Plus3TLCEC1} is relevant for this model---the master equations are [(\ref{eq:Case2Plus3TLCEC1Eq02a})-(\ref{eq:Case2Plus3TLCEC1Eq03})] (with a complex unknown $G$ and real nonnegative unknown \smash{$\tilde{h}$}), which upon inserting (\ref{eq:TM3DefinitionEq02}) into them and using (\ref{eq:TM3DefinitionEq03}) take on the form
\begin{subequations}
\begin{align}
F_{1} ( G , \tilde{h} ) &= 0 ,\label{eq:TM3TLCEEq01a}\\
F_{2} ( G , \tilde{h} ) &= z ,\label{eq:TM3TLCEEq01b}
\end{align}
\end{subequations}
where for short,
\begin{subequations}
\begin{align}
F_{1} ( G , \tilde{h} ) &\equiv r A_{1}^{2} \frac{1}{2 \pi \ii} \ointctrclockwise_{C ( 0 , 1 )} \dd u \frac{u^{t}}{W ( u )} - \frac{1}{| G |^{2} + \tilde{h}} ,\label{eq:TM3TLCEEq02a}\\
F_{2} ( G , \tilde{h} ) &\equiv A_{1} \frac{1}{2 \pi \ii} \ointctrclockwise_{C ( 0 , 1 )} \dd u \frac{u^{2 t - 1} \left( - u^{2} + A_{2} u - 1 \right)}{W ( u )} ,\label{eq:TM3TLCEEq02b}
\end{align}
\end{subequations}
where $W ( u )$ is a polynomial in $u$ of order $2 ( t + 1 )$ (i.e., $4$, $6$, $8$, etc., for $t = 1$, $2$, $3$, etc.), the leading coefficient \smash{$( r A_{1} G + \delta_{t 1} )$}, and a property that its reciprocal coefficients are mutually complex conjugate,
\begin{equation}
\begin{split}\label{eq:TM3TLCEEq03}
W ( u ) &\equiv\\
&\equiv r A_{1} \overline{G} - r A_{1} A_{2} \overline{G} u + r A_{1} \overline{G} u^{2} +\\
&+ u^{t - 1} - 2 A_{2} u^{t} +\\
&+ \left( 2 + A_{2}^{2} + r^{2} A_{1}^{2} \left( | G |^{2} + \tilde{h} \right) \right) u^{t + 1} -\\
&- 2 A_{2} u^{t + 2} + u^{t + 3} +\\
&+ r A_{1} G u^{2 t} - r A_{1} A_{2} G u^{2 t + 1} + r A_{1} G u^{2 t + 2} .
\end{split}
\end{equation}
This polynomial has too high an order for it to be factorized analytically (thereby calculating the residues and consequently the two integrals), which is the main reason behind the difficulties in solving this set of integral equations---one has to resort to numerical methods.

%%%%%%%%%%%%%%%%%%%%%%%%%%%%%%%%%%%%%%%%%%%%%%%%%%%%%%%%%%%%%%%%%%%%%%

\subsubsection{Borderline of the mean spectral domain}
\label{sss:TM3TLCEBorderline}

The equation of the borderline is given by the above set with \smash{$\tilde{h} = 0$}. Equation~(\ref{eq:TM3TLCEEq01a}) (a real equation for a complex unknown $G \equiv X + \ii Y$) gives thus a curve in the $( X , Y )$-plane, which is then mapped to the $( x , y )$-plane by Eq.~(\ref{eq:TM3TLCEEq01b}). As evident from (\ref{eq:Case2Plus3TLCEC1Eq03}), the denominator (\ref{eq:TM3TLCEEq03}) factorizes into two polynomials of order $( t + 1 )$ each,
\begin{equation}
\begin{split}\label{eq:TM3TLCEEq04}
W ( u ) &=\\
&= \left( 1 - A_{2} u + u^{2} + r A_{1} G u^{t + 1} \right) \cdot\\
&\cdot \left( r A_{1} \overline{G} + u^{t - 1} - A_{2} u^{t} + u^{t + 1} \right) ,
\end{split}
\end{equation}
whose roots (related by \smash{$u \to 1 / \overline{u}$}) should be found in order to evaluate the integrals [(\ref{eq:TM3TLCEEq02a}), (\ref{eq:TM3TLCEEq02b})].

\emph{Borderline for $t = 1$.} For $t = 1$, one is faced with two quadratic equations, which therefore allows for considerable analytical progress: The roots are easily derived, \smash{$u_{1 \pm} \equiv A_{2} / 2 \pm ( A_{2}^{2} / 4 - 1 - r A_{1} \overline{G} )^{1 / 2}$} and \smash{$u_{2 \pm} = 1 / \overline{u_{1 \mp}}$}, and so are the integrals in question. However, one has to carefully determine which two of these four roots lie inside the circle $C ( 0 , 1 )$ and which two outside, for $G$ satisfying (\ref{eq:TM3TLCEEq01a}). A case-by-case analysis (performed in the most relevant situation of $r < 1$) reveals what follows:

(i) The external borderline originates from \smash{$u_{1 -}$} and \smash{$u_{2 -}$} inside $C ( 0 , 1 )$. Its equation in the $( X , Y )$-plane (\ref{eq:TM3TLCEEq01a}) reads
\begin{subequations}
\begin{align}
&U^{4} + \left( 4 - A_{2}^{2} + 4 r A_{1} X \right) U^{2} - 4 r^{2} A_{1}^{2} Y^{2} = 0 ,\label{eq:TM3TLCEEq05a}\\
&\Big( 4 - A_{2}^{2} + 4 r A_{1} X + 2 r^{2} A_{1}^{2} \left( X^{2} + Y^{2} \right) \Big) U^{2} +\nonumber\\
&+ r A_{1}^{2} A_{2} \left( X^{2} + Y^{2} \right) U +\nonumber\\
&+ r^{2} A_{1}^{2} \Big( \left( 4 - A_{2}^{2} + 4 r A_{1} X \right) X^{2} +\nonumber\\
&+ \left( - 4 - A_{2}^{2} + 4 r A_{1} X \right) Y^{2} \Big) = 0 ,\label{eq:TM3TLCEEq05b}
\end{align}
\end{subequations}
where $U$ is an auxiliary real unknown (which can easily be solved out, which would however leave a very lengthy expression), and where the correct solution is such that the roots \smash{$u_{1 -}$}, \smash{$u_{2 -}$} indeed lie inside $C ( 0 , 1 )$.

The map to the $( x , y )$-plane (\ref{eq:TM3TLCEEq01b}) is
\begin{subequations}
\begin{align}
x &= \frac{1}{4 \left( X^{2} + Y^{2} \right) \left( \left( 1 + r A_{1} X \right)^{2} + r^{2} A_{1}^{2} Y^{2} \right)} \cdot\nonumber\\
&\cdot \bigg( \Big( 2 / ( r A_{1} ) + 2 X - r A_{1} \left( X^{2} + Y^{2} \right) \Big) U^{2} +\nonumber\\
&+ X \Big( 4 + A_{1} \left( - 2 ( 1 - r ) + r A_{2}^{2} \right) X -\nonumber\\
&- 2 r ( 1 + r ) A_{1}^{2} X^{2} \Big) +\nonumber\\
&+ A_{1} \Big( - 2 ( 1 + r ) + r A_{2}^{2} - 2 r ( 1 + r ) A_{1} X \Big) Y^{2} \bigg) ,\label{eq:TM3TLCEEq06a}\\
y &= \frac{1}{4 Y \left( X^{2} + Y^{2} \right) \left( \left( 1 + r A_{1} X \right)^{2} + r^{2} A_{1}^{2} Y^{2} \right)} \cdot\nonumber\\
&\cdot \bigg( - \Big( X^{2} \left( 1 + r A_{1} X \right) + \left( 3 + r A_{1} X \right) Y^{2} \Big) U^{2} +\nonumber\\
&+ X^{2} \left( 1 + r A_{1} X \right) \left( - 4 + A_{2}^{2} - 4 r A_{1} X \right) +\nonumber\\
&+ \Big( - 4 + A_{2}^{2} + r A_{1} \left( - 12 + A_{2}^{2} \right) X +\nonumber\\
&+ 2 r ( 1 - 3 r ) A_{1}^{2} X^{2} \Big) Y^{2} +\nonumber\\
&+ 2 r ( 1 - r ) A_{1}^{2} Y^{4} \bigg) .\label{eq:TM3TLCEEq06b}
\end{align}
\end{subequations}

On the one hand, an algebraic form of these equations (in contrast to an integral form for $t > 1$) allows for quite easy plotting, as well as some analytical insight into the properties of the external borderline; on the other hand, any investigation of [(\ref{eq:TM3TLCEEq05a})-(\ref{eq:TM3TLCEEq06b})] remains complicated. For instance, one may infer that the curve is symmetric with respect to the abscissa axis, in both coordinate systems ($Y \to - Y$, which corresponds to $y \to - y$), and that it crosses this axis ($Y = 0$, which corresponds to $y = 0$) in two points (this number of crossings is a numerical observation and requires a proof); the crossings with the $X$-axis are given by a quintic polynomial equation,
\begin{equation}
\begin{split}\label{eq:TM3TLCEEq07}
&4 r^{5} A_{1}^{5} X_{\star}^{5} +\\
&+ r^{2} A_{1}^{4} \left( A_{2}^{2} + 36 r^{2} - A_{2}^{2} r^{2} \right) X_{\star}^{4} +\\
&+ 16 r^{3} A_{1}^{3} \left( 8 - A_{2}^{2} \right) X_{\star}^{3} +\\
&+ 2 r^{2} A_{1}^{2} \left( 4 - A_{2}^{2} \right) \left( 28 - A_{2}^{2} \right) X_{\star}^{2} +\\
&+ 12 r A_{1} \left( 4 - A_{2}^{2} \right)^{2} X_{\star} +\\
&+ \left( 4 - A_{2}^{2} \right)^{3} = 0
\end{split}
\end{equation}
[one should select its real roots with the property that \smash{$u_{1 -}$}, \smash{$u_{2 -}$} lie inside $C ( 0 , 1 )$], which yield the crossings with the $x$-axis to be
\begin{equation}\label{eq:TM3TLCEEq08}
x_{\star} = - \frac{4 - A_{2}^{2} + 2 r A_{1} X_{\star} + r ( 1 - r ) A_{1}^{2} X_{\star}^{2}}{2 r A_{1} \left( 1 +  r A_{1} X_{\star} \right) X_{\star}^{2}}
\end{equation}
[\smash{$x_{\star}$} may be checked to also obey a quintic polynomial equation, albeit lengthier than (\ref{eq:TM3TLCEEq07}), and thus not printed]. In Fig.~\ref{fig:TM3TLCEEIGEdges}, one sees plots of \smash{$x_{\star , \min}$} and \smash{$x_{\star , \max}$} versus $\tau$ for $r = 0.1 , 0.9$, combined with the plots of the edges of the mean spectrum of the ETCE (cf.~App.~\ref{aa:TM3ETCE}), solving the sextic equation (\ref{eq:TM3ETCEEq03}). [One can verify that---as may be suspected from Fig.~\ref{fig:TM3TLCEEIGEdges}---in the limit $\delta t \to 0$ (in which $\tau \sim 1 / \delta t \to \infty$ and $r \sim \delta t \to 0$), the edges for both the ETCE and TLCE coincide; at the leading order, they both solve a certain quartic polynomial equation.] These graphs may be used as a genuine tool to determine whether a given set of data has intrinsic temporal covariances, and of what $\tau$.

(ii) The internal borderline arises when \smash{$u_{2 \pm}$} are inside $C ( 0 , 1 )$. It may be verified to exist only for
\begin{equation}\label{eq:TM3TLCEEq09}
r_{\textrm{c}} \equiv \frac{1}{1 + \ee^{- 2 / \tau}} < r < 1
\end{equation}
(as mentioned, $r \geq 1$ has not been analyzed), which means that there occurs a ``topological phase transition'' (forming of a hole inside the domain) as one increases $r$ past the critical value \smash{$r_{\textrm{c}}$}. This \smash{$r_{\textrm{c}} ( \tau )$} is plotted in Fig.~\ref{fig:TM3TLCEEIGrc}, and remark that it decreases from $1$ to $1 / 2$ when $\tau$ grows from zero to infinity---this graph may also be a help in searches for true temporal covariances in multivariate time series, yet only for $r > 1 / 2$.

If (\ref{eq:TM3TLCEEq09}) holds, the equation of the internal curve in the $( X , Y )$-plane reads
\begin{equation}
\begin{split}\label{eq:TM3TLCEEq10}
&X^{2} \Big( - 4 + A_{2}^{2} + 2 ( 1 - 2 r ) A_{1} X + r ( 1 - r ) A_{1}^{2} X^{2} \Big) +\\
&+ \Big( A_{2}^{2} + 2 ( 1 - 2 r ) A_{1} X + 2 r ( 1 - r ) A_{1}^{2} X^{2} \Big) Y^{2} +\\
&+ r ( 1 - r ) A_{1}^{2} Y^{4} = 0 ,
\end{split}
\end{equation}
for \smash{$X_{-} \leq X \leq X_{+}$}, with
\begin{equation}\label{eq:TM3TLCEEq11}
X_{\pm} \equiv \frac{- 1 + 2 r \pm \sqrt{1 - r ( 1 - r ) A_{2}^{2}}}{r ( 1 - r ) A_{1}} .
\end{equation}
[The defining conditions of \smash{$u_{2 \pm}$} inside $C ( 0 , 1 )$ are then automatically satisfied.]

It is mapped to the $( x , y )$-plane by
\begin{subequations}
\begin{align}
x &= \frac{1}{\left( X^{2} + Y^{2} \right) \left( \left( 1 + r A_{1} X \right)^{2} + r^{2} A_{1}^{2} Y^{2} \right)} \cdot\nonumber\\
&\cdot \frac{1}{2 X + r A_{1} \left( X^{2} + Y^{2} \right)} \cdot\nonumber\\
&\cdot \bigg( - X^{2} \left( 1 + r A_{1} X \right) \left( 2 - A_{2}^{2} + r A_{1} X \right) +\nonumber\\
&+ \Big( A_{2}^{2} + r A_{1} \left( - 3 + A_{2}^{2} \right) X - 2 r^{2} A_{1}^{2} X^{2} \Big) Y^{2} -\nonumber\\
&- r^{2} A_{1}^{2} Y^{4} \bigg) ,\label{eq:TM3TLCEEq12a}\\
y &= \frac{- Y}{\left( X^{2} + Y^{2} \right) \left( \left( 1 + r A_{1} X \right)^{2} + r^{2} A_{1}^{2} Y^{2} \right)} \cdot\nonumber\\
&\cdot \frac{1}{2 X + r A_{1} \left( X^{2} + Y^{2} \right)} \cdot\nonumber\\
&\cdot \bigg( X \left( 2 + r A_{1} \left( 1 + A_{2}^{2} \right) X \right) + r A_{1} \left( 1 + A_{2}^{2} \right) Y^{2} \bigg) .\label{eq:TM3TLCEEq12b}
\end{align}
\end{subequations}

Figure~\ref{fig:TM3TLCEEIGt1} presents the numerically generated eigenvalues of the TLCE and the numerically solved equations of the borderline(s), [(\ref{eq:TM3TLCEEq05a})-(\ref{eq:TM3TLCEEq06b})] and (for \smash{$r > r_{\textrm{c}}$}) [(\ref{eq:TM3TLCEEq10})-(\ref{eq:TM3TLCEEq12b})], for the parameters $t = 1$, and $r = 0.1$ (left column) or $r = 0.9$ (right column), as well as $\tau = 0.5$ (top row), $2.5$ (middle row) or $12.5$ (bottom row), with very good agreement between the two, modulo the finite-size leaking out of the eigenvalues. The mean spectral domain is ``bullet-shaped''; with growing $\tau$, it gets thinner in the $y$-direction (the more the smaller $r$) and longer in the positive part of the $x$-direction (the more the larger $r$). Moreover, Figs.~[(e), (f)] show the phase with a hole inside the domain.

\emph{Borderline for $t > 1$.} In this case, one is left only with numerical methods to evaluate the borderline of the mean spectral domain, as the factors in (\ref{eq:TM3TLCEEq04}) have too high an order for an effective analytical treatment. The domain may have more complicated and very interesting shapes, possibly with several holes, as indicated by Figs.~\ref{fig:TM3TLCEEIGt5} and~\ref{fig:TM3TLCEEIGt15}, which correspond respectively to $t = 5$ or $t = 15$, with other parameters as in Fig.~\ref{fig:TM3TLCEEIGt1}.

All these graphs reveal that the spectrum of the TLCE is a sensitive tool for unveiling true temporal correlations present in the system, unlike the ETCE (cf.~App.~\ref{aa:TM3ETCE} and Fig.~\ref{fig:TM3ETCE}), whose spectrum is ``MP-shaped'' and does not have enough conspicuous features to be an effective probe of autocorrelations.

The numerical technique used to find these curves is very simple and could certainly be improved in order to obtain their more accurate approximations: (i) By trial and error, the extension of the borderline in the $( X , Y )$-plane has been estimated. (ii) The resulting rectangle in the $( X , Y )$-plane has been divided into a grid of dimensions $1000 \times 1000$, and the value of \smash{$F_{1} ( G , \tilde{h} = 0 )$} (\ref{eq:TM3TLCEEq02a}) at each point has been calculated by the residues. (iii) Only the values close to zero (with some assumed precision) have been selected (\ref{eq:TM3TLCEEq01a}). (iv) These small values have then been mapped to the $( x , y )$-values through [(\ref{eq:TM3TLCEEq01b}), (\ref{eq:TM3TLCEEq02b})], yielding an approximation of the desired borderline.

Remark (by inspecting Figs.~\ref{fig:TM3TLCEEIGt5} and~\ref{fig:TM3TLCEEIGt15}) that the central region of the borderline [which may be checked to originate from its outer part in the $( X , Y )$-plane]---which is also a region of the largest MSD (cf.~Fig.~\ref{fig:TM3TLCEMSD})---is numerically reproduced much better than the peripheral region of the borderline [corresponding to its central part in the $( X , Y )$-plane]. Hence, one may obtain a better approximation to the borderline by using an $( X , Y )$-grid denser in the central part.

%%%%%%%%%%%%%%%%%%%%%%%%%%%%%%%%%%%%%%%%%%%%%%%%%%%%%%%%%%%%%%%%%%%%%%

\subsubsection{Mean spectral density}
\label{sss:TM3TLCEMSD}

Calculating the MSD is even more involved than the borderline as one needs to solve the master equations [(\ref{eq:TM3TLCEEq01a})-(\ref{eq:TM3TLCEEq01b})] for both unknowns, $G$ and \smash{$\tilde{h}$}---the numerical algorithm is outlined in App.~\ref{a:NumericalEvaluationOfTheMSDOfTheTLCEForToyModel3}, and Fig.~\ref{fig:TM3TLCEMSD} visualizes the resulting MSD in 3D.

%%%%%%%%%%%%%%%%%%%%%%%%%%%%%%%%%%%%%%%%%%%%%%%%%%%%%%%%%%%%%%%%%%%%%%
%%%%%%%%%%%%%%%%%%%%%%%%%%%%%%%%%%%%%%%%%%%%%%%%%%%%%%%%%%%%%%%%%%%%%%

\subsection{Toward more realistic toy models}
\label{ss:TM4}

\begin{figure*}[t]
\includegraphics[width=\columnwidth]{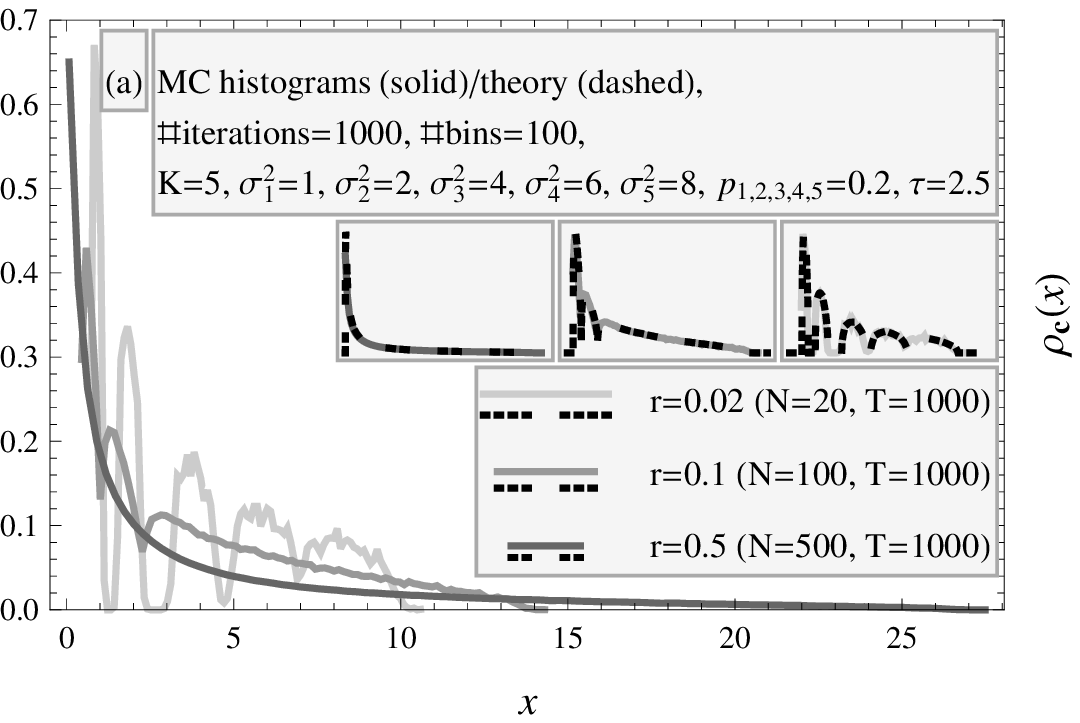}
\includegraphics[width=\columnwidth]{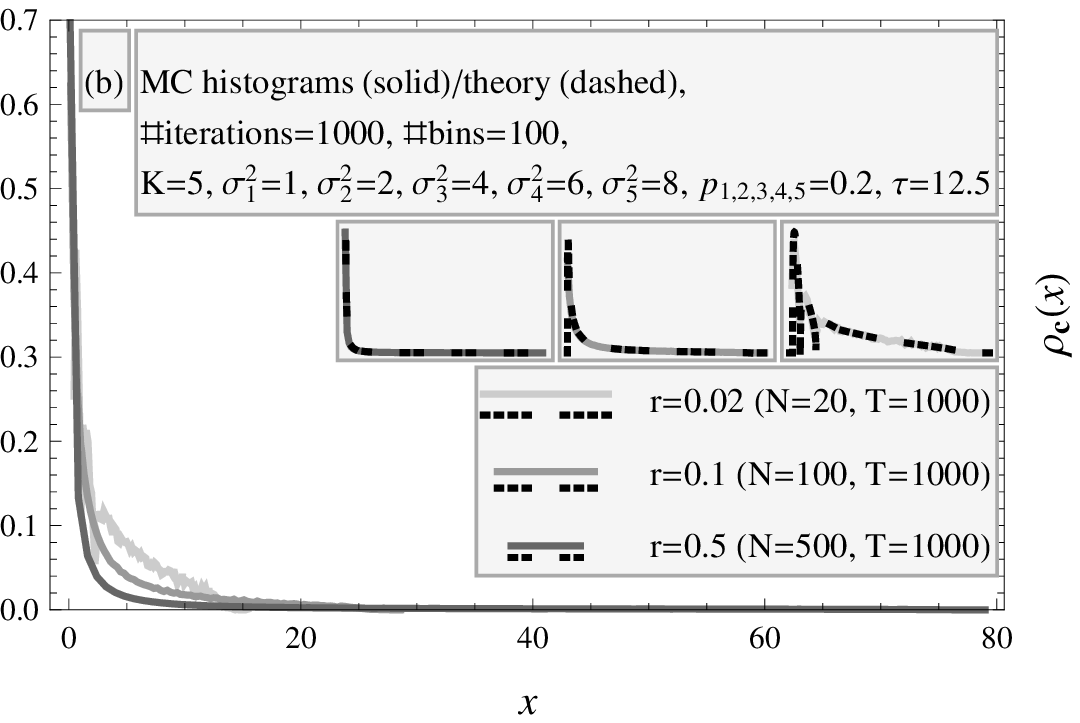}
\caption{Monte Carlo eigenvalues versus theory for the ETCE for Toy Model 4a. The graphs in the main figures are hardly distinguishable, hence the insets show them separately.}
\label{fig:TM4aETCE}
\end{figure*}

\begin{figure*}[t]
\includegraphics[width=\columnwidth]{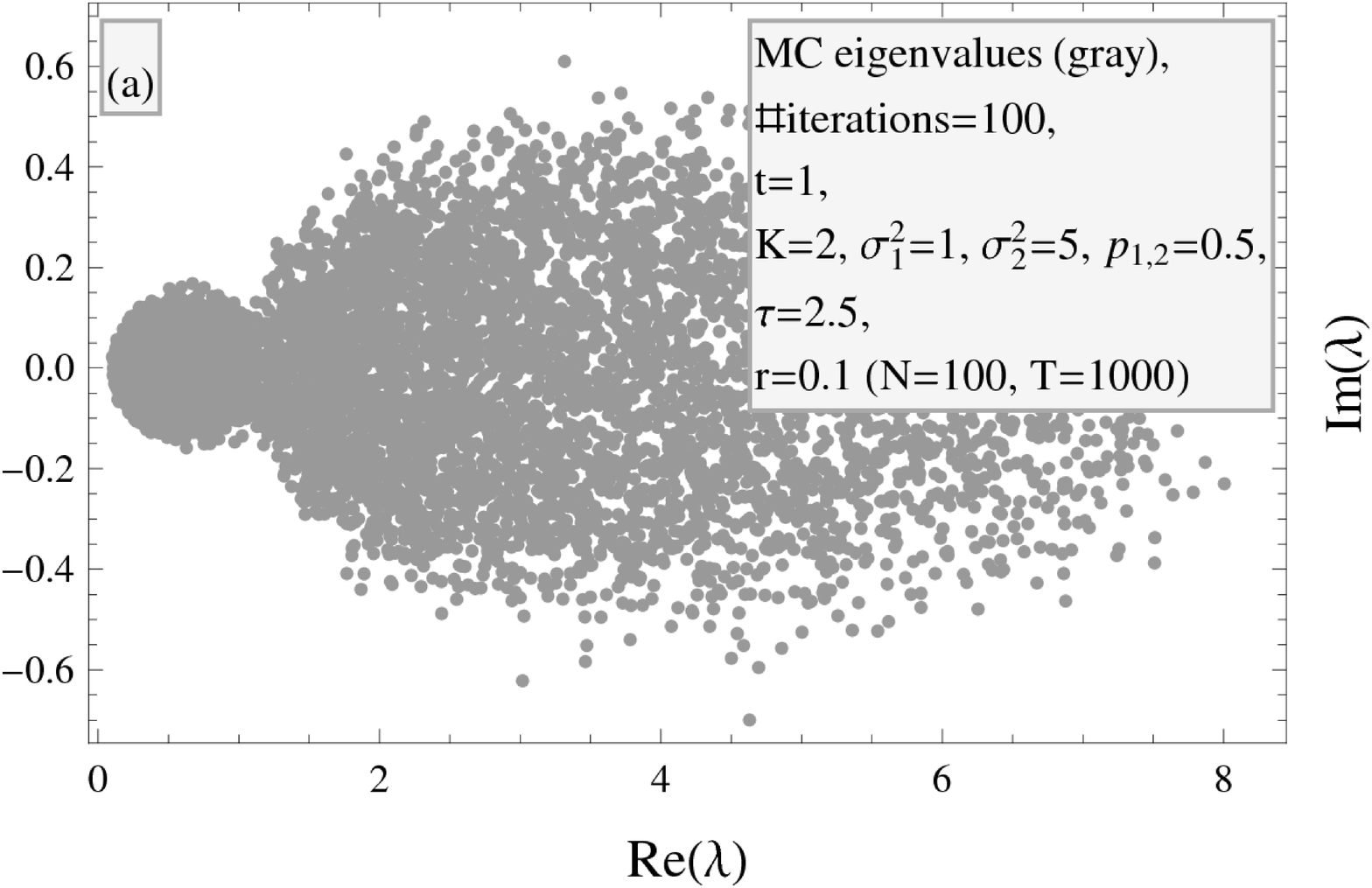}
\includegraphics[width=\columnwidth]{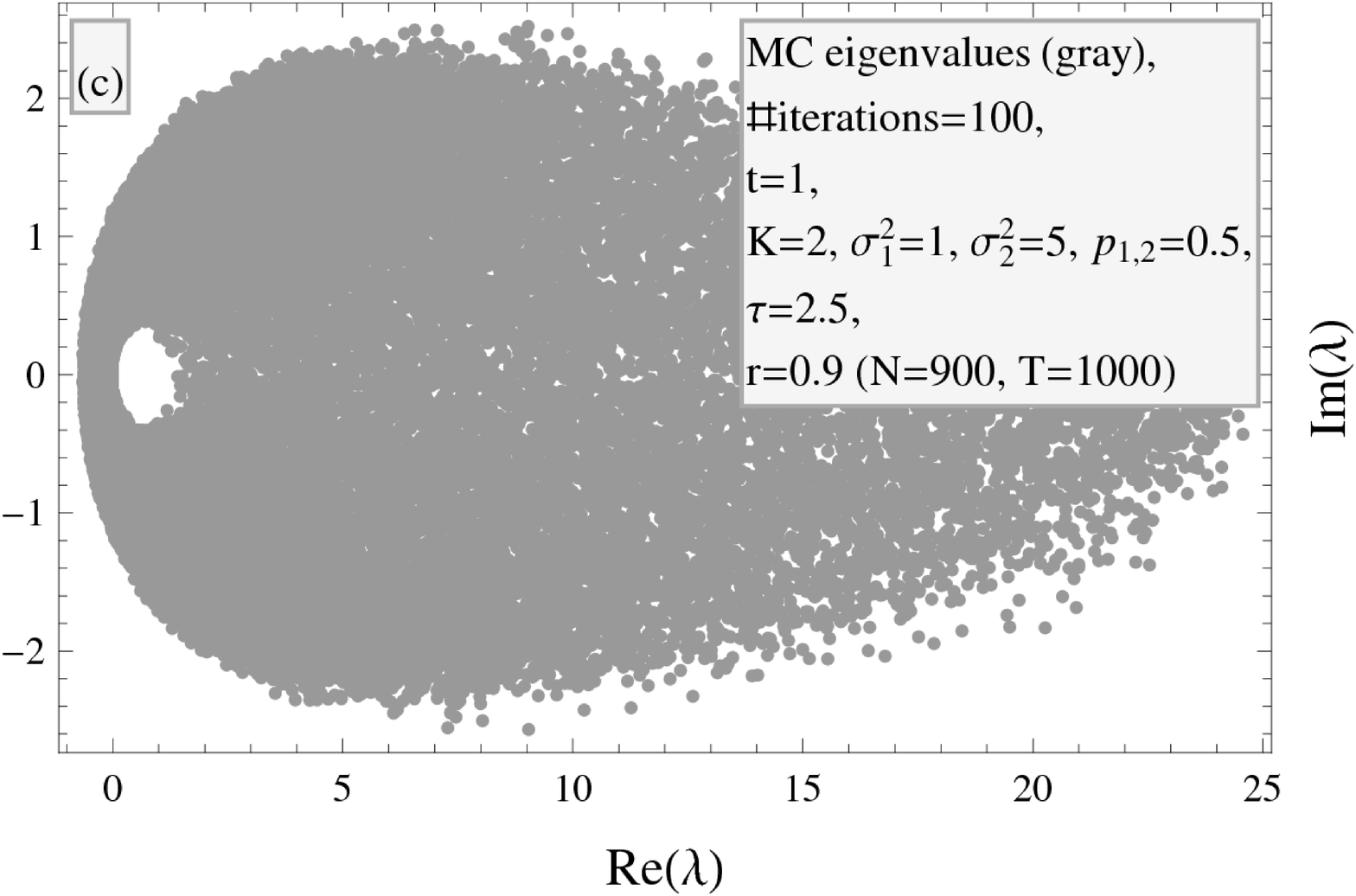}
\includegraphics[width=\columnwidth]{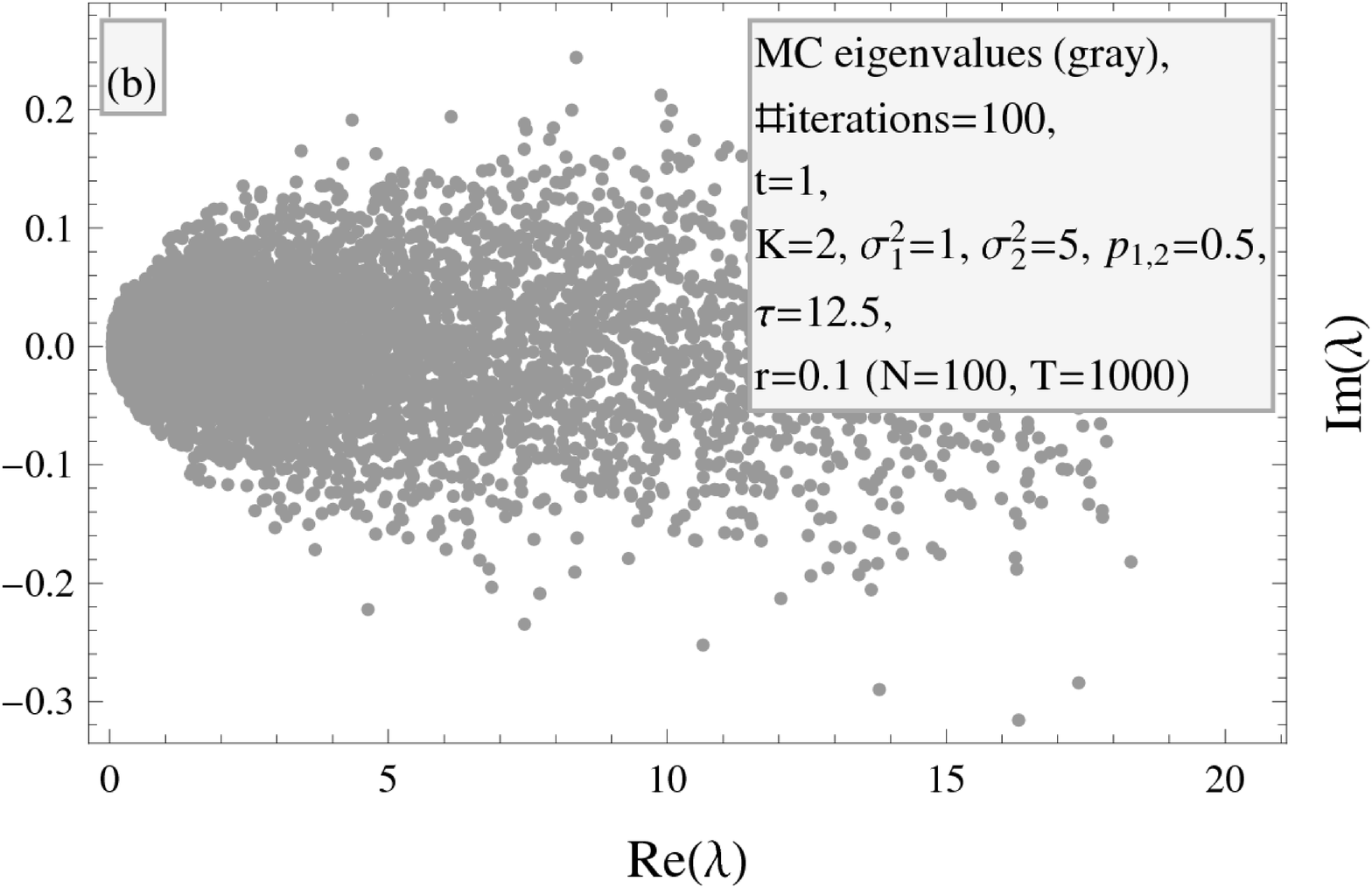}
\includegraphics[width=\columnwidth]{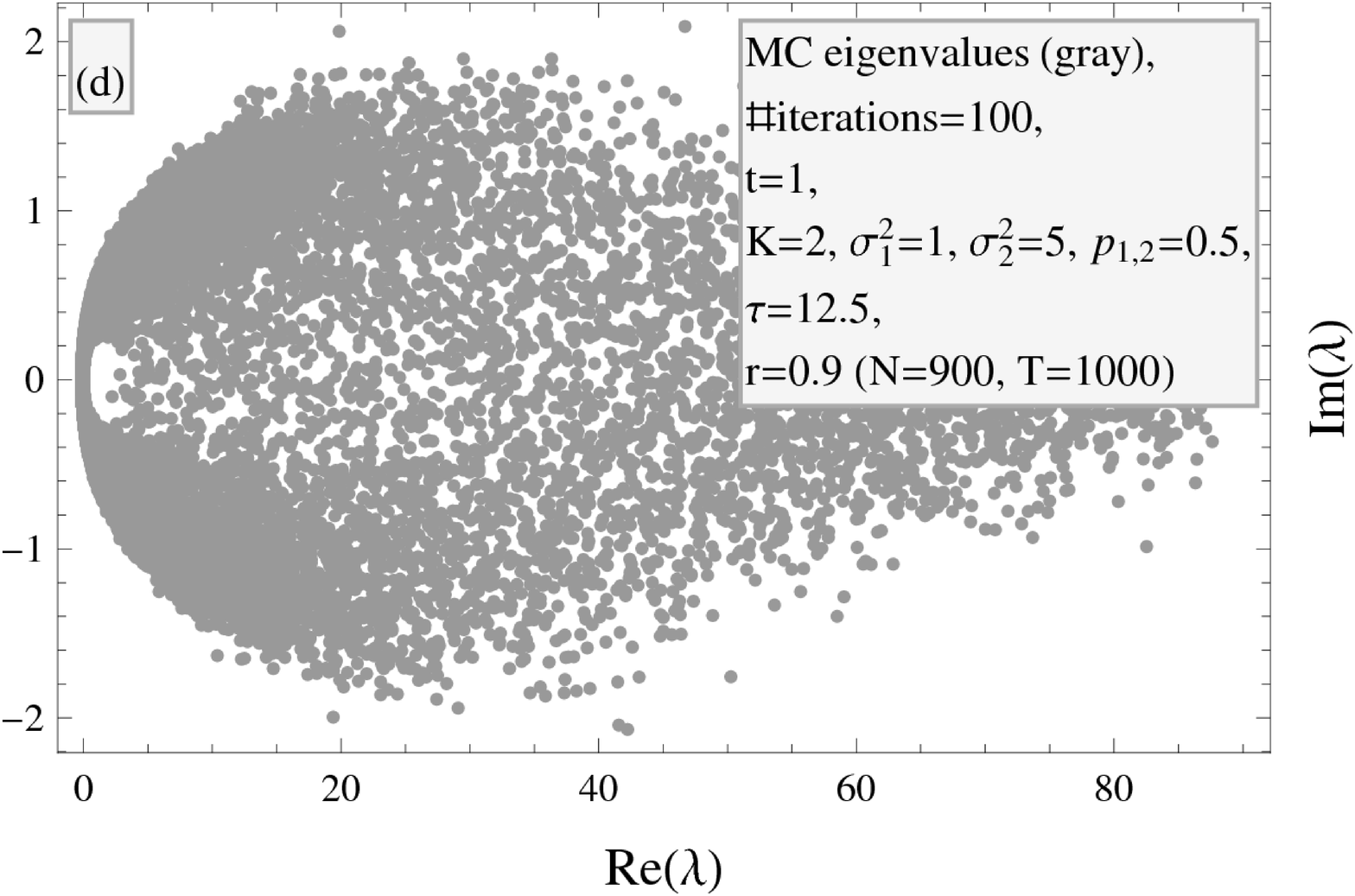}
\caption{Monte Carlo eigenvalues for the TLCE for Toy Model 4a.}
\label{fig:TM4aTLCE}
\end{figure*}

\begin{figure*}[t]
\includegraphics[width=\columnwidth]{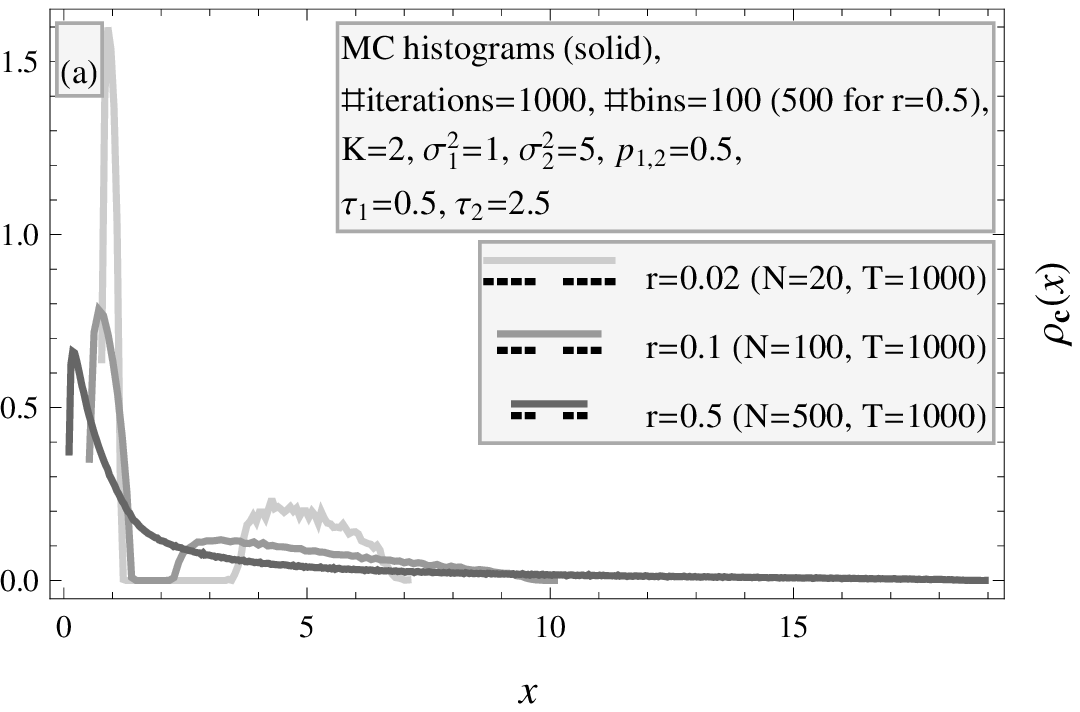}
\includegraphics[width=\columnwidth]{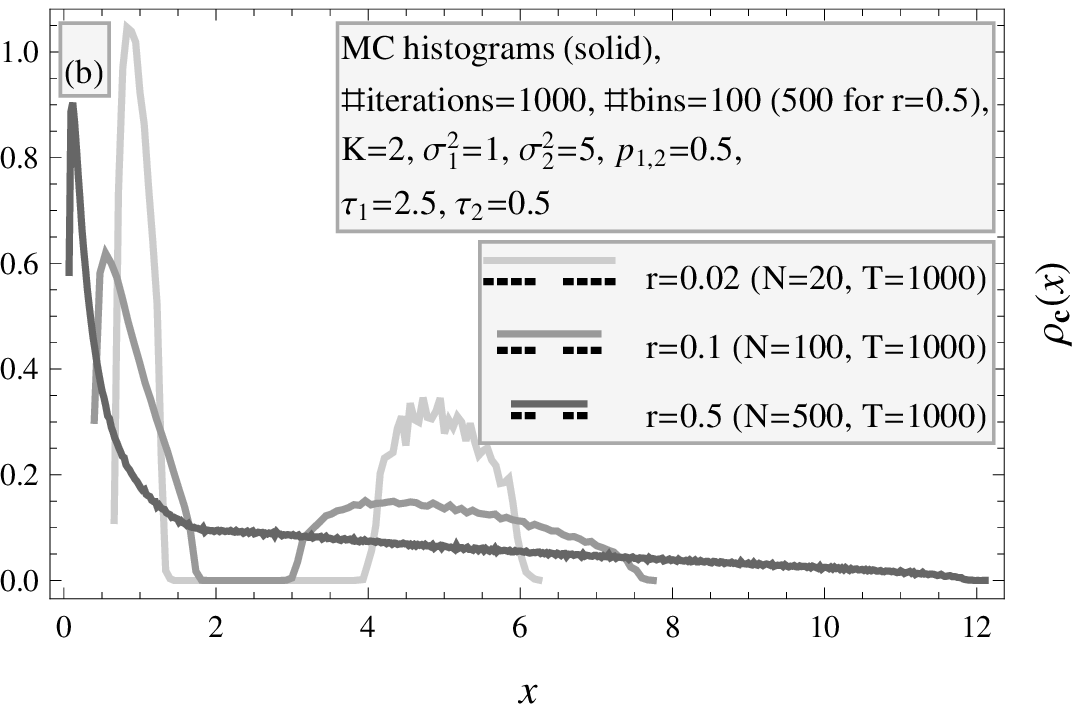}
\caption{Monte Carlo eigenvalues versus theory for the ETCE for Toy Model 4b.}
\label{fig:TM4bETCE}
\end{figure*}

\begin{figure*}[t]
\includegraphics[width=\columnwidth]{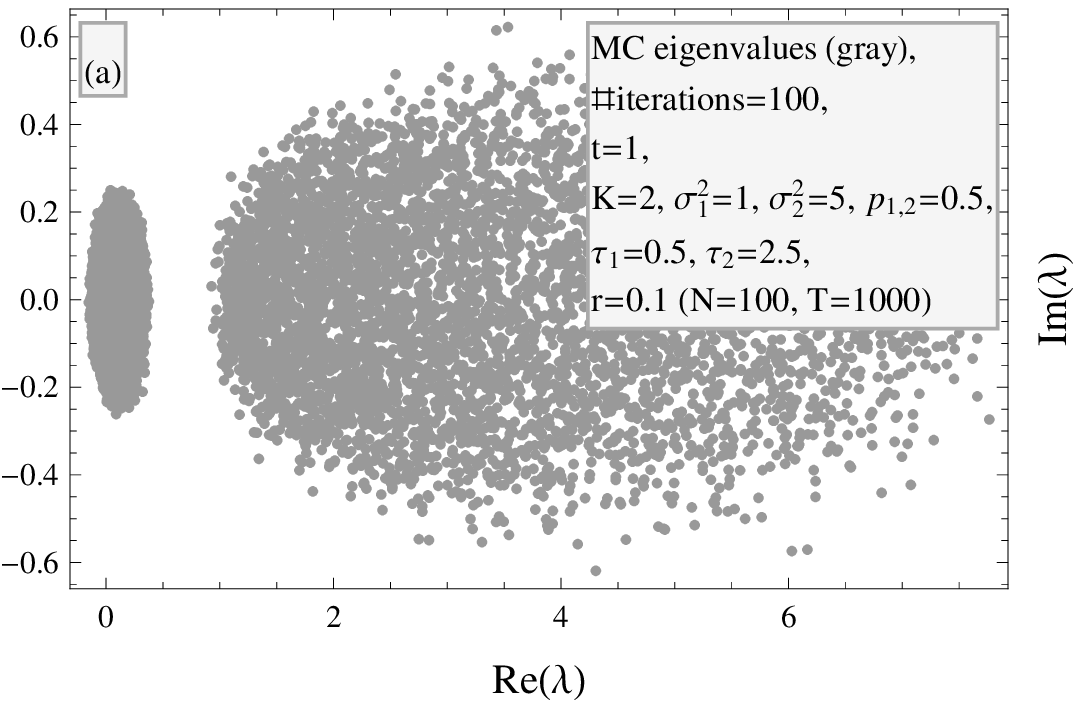}
\includegraphics[width=\columnwidth]{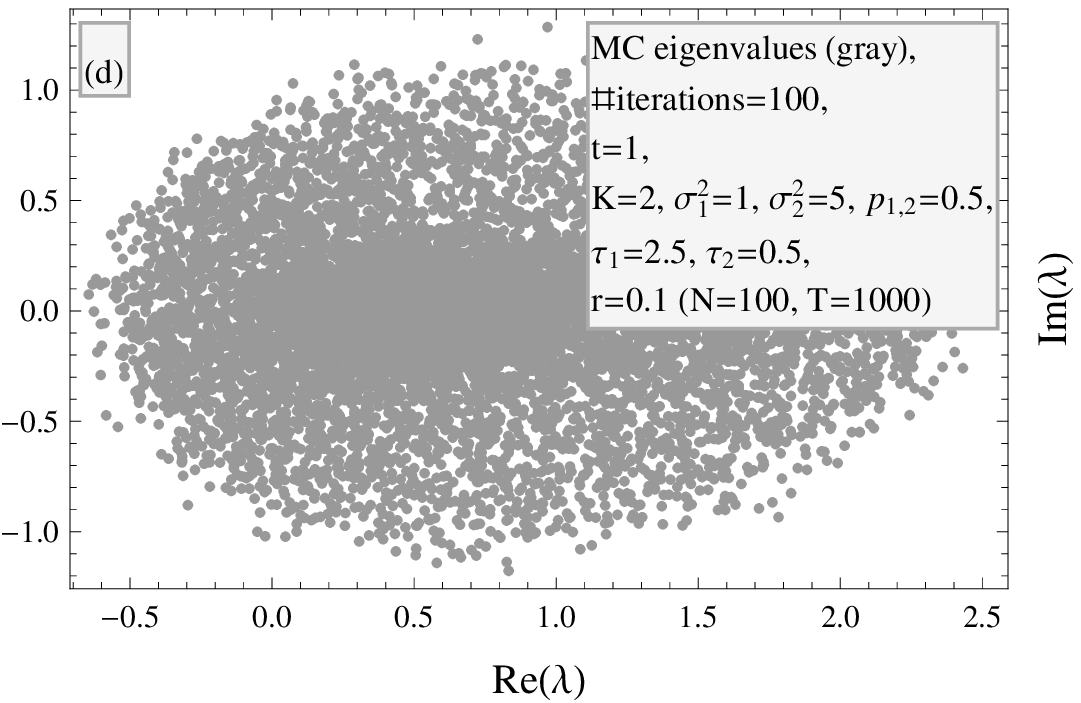}
\includegraphics[width=\columnwidth]{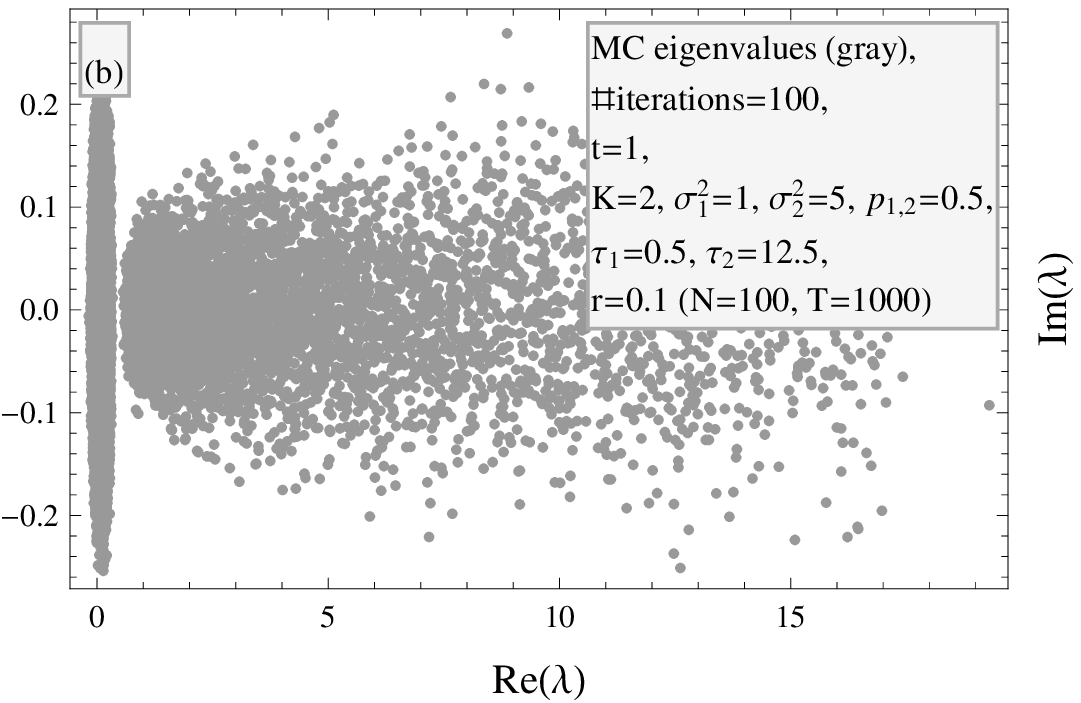}
\includegraphics[width=\columnwidth]{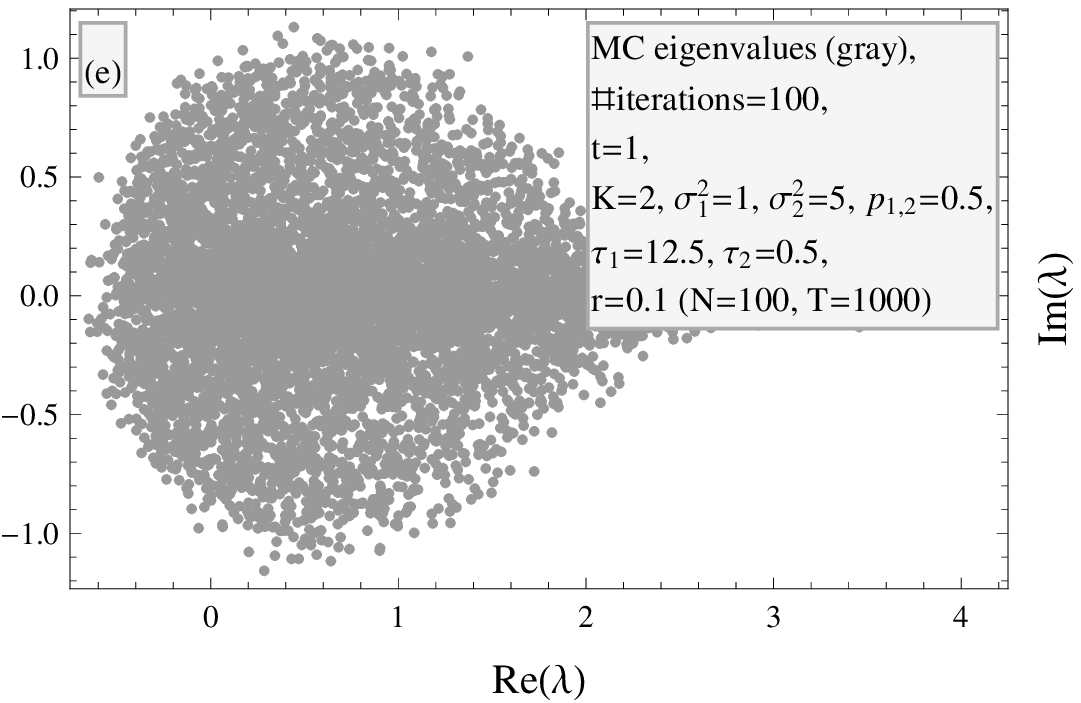}
\includegraphics[width=\columnwidth]{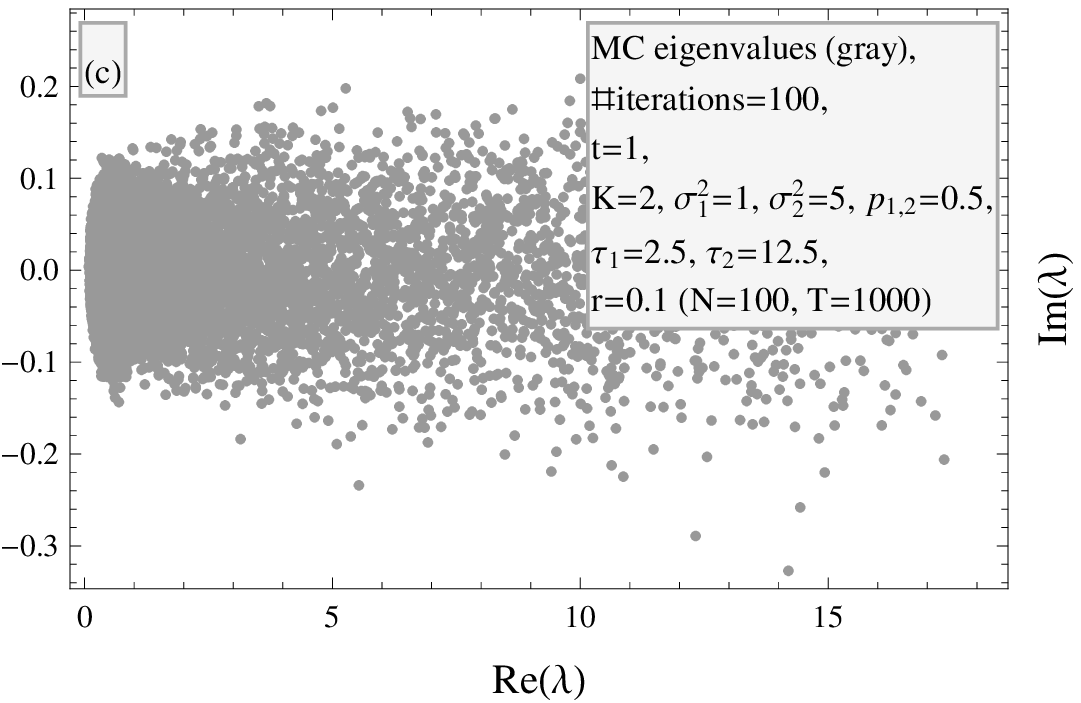}
\includegraphics[width=\columnwidth]{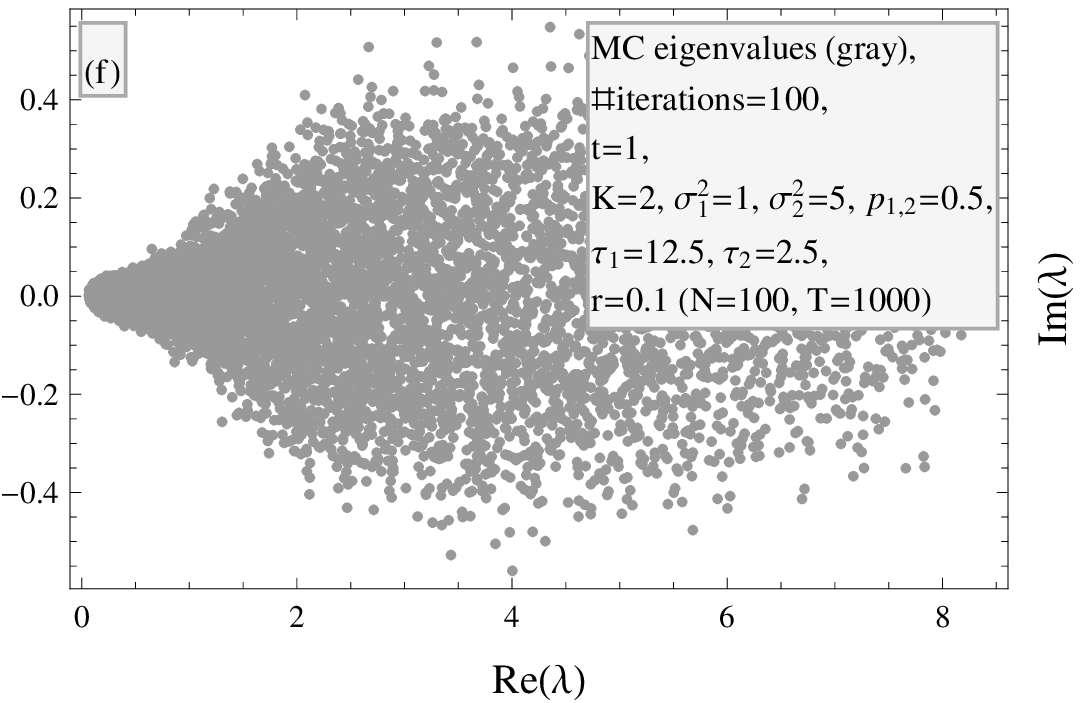}
\caption{Monte Carlo eigenvalues for the TLCE for Toy Model 4b.}
\label{fig:TM4bTLCE}
\end{figure*}

\begin{figure*}[t]
\includegraphics[width=\columnwidth]{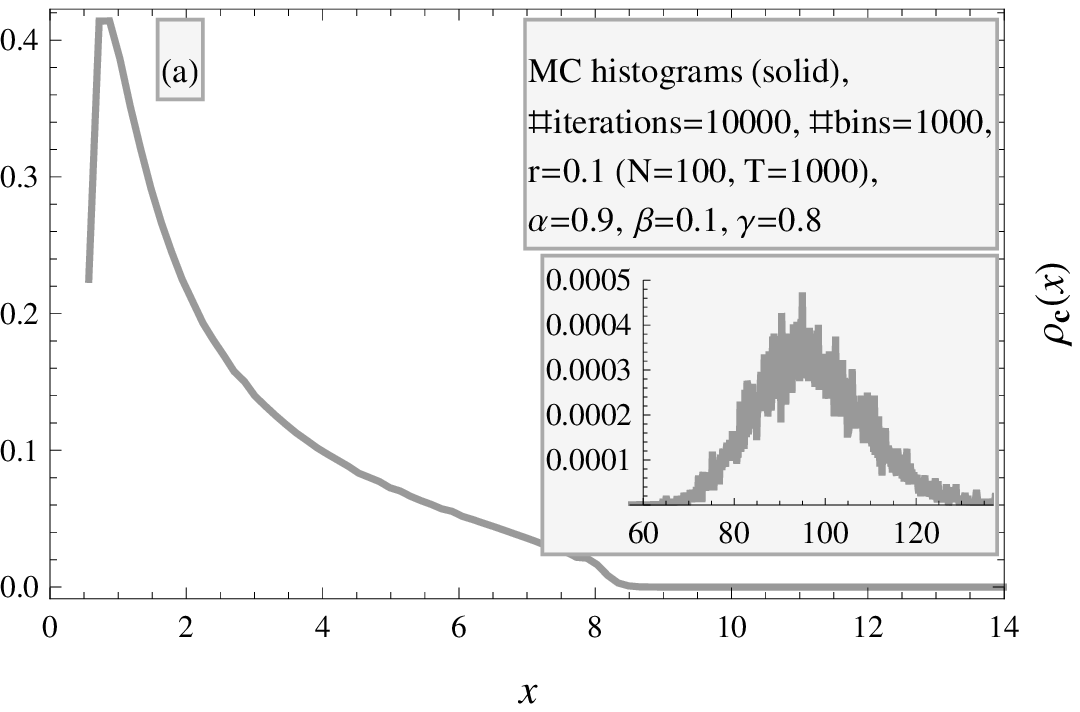}
\includegraphics[width=\columnwidth]{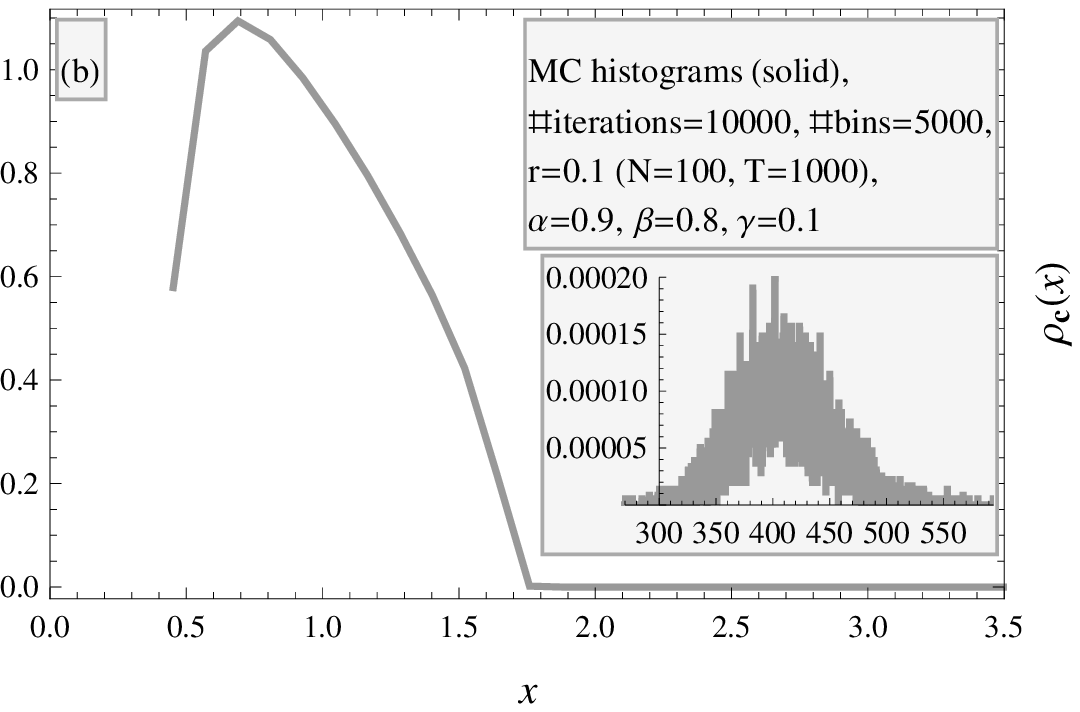}
\caption{Monte Carlo eigenvalues versus theory for the ETCE for Toy Model 4c. The insets magnify a small peak originating from the ``large'' eigenvalue \smash{$\lambda_{3} ( c )$}, while the peak around zero corresponds from the two ``small'' eigenvalues \smash{$\lambda_{1 , 2} ( c )$}.}
\label{fig:TM4cETCE}
\end{figure*}

\begin{figure*}[t]
\includegraphics[width=\columnwidth]{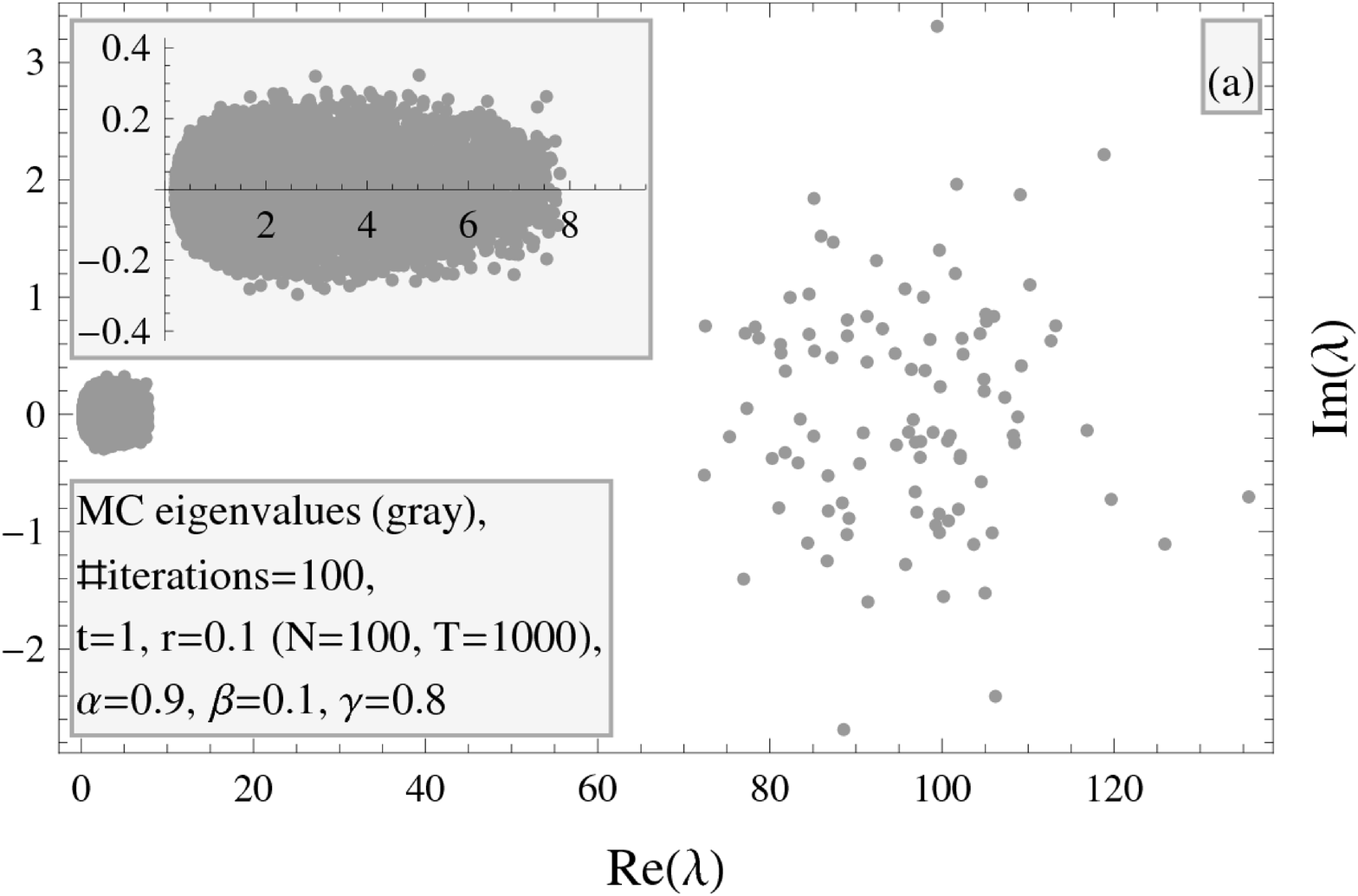}
\includegraphics[width=\columnwidth]{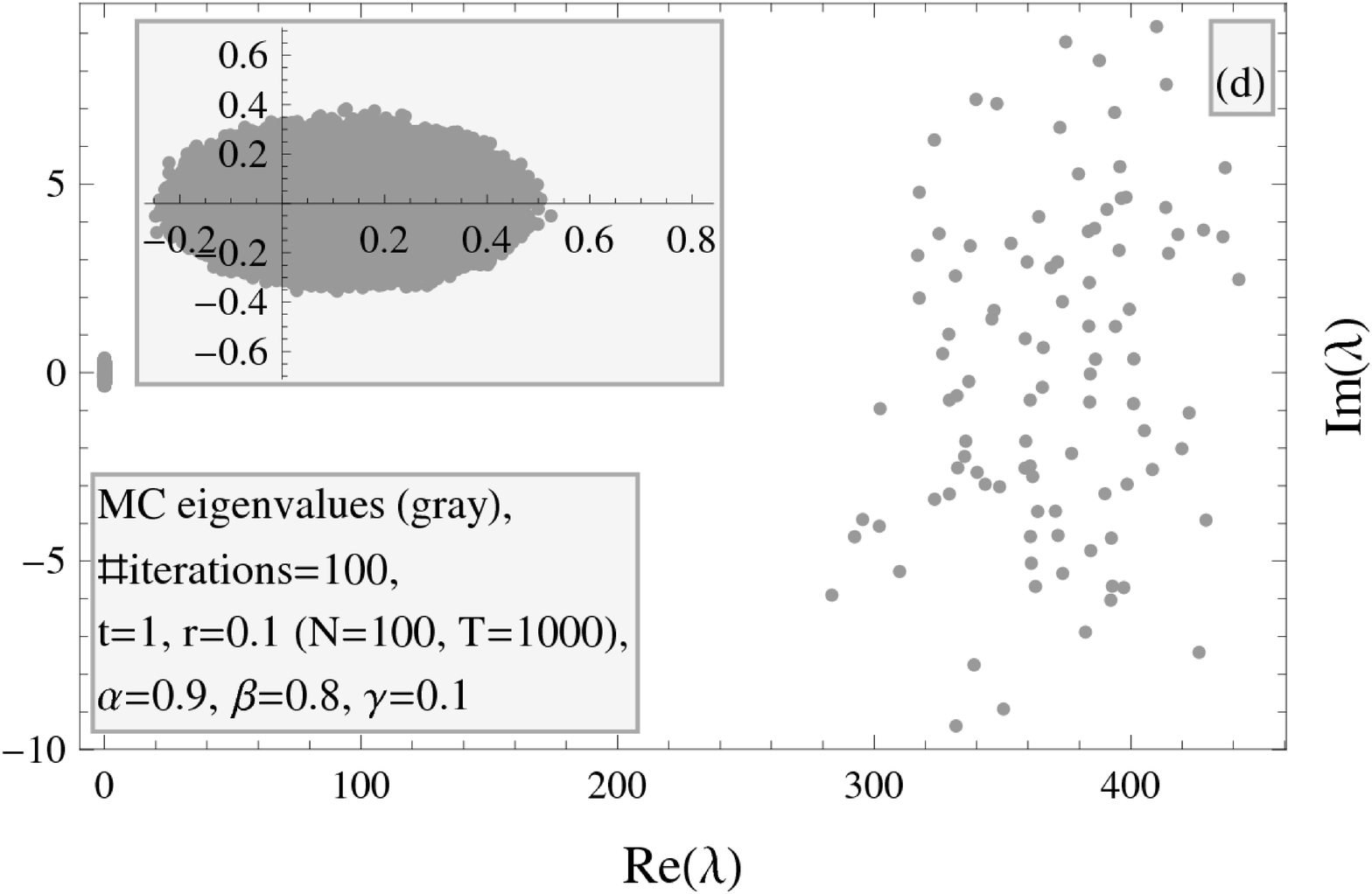}
\includegraphics[width=\columnwidth]{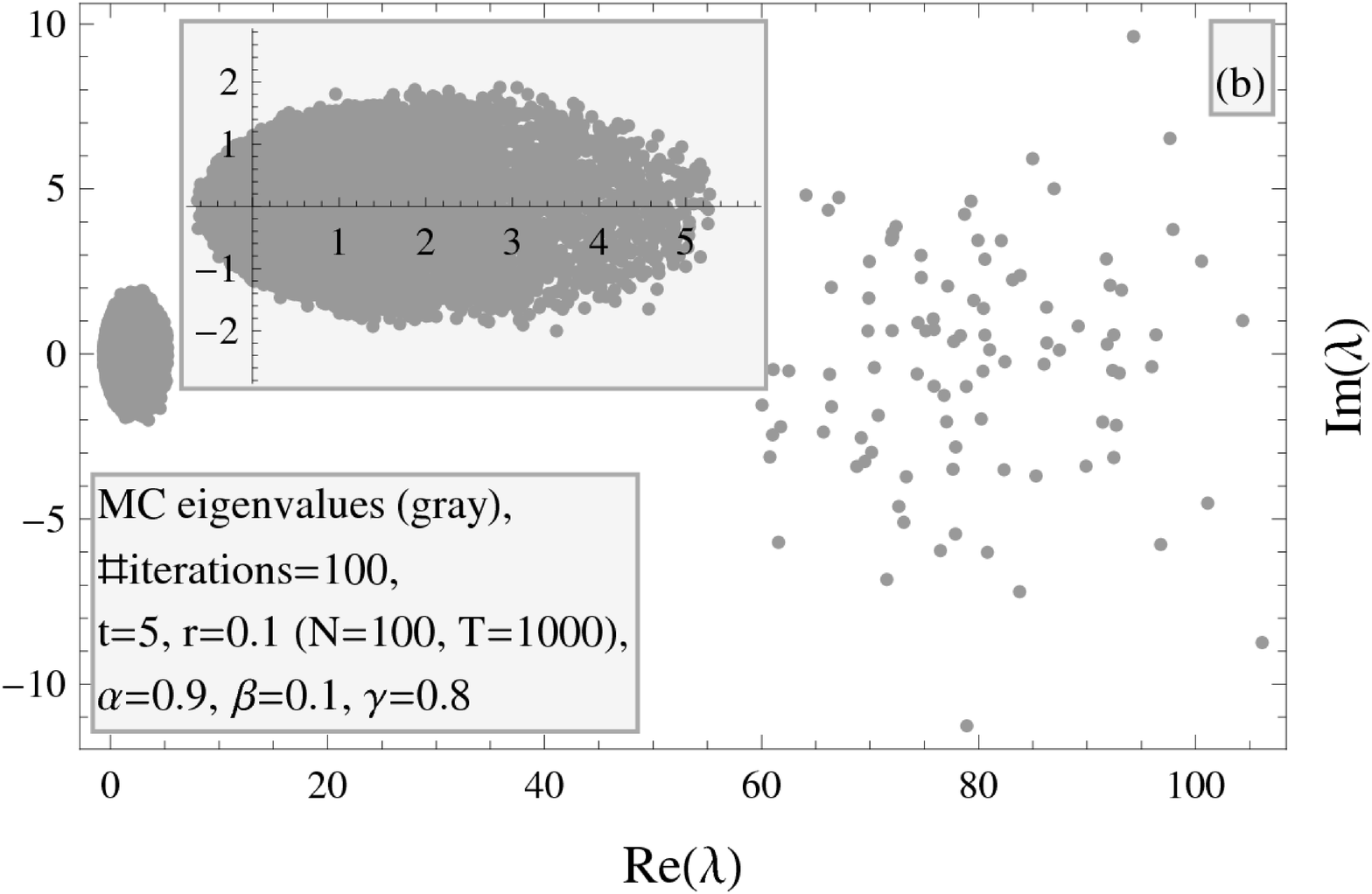}
\includegraphics[width=\columnwidth]{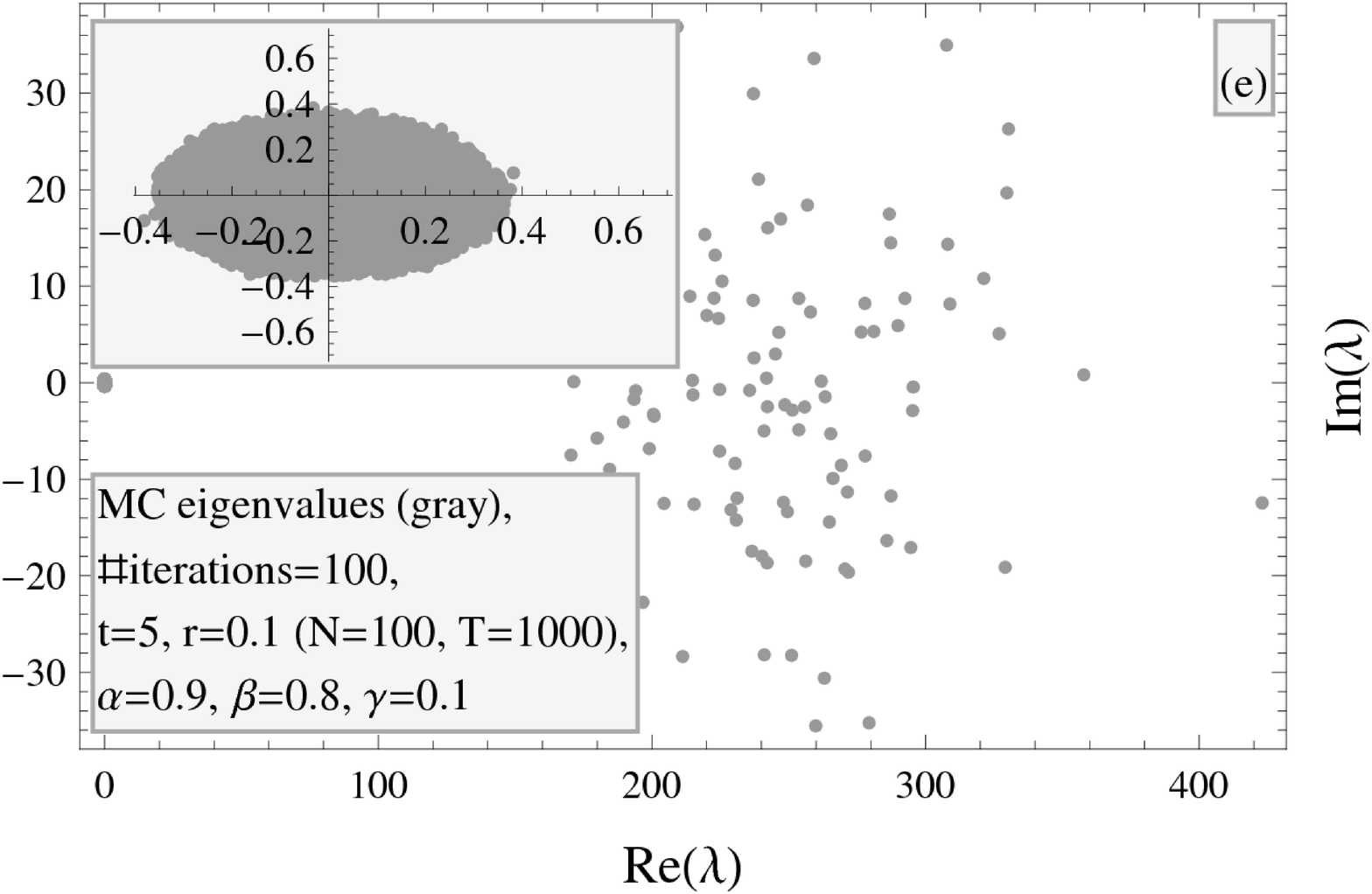}
\includegraphics[width=\columnwidth]{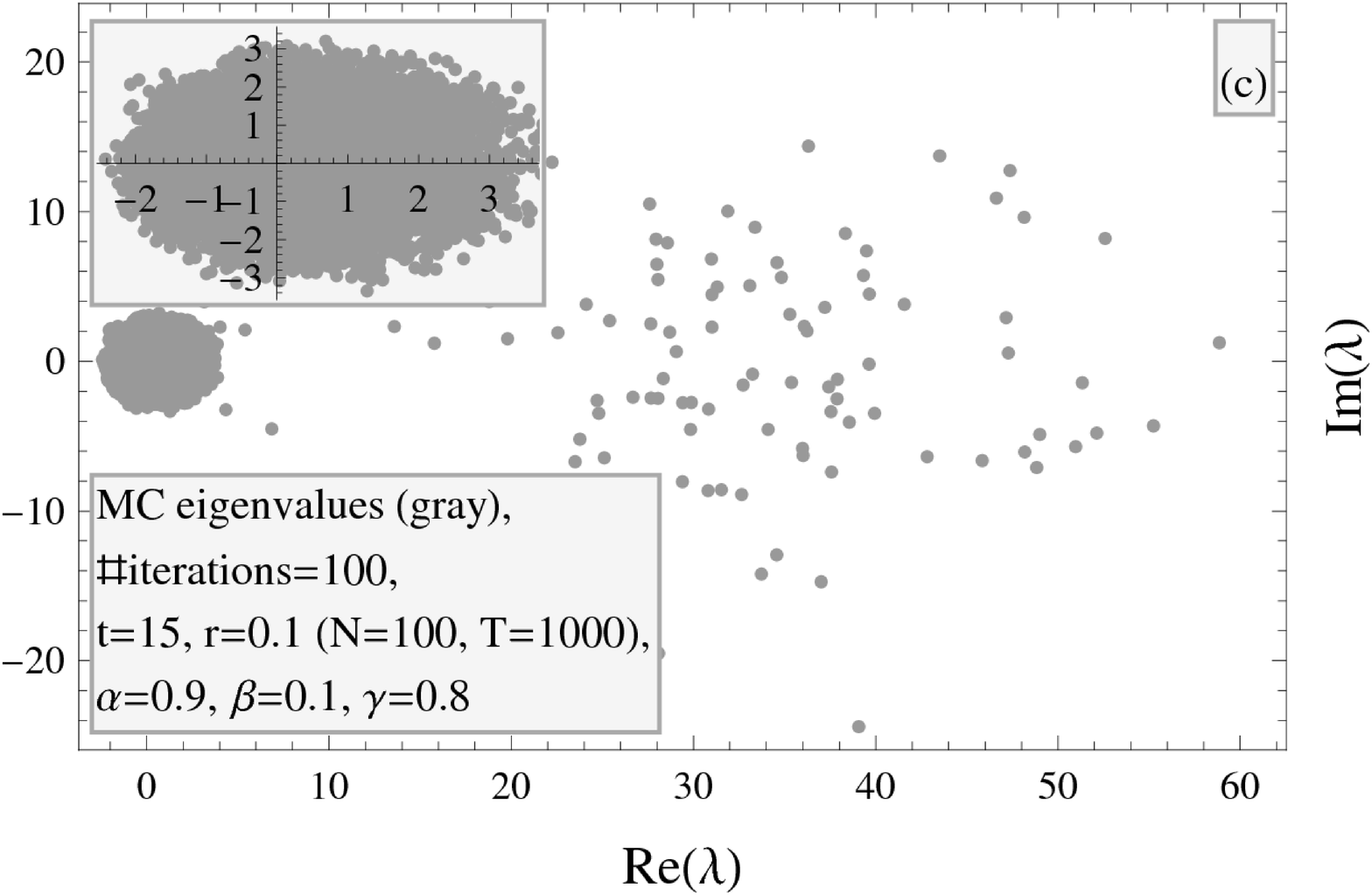}
\includegraphics[width=\columnwidth]{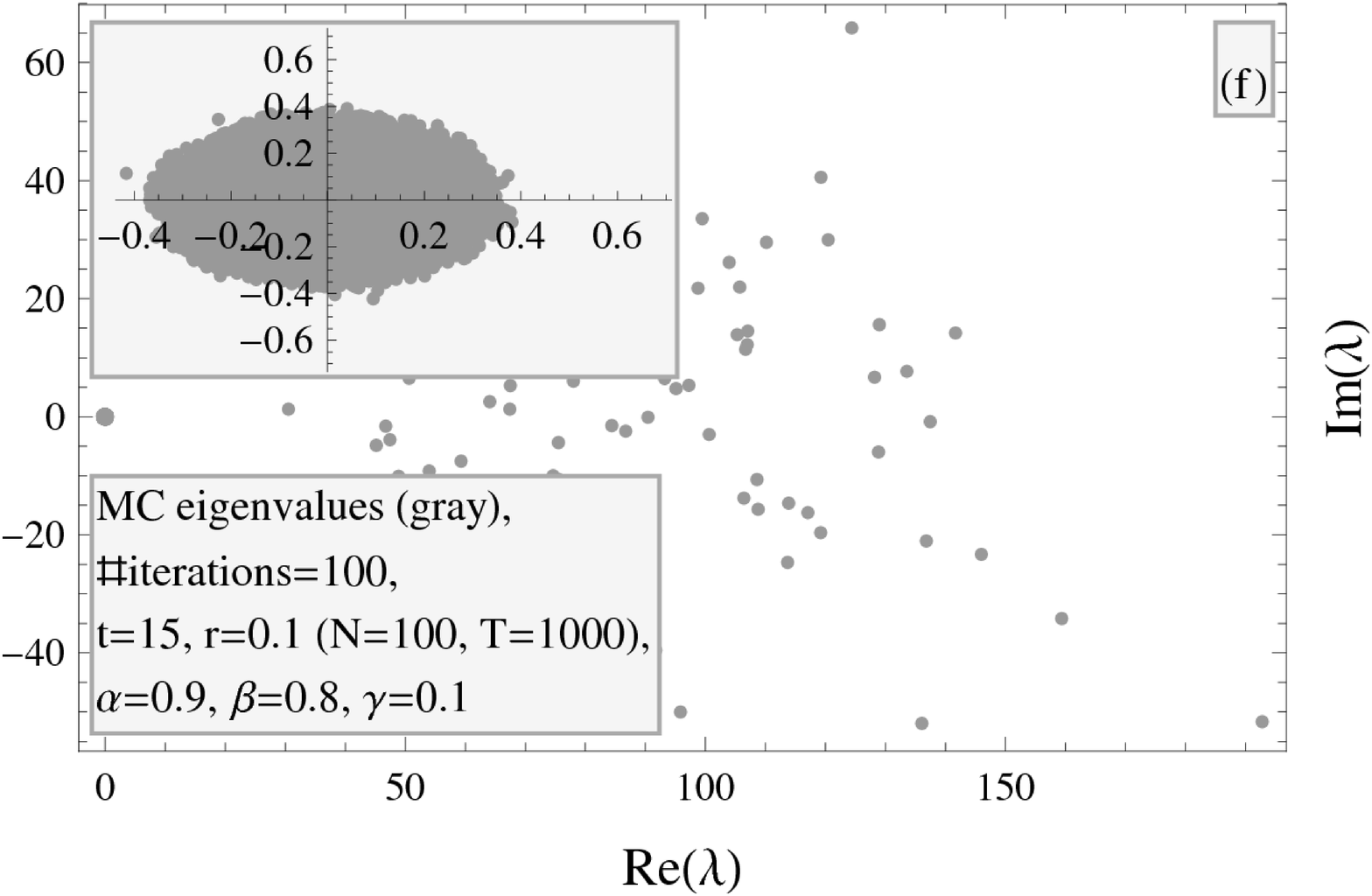}
\caption{Monte Carlo eigenvalues for the TLCE for Toy Model 4c. The main figures present the complete spectrum, while the insets zoom into the central region, corresponding to \smash{$\lambda_{1 , 2} ( c )$}. There are $100$ (= number of iterations) points in the distant region, corresponding to \smash{$\lambda_{3} ( c )$}.}
\label{fig:TM4cTLCE}
\end{figure*}

In the above examples, the effects of either the spatial or temporal intrinsic covariances on the MSD of the TLCE have been separately analyzed. However, in a description of complex systems such as financial markets one will certainly encounter spatial and temporal covariances both present and mixed (cf.~App.~\ref{aa:RandomVolatilityModels}), such as in the VAR models (cf.~Sec.~\ref{sss:VAR}). Such situations pose a much greater challenge for an analytical approach to the MSD, and another publication should be devoted to them, but in this Section, some hints will be given as to the directions of generalizations, as well as a few preliminary results. Moreover, the ETCE will also be discussed since its MSD for the below models have as yet been unknown.

%%%%%%%%%%%%%%%%%%%%%%%%%%%%%%%%%%%%%%%%%%%%%%%%%%%%%%%%%%%%%%%%%%%%%%

\subsubsection{Toy Model 4a: Different variances and identical temporal exponential decay}
\label{sss:TM4a}

\emph{Definition.} A first step toward more meaningful models could be to combine the above Toy Models 2a and 3, i.e., suppose that the assets are divided into $K$ sectors of distinct variances \smash{$\sigma_{k}^{2}$} (hence, the different assets are still independent), all of them exhibiting exponentially decaying temporal autocorrelations, with the same autocorrelation time $\tau$. This is the SVAR($1$) model (cf.~Sec.~\ref{sss:VAR}) with $\mathbf{B}$ proportional to the identity matrix and a finite number of distinct idiosyncratic variances. The covariance function (\ref{eq:SVAR1CovarianceFunction}) is factorized and translationally symmetric in time (\ref{eq:Case2Plus3Definition}) with
\begin{equation}
\begin{split}\label{eq:TM4aDefinitionEq01}
\mathbf{C} &= \diag ( \underbrace{\sigma_{1}^{2}}_{N_{1} \textrm{ times}} , \underbrace{\sigma_{2}^{2}}_{N_{2} \textrm{ times}} , \ldots , \underbrace{\sigma_{K}^{2}}_{N_{K} \textrm{ times}} ) ,\\
A ( b - a ) &= \ee^{- \frac{| b - a |}{\tau}} .
\end{split}
\end{equation}

\emph{ETCE.} The master equation for \smash{$M \equiv M_{\mathbf{c}} ( z )$} is (\ref{eq:Case2ETCEEq06}) with (\ref{eq:TM2aEq02}) and (\ref{eq:TM3ETCEEq01}), that is
\begin{subequations}
\begin{align}
M &= \sum_{k = 1}^{K} \frac{p_{k}}{\frac{z}{r \sigma_{k}^{2} M ( \chi + s )} - 1} ,\label{eq:TM4aETCEEq01a}\\
s^{2} &= \frac{1}{r^{2} M^{2}} + \chi^{2} - 1 ,\label{eq:TM4aETCEEq01b}
\end{align}
\end{subequations}
where $s$ is an auxiliary complex unknown, and recall \smash{$p_{k} = N_{k} / N$}, \smash{$\chi = \coth ( 1 / \tau )$}. By solving out $s$, one obtains a (lengthy) polynomial equation for $M$ of order $2 ( K + 1 )$, which is easily appropriated numerically. [Remark that switching on exponentially-decaying autocorrelations doubles the polynomial order of the master equation for the ETCE; compare the current situation with Toy Model 2a (\ref{eq:TM2ETCEEq01}), and Toy Model 3 (\ref{eq:TM3ETCEEq02}) with Toy Model 1 (\ref{eq:TM1ETCEEq01}).]

In Fig.~\ref{fig:TM4aETCE}, the MSD from Monte Carlo simulations is compared to the numerical solutions of [(\ref{eq:TM4aETCEEq01a}), (\ref{eq:TM4aETCEEq01b})], for the values of the parameters $K = 5$, \smash{$\sigma_{1}^{2} = 1$}, \smash{$\sigma_{2}^{2} = 2$}, \smash{$\sigma_{3}^{2} = 4$}, \smash{$\sigma_{4}^{2} = 6$}, \smash{$\sigma_{5}^{2} = 8$}, equal relative multiplicities $0.2$, and three values of $r = 0.02 , 0.1 , 0.5$, all for either $\tau = 2.5$ (a) or $\tau = 12.5$ (b), and all with perfect agreement. One observes that the ETCE could still be an efficient tool in distinguishing the variance sectors (the distinct ``lumps''), the better the smaller $r$, although the effect of the autocorrelations is to smear these lumps out, the more severely the greater $\tau$, making this analysis more difficult.

\emph{TLCE.} The master equations are [(\ref{eq:Case2Plus3TLCEEq11a})-(\ref{eq:Case2Plus3TLCEEq11c})], with [(\ref{eq:Case2Plus3TLCEEq06a}), (\ref{eq:Case2Plus3TLCEEq06b}), (\ref{eq:Case2Plus3TLCEEq04}), (\ref{eq:Case2Plus3TLCEEq10a}), (\ref{eq:Case2Plus3TLCEEq10b}), (\ref{eq:Case2Plus3TLCEEq08})], and a numerical method of solution remains to be developed.

Figure~\ref{fig:TM4aTLCE} only presents the numerical eigenvalues of the TLCE, for $t = 1$, $K = 2$, \smash{$\sigma_{1}^{2} = 1$}, \smash{$\sigma_{2}^{2} = 5$}, equal relative multiplicities $0.5$, as well as $r = 0.1$ (left column) or $r = 0.9$ (right column), and $\tau = 2.5$ (top row) or $\tau = 12.5$ (bottom row). The domain is ``bullet-shaped,'' with a possible hole inside, similarly as for Toy Model 3. However, for sufficiently small $r$ and $\tau$, one recognizes its parts corresponding to the different variances; this shape might be exploited in determining the parameters of the model from a given set of data, once the theoretical master equations become numerically solvable. Also, just like for the ETCE, increasing $r$ or $\tau$ makes these parts less distinguishable.

%%%%%%%%%%%%%%%%%%%%%%%%%%%%%%%%%%%%%%%%%%%%%%%%%%%%%%%%%%%%%%%%%%%%%%

\subsubsection{Toy Model 4b: Different variances and different temporal exponential decays}
\label{sss:TM4b}

\emph{Definition.} A reasonable way to further generalize the above model would be to assume that each variance sector has also a different autocorrelation time,
\begin{equation}\label{eq:TM4bDefinitionEq01}
\mathbf{C} ( c ) = \diag \Big( \underbrace{\sigma_{1}^{2} \ee^{- \frac{| c |}{\tau_{1}}}}_{N_{1} \textrm{ times}} , \underbrace{\sigma_{2}^{2} \ee^{- \frac{| c |}{\tau_{2}}}}_{N_{2} \textrm{ times}} , \ldots , \underbrace{\sigma_{K}^{2} \ee^{- \frac{| c |}{\tau_{K}}}}_{N_{K} \textrm{ times}} \Big) .
\end{equation}
This is the SVAR($1$) model with $\mathbf{B}$ diagonal with a finite number of distinct beta coefficients corresponding to distinct idiosyncratic variances. The covariance function (\ref{eq:SVAR1CovarianceFunction}) is translationally symmetric in time but not factorized into a spatial and temporal parts, being the first instance of such mixing in this paper (it is Case 3; cf.~Sec.~\ref{ss:Case3}).

The Fourier transform (\ref{eq:Case3FourierEq03a}) of (\ref{eq:TM4bDefinitionEq01}) in the variable $u$ (\ref{eq:TM3DefinitionEq03}) reads [cf.~(\ref{eq:TM3DefinitionEq02})],
\begin{equation}\label{eq:TM4bDefinitionEq02}
\hat{\mathbf{C}} ( u ) = \diag \bigg( \underbrace{\frac{A_{1 , 1}}{A_{2 , 1} - u - \frac{1}{u}}}_{N_{1} \textrm{ times}} , \ldots , \underbrace{\frac{A_{1 , K}}{A_{2 , K} - u - \frac{1}{u}}}_{N_{K} \textrm{ times}} \bigg) ,
\end{equation}
with \smash{$A_{1 , k} \equiv 2 \sigma_{k}^{2} \sinh ( 1 / \tau_{k} )$} and \smash{$A_{2 , k} \equiv 2 \cosh ( 1 / \tau_{k} )$}.

\emph{ETCE.} There is a common pattern in solving the master equations (for both the ETCE and TLCE) in the situation of mixed spatial and temporal covariances, which will be sketched on the current example: Firstly, the matrix structure of \smash{$\hat{\mathbf{C}} ( u )$} determines the matrix structure of \smash{$\boldsymbol{\Sigma}^{N N}$}, as visible from (\ref{eq:Case3ETCEEq01a}), i.e., \smash{$\boldsymbol{\Sigma}^{N N}$} is diagonal, with some elements \smash{$\Sigma_{k}$} with multiplicities \smash{$N_{k}$}, for $k = 1 , 2 , \ldots , K$; these constitute the unknowns. Secondly, (\ref{eq:Case3ETCEEq02a}) implies that the same is true of \smash{$\mathbf{G}^{N N}$}, with \smash{$G_{k} = 1 / ( z - \Sigma_{k} )$}. Thirdly, (\ref{eq:Case3ETCEEq01b}) becomes
\begin{equation}\label{eq:TM4bETCEEq01}
\hat{\Sigma}^{T T} ( u ) = r \sum_{l = 1}^{K} p_{l} \frac{A_{1 , l}}{A_{2 , l} - u - \frac{1}{u}} \frac{1}{z - \Sigma_{l}} .
\end{equation}
Fourthly, one plugs this into (\ref{eq:Case3ETCEEq02b}), and fifthly, further into (\ref{eq:Case3ETCEEq01a}), which gives the final $K$ equations for the $K$ unknowns,
\begin{equation}\label{eq:TM4bETCEEq02}
\Sigma_{k} = \frac{1}{2 \pi \ii} \ointctrclockwise_{C ( 0 , 1 )} \dd u \frac{1}{u} \frac{A_{1 , k}}{A_{2 , k} - u - \frac{1}{u}} \frac{1}{z - \hat{\Sigma}^{T T} ( u )} .
\end{equation}
Solving this system requires computing the residues of the integrands inside the centered unit circle, which is equivalent to finding the zeros in $u$ of their denominators, and these are all polynomials of order $2 K$; therefore, it poses a challenge to construct even a numerical algorithm able to calculate the values \smash{$\Sigma_{k}$} from (\ref{eq:TM4bETCEEq02}).

In Fig.~\ref{fig:TM4bETCE}, there are examples of Monte Carlo histograms of the MSD of the ETCE, for $K = 2$, \smash{$\sigma_{1}^{2} = 1$}, \smash{$\sigma_{2}^{2} = 5$}, equal relative multiplicities $0.5$, three values of $r = 0.02 , 0.1 , 0.5$, and respectively \smash{$\tau_{1} = 0.5$}, \smash{$\tau_{2} = 2.5$} (a) or \smash{$\tau_{1} = 2.5$}, \smash{$\tau_{2} = 0.5$} (b). As with Toy Model 4a (cf.~Fig.~\ref{fig:TM4aETCE}), the variance sectors are visible for small enough $r$, but larger autocorrelation time \smash{$\tau_{k}$} broadens the peak corresponding to \smash{$\sigma_{k}$}.

\emph{TLCE.} The master equations are [(\ref{eq:Case3TLCEEq01a})-(\ref{eq:Case3TLCEEq04})], and a similar method as for the ETCE should be used to recast them in a form appropriate for a yet to be devised numerical procedure.

In Fig.~\ref{fig:TM4bTLCE}, the Monte Carlo eigenvalues of the TLCE are shown, for $t = 1$, $K = 2$, \smash{$\sigma_{1}^{2} = 1$}, \smash{$\sigma_{2}^{2} = 5$}, equal relative multiplicities $0.5$, and $r = 0.1$, as well as six combinations of the values $0.5$, $2.5$, $12.5$ for \smash{$\tau_{1}$}, \smash{$\tau_{2}$}. One again recognizes that increasing any \smash{$\tau_{k}$} causes the sub-domain corresponding to \smash{$\sigma_{k}^{2}$} to expand, with rate depending on the value of the variance (it seems it has a stronger impact on smaller variances); for sufficiently small autocorrelation times (especially for smaller variances), and of course for small enough $r$, the sub-domains may separate (topological phase transition), which has not been observed for Toy Model 4a.

%%%%%%%%%%%%%%%%%%%%%%%%%%%%%%%%%%%%%%%%%%%%%%%%%%%%%%%%%%%%%%%%%%%%%%

\subsubsection{Toy Model 4c: A simple vector autoregression model with a market mode}
\label{sss:TM4c}

\emph{Definition.} Continuing the above progression of SVAR($1$) models, a next reasonable step could be to include genuine correlations between various assets. A simplest such example is to consider the market mode which depends on itself in the previous moment of time, while any other asset depends both on itself and the market in the previous moment, as outlined at the end of Sec.~\ref{sss:VAR}. This is the most advanced model in this article, with the time-dependent covariance matrix [(\ref{eq:SVAR1BMarketCovarianceFunction1})-(\ref{eq:SVAR1BMarketCovarianceFunction5})] (belonging to Case 3).

The Fourier transform (\ref{eq:Case3FourierEq03a}) of the covariance function (\ref{eq:VARCovarianceFunction}) of an arbitrary VAR model reads
\begin{equation}\label{eq:TM4cDefinitionEq01}
\hat{\mathbf{C}} ( p ) = \frac{1}{1 - \hat{\mathbf{K}} ( p )} \hat{\mathbf{C}}^{\textrm{id.}} ( p ) \frac{1}{1 - \hat{\mathbf{K}} ( p )^{\dagger}} ,
\end{equation}
which in the considered simplified version turns into (note, \smash{$\hat{\mathbf{K}} ( u ) = u \mathbf{B}$}),
\begin{subequations}
\begin{align}
\hat{C}_{1 1} ( u ) &\equiv c_{1} ( u ) =\nonumber\\
&= \frac{u \left( - \gamma u^{2} + \left( 1 + \gamma^{2} + ( N - 1 ) \beta^{2} \right) u - \gamma \right)}{( u - \alpha ) ( u \alpha - 1 ) ( u - \gamma ) ( u \gamma - 1 )} ,\label{eq:TM4cDefinitionEq02a}\\
\hat{C}_{i i} ( u ) &\equiv c_{2} ( u ) = \frac{- u}{( u - \gamma ) ( u \gamma - 1 )} ,\label{eq:TM4cDefinitionEq02b}\\
\hat{C}_{1 i} ( u ) &\equiv c_{3} ( u ) = \frac{- \beta u}{( u - \alpha ) ( u - \gamma ) ( u \gamma - 1 )} ,\label{eq:TM4cDefinitionEq02c}\\
\hat{C}_{i 1} ( u ) &\equiv c_{4} ( u ) = \frac{\beta u^{2}}{( u \alpha - 1 ) ( u - \gamma ) ( u \gamma - 1 )} ,\label{eq:TM4cDefinitionEq02d}
\end{align}
\end{subequations}
for $i > 1$, with the remaining matrix elements zero (i.e., there are four distinct matrix entries on the diagonal, first row and first column). This constitutes an input data for the master equations [(\ref{eq:Case3ETCEEq01a})-(\ref{eq:Case3ETCEEq02b})] (for the ETCE) or [(\ref{eq:Case3TLCEEq01a})-(\ref{eq:Case3TLCEEq04})] (for the TLCE).

\emph{ETCE.} Following the same logic as in Sec.~\ref{sss:TM4b}, one starts from noting that according to (\ref{eq:Case3ETCEEq01a}) the matrix structure of \smash{$\boldsymbol{\Sigma}^{N N}$} mimics that of \smash{$\hat{\mathbf{C}} ( u )$}, i.e., it is determined by four complex parameters, \smash{$\Sigma^{N N}_{1 1} \equiv \Sigma_{1}$}, \smash{$\Sigma^{N N}_{i i} \equiv \Sigma_{2}$}, \smash{$\Sigma^{N N}_{1 i} \equiv \Sigma_{3}$}, \smash{$\Sigma^{N N}_{i 1} \equiv \Sigma_{4}$}, for all $i > 1$, which are the unknowns. One substitutes this form into (\ref{eq:Case3ETCEEq02a}), where the matrix inversion is explicitly doable, and further into (\ref{eq:Case3ETCEEq01b}),
\begin{equation}
\begin{split}\label{eq:TM4cETCEEq01}
\hat{\Sigma}^{T T} ( u ) &=\\
&= \frac{r}{\frac{\Sigma_{3} \Sigma_{4}}{z - \Sigma_{2}} - ( z - \Sigma_{1} ) \frac{1}{N - 1}} \cdot\\
&\cdot \bigg( \frac{c_{2} ( u ) \Sigma_{3} \Sigma_{4}}{( z - \Sigma_{2} )^{2}} \frac{N - 2}{N} -\\
&- \frac{c_{2} ( u ) ( z - \Sigma_{1} ) + c_{4} ( u ) \Sigma_{3} + c_{3} ( u ) \Sigma_{4}}{z - \Sigma_{2}} \frac{1}{N} -\\
&- c_{1} ( u ) \frac{1}{N ( N - 1 )} \bigg) .
\end{split}
\end{equation}
[Although $N$ is always infinite in this paper (\ref{eq:ThermodynamicLimit}), here one should treat it as finite, governing the number of assets relative to the market size.] Plugging this result into (\ref{eq:Case3ETCEEq02b}), and then into (\ref{eq:Case3ETCEEq01a}), one arrives at four integral equations for the four unknowns,
\begin{equation}\label{eq:TM4cETCEEq02}
\Sigma_{s} = \frac{1}{2 \pi \ii} \ointctrclockwise_{C ( 0 , 1 )} \dd u \frac{1}{u} c_{s} ( u ) \frac{1}{z - \hat{\Sigma}^{T T} ( u )} ,
\end{equation}
for $s = 1 , 2 , 3 , 4$. One finds that the four integrands are rational functions of $u$, with the same denominator being a quartic polynomial in $u$. For that reason, analytical solution is impossible, and an effective numerical method needs to be devised.

Figure~\ref{fig:TM4cETCE} presents the Monte Carlo histograms of the eigenvalues of the ETCE, for $r = 0.1$, $\alpha = 0.9$ and respectively $\beta = 0.1$, $\gamma = 0.8$ (a) or $\beta = 0.8$, $\gamma = 0.1$ (b). The eigenvalues \smash{$\lambda_{1 , 2 , 3} ( c = 0 )$} [(\ref{eq:SVAR1BMarketCovarianceFunctionEigenvalue1}), (\ref{eq:SVAR1BMarketCovarianceFunctionEigenvalue2})] of the true covariance matrix acquire then the approximate values $2.7778$, $2.1005$, $94.8502$ (a) or $1.0101$, $1.0082$, $408.735$ (b), which are seen to be broadened by the measurement noise in the form of a large peak corresponding to the two smallest eigenvalues (they are close to each other and thus indistinguishable from this spectrum) and a small distant peak originating from the one large eigenvalue.

\emph{TLCE.} The master equations in a form suitable for numerical evaluation are derived analogously to the ETCE case above; they will not be explicitly written down, only some examples of the Monte Carlo spectra of the TLCE are depicted in Fig.~\ref{fig:TM4cTLCE}, for the same values of $r$ and the beta coefficients as above, as well as respectively $t = 1$ (top row), $t = 5$ (middle row) or $t = 15$ (bottom row). Again, there is a bullet-shaped central region corresponding to the two smallest eigenvalues and a distant region from the noise-broadened large one. In other words, one not only learns from these plots about the presence of the market mode, but by investigating the shapes of the domain for various $t$ one may infer about the autocorrelations in the system.

%%%%%%%%%%%%%%%%%%%%%%%%%%%%%%%%%%%%%%%%%%%%%%%%%%%%%%%%%%%%%%%%%%%%%%
%%%%%%%%%%%%%%%%%%%%%%%%%%%%%%%%%%%%%%%%%%%%%%%%%%%%%%%%%%%%%%%%%%%%%%
%%%%%%%%%%%%%%%%%%%%%%%%%%%%%%%%%%%%%%%%%%%%%%%%%%%%%%%%%%%%%%%%%%%%%%

\section{Conclusions}
\label{s:Conclusions}

%%%%%%%%%%%%%%%%%%%%%%%%%%%%%%%%%%%%%%%%%%%%%%%%%%%%%%%%%%%%%%%%%%%%%%
%%%%%%%%%%%%%%%%%%%%%%%%%%%%%%%%%%%%%%%%%%%%%%%%%%%%%%%%%%%%%%%%%%%%%%

\subsection{Summary}
\label{ss:Summary}

The focus of this article is on an important problem of assessing from historical data correlations between various objects of a complex system (such as a financial market) at identical or distinct time moments. In the first case, a useful tool is the spectral analysis of the equal-time covariance estimator, which is a Hermitian random matrix and which has been extensively studied in the literature. It has been shown for various models to reproduce the true equal-time correlations, albeit with a measurement noise, which nevertheless diminishes with decreasing rectangularity ratio $r$. Unfortunately but quite understandably, its spectrum practically fails to be sensitive enough to true time-lagged correlations, a property perhaps even more important, reflecting rich, complicated and relatively little known temporal dynamics of the markets (e.g., heteroscedasticity, leverage effects, Epps effect etc.).

It is the time-lagged covariance estimator which constitutes a sensitive probe of these temporal correlations. It is a non-Hermitian random matrix, and as such is much more involved from the analytical point of view than Hermitian matrices, which explains why it has been largely neglected. The only analytical result to date has been~\cite{BielyThurner2006,ThurnerBiely2007}, whose main result is shown here to be incorrect while the right one is given.

This paper is a comprehensive treatment of both these estimators. The method of planar diagrammatic expansion, adjusted to the non-Hermitian world (explained at length in App.~\ref{a:GaussianPlanarDiagrammaticExpansion}), is applied in the most general Gaussian case to derive the master equations (cf.~Sec.~\ref{s:MasterEquations} and Tab.~\ref{tab:MasterEquations}) from which the mean spectral density can be calculated.

These general results are then used for a detailed spectral analysis (cf.~Sec.~\ref{s:TM} and Tab.~\ref{tab:TM})---i.e., finding the density, comparing it to Monte Carlo simulations and discussing it with respect to the effectiveness in unveiling true covariances---of the time-lagged (and sometimes also equal-time) covariance estimator for four toy models of the market, each designed to approximate its certain aspect. The simplest Toy Model 1 has zero true correlations and may be treated as the null hypothesis. Toy Model 2 assumes zero temporal correlations and investigates nontrivial variances, either their finite number (2a) or a power-law distribution (2b), which both model the presence of industrial sectors. They are recognized to be much more accurately reflected in the spectrum of the equal-time estimator. Toy Model 3, on the contrary, has zero correlations between distinct assets but nontrivial autocorrelations; it is a simplest example of the class of vector autoregression models, with exponentially-decaying temporal correlations identical for all the assets. The spectral analysis is quite difficult both analytically and numerically but leads to very intricate shapes of mean spectral domains which may serve as a sensitive probe of true temporal correlations in the system, unlike the equal-time estimator.

Since these models are both already complicated and yet quite crude approximations, three more realistic models are further investigated---due to analytical difficulties, both the equal-time and time-lagged estimators are considered but it is only sketched how to write down their master equation in a form suitable for numerical evaluation and Monte Carlo simulations are presented. Toy Model 4a has a finite number of variances and exponentially-decaying autocorrelations identical for all the assets; Toy Model 4b differs from it by assuming distinct autocorrelation times for all the sectors; Toy Model 4c is the most advanced example with a nondiagonal time-dependent covariance matrix originating from the presence of the market mode such that the evolution of each asset depends on the value of this asset and of the market in a previous moment of time.

Due to analytical tractability, all the above calculations are based on the Gaussian probability distribution of the assets, which is an assumption universally made both within the scientific approach and financial practice, yet certainly quite far from the truth (cf.~App.~\ref{a:NonGaussianProbabilityDistributionsForFinancialInstruments}; however, this can be a basis for non-Gaussian generalizations through random volatility models). Nevertheless, four extensions of Toy Model 1 beyond this realm are also accomplished: (i) Gaussian assets but with the EWMA estimator; (ii) Student t distribution of the assets, obtained either from a common random volatility factor or, more realistically, IID temporal volatilities; (iii) free L\'{e}vy distribution, which is a more tractable approximation to the more realistic Wigner-L\'{e}vy distribution of the assets. It is the first step in describing how fat tails impact the spectrum of the time-lagged estimator; applications to more complicated toy models constitute a task for another project.

An addition to the above results is a formulation and numerical tests of two mathematical hypotheses valid for non-Hermitian random matrix models with rotational symmetry in their mean spectrum: (i) the $N$-transform conjecture, which relates the eigenvalues and singular values, thereby being a bridge between Hermitian and non-Hermitian worlds; (ii) the form-factor using the complementary error function which describes the universal decline of the density close to the borderline of its domain.

%%%%%%%%%%%%%%%%%%%%%%%%%%%%%%%%%%%%%%%%%%%%%%%%%%%%%%%%%%%%%%%%%%%%%%
%%%%%%%%%%%%%%%%%%%%%%%%%%%%%%%%%%%%%%%%%%%%%%%%%%%%%%%%%%%%%%%%%%%%%%

\subsection{Open problems}
\label{ss:OpenProblems}

Several paths may be taken from where this paper ends. One major research program would be to repeat the analysis of the mean spectral density for more realistic (say, from the financial point of view) toy models; this may include: (i) Development of numerical methods able to handle the master equations in the Gaussian case but for e.g. the vector autoregression (VAR) models with more involved structure of cross-covariances; the ``vector moving average'' (VMA) models; or a combination of the two called the ``VARMA''~\cite{BurdaJaroszNowakSnarska2010}. (ii) Generalization to weighted estimators, such as the EWMA or with logarithmic weights. (iii) Generalization to non-Gaussian monovariate probability distributions of the assets (e.g., Student or L\'{e}vy) for models with nontrivial covariance structure. (iv) Attempts of incorporating more involved temporal effects, such as the heteroscedasticity, monovariate or index leverage effects, Epps effect etc.

An important part yet to be accomplished is a comparison of the results stemming from some selected toy models with real-world financial data, and learning how to validate them and assess their parameters by fitting. This method may give a profound insight into the structure of intrinsic correlations between financial objects.

Besides, several mathematical problems are worth investigating with a view for a broader understanding of the time-lagged estimator; one could compute: (i) finite-size effects; (ii) statistical properties of the eigenvectors; (iii) the statistics of the ``largest'' (most distant) eigenvalue, such as in Toy Model 4c; (iv) a more detailed analysis of the topological phase transitions in the mean spectral domains.

%%%%%%%%%%%%%%%%%%%%%%%%%%%%%%%%%%%%%%%%%%%%%%%%%%%%%%%%%%%%%%%%%%%%%%
%%%%%%%%%%%%%%%%%%%%%%%%%%%%%%%%%%%%%%%%%%%%%%%%%%%%%%%%%%%%%%%%%%%%%%
%%%%%%%%%%%%%%%%%%%%%%%%%%%%%%%%%%%%%%%%%%%%%%%%%%%%%%%%%%%%%%%%%%%%%%

\begin{acknowledgments}
I express my deep gratitude to Zdzis{\l}aw Burda from Jagiellonian University, Krak\'{o}w, who has suggested to me the topic of the time-lagged covariance estimator and spent many hours discussing these matters with me. I thank Piotr Bo\.{z}ek, Zdzis{\l}aw Burda, Giacomo Livan and Artur \'{S}wi\textpolhook{e}ch for valuable discussions and input in the initial stages of this project. My work has been partially supported by the Polish Ministry of Science and Higher Education Grant ``Iuventus Plus'' No.~0148/H03/2010/70. I acknowledge the financial support of Clico Ltd., Oleandry 2, 30-063 Krak\'{o}w, Poland, during the work on this paper.
\end{acknowledgments}

%%%%%%%%%%%%%%%%%%%%%%%%%%%%%%%%%%%%%%%%%%%%%%%%%%%%%%%%%%%%%%%%%%%%%%
%%%%%%%%%%%%%%%%%%%%%%%%%%%%%%%%%%%%%%%%%%%%%%%%%%%%%%%%%%%%%%%%%%%%%%
%%%%%%%%%%%%%%%%%%%%%%%%%%%%%%%%%%%%%%%%%%%%%%%%%%%%%%%%%%%%%%%%%%%%%%

\appendix

%%%%%%%%%%%%%%%%%%%%%%%%%%%%%%%%%%%%%%%%%%%%%%%%%%%%%%%%%%%%%%%%%%%%%%
%%%%%%%%%%%%%%%%%%%%%%%%%%%%%%%%%%%%%%%%%%%%%%%%%%%%%%%%%%%%%%%%%%%%%%
%%%%%%%%%%%%%%%%%%%%%%%%%%%%%%%%%%%%%%%%%%%%%%%%%%%%%%%%%%%%%%%%%%%%%%

\section{Non-Gaussian probability distributions for financial instruments}
\label{a:NonGaussianProbabilityDistributionsForFinancialInstruments}

This appendix is a short outline of the main features of the probability distributions encountered in finances; they are non-Gaussian and reflect an involved structure of correlations present on the market. On the contrary, in the majority of this paper the Gaussian distribution is assumed---with notable exceptions of Secs.~\ref{sss:TM1TLCEStudent} (Student) and~\ref{sss:TM1TLCELevy} (free L\'{e}vy)---with relatively simple structure of the true correlations. This is because the calculations in this approximation are already very complicated, yet much less than with any of the additional features taken into account; also, Gaussian results can serve as a basis to construct more realistic random volatility models (cf.~Sec.~\ref{aa:RandomVolatilityModels}). This appendix explains what sorts of errors are committed by using the Gaussian distribution, and the necessary means to amend them. For a more thorough introduction to the subject, cf.~especially the textbook~\cite{BouchaudPotters2003}, as well as the reviews~\cite{BouchaudPotters2009,BurdaJaroszJurkiewiczNowakPappZahed2010}.

%%%%%%%%%%%%%%%%%%%%%%%%%%%%%%%%%%%%%%%%%%%%%%%%%%%%%%%%%%%%%%%%%%%%%%
%%%%%%%%%%%%%%%%%%%%%%%%%%%%%%%%%%%%%%%%%%%%%%%%%%%%%%%%%%%%%%%%%%%%%%

\subsection{Student t and free L\'{e}vy distributions}
\label{aa:StudenttAndFreeLevyDistributions}

%%%%%%%%%%%%%%%%%%%%%%%%%%%%%%%%%%%%%%%%%%%%%%%%%%%%%%%%%%%%%%%%%%%%%%

\subsubsection{Weak stability for covariances}
\label{aaa:WeakStabilityForCovariances}

In the statistical approach, one typically makes an assumption of ``weak stability,'' i.e., that the mechanisms governing the market (thus, the probability distributions) do not significantly vary with time over the considered periods, so that historical data could be used to assess future behavior (cf.~Sec.~\ref{sss:MeasurementNoise}). This is rather untrue during the crisis times, and also in attempts to predict future returns (which have highly non-stationary distributions), but is typically justified for covariances (thus, risks), which are the subject of this article. Therefore, the mean trend (``drift'') will be subtracted in all the considerations and fluctuations will be the only focus,
\begin{equation}\label{eq:ZeroMean}
\la R_{i a} \ra = 0 .
\end{equation}

A further comment on these disregarded mean values of the returns: First of all, for short time periods (less than several weeks), the mean return is negligible as compared to its ``volatility'' (a financial term for the standard deviation; e.g., for stocks over one business day, a fraction of a percent versus several percent). Moreover, the returns should have no temporal autocorrelations, since their nonzero value would enable a profitable trading strategy (``arbitrage'') until these autocorrelations would disappear, according to the ``efficient market hypothesis.'' On the other hand, there exist riskless assets (bonds), hence stocks should earn on average at least the riskless rate of return; furthermore, there are factors impacting the mean returns which are not random and quite predictable, such as short-term interest rates. For these reasons, the autocorrelations of returns are actually nonzero, yet weak and over small time lags (less than about $30$ minutes for liquid assets). Note that this does not contradict the efficient market hypothesis---as it may seem, profits could be obtained using the high-frequency (HF) trading---because of transaction costs. They also have an important implication, being one reason for the ``Epps effect'' (cf.~the end of~Sec.~\ref{aaa:MultivariateRandomVolatilityModels}). Some risk assessment methodologies (such as RiskMetrics~2006~\cite{RiskMetrics2006}) attempt to incorporate these effects; however, they will henceforth be neglected.

%%%%%%%%%%%%%%%%%%%%%%%%%%%%%%%%%%%%%%%%%%%%%%%%%%%%%%%%%%%%%%%%%%%%%%

\subsubsection{Fat tails}
\label{aaa:FatTails}

In order to make the discussion from this paper truly practically applicable, one will need to face the fact that the returns are strongly non-Gaussian. It is an observed fact that their unconditional probability distribution over all the time scales between several minutes and several days has ``fat (heavy/Paretian/power-law) tails,''
\begin{equation}\label{eq:FatTails}
\mathcal{P}_{\textrm{unconditional}} \left( R_{\bullet \bullet} \right) \sim \frac{1}{\left| R_{\bullet \bullet} \right|^{1 + \mu}} , \quad R_{\bullet \bullet} \to \pm \infty ,
\end{equation}
where the experimental value of the exponent is $\mu \sim 3 \div 5$~\cite{PlerouGopikrishnanAmaralMeyerStanley1999,GopikrishnanPlerouAmaralMeyerStanley1999}. This reflects considerable chances for extreme events, and also signifies the presence of multi-scale phenomena (i.e., when both very small and very large values appear). For any such distribution, all the moments \smash{$\langle R_{\bullet \bullet}^{n} \rangle$} for $n \geq \mu$ diverge.

%%%%%%%%%%%%%%%%%%%%%%%%%%%%%%%%%%%%%%%%%%%%%%%%%%%%%%%%%%%%%%%%%%%%%%

\subsubsection{Student t distribution}
\label{aaa:StudenttDistribution}

The financial data can be well fitted with the Student t-distribution (cf.~e.g.~\cite{BouchaudPotters2003,BiroliBouchaudPotters2007-1,BurdaGorlichWaclaw2006}),
\begin{equation}\label{eq:MonovariateStudenttDistributionDefinition}
\mathcal{P}_{\textrm{unconditional}} \left( R_{\bullet \bullet} \right) = \frac{1}{\sqrt{\pi}} \frac{\Gamma \left( \frac{1 + \mu}{2} \right)}{\Gamma \left( \frac{\mu}{2} \right)} \frac{\theta^{\mu}}{\left( \theta^{2} + R_{\bullet \bullet}^{2} \right)^{\frac{1 + \mu}{2}}} ,
\end{equation}
where $\Gamma ( \cdot )$ here is the gamma function, and the variance reads \smash{$\sigma^{2} = \theta^{2} / ( \mu - 2 )$}.

%%%%%%%%%%%%%%%%%%%%%%%%%%%%%%%%%%%%%%%%%%%%%%%%%%%%%%%%%%%%%%%%%%%%%%

\subsubsection{Free L\'{e}vy distribution}
\label{aaa:FreeLevyDistribution}

\emph{Wigner-L\'{e}vy random matrices.} Another candidate for a fat-tailed probability distribution of the returns is the L\'{e}vy stable law. Recall that it is described by four parameters, the stability index $\alpha \in ( 0 , 2 ]$, skewness $\beta \in [ - 1 , 1 ]$, range $\gamma > 0$ and location $\delta \in \mathbb{R}$ (which will always besides this Section be set to zero). Its characteristic function (related to the PDF through \smash{$L_{\alpha , \beta , \gamma , \delta} ( x ) = \frac{1}{2 \pi} \int_{- \infty}^{+ \infty} \dd k \ee^{- \ii k x} \hat{L}_{\alpha , \beta , \gamma , \delta} ( k )$}) reads
\begin{equation}
\begin{split}\label{eq:LevyStableDistributionCharacteristicFunction}
&\log \hat{L}_{\alpha , \beta , \gamma , \delta} ( k ) =\\
&= \left\lbrace \begin{array}{ll} - \gamma | k |^{\alpha} \left( 1 - \textrm{sign} ( k ) \ii \beta \tan \frac{\pi \alpha}{2} \right) + \ii \delta k , & \textrm{for } \alpha \neq 1 , \\ - \gamma | k | \left( 1 + \textrm{sign} ( k ) \ii \beta \frac{2}{\pi} \log | k | \right) + \ii \delta k , & \textrm{for } \alpha = 1 . \end{array} \right.
\end{split}
\end{equation}
It has: (i) fat tails, \smash{$L_{\alpha , \beta , \gamma , \delta} ( x ) \sim A_{\pm} / | x |^{1 + \alpha}$} for $x \to \pm \infty$; (ii) all the positive moments divergent for $\alpha \leq 1$, and only the mean finite ($= \delta$) for $\alpha > 1$; (iii) the relative ``weight'' of the right and left tail quantified by skewness, \smash{$( A_{+} - A_{-} ) / ( A_{+} + A_{-} ) = \beta$}; (iv) the scale \smash{$\gamma^{1 / \alpha}$}. Besides, it possesses a host of mathematically interesting features, in particular: (i) It is a fixed point of convolution modulo an affine transformation (dilation and translation), i.e., the PDF of a sum of a large number $T$ of IID random variables is identical to the PDF of the single variable modulo an affine transformation if and only if the variables are Gaussian or L\'{e}vy. The dilation is ``super-diffusive,'' i.e., this sum should be rescaled by \smash{$T^{- 1 / \alpha}$}. (ii) It is $\alpha$-stable, i.e., a sum of $\alpha$-L\'{e}vy (not necessarily independent) remains $\alpha$-L\'{e}vy. (iii) It constitutes the basin of attraction for summing a large number $T$ of random variables (not necessarily independent or identical), rescaled super-diffusively, which have fat tails with the exponent in $( 0 , 2 ]$ [a generalized central limit theorem (CLT)].

In RMT one wants to mimic the above scalar construction while preserving its foundational properties. The most obvious way~\cite{CizeauBouchaud1994} (cf.~also~\cite{BurdaJurkiewicz2009,BurdaJurkiewiczNowakPappZahed2007}) would be to fill a real symmetric $T \times T$ matrix with IID (for $a \leq b$) entries \smash{$h_{a b}$} sampled from some L\'{e}vy stable distribution, rescaled super-diffusively to ensure a finite thermodynamic limit,
\begin{equation}\label{eq:WignerLevyRandomMatrixDefinition}
H^{\mathrm{WL}}_{a b} \equiv \frac{h_{a b}}{T^{1 / \alpha}} ,
\end{equation}
which is dubbed a ``Wigner-L\'{e}vy'' (WL) or ``Bouchaud-Cizeau'' matrix. This may then be extended to complex or rectangular matrices.

\emph{Free L\'{e}vy random matrices.} There exists however another matrix generalization of the L\'{e}vy stable laws in which one asks about matrix ensembles whose MSD is preserved modulo an affine transformation upon summing IID random matrices. To be precise, the notion of statistical independence has been generalized to the noncommuting realm, and called freeness, in Voiculescu-Speicher free probability theory~\cite{VoiculescuDykemaNica1992,Speicher1994}. Moreover, in analogy to the prescription in classical probability theory for summing IID random variables (one should sum the logarithms of the characteristic functions), in order to sum free Hermitian random matrices \smash{$\mathbf{H}_{1}$} and \smash{$\mathbf{H}_{2}$}, one should invert functionally their holomorphic Green functions,
\begin{equation}\label{eq:HolomorphicBlueFunctionDefinition}
G_{\mathbf{H}} \left( B_{\mathbf{H}} ( z ) \right) = B_{\mathbf{H}} \left( G_{\mathbf{H}} ( z ) \right) = z ,
\end{equation}
which is called the ``Blue function'' and which in turn obeys the ``addition law,''
\begin{equation}\label{eq:FreeProbabilityAdditionLaw}
B_{\mathbf{H}_{1} + \mathbf{H}_{2}} ( z ) = B_{\mathbf{H}_{1}} ( z ) + B_{\mathbf{H}_{2}} ( z ) - \frac{1}{z} .
\end{equation}
[Note a close relationship with the multiplication law (\ref{eq:FreeProbabilityMultiplicationLaw}), outlined in Sec.~\ref{sss:Case2ETCE}, which uses the $N$-transform (\ref{eq:HolomorphicNTransformDefinition}) being the functional inverse of the holomorphic $M$-transform. Cf.~also the end of App.~\ref{aaa:NonHermitianGreenFunctions} for its non-Hermitian generalization.]

This new stability condition can then be solved with aid of (\ref{eq:FreeProbabilityAdditionLaw}) leading to the ``free L\'{e}vy stable laws''~\cite{BercoviciVoiculescu1993},
\begin{equation}
\begin{split}\label{eq:LevyStableDistributionCharacteristicFunction}
&B_{\mathbf{L}_{\alpha , \beta , \gamma , \delta}} ( z ) =\\
&= \left\lbrace \begin{array}{ll} \delta + b z^{\alpha - 1} + \frac{1}{z} , & \textrm{for } \alpha \neq 1 , \\ \delta - \ii \gamma ( 1 + \beta ) - \frac{2 \beta \gamma}{\pi} \log ( \gamma z ) + \frac{1}{z} , & \textrm{for } \alpha = 1 , \end{array} \right.
\end{split}
\end{equation}
where the branch structure is chosen such that the upper complex half-plane is mapped to itself, and where for short,
\begin{equation}\label{eq:bDefinition}
b \equiv \left\lbrace \begin{array}{ll} \gamma \ee^{\ii \pi \left( 1 + \frac{\alpha}{2} ( 1 + \beta ) \right)} , & \textrm{for } \alpha < 1 , \\ \gamma \ee^{\ii \pi \left( \frac{\alpha}{2} - 1 \right) ( 1 + \beta )} , & \textrm{for } \alpha > 1 . \end{array} \right.
\end{equation}

The free L\'{e}vy laws are convenient to work with due to their free probability framework, and actually do not substantially differ from the WL distributions~\cite{BurdaJurkiewiczNowakPappZahed2007}. Therefore, in a series of works (cf.~e.g.~\cite{BurdaJanikJurkiewiczNowakPappZahed2002,BurdaJurkiewiczNowakPappZahed2003,BurdaJurkiewiczNowakPappZahed2004,BurdaJurkiewiczNowakPappZahed2001,BenArousGuionnet2008,BelinschiDemboGuionnet2009}), they have been chosen to model the behavior of financial assets. E.g. in~\cite{BurdaJurkiewiczNowakPappZahed2004}, the MSD of the ETCE observed for $N = 406$ stocks from the S\&P 500 market over $T = 1309$ days (i.e., $r \approx 0.31$) has been well fitted with the free L\'{e}vy distribution with $\alpha = 3 / 2$ and $\beta = 0$.

%%%%%%%%%%%%%%%%%%%%%%%%%%%%%%%%%%%%%%%%%%%%%%%%%%%%%%%%%%%%%%%%%%%%%%
%%%%%%%%%%%%%%%%%%%%%%%%%%%%%%%%%%%%%%%%%%%%%%%%%%%%%%%%%%%%%%%%%%%%%%

\subsection{Random volatility models}
\label{aa:RandomVolatilityModels}

%%%%%%%%%%%%%%%%%%%%%%%%%%%%%%%%%%%%%%%%%%%%%%%%%%%%%%%%%%%%%%%%%%%%%%

\subsubsection{Monovariate random volatility models}
\label{aaa:MonovariateRandomVolatilityModels}

However, the above non-Gaussian probability laws are only a part of the picture, since it is well known that the returns at different time moments are not IID, but have different distributions and are correlated with each other. Otherwise, their sum would too quickly converge toward a Gaussian fixed point (according to the CLT for IID variables), contrary to the observation (vide: market crashes). A convenient way to express this is in terms of ``random (stochastic) volatility models''~\cite{BouchaudPotters2003} (a.k.a. ``deformations''~\cite{ToscanoVallejosTsallis2004,BertuolaBohigasPato2004,BohigasDeCarvalhoPato2008,MuttalibKlauder2005,BiroliBouchaudPotters2007-2}), where the return at time $a$,
\begin{equation}\label{eq:MonovariateRandomVolatilityModelDefinition}
R_{a} = \sigma_{a} \epsilon_{a}
\end{equation}
(the index $i$ is removed in this Section), where the standardized returns (``residuals'') \smash{$\epsilon_{a}$} are assumed IID Gaussian $\textrm{N} ( 0 , 1 )$, while the volatilities \smash{$\sigma_{a}$} are also taken to be random, where the JPDF \smash{$\mathcal{P} ( \{ \epsilon_{a} \}, \{ \sigma_{a} \} )$} should reproduce the following ``stylized facts'':

(i) It should yield the proper marginal distribution of the returns. For instance, if all the \smash{$\sigma_{a}$} were independent from each other and from all the \smash{$\epsilon_{a}$} (which is not true, but may be considered a simple toy model), then choosing
\begin{equation}\label{eq:SquareRootInverseGammaDistributionDefinition}
\mathcal{P} ( \sigma ) = \frac{1}{2^{\mu / 2 - 1} \Gamma \left( \frac{\mu}{2} \right)} \frac{\theta^{\mu}}{\sigma^{\mu + 1}} \ee^{- \frac{\theta^{2}}{2 \sigma^{2}}}
\end{equation}
(i.e., the inverse variance has the gamma distribution, which is a two-parameter generalization of the chi-squared distribution) would lead to the Student t-distribution (\ref{eq:MonovariateStudenttDistributionDefinition}).

(ii) The volatility depends on time (in a random way) and displays time-lagged correlations, which are quite small (a few percent; this value depends on the volatility proxy used---e.g., the ``HF proxy,'' being the daily average of HF volatilities---and the time lag), but persisting over long periods of time (``long memory'')---i.e., that high volatilities persist in crisis times and low ones in calm periods---which may be described by a power-law (slow) decay,
\begin{equation}\label{eq:Heteroscedasticity}
\frac{\la \sigma_{a}^{2} \sigma_{b}^{2} \ra - \la \sigma_{a}^{2} \ra \la \sigma_{b}^{2} \ra}{\la \sigma_{a}^{2} \ra \la \sigma_{b}^{2} \ra} \equiv g ( | b - a | ) \sim | b - a |^{- \nu} ,
\end{equation}
where the averaging uses the volatility JPDF, and the exponent $\nu \sim 0.2 \div 0.4$. This phenomenon is called ``heteroscedasticity,'' ``volatility clustering'' or ``intermittence,'' in analogy with the motion of liquids, where also periods of laminar and turbulent flow persist consecutively (cf~e.g.~\cite{BouchaudPotters2003,Zumbach2004,Zumbach2009}).

\emph{Remark 1:} In particular, it implies that estimation based on historical time series is a valid means to assess future risks.

\emph{Remark 2:} Even if the residuals were uncorrelated with the volatilities (which is not true; see below), then the returns would also be uncorrelated, \smash{$\langle R_{a} R_{b} \rangle =$} \smash{$\delta_{a b} \langle \sigma_{a} \sigma_{b} \rangle$}, but they would not be independent, hence there would be nontrivial higher-order correlations, such as \smash{$\langle R_{a}^{2} R_{b}^{2} \rangle -$} \smash{$\langle R_{a}^{2} \rangle \langle R_{b}^{2} \rangle =$} \smash{$g ( | b - a | )$}, for $a \neq b$ (from the Wick theorem).

\emph{Remark 3:} With the heteroscedasticity (\ref{eq:Heteroscedasticity}) present, a sum of a large number $T$ of returns (aggregation with time) still tends to a Gaussian fixed point (CLT), but much slower than for independent variables. One may check this by computing the kurtosis \smash{$\kappa_{T}$} (the normalized fourth moment; it measures a distance of a given distribution from the Gaussian) of this sum, which for large $T$ behaves as
\begin{equation}\label{eq:KurtosisOfTheAggregationOfReturnsWithTheHeteroscedasticityPresent}
\kappa_{T} \sim \left\{ \begin{array}{ll} \frac{1}{T} , & \textrm{for } \nu > 1 , \\ \frac{1}{T^{\nu}} , & \textrm{for } \nu < 1 , \end{array} \right.
\end{equation}
which means that the convergence is slow (several weeks) for long-memory volatility autocorrelations (``anomalous kurtosis'').

\emph{Remark 4:} A common approach to modeling the volatility dynamics is by a stochastic process. Even though a proper way would be to search for models of the mechanisms underlying the volatility process, one often selects a process according to the criterium of computability, which in practice restricts their class to the so-called (integrated) ``auto-regressive conditional heteroscedasticity'' (ARCH) processes~\cite{Engle1982,EngleBollerslev1986,Bollerslev1986}, in which the variance depends linearly on the past squared returns,
\begin{equation}\label{eq:ARCHDefinition}
\sigma_{a}^{2} = \sum_{b = 1}^{T} w_{b} R_{a - b}^{2} ,
\end{equation}
where $T$ here is a lower cutoff, and \smash{$w_{b}$} are some positive weights, satisfying the sum rule, \smash{$\sum_{b = 1}^{T} w_{b} = 1$}. An example broadly used by the financial industry (introduced in RiskMetrics~1994~\cite{RiskMetrics1996,MinaXiao2001}) was the EWMA (\ref{eq:EWMADefinition}) (cf.~e.g.~\cite{PafkaPottersKondor2004,Svensson2007}), \smash{$w_{b} \propto$} \smash{$\exp ( - b / \tau )$}, where one parameter is used for all the assets, \smash{$\tau \delta t = 16.2$} business days. In RiskMetrics~2006, it was argued that a better fit is given by a logarithmic decay, \smash{$w_{b} \propto$} \smash{$1 - \log b / \log \tau$}, with \smash{$\tau \delta t \sim 3 \div 6$} years.

(iii) The ``leverage effect''~\cite{GlostenJagannathanRunkle1993,BekaertWu2000,BouchaudMataczPotters2001}, i.e., the empirical fact that negative past returns typically increase future volatilities, which means that \smash{$\epsilon_{a}$} and \smash{$\sigma_{a}$} are correlated in such a way that
\begin{equation}\label{eq:LeverageEffect}
\la \epsilon_{a} \sigma_{b} \ra < 0 , \quad \textrm{for } b > a .
\end{equation}

%%%%%%%%%%%%%%%%%%%%%%%%%%%%%%%%%%%%%%%%%%%%%%%%%%%%%%%%%%%%%%%%%%%%%%

\subsubsection{Multivariate random volatility models}
\label{aaa:MultivariateRandomVolatilityModels}

The above experimental constraints on the statistics of a single return are both very nontrivial to implement in such a way that the resulting model would be mathematically tractable, and also leave plenty of freedom for model specification. To make things even tougher, this is still an incomplete description, as there are many correlated assets on the market,
\begin{equation}\label{eq:MultivariateRandomVolatilityModelDefinition}
R_{i a} = \sigma_{i a} \epsilon_{i a}
\end{equation}
(``multivariate random volatility model''), and one has to select a JPDF which will additionally reproduce their observed dependence structure.

\emph{Multivariate Gaussian returns.} This task is straightforward only if the returns are Gaussian (which, as explained above, is quite far from truth), since the most general (real) multivariate Gaussian distribution (of zero mean) is fully described by the two-point covariance function,
\begin{equation}\label{eq:RealMultivariateGaussianCovarianceFunctionDefinition}
\mathcal{C}_{i j , a b} \equiv \la R_{i a} R_{j b} \ra .
\end{equation}

\emph{Pseudo-elliptical and elliptical models; multivariate Student returns.} This suggests a way to introduce a dependence structure between the assets using the form (\ref{eq:MultivariateRandomVolatilityModelDefinition}) (the index $a$ is removed in the following discussion, as it pertains to the cross-correlations only)---by supposing that the Gaussian residuals are cross-correlated, \smash{$\langle \epsilon_{i} \epsilon_{j} \rangle =$} \smash{$C_{i j}$}, and multiplying them with some random volatilities \smash{$\sigma_{i}$}, perhaps also cross-correlated, but independent of the residuals---which leads to the class of ``pseudo-elliptical models.''

A commonly used example (``elliptical models'') is to take a single volatility for all the assets, \smash{$R_{i} = \sigma \epsilon_{i}$}. For instance, if the volatility is distributed according to (\ref{eq:SquareRootInverseGammaDistributionDefinition}) (with \smash{$\theta^{2} = 2 \mu$}), this yields the ``multivariate Student t-distribution'' of the (simultaneous) returns,
\begin{equation}
\begin{split}\label{eq:MultivariateStudenttDistributionDefinition}
\mathcal{P} ( \{ R_{i} \} ) &= \frac{1}{\sqrt{( \pi \mu )^{N} \Det \mathbf{C}}} \frac{\Gamma \left( \frac{N + \mu}{2} \right)}{\Gamma \left( \frac{\mu}{2} \right)} \cdot\\
&\cdot \frac{1}{\left( 1 + \frac{1}{\mu} \sum_{i , j = 1}^{N} R_{i} [ \mathbf{C}^{- 1} ]_{i j} R_{j} \right)^{\frac{N + \mu}{2}}} .
\end{split}
\end{equation}
In this case, the matrix $\mathbf{C}$ is sufficient to describe all the correlation functions: the two-point one, \smash{$\langle R_{i} R_{j} \rangle =$} \smash{$\frac{\mu}{\mu - 2} C_{i j}$}, as well as the higher ones through the Wick theorem. However, it has been argued~\cite{ChicheporticheBouchaud2010} that the JPDF of returns is not elliptical, and several volatility factors need to be exploited.

\emph{Copulas.} Let it also be added that for more general non-Gaussian distributions, the two-point covariance function is typically insufficient (because of higher-order correlations), invalid (because it diverges), or uninteresting (e.g., because one is more concerned about correlations of extreme events). There is a method---based on the Sklar theorem~\cite{Sklar1959}---which factorizes the problem of defining multivariate non-Gaussian distributions into a specification of marginal distributions [such as (\ref{eq:MonovariateStudenttDistributionDefinition})] and a description of dependence: (i) One transforms the returns \smash{$R_{i}$} into \smash{$U_{i} \equiv$} \smash{$\mathcal{P}_{< , i} ( R_{i} )$}, where \smash{$\mathcal{P}_{< , i} ( \cdot )$} is the marginal cumulative distribution function (CDF) of asset $i$; these new random variables are uniformly distributed on $[ 0 , 1 ]$. (ii) Their joint CDF,
\begin{equation}\label{eq:CopulaDefinition}
\mathcal{C} \left( \{ u_{i} \} \right) \equiv \textrm{Prob} \left( \{ U_{i} \leq u_{i} \} \right) ,
\end{equation}
is named the ``copula,'' and is responsible for the dependence of the returns. This framework includes a number of known measures of dependence used in the presence of fat tails, such as the Kendall tau, Spearman rho or tail dependence. A copula has to be chosen so as to reproduce the empirical two-point covariance function \smash{$\langle R_{i} R_{j} \rangle$}, which again is a greatly underconstrained task (``copula specification problem''), leading to the choices based rather on mathematical computability, and not founded upon a solid financial reality (cf.~e.g.~\cite{EmbrechtsLindskogMcNeil2001,DoreyJoubert2007}).

\emph{Temporal cross-correlations.} From the above discussion, it is quite obvious that disregarding temporal correlations---not only the autocorrelations of any single asset but between distinct assets at different time moments---is a serious simplification. There are at least two important stylized facts which require incorporation of the temporal cross-correlations:

(i) The ``index (correlation) leverage effect''~\cite{BaloghSimonsenNagyNeda2010,BorlandHassid2010,ReigneronAllezBouchaud2010}---not only do the volatilities increase upon large negative past returns [cf.~point (iii) in Sec.~\ref{aaa:MonovariateRandomVolatilityModels}], but so do the cross-correlations (``cross-correlations jump to one in crisis periods''). This is reflected by the fact that the leverage correlation (\ref{eq:LeverageEffect}) for market indices (i.e., the averages of the returns of all the stocks) is stronger than for separate stocks---the additional contribution originating precisely from the cross-correlations.

(ii) The ``Epps effect''~\cite{Epps1979}---the equal-time cross-covariance \smash{$\langle R_{i a} R_{j a} \rangle$} grows significantly as one increases the observation time lag $\delta t$ between about $5$ and $30$ minutes, then more slowly, finally to saturate after a few days (cf.~also~\cite{BonannoVandewalleMantegna2000}). It can be explained~\cite{TothTothKertesz2007,TothKertesz2007} by relating \smash{$\langle R_{i a} R_{j a} \rangle$} to the time-lagged cross-covariances \smash{$\langle R_{i a} R_{j , a + t} \rangle$} on shorter time scales $\delta t$, and proposing reasonable toy models of the latter, which reproduces the ``Epps curve,'' in agreement with the experiment. It is an important feature of the structure of cross-covariances, and also from the point of view of their historical measurement (cf.~Sec.~\ref{sss:MeasurementNoise}).

%%%%%%%%%%%%%%%%%%%%%%%%%%%%%%%%%%%%%%%%%%%%%%%%%%%%%%%%%%%%%%%%%%%%%%
%%%%%%%%%%%%%%%%%%%%%%%%%%%%%%%%%%%%%%%%%%%%%%%%%%%%%%%%%%%%%%%%%%%%%%
%%%%%%%%%%%%%%%%%%%%%%%%%%%%%%%%%%%%%%%%%%%%%%%%%%%%%%%%%%%%%%%%%%%%%%

\section{Gaussian planar diagrammatic expansion}
\label{a:GaussianPlanarDiagrammaticExpansion}

This appendix contains (i) an introduction of the basic notions of random matrix theory in whose language the results of this paper are expressed (Hermitian case in App.~\ref{aaa:HermitianGreenFunction}, non-Hermitian in App.~\ref{aaa:NonHermitianGreenFunctions}); (ii) a sketch of the main technique used to obtain these results, i.e., the Gaussian planar diagrammatic expansion (Hermitian case in App.~\ref{aaa:DSAndGUE}, non-Hermitian in App.~\ref{aaa:DSAndGinUE}).

%%%%%%%%%%%%%%%%%%%%%%%%%%%%%%%%%%%%%%%%%%%%%%%%%%%%%%%%%%%%%%%%%%%%%%
%%%%%%%%%%%%%%%%%%%%%%%%%%%%%%%%%%%%%%%%%%%%%%%%%%%%%%%%%%%%%%%%%%%%%%

\subsection{Hermitian random matrices}
\label{aa:HermitianRandomMatrices}

%%%%%%%%%%%%%%%%%%%%%%%%%%%%%%%%%%%%%%%%%%%%%%%%%%%%%%%%%%%%%%%%%%%%%%

\subsubsection{Hermitian Green function}
\label{aaa:HermitianGreenFunction}

Consider an $N \times N$ Hermitian random matrix $\mathbf{H}$, with a corresponding probability measure $\dd \mu ( \mathbf{H} )$, in the large-$N$ limit [cf.~(\ref{eq:ThermodynamicLimit})]. One of the basic characteristics of this model is the ``mean spectral density'' (MSD), i.e., the distribution of its (real) eigenvalues \smash{$\lambda_{i}$}, averaged according to $\langle \ldots \rangle \equiv \int ( \ldots ) \dd \mu ( \mathbf{H} )$,
\begin{equation}\label{eq:HermitianMSDDefinition}
\rho_{\mathbf{H}} ( x ) \equiv \frac{1}{N} \sum_{i = 1}^{N} \la \delta \left( x - \lambda_{i} \right) \ra ,
\end{equation}
where $\delta ( x )$ is the real Dirac delta function.

However, there exists an equivalent but handier object which comes in three versions:

(i) The $N \times N$ ``holomorphic Green function matrix,''
\begin{equation}\label{eq:HolomorphicGreenFunctionMatrixDefinition}
\mathbf{G}_{\mathbf{H}} ( z ) \equiv \la \frac{1}{\mathbf{Z} - \mathbf{H}} \ra , \quad \textrm{where} \quad \mathbf{Z} \equiv z \Id_{N} ,
\end{equation}
where $z \equiv x + \ii y$ is a complex argument.

(ii) Its normalized trace is known as the ``holomorphic Green function'' (other names: ``resolvent,'' ``Stjeltjes transform''),
\begin{equation}\label{eq:HolomorphicGreenFunctionDefinition}
G_{\mathbf{H}} ( z ) \equiv \frac{1}{N} \Tr \mathbf{G}_{\mathbf{H}} ( z ) = \frac{1}{N} \sum_{i = 1}^{N} \la \frac{1}{z - \lambda_{i}} \ra .
\end{equation}

(iii) In some situations an even more convenient quantity is the ``holomorphic $M$-transform,''
\begin{equation}\label{eq:HolomorphicMTransformDefinition}
M_{\mathbf{H}} ( z ) \equiv z G_{\mathbf{H}} ( z ) - 1 ,
\end{equation}
whose expansion around $z = \infty$ (if it exists) is a power series of the (positive) ``moments,''
\begin{subequations}
\begin{align}
M_{\mathbf{H}} ( z ) &= \sum_{n \geq 1} \frac{m_{\mathbf{H} , n}}{z^{n}} ,\label{eq:HolomorphicMTransformMomentExpansion}\\
m_{\mathbf{H} , n} &\equiv \frac{1}{N} \Tr \la \mathbf{H}^{n} \ra = \int_{\textrm{cuts}} \dd x \rho_{\mathbf{H}} ( x ) x^{n} .\label{eq:MomentsDefinition}
\end{align}
\end{subequations}

The relationship of the holomorphic Green function to the MSD is the following:

(i) For any finite $N$, the former is a meromorphic function, with poles at the mean eigenvalues; but when $N$ grows to infinity, these poles merge into continuous intervals on the $x$-axis, distributed according to \smash{$\rho_{\mathbf{H}} ( x )$}, and thus \smash{$G_{\mathbf{H}} ( z )$} becomes holomorphic on the whole complex plane except these real ``cuts.'' Then, the MSD can be used to calculate the holomorphic Green function through
\begin{equation}\label{eq:HolomorphicGreenFunctionFromMSD}
G_{\mathbf{H}} ( z ) = \int_{\textrm{cuts}} \dd x \rho_{\mathbf{H}} ( x ) \frac{1}{z - x} .
\end{equation}

\begin{figure}[ht]
\includegraphics[width=\columnwidth]{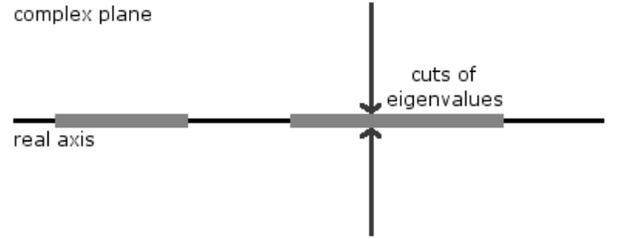}
\caption{For Hermitian random matrices, the real MSD is evaluated from the complex Green function of a complex argument in the vicinity of the real axis (\ref{eq:MSDFromHolomorphicGreenFunction}).}
\label{fig:DensityFromGreenFunctionForRealSpectrum}
\end{figure}

(ii) On the other hand, since the real Dirac delta is known to have a representation (the ``Sokhotsky-Weierstrass formula''),
\begin{equation}\label{eq:SokhotskyWeierstrassFormula}
\delta ( x ) = - \frac{1}{2 \pi \ii} \lim_{\epsilon \to 0^{+}} \left( \frac{1}{x + \ii \epsilon} - \frac{1}{x - \ii \epsilon} \right) ,
\end{equation}
one infers also an opposite link,
\begin{equation}\label{eq:MSDFromHolomorphicGreenFunction}
\rho_{\mathbf{H}} ( x ) = - \frac{1}{2 \pi \ii} \lim_{\epsilon \to 0^{+}} \left( G_{\mathbf{H}} ( x + \ii \epsilon ) - G_{\mathbf{H}} ( x - \ii \epsilon ) \right) ,
\end{equation}
i.e., the MSD is the jump of the holomorphic Green function across the cuts. This is schematically illustrated in Fig.~\ref{fig:DensityFromGreenFunctionForRealSpectrum}.

%%%%%%%%%%%%%%%%%%%%%%%%%%%%%%%%%%%%%%%%%%%%%%%%%%%%%%%%%%%%%%%%%%%%%%

\subsubsection{Dyson-Schwinger equations and Gaussian unitary ensemble}
\label{aaa:DSAndGUE}

There exist various techniques of calculating the holomorphic Green function of a given Hermitian random matrix model $\mathbf{H}$. In the following, the method of ``planar diagrammatic expansion'' and ``Dyson-Schwinger (DS) equations''~\cite{FeinbergZee1997-03,MitraSengupta1999} will be introduced and applied as an example to solve a simple model, the ``Gaussian unitary ensemble'' (GUE),
\begin{equation}\label{eq:GUEMeasureDefinition}
\dd \mu ( \mathbf{H} ) \propto \ee^{- \frac{N}{2 \sigma^{2}} \Tr \mathbf{H}^{2}} \dd \mathbf{H} ,
\end{equation}
i.e., such that all \smash{$\re H_{i j}$}, \smash{$\im H_{i j}$} for $i \leq j$ are independent Gaussian of zero mean and variance \smash{$\sigma^{2} / N$} (for $i = j$) or \smash{$\sigma^{2} / ( 2 N )$} (for $i < j$); moreover \smash{$\dd \mathbf{H} \equiv \prod_{i} \dd H_{i i} \prod_{i < j} \dd ( \re H_{i j} ) \dd ( \im H_{i j} )$} is the flat measure.

To begin with, the holomorphic Green function matrix (\ref{eq:HolomorphicGreenFunctionMatrixDefinition}) should be expanded around $z = \infty$,
\begin{equation}
\begin{split}\label{eq:HolomorphicGreenFunctionMatrixExpansion}
\mathbf{G}_{\mathbf{H}} ( z ) &= \mathbf{Z}^{- 1} + \la \mathbf{Z}^{- 1} \mathbf{H} \mathbf{Z}^{- 1} \mathbf{H} \mathbf{Z}^{- 1} \ra +\\
&+ \la \mathbf{Z}^{- 1} \mathbf{H} \mathbf{Z}^{- 1} \mathbf{H} \mathbf{Z}^{- 1} \mathbf{H} \mathbf{Z}^{- 1} \mathbf{H} \mathbf{Z}^{- 1} \ra + \ldots ,
\end{split}
\end{equation}
where only the even moments have been kept due to the evenness of the GUE measure (\ref{eq:GUEMeasureDefinition}).

Therefore, in order to sum up (\ref{eq:HolomorphicGreenFunctionMatrixExpansion}), one needs to calculate at each (even) order $n$ the averages ($n$-point correlation functions) \smash{$\langle [ \mathbf{H} ]_{k_{1} k_{2}} [ \mathbf{H} ]_{k_{3} k_{4}} \ldots [ \mathbf{H} ]_{k_{2 n - 1} k_{2 n}} \rangle$}. According to the Wick theorem, this is done by making all the contractions, i.e., forming all the possible pairings of the $\mathbf{H}$-factors, and replacing each of them by an appropriate $2$-point correlation function (propagator), \smash{$\langle [ \mathbf{H} ]_{k_{p} k_{p + 1}} [ \mathbf{H} ]_{k_{q} k_{q + 1}} \rangle$}. The GUE propagator is inferred from (\ref{eq:GUEMeasureDefinition}) to be
\begin{equation}\label{eq:GUEPropagator}
\la [ \mathbf{H} ]_{i j} [ \mathbf{H} ]_{k l} \ra = \frac{\sigma^{2}}{N} \delta_{i l} \delta_{j k} .
\end{equation}
This procedure is best explained using a pictorial language,

\setlength{\unitlength}{0.8pt}
\begin{picture}(250,80)(0,-40)
\thicklines
\put(5,0){\circle*{6}}
\put(3,-13){$i$}
\put(45,0){\circle*{6}}
\put(43,-13){$j$}
\put(5,0){\line(1,0){40}}
\put(55,0){$\equiv \left[ \mathbf{Z}^{- 1} \right]_{i j} = \frac{1}{z} \delta_{i j}$,}
\end{picture}

\begin{picture}(250,80)(0,-40)
\thicklines
\put(5,0){\circle*{6}}
\put(3,-13){$i$}
\put(15,0){\circle*{6}}
\put(13,-13){$j$}
\put(75,0){\circle*{6}}
\put(73,-13){$k$}
\put(85,0){\circle*{6}}
\put(83,-13){$l$}
\put(45,0){\arc{80}{3.1416}{6.2832}}
\put(45,0){\arc{60}{3.1416}{6.2832}}
\put(95,20){$\equiv \la [ \mathbf{H} ]_{i j} [ \mathbf{H} ]_{k l} \ra = \frac{\sigma^{2}}{N} \delta_{i l} \delta_{j k}$,}
\end{picture}
\\since then at each order in (\ref{eq:HolomorphicGreenFunctionMatrixExpansion}), written in terms of matrix indices, one obtains a set of alternating horizontal segments (\smash{$\mathbf{Z}^{- 1}$}) and double dots ($\mathbf{H}$), where the pairs of double dots should be connected with double arcs (propagators)---this series for $n = 0 , 2 , 4$ reads

\begin{picture}(250,80)(0,-40)
\thicklines
\put(5,0){\circle*{6}}
\put(3,-13){$i$}
\put(85,0){\circle*{6}}
\put(83,-13){$j$}
\put(5,0){\line(1,0){20}}
\put(85,0){\line(-1,0){20}}
\filltype{shade}
\put(45,0){\circle*{40}}
\filltype{black}
\put(41,-3){{\color{white}$\mathbf{G}$}}
\put(25,-35){$[ \mathbf{G}_{\mathbf{H}} ( z ) ]_{i j}$}
\put(97,0){$=$}
\put(115,0){\circle*{6}}
\put(113,-13){$i$}
\put(155,0){\circle*{6}}
\put(153,-13){$j$}
\put(115,0){\line(1,0){40}}
\put(167,0){$+$}
\end{picture}

\begin{picture}(250,80)(0,-40)
\thicklines
\put(5,0){$+$}
\put(23,0){\circle*{6}}
\put(21,-13){$i$}
\put(63,0){\circle*{6}}
\put(60,-13){$k_{1}$}
\put(73,0){\circle*{6}}
\put(70,-13){$k_{2}$}
\put(113,0){\circle*{6}}
\put(110,-13){$k_{3}$}
\put(123,0){\circle*{6}}
\put(120,-13){$k_{4}$}
\put(163,0){\circle*{6}}
\put(161,-13){$j$}
\put(23,0){\line(1,0){40}}
\put(73,0){\line(1,0){40}}
\put(123,0){\line(1,0){40}}
\put(93,0){\arc{40}{3.1416}{6.2832}}
\put(93,0){\arc{60}{3.1416}{6.2832}}
\put(53,-35){$\la [ \rnode{1}{\mathbf{H}} ]_{k_{1} k_{2}} [ \rnode{2}{\mathbf{H}} ]_{k_{3} k_{4}} \ra \ncbar[linewidth=0.01,nodesep=2pt,arm=0.15,angle=90]{-}{1}{2}$}
\put(175,0){$+$}
\end{picture}

\begin{picture}(250,80)(0,-40)
\thicklines
\put(5,0){$+$}
\put(23,0){\circle*{6}}
\put(21,-13){$i$}
\put(63,0){\circle*{6}}
\put(60,-13){$k_{1}$}
\put(73,0){\circle*{6}}
\put(70,-13){$k_{2}$}
\put(113,0){\circle*{6}}
\put(110,-13){$k_{3}$}
\put(123,0){\circle*{6}}
\put(120,-13){$k_{4}$}
\put(163,0){\circle*{6}}
\put(160,-13){$k_{5}$}
\put(173,0){\circle*{6}}
\put(170,-13){$k_{6}$}
\put(213,0){\circle*{6}}
\put(210,-13){$k_{7}$}
\put(223,0){\circle*{6}}
\put(220,-13){$k_{8}$}
\put(263,0){\circle*{6}}
\put(261,-13){$j$}
\put(23,0){\line(1,0){40}}
\put(73,0){\line(1,0){40}}
\put(123,0){\line(1,0){40}}
\put(173,0){\line(1,0){40}}
\put(223,0){\line(1,0){40}}
\put(93,0){\arc{40}{3.1416}{6.2832}}
\put(93,0){\arc{60}{3.1416}{6.2832}}
\put(193,0){\arc{40}{3.1416}{6.2832}}
\put(193,0){\arc{60}{3.1416}{6.2832}}
\put(78,-35){$\la [ \rnode{1}{\mathbf{H}} ]_{k_{1} k_{2}} [ \rnode{2}{\mathbf{H}} ]_{k_{3} k_{4}} [ \rnode{3}{\mathbf{H}} ]_{k_{5} k_{6}} [ \rnode{4}{\mathbf{H}} ]_{k_{7} k_{8}} \ra \ncbar[linewidth=0.01,nodesep=2pt,arm=0.15,angle=90]{-}{1}{2} \ncbar[linewidth=0.01,nodesep=2pt,arm=0.15,angle=90]{-}{3}{4}$}
\put(275,0){$+$}
\end{picture}

\begin{picture}(250,130)(0,-40)
\thicklines
\put(5,0){$+$}
\put(23,0){\circle*{6}}
\put(21,-13){$i$}
\put(63,0){\circle*{6}}
\put(60,-13){$k_{1}$}
\put(73,0){\circle*{6}}
\put(70,-13){$k_{2}$}
\put(113,0){\circle*{6}}
\put(110,-13){$k_{3}$}
\put(123,0){\circle*{6}}
\put(120,-13){$k_{4}$}
\put(163,0){\circle*{6}}
\put(160,-13){$k_{5}$}
\put(173,0){\circle*{6}}
\put(170,-13){$k_{6}$}
\put(213,0){\circle*{6}}
\put(210,-13){$k_{7}$}
\put(223,0){\circle*{6}}
\put(220,-13){$k_{8}$}
\put(263,0){\circle*{6}}
\put(261,-13){$j$}
\put(23,0){\line(1,0){40}}
\put(73,0){\line(1,0){40}}
\put(123,0){\line(1,0){40}}
\put(173,0){\line(1,0){40}}
\put(223,0){\line(1,0){40}}
\put(143,0){\arc{140}{3.1416}{6.2832}}
\put(143,0){\arc{160}{3.1416}{6.2832}}
\put(143,0){\arc{40}{3.1416}{6.2832}}
\put(143,0){\arc{60}{3.1416}{6.2832}}
\put(78,-35){$\la [ \rnode{1}{\mathbf{H}} ]_{k_{1} k_{2}} [ \rnode{2}{\mathbf{H}} ]_{k_{3} k_{4}} [ \rnode{3}{\mathbf{H}} ]_{k_{5} k_{6}} [ \rnode{4}{\mathbf{H}} ]_{k_{7} k_{8}} \ra \ncbar[linewidth=0.01,nodesep=2pt,arm=0.25,angle=90]{-}{1}{4} \ncbar[linewidth=0.01,nodesep=2pt,arm=0.15,angle=90]{-}{2}{3}$}
\put(275,0){$+$}
\end{picture}

\begin{picture}(250,130)(0,-65)
\thicklines
\put(5,0){$+$}
\put(23,0){\circle*{6}}
\put(21,-13){$i$}
\put(63,0){\circle*{6}}
\put(60,-13){$k_{1}$}
\put(73,0){\circle*{6}}
\put(70,-13){$k_{2}$}
\put(113,0){\circle*{6}}
\put(110,-13){$k_{3}$}
\put(123,0){\circle*{6}}
\put(120,-13){$k_{4}$}
\put(163,0){\circle*{6}}
\put(160,-13){$k_{5}$}
\put(173,0){\circle*{6}}
\put(170,-13){$k_{6}$}
\put(213,0){\circle*{6}}
\put(210,-13){$k_{7}$}
\put(223,0){\circle*{6}}
\put(220,-13){$k_{8}$}
\put(263,0){\circle*{6}}
\put(261,-13){$j$}
\put(23,0){\line(1,0){40}}
\put(73,0){\line(1,0){40}}
\put(123,0){\line(1,0){40}}
\put(173,0){\line(1,0){40}}
\put(223,0){\line(1,0){40}}
\put(113,0){\arc{100}{3.1416}{6.2832}}
\put(123,0){\arc{100}{3.1416}{6.2832}}
\put(163,0){\arc{100}{3.1416}{6.2832}}
\put(173,0){\arc{100}{3.1416}{6.2832}}
\put(78,-35){$\la [ \rnode{1}{\mathbf{H}} ]_{k_{1} k_{2}} [ \rnode{2}{\mathbf{H}} ]_{k_{3} k_{4}} [ \rnode{3}{\mathbf{H}} ]_{k_{5} k_{6}} [ \rnode{4}{\mathbf{H}} ]_{k_{7} k_{8}} \ra \ncbar[linewidth=0.01,nodesep=2pt,arm=0.15,angle=90]{-}{1}{3} \ncbar[linewidth=0.01,nodesep=2pt,arm=0.25,angle=90]{-}{2}{4}$}
{\color{gray}
\put(0,0){\path(25,60)(256,-60)}
\put(0,0){\path(25,-60)(256,60)}
\put(40,-60){non-planar}}
\put(275,0){$+$}
\end{picture}

\begin{picture}(250,80)(0,-40)
\thicklines
\put(5,0){$+ \ldots$.}
\end{picture}
\\(The $k$-indices are summed over.) Hence, (i) at $n = 0$, there is just one \smash{$\mathbf{Z}^{- 1}$} (one horizontal line); (ii) at $n = 2$, there are three \smash{$\mathbf{Z}^{- 1}$} (three horizontal lines) and two $\mathbf{H}$ (two double dots), so there is one possible contraction into one propagator (one double arc); (iii) at $n = 4$, one finds five \smash{$\mathbf{Z}^{- 1}$} and four $\mathbf{H}$, so there are three possible contractions into two propagators (two double arcs); one of them yields a non-planar graph; (iv) etc.

One might worry that all the above infinite series of graphs would have to be computed in order to evaluate (\ref{eq:HolomorphicGreenFunctionMatrixExpansion}), but in the large-$N$ limit there exists a powerful alternative---infinite $N$ is the 't Hooft limit in which only the planar diagrams contribute to the sum~\cite{tHooft1974}, and these can be effectively summed as follows:

Consider the ``one-line-irreducible'' (1LI) diagrams (i.e., those that cannot by disconnected into two sub-graphs by cutting a single horizontal line), and let \smash{$\boldsymbol{\Sigma}_{\mathbf{H}} ( z )$} (the $N \times N$ ``self-energy matrix'') be their generating function,

\begin{picture}(250,80)(0,-40)
\thicklines
\filltype{shade}
\put(45,0){\ellipse*{80}{40}}
\filltype{black}
\put(5,0){\circle*{6}}
\put(85,0){\circle*{6}}
\put(41,-3){{\color{white}$\boldsymbol{\Sigma}$}}
\put(97,0){$\equiv$}
\put(115,0){\circle*{6}}
\put(125,0){\circle*{6}}
\put(165,0){\circle*{6}}
\put(175,0){\circle*{6}}
\put(125,0){\line(1,0){40}}
\put(145,0){\arc{40}{3.1416}{6.2832}}
\put(145,0){\arc{60}{3.1416}{6.2832}}
\put(187,0){$+$}
\end{picture}

\begin{picture}(250,130)(0,-40)
\thicklines
\put(5,0){$+$}
\put(23,0){\circle*{6}}
\put(33,0){\circle*{6}}
\put(73,0){\circle*{6}}
\put(83,0){\circle*{6}}
\put(123,0){\circle*{6}}
\put(133,0){\circle*{6}}
\put(173,0){\circle*{6}}
\put(183,0){\circle*{6}}
\put(33,0){\line(1,0){40}}
\put(83,0){\line(1,0){40}}
\put(133,0){\line(1,0){40}}
\put(103,0){\arc{40}{3.1416}{6.2832}}
\put(103,0){\arc{60}{3.1416}{6.2832}}
\put(103,0){\arc{140}{3.1416}{6.2832}}
\put(103,0){\arc{160}{3.1416}{6.2832}}
\put(195,0){$+ \ldots$.}
\end{picture}

This subset is related to all the diagrams in a twofold way:

(i) On one hand, any planar graph has necessarily a form of a number of 1LI graphs connected with horizontal segments,

\begin{picture}(250,60)(0,-30)
\thicklines
\put(5,0){\circle*{6}}
\put(85,0){\circle*{6}}
\put(5,0){\line(1,0){20}}
\put(85,0){\line(-1,0){20}}
\filltype{shade}
\put(45,0){\circle*{40}}
\filltype{black}
\put(41,-3){{\color{white}$\mathbf{G}$}}
\put(97,0){$=$}
\put(115,0){\circle*{6}}
\put(155,0){\circle*{6}}
\put(115,0){\line(1,0){40}}
\put(167,0){$+$}
\end{picture}

\begin{picture}(250,60)(0,-30)
\thicklines
\put(5,0){$+$}
\filltype{shade}
\put(103,0){\ellipse*{80}{40}}
\filltype{black}
\put(23,0){\circle*{6}}
\put(63,0){\circle*{6}}
\put(143,0){\circle*{6}}
\put(183,0){\circle*{6}}
\put(23,0){\line(1,0){40}}
\put(143,0){\line(1,0){40}}
\put(99,-3){{\color{white}$\boldsymbol{\Sigma}$}}
\put(195,0){$+$}
\end{picture}

\begin{picture}(250,60)(0,-30)
\thicklines
\put(5,0){$+$}
\filltype{shade}
\put(103,0){\ellipse*{80}{40}}
\put(223,0){\ellipse*{80}{40}}
\filltype{black}
\put(23,0){\circle*{6}}
\put(63,0){\circle*{6}}
\put(143,0){\circle*{6}}
\put(183,0){\circle*{6}}
\put(263,0){\circle*{6}}
\put(303,0){\circle*{6}}
\put(23,0){\line(1,0){40}}
\put(143,0){\line(1,0){40}}
\put(263,0){\line(1,0){40}}
\put(99,-3){{\color{white}$\boldsymbol{\Sigma}$}}
\put(219,-3){{\color{white}$\boldsymbol{\Sigma}$}}
\end{picture}

\begin{picture}(250,60)(0,-30)
\put(5,0){$+ \ldots$,}
\end{picture}
\\i.e., symbolically,
\begin{equation}
\begin{split}\label{eq:DS1}
\mathbf{G}_{\mathbf{H}} ( z ) &=\\
&= \mathbf{Z}^{- 1} + \mathbf{Z}^{- 1} \boldsymbol{\Sigma}_{\mathbf{H}} ( z ) \mathbf{Z}^{- 1} +\\
&+ \mathbf{Z}^{- 1} \boldsymbol{\Sigma}_{\mathbf{H}} ( z ) \mathbf{Z}^{- 1} \boldsymbol{\Sigma}_{\mathbf{H}} ( z ) \mathbf{Z}^{- 1} + \ldots =\\
&= \frac{1}{\mathbf{Z} - \boldsymbol{\Sigma}_{\mathbf{H}} ( z )} ,
\end{split}
\end{equation}
which is the ``first Dyson-Schwinger equation.''

(ii) On the other hand, for Gaussian random matrices and in the planar limit, it is true that any 1LI diagram originates from some other diagram by adding to it an external double arc,

\begin{picture}(250,100)(0,-35)
\thicklines
\filltype{shade}
\put(45,0){\ellipse*{80}{40}}
\filltype{black}
\put(5,0){\circle*{6}}
\put(3,-13){$i$}
\put(85,0){\circle*{6}}
\put(83,-13){$j$}
\put(41,-3){{\color{white}$\boldsymbol{\Sigma}$}}
\put(97,0){$=$}
\put(115,0){\circle*{6}}
\put(113,-13){$i$}
\put(125,0){\circle*{6}}
\put(122,-13){$k_{1}$}
\put(205,0){\circle*{6}}
\put(202,-13){$k_{2}$}
\put(215,0){\circle*{6}}
\put(213,-13){$j$}
\put(125,0){\line(1,0){20}}
\put(205,0){\line(-1,0){20}}
\filltype{shade}
\put(165,0){\circle*{40}}
\filltype{black}
\put(161,-3){{\color{white}$\mathbf{G}$}}
\put(165,0){\arc{80}{3.1416}{6.2832}}
\put(165,0){\arc{100}{3.1416}{6.2832}}
\end{picture}
\\i.e.,
\begin{equation}\label{eq:DS2}
[ \boldsymbol{\Sigma}_{\mathbf{H}} ( z ) ]_{i j} = \sum_{k_{1} , k_{2} = 1}^{N} [ \mathbf{G}_{\mathbf{H}} ( z ) ]_{k_{1} k_{2}} \la [ \mathbf{H} ]_{i k_{1}} [ \mathbf{H} ]_{k_{2} j} \ra ,
\end{equation}
which is the ``second Dyson-Schwinger equation.'' For instance, inserting here the GUE propagator (\ref{eq:GUEPropagator}),
\begin{equation}\label{eq:DS2ForGUE}
\boldsymbol{\Sigma}_{\mathbf{GUE}} ( z ) = \sigma^{2} G_{\mathbf{GUE}} ( z ) \Id_{N} .
\end{equation}
Analogously to the holomorphic Green function (\ref{eq:HolomorphicGreenFunctionDefinition}), one defines the ``self-energy'' as the normalized trace,
\begin{equation}\label{eq:SelfEnergyDefinition}
\Sigma_{\mathbf{H}} ( z ) \equiv \frac{1}{N} \Tr \boldsymbol{\Sigma}_{\mathbf{H}} ( z ) ,
\end{equation}
and for the GUE it reads \smash{$\Sigma_{\mathbf{GUE}} ( z ) = \sigma^{2} G_{\mathbf{GUE}} ( z )$}.

The first (\ref{eq:DS1}) and second (\ref{eq:DS2}) DS equations are a set of two matrix equations for \smash{$\mathbf{G}_{\mathbf{H}} ( z )$} and \smash{$\boldsymbol{\Sigma}_{\mathbf{H}} ( z )$}---this is the fundamental tool exploited in this paper. For example, for the GUE, plugging (\ref{eq:DS2ForGUE}) into (\ref{eq:DS1}), and taking the normalized trace of both sides gives a quadratic equation for the holomorphic Green function whose solution with the proper asymptotics at infinity [i.e., \smash{$G_{\mathbf{H}} ( z \to \infty ) \sim 1 / z$}, which is equivalent to the normalization of the MSD, \smash{$\int_{\textrm{cuts}} \dd x \rho_{\mathbf{H}} ( x ) = 1$}, cf.~ (\ref{eq:HolomorphicGreenFunctionFromMSD})] is
\begin{equation}\label{eq:HolomorphicGreenFunctionForGUE}
G_{\mathbf{GUE}} ( z ) = \frac{1}{2 \sigma^{2}} \left( z - \sqrt{z - 2 \sigma} \sqrt{z + 2 \sigma} \right) ,
\end{equation}
where the square roots are principal. Applying (\ref{eq:MSDFromHolomorphicGreenFunction}), one arrives at the famous ``Wigner semicircle law,''
\begin{equation}\label{eq:MSDForGUE}
\rho_{\mathbf{GUE}} ( x ) = \left\{ \begin{array}{ll} \frac{1}{2 \pi \sigma^{2}} \sqrt{4 \sigma^{2} - x^{2}} , & \textrm{for } x \in [ - 2 \sigma , 2 \sigma ] , \\ 0 , & \textrm{otherwise} . \end{array} \right.
\end{equation}

This completes an exposition of the DS method in the Hermitian world; below it will be extended to the more involved non-Hermitian case.

%%%%%%%%%%%%%%%%%%%%%%%%%%%%%%%%%%%%%%%%%%%%%%%%%%%%%%%%%%%%%%%%%%%%%%
%%%%%%%%%%%%%%%%%%%%%%%%%%%%%%%%%%%%%%%%%%%%%%%%%%%%%%%%%%%%%%%%%%%%%%

\subsection{Non-Hermitian random matrices}
\label{aa:NonHermitianRandomMatrices}

%%%%%%%%%%%%%%%%%%%%%%%%%%%%%%%%%%%%%%%%%%%%%%%%%%%%%%%%%%%%%%%%%%%%%%

\subsubsection{Non-Hermitian Green functions}
\label{aaa:NonHermitianGreenFunctions}

A fundamental difference between Hermitian and non-Hermitian random matrices which makes the latter (denote an example by $\mathbf{X}$) both richer and more difficult is that the eigenvalues \smash{$\lambda_{i}$} are generically complex instead of real, thus in the large-$N$ limit coalesce not into real cuts but some two-dimensional ``domains'' $\mathcal{D}$ on the $z$-plane with the MSD,
\begin{equation}\label{eq:NonHermitianMSDDefinition}
\rho_{\mathbf{X}} ( z , \overline{z} ) \equiv \frac{1}{N} \sum_{i = 1}^{N} \la \delta^{( 2 )} \left( z - \lambda_{i} , \overline{z - \lambda_{i}} \right) \ra .
\end{equation}

\emph{Nonholomorphic Green function.} The above definition is analogous to (\ref{eq:HermitianMSDDefinition}) except that the Dirac delta function is now complex instead of real. Consequently, the Sokhotsky-Weierstrass formula (\ref{eq:SokhotskyWeierstrassFormula}) can no longer be used as its representation, and therefore, the concept of the Green function (\ref{eq:HolomorphicGreenFunctionDefinition}) needs to be reestablished according to another valid representation,
\begin{equation}\label{eq:ComplexDiracDeltaRepresentation}
\delta^{( 2 )} ( z , \overline{z} ) = \frac{1}{\pi} \lim_{\epsilon \to 0} \frac{\epsilon^{2}}{\left( | z |^{2} + \epsilon^{2} \right)^{2}} = \frac{1}{\pi} \frac{\partial}{\partial \overline{z}} \lim_{\epsilon \to 0} \frac{\overline{z}}{| z |^{2} + \epsilon^{2}} .
\end{equation}
Inspired by this form, one introduces the ``nonholomorphic Green function''~\cite{SommersCrisantiSompolinskyStein1988,HaakeIzrailevLehmannSaherSommers1992,LehmannSaherSokolovSommers1995,FyodorovSommers1997,FyodorovKhoruzhenkoSommers1997},
\begin{equation}
\begin{split}\label{eq:NonHolomorphicGreenFunctionDefinition}
&G_{\mathbf{X}} ( z , \overline{z} ) \equiv \lim_{\epsilon \to 0} \lim_{N \to \infty} \frac{1}{N} \sum_{i = 1}^{N} \la \frac{\overline{z} - \overline{\lambda_{i}}}{\left| z - \lambda_{i} \right|^{2} + \epsilon^{2}} \ra =\\
&= \lim_{\epsilon \to 0} \lim_{N \to \infty} \frac{1}{N} \Tr \la \frac{\overline{z} \Id_{N} - \mathbf{X}^{\dagger}}{\left( z \Id_{N} - \mathbf{X} \right) \left( \overline{z} \Id_{N} - \mathbf{X}^{\dagger} \right) + \epsilon^{2} \Id_{N}} \ra
\end{split}
\end{equation}
(the matrix inversion will always be understood as \smash{$\mathbf{A} / \mathbf{B} \equiv \mathbf{A} \mathbf{B}^{- 1}$}), since then the MSD is simply proportional to its Wirtinger derivative with respect to \smash{$\overline{z}$},
\begin{equation}\label{eq:MSDFromNonHolomorphicGreenFunction}
\rho_{\mathbf{X}} ( z , \overline{z} ) = \frac{1}{\pi} \frac{\partial}{\partial \overline{z}} G_{\mathbf{X}} ( z , \overline{z} ) , \quad \textrm{for} \quad z \in \mathcal{D} .
\end{equation}

Remark that, strictly speaking, the order of limits in (\ref{eq:NonHolomorphicGreenFunctionDefinition}) is opposite to the right one: One should take $\epsilon \to 0$ for any finite $N$ first, which can be directly performed everywhere except a finite number of points \smash{$\lambda_{i}$}, but this would reduce the nonholomorphic Green function to the holomorphic one (\ref{eq:HolomorphicGreenFunctionDefinition}). Only for $N$ already infinite, and inside the (then continuous) mean spectral domain $\mathcal{D}$, the limit $\epsilon \to 0$ produces a nontrivial quantity. This intricacy will be disregarded in the following and an assumption made that for the matrix models investigated in this paper the limits do commute.

Moreover, outside $\mathcal{D}$, the regulator $\epsilon$ is unnecessary, and one is left with the holomorphic Green function,
\begin{equation}\label{eq:NonHolomorphicGreenFunctionOutsideD}
G_{\mathbf{X}} ( z , \overline{z} ) = G_{\mathbf{X}} ( z ) , \quad \textrm{for} \quad z \notin \mathcal{D} ,
\end{equation}
which however---contrary to the Hermitian case---does not carry any information about the MSD.

Alternatively to (\ref{eq:NonHolomorphicGreenFunctionDefinition}), it is sometimes more natural to use the ``nonholomorphic $M$-transform,'' defined analogously to (\ref{eq:HolomorphicMTransformDefinition}),
\begin{equation}\label{eq:NonHolomorphicMTransformDefinition}
M_{\mathbf{X}} ( z , \overline{z} ) \equiv z G_{\mathbf{X}} ( z , \overline{z} ) - 1 .
\end{equation}

\emph{Matrix-valued Green function.} The nonholomorphic Green function (\ref{eq:NonHolomorphicGreenFunctionDefinition}) encodes the MSD (\ref{eq:MSDFromNonHolomorphicGreenFunction}) but still remains inconvenient to evaluate due to the denominator quadratic in $\mathbf{X}$ unlike the linear denominator in the holomorphic version (\ref{eq:HolomorphicGreenFunctionDefinition}). This problem has been resolved by ``duplication''~\cite{JanikNowakPappWambachZahed1997,JanikNowakPappZahed1997-01} (cf.~also~\cite{FeinbergZee1997-01,FeinbergZee1997-02} for a slightly different approach), i.e., introducing a $2 N \times 2 N$ matrix,
\begin{equation}\label{eq:MatrixValuedGreenFunctionMatrixDefinition1}
\mathbf{G}^{\dupl}_{\mathbf{X}} ( z , \overline{z} ) \equiv \lim_{\epsilon \to 0} \lim_{N \to \infty} \la \frac{1}{\mathbf{Z}^{\dupl}_{\epsilon} - \mathbf{X}^{\dupl}} \ra ,
\end{equation}
where for short,
\begin{equation}\label{eq:MatrixValuedGreenFunctionMatrixDefinition2}
\mathbf{Z}^{\dupl}_{\epsilon} \equiv \left( \begin{array}{c|c} z \Id_{N} & \ii \epsilon \Id_{N} \\ \hline \ii \epsilon \Id_{N} & \overline{z} \Id_{N} \end{array} \right) , \quad \mathbf{X}^{\dupl} \equiv \left( \begin{array}{c|c} \mathbf{X} & \Zero_{N} \\ \hline \Zero_{N} & \mathbf{X}^{\dagger} \end{array} \right) .
\end{equation}
This form exactly parallels the holomorphic one (\ref{eq:HolomorphicGreenFunctionMatrixDefinition}), with a denominator linear in the random matrix, for which the cost is the doubling of the matrix dimensions. The four $N \times N$ blocks of this and alike matrices will be distinguished by the following superscripts,
\begin{equation}\label{eq:MatrixValuedGreenFunctionMatrixDefinition3}
\mathbf{G}^{\dupl}_{\mathbf{X}} \equiv \left( \begin{array}{c|c} \mathbf{G}^{z z}_{\mathbf{X}} & \mathbf{G}^{z \overline{z}}_{\mathbf{X}} \\ \hline \mathbf{G}^{\overline{z} z}_{\mathbf{X}} & \mathbf{G}^{\overline{z} \overline{z}}_{\mathbf{X}} \end{array} \right) .
\end{equation}

Taking the normalized ``block-trace,''
\begin{equation}\label{eq:BlockTraceDefinition}
\bTr \left( \begin{array}{c|c} \mathbf{A} & \mathbf{B} \\ \hline \mathbf{C} & \mathbf{D} \end{array} \right) \equiv \left( \begin{array}{c|c} \Tr \mathbf{A} & \Tr \mathbf{B} \\ \hline \Tr \mathbf{C} & \Tr \mathbf{D} \end{array} \right) ,
\end{equation}
one obtains from (\ref{eq:MatrixValuedGreenFunctionMatrixDefinition1}) the ($2 \times 2$) ``matrix-valued Green function,''
\begin{equation}\label{eq:MatrixValuedGreenFunctionDefinition}
\mathcal{G}_{\mathbf{X}} \equiv \left( \begin{array}{c|c} \mathcal{G}^{z z}_{\mathbf{X}} & \mathcal{G}^{z \overline{z}}_{\mathbf{X}} \\ \hline \mathcal{G}^{\overline{z} z}_{\mathbf{X}} & \mathcal{G}^{\overline{z} \overline{z}}_{\mathbf{X}} \end{array} \right) \equiv \frac{1}{N} \bTr \mathbf{G}^{\dupl}_{\mathbf{X}} .
\end{equation}

It is a fundamental object in the non-Hermitian world for two reasons:

(i) Its upper left corner is precisely the desired nonholomorphic Green function (\ref{eq:NonHolomorphicGreenFunctionDefinition}),
\begin{equation}\label{eq:MatrixValuedGreenFunctionBlockZZ}
\mathcal{G}^{z z}_{\mathbf{X}} ( z , \overline{z} ) = G_{\mathbf{X}} ( z , \overline{z} ) .
\end{equation}
In other words, the MSD has been encoded as (a derivative of) a block of an object analogous to the holomorphic Green function (\ref{eq:HolomorphicGreenFunctionDefinition}), thus opening the door for methods from Hermitian RMT (such as planar diagrams and DS equations; cf.~Sec.~\ref{aaa:DSAndGinUE}) to be applied in the non-Hermitian case as well. [The lower right block is simply the complex conjugate of (\ref{eq:MatrixValuedGreenFunctionBlockZZ}); it does not bring any new information.]

(ii) Furthermore, the off-diagonal blocks of (\ref{eq:MatrixValuedGreenFunctionDefinition}),
\begin{subequations}
\begin{align}
&\mathcal{G}^{z \overline{z}}_{\mathbf{X}} ( z , \overline{z} ) =\nonumber\\
&= \lim_{\epsilon \to 0} \lim_{N \to \infty} \frac{1}{N} \Tr \la \frac{- \ii \epsilon}{\left( \overline{z} \Id_{N} - \mathbf{X}^{\dagger} \right) \left( z \Id_{N} - \mathbf{X} \right) + \epsilon^{2} \Id_{N}} \ra ,\label{eq:MatrixValuedGreenFunctionBlockZZBar}\\
&\mathcal{G}^{\overline{z} z}_{\mathbf{X}} ( z , \overline{z} ) =\nonumber\\
&= \lim_{\epsilon \to 0} \lim_{N \to \infty} \frac{1}{N} \Tr \la \frac{- \ii \epsilon}{\left( z \Id_{N} - \mathbf{X} \right) \left( \overline{z} \Id_{N} - \mathbf{X}^{\dagger} \right) + \epsilon^{2} \Id_{N}} \ra ,\label{eq:MatrixValuedGreenFunctionBlockZBarZ}
\end{align}
\end{subequations}
which are equal and purely imaginary, so it is convenient to use their negated product,
\begin{equation}\label{eq:BDefinition}
\mathcal{B}_{\mathbf{X}} ( z , \overline{z} ) \equiv - \mathcal{G}^{z \overline{z}}_{\mathbf{X}} ( z , \overline{z} ) \mathcal{G}^{\overline{z} z}_{\mathbf{X}} ( z , \overline{z} ) \in \mathbb{R}_{+} \cup \{ 0 \} ,
\end{equation}
which has a twofold meaning: Firstly, it speaks about the correlation between the left and right eigenvectors of $\mathbf{X}$~\cite{ChalkerMehlig1998}, a property not investigated in this article. Secondly, it is an ``order parameter'': It vanishes outside the mean spectral domain $\mathcal{D}$, since the regulator is unnecessary there, $\epsilon = 0$, and strictly positive inside this domain. Hence, once (\ref{eq:BDefinition}) has been calculated inside $\mathcal{D}$ (i.e., in the nonholomorphic sector), then
\begin{equation}\label{eq:BorderlineEquation}
\mathcal{B}_{\mathbf{X}} ( z , \overline{z} ) = 0
\end{equation}
is the equation of the borderline of $\mathcal{D}$ in the Cartesian coordinates $( x , y )$, i.e., $z = x + \ii y \in \partial \mathcal{D}$.

\begin{figure}[ht]
\includegraphics[width=\columnwidth]{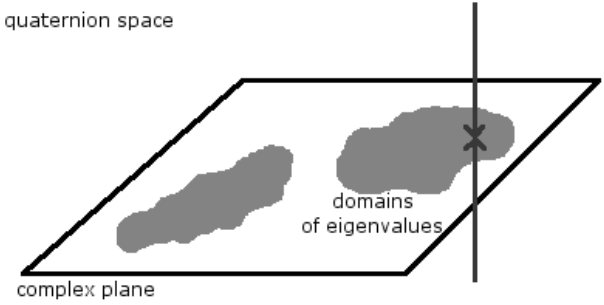}
\caption{For non-Hermitian random matrices, the complex MSD is evaluated from the quaternion Green function of a quaternion argument in the vicinity of the complex plane, i.e., from the matrix-valued Green function [(\ref{eq:MatrixValuedGreenFunctionMatrixDefinition1})-(\ref{eq:MatrixValuedGreenFunctionMatrixDefinition3}), (\ref{eq:MatrixValuedGreenFunctionDefinition})].}
\label{fig:DensityFromGreenFunctionForComplexSpectrum}
\end{figure}

\emph{Quaternion Green function.} Although not used in this paper, remark for completeness that the matrix-valued Green function [(\ref{eq:MatrixValuedGreenFunctionMatrixDefinition1})-(\ref{eq:MatrixValuedGreenFunctionDefinition})] may be naturally generalized by replacing the infinitesimal parameter $\epsilon$ with a complex (actually, real nonnegative is sufficient) argument $b$ (plus $z$ is renamed $a$),
\begin{equation}\label{eq:QuaternionGreenFunctionDefinition}
\mathcal{G}_{\mathbf{X}} ( \mathcal{Q} ) \equiv \frac{1}{N} \bTr \la \frac{1}{\mathcal{Q} \otimes \mathbf{1}_{N} - \mathbf{X}^{\dupl}} \ra ,
\end{equation}
where the argument is a quaternion,
\begin{equation}\label{eq:QuaternionDefinition}
\mathcal{Q} \equiv \left( \begin{array}{c|c} a & \ii \overline{b} \\ \hline \ii b & \overline{a} \end{array} \right) , \quad a , b \in \mathbb{C} ,
\end{equation}
and the tensor product means that each of its four blocks is multiplied by \smash{$\mathbf{1}_{N}$}. It is known as the ``quaternion Green function''~\cite{JaroszNowak2004,JaroszNowak2006} (cf.~\cite{Jarosz2011-01,Jarosz2012-01} for its recent application). This extension allows to define, in analogy to (\ref{eq:HolomorphicBlueFunctionDefinition}), the functional inverse (in the quaternion space) of (\ref{eq:QuaternionGreenFunctionDefinition}) (the ``quaternion Blue function''), which in turn leads to an extension of the free probability addition law (\ref{eq:FreeProbabilityAdditionLaw}) from the Hermitian to non-Hermitian world. It also shows that just as in the Hermitian case, a real MSD requires a complex Green function in the vicinity of the real axis (cf.~Fig.~\ref{fig:DensityFromGreenFunctionForRealSpectrum}), so in the non-Hermitian case, a quaternion Green function in the vicinity of the complex plane (cf.~Fig.~\ref{fig:DensityFromGreenFunctionForComplexSpectrum}) needs to be employed in order to reproduce a complex density.

%%%%%%%%%%%%%%%%%%%%%%%%%%%%%%%%%%%%%%%%%%%%%%%%%%%%%%%%%%%%%%%%%%%%%%

\subsubsection{Dyson-Schwinger equations and Ginibre unitary ensemble}
\label{aaa:DSAndGinUE}

In order to get better acquainted with the DS technique in the non-Hermitian case, the steps from Sec.~\ref{aaa:DSAndGUE} will now be applied to an example, the (square) ``Ginibre unitary ensemble'' (GinUE)~\cite{Ginibre1965,Girko1984,Girko1985,TaoVu2007},
\begin{equation}\label{eq:GinUEMeasureDefinition}
\dd \mu ( \mathbf{X} ) \propto \ee^{- \frac{N}{\sigma^{2}} \Tr ( \mathbf{X}^{\dagger} \mathbf{X} )} \dd \mathbf{X} ,
\end{equation}
i.e., all the \smash{$\re X_{i j}$}, \smash{$\im X_{i j}$} are IID Gaussian random numbers of zero mean and variance \smash{$\sigma^{2} / ( 2 N )$}, for some $\sigma$.

One starts from writing down the nonzero propagators, immediately stemming from (\ref{eq:GinUEMeasureDefinition}),
\begin{equation}\label{eq:GinUEPropagator}
\la [ \mathbf{X} ]_{i j} [ \mathbf{X}^{\dagger} ]_{k l} \ra = \frac{\sigma^{2}}{N} \delta_{i l} \delta_{j k} .
\end{equation}
However, here one needs the propagator of the duplicated random matrix (\ref{eq:MatrixValuedGreenFunctionMatrixDefinition2}),
\begin{equation}\label{eq:DuplicatedGinUEPropagator}
\la [ \mathbf{X}^{\dupl} ]_{i j} [ \mathbf{X}^{\dupl} ]_{\overline{k} \overline{l}} \ra = \frac{\sigma^{2}}{N} \delta_{i \overline{l}} \delta_{j \overline{k}} ,
\end{equation}
where a bar over an index or its lack determines the block to which the index belongs, according to the convention (\ref{eq:MatrixValuedGreenFunctionMatrixDefinition3}) (so here the nonzero propagator is between the upper left and lower right blocks).

This is the necessary ingredient of the DS equations. The self-energy matrix is also duplicated,
\begin{equation}\label{eq:DuplicatedSelfEnergyMatrixDefinition}
\mathbf{\Sigma}^{\dupl} \equiv \left( \begin{array}{c|c} \mathbf{\Sigma}^{z z} & \mathbf{\Sigma}^{z \overline{z}} \\ \hline \mathbf{\Sigma}^{\overline{z} z} & \mathbf{\Sigma}^{\overline{z} \overline{z}} \end{array} \right) .
\end{equation}
The first DS equation is analogous to (\ref{eq:DS1}),
\begin{equation}\label{eq:GinUEDysonSchwingerEquation1}
\mathbf{G}^{\dupl} = \frac{1}{\mathbf{Z}^{\dupl} - \mathbf{\Sigma}^{\dupl}} ,
\end{equation}
where for short, \smash{$\mathbf{Z}^{\dupl} \equiv \mathbf{Z}^{\dupl}_{\epsilon = 0}$} (\ref{eq:MatrixValuedGreenFunctionMatrixDefinition2}) (one may set $\epsilon = 0$ here as it will be seen that the DS equations by themselves regularize the singularities inside the mean spectral domain). The second DS equation is analogous to (\ref{eq:DS2})
\begin{subequations}
\begin{align}
\left[ \mathbf{\Sigma}^{\dupl} \right]_{i \overline{j}} &= \sum_{k_{1} , \overline{k_{2}} = 1}^{N} \left[ \mathbf{G}^{\dupl} \right]_{k_{1} \overline{k_{2}}} \cdot \nonumber\\
&\cdot \la \left[ \mathbf{X}^{\dupl} \right]_{i k_{1}} \left[ \mathbf{X}^{\dupl} \right]_{\overline{k_{2}} \overline{j}} \ra = \nonumber\\
&= \sigma^{2} \mathcal{G}^{z \overline{z}} \delta_{i \overline{j}} ,\label{eq:GinUEDysonSchwingerEquation2BlockZZBar}\\
\left[ \mathbf{\Sigma}^{\dupl} \right]_{\overline{i} j} &= \sigma^{2} \mathcal{G}^{\overline{z} z} \delta_{\overline{i} j} ,\label{eq:GinUEDysonSchwingerEquation2BlockZBarZ}\\
\left[ \mathbf{\Sigma}^{\dupl} \right]_{i j} &= \left[ \mathbf{\Sigma}^{\dupl} \right]_{\overline{i} \overline{j}} = 0 ,\label{eq:GinUEDysonSchwingerEquation2BlockZZAndZBarZBar}
\end{align}
\end{subequations}
i.e.,
\begin{equation}\label{eq:GinUEDysonSchwingerEquation2}
\mathbf{\Sigma}^{\dupl} = \left( \begin{array}{c|c} \Zero_{N} & \sigma^{2} \mathcal{G}^{z \overline{z}} \Id_{N} \\ \hline \sigma^{2} \mathcal{G}^{\overline{z} z} \Id_{N} & \Zero_{N} \end{array} \right) .
\end{equation}

Inserting (\ref{eq:GinUEDysonSchwingerEquation2}) into (\ref{eq:GinUEDysonSchwingerEquation1}), one finds a $2 \times 2$ matrix equation for the elements of the matrix-valued Green function of the GinUE,
\begin{equation}
\begin{split}\label{eq:MatrixValuedGreenFunctionForGinUE}
\left( \begin{array}{c|c} \mathcal{G}^{z z} & \mathcal{G}^{z \overline{z}} \\ \hline \mathcal{G}^{\overline{z} z} & \mathcal{G}^{\overline{z} \overline{z}} \end{array} \right) &= \left( \begin{array}{c|c} z & - \sigma^{2} \mathcal{G}^{z \overline{z}} \\ \hline - \sigma^{2} \mathcal{G}^{\overline{z} z} & \overline{z} \end{array} \right)^{- 1} =\\
&= \frac{1}{| z |^{2} - \sigma^{4} \mathcal{G}^{z \overline{z}} \mathcal{G}^{\overline{z} z}} \left( \begin{array}{c|c} \overline{z} & \sigma^{2} \mathcal{G}^{z \overline{z}} \\ \hline \sigma^{2} \mathcal{G}^{\overline{z} z} & z \end{array} \right) .
\end{split}
\end{equation}
Firstly, the equation for the order parameter \smash{$\mathcal{B} \equiv \mathcal{B}_{\mathbf{X}} ( z , \overline{z} )$} (\ref{eq:BDefinition}) reads therefore,
\begin{equation}\label{eq:MatrixValuedGreenFunctionForGinUEOffDiagonal}
\mathcal{B} = \frac{\sigma^{4} \mathcal{B}}{\left( | z |^{2} + \sigma^{4} \mathcal{B} \right)^{2}} ,
\end{equation}
which has the ``holomorphic solution'' $\mathcal{B} = 0$, valid outside of $\mathcal{D}$, and a relevant ``nonholomorphic solution,''
\begin{equation}\label{eq:MatrixValuedGreenFunctionForGinUEOffDiagonalNonholomorphicSolution}
\mathcal{B} = \frac{1}{\sigma^{2}} \left( 1 - \frac{| z |^{2}}{\sigma^{2}} \right) .
\end{equation}
It implies (\ref{eq:BorderlineEquation}) that the mean spectral domain $\mathcal{D}$ for GinUE is the centered circle of radius
\begin{equation}\label{eq:GinUEBoundary}
| z | = \sigma .
\end{equation}
Secondly, the upper left block of (\ref{eq:MatrixValuedGreenFunctionForGinUE}) yields
\begin{equation}\label{eq:MatrixValuedGreenFunctionForGinUEDiagonalNonholomorphicSolution}
\mathcal{G}^{z z} = \frac{\overline{z}}{| z |^{2} + \sigma^{4} \mathcal{B}} = \left\{ \begin{array}{ll} \frac{\overline{z}}{\sigma^{2}} , & \textrm{for } | z | \leq \sigma , \\ \frac{1}{z} , & \textrm{for } | z | > \sigma . \end{array} \right.
\end{equation}
Hence, one recognizes the (trivial) holomorphic Green function $1 / z$ outside $\mathcal{D}$, and the nonholomorphic one, \smash{$G_{\mathbf{X}} ( z , \overline{z} ) = \overline{z} / \sigma^{2}$}, inside $\mathcal{D}$, which leads (\ref{eq:MSDFromNonHolomorphicGreenFunction}) to the constant MSD,
\begin{equation}\label{eq:MSDForGinUE}
\rho_{\mathbf{X}} ( z , \overline{z} ) = \left\{ \begin{array}{ll} \frac{1}{\pi \sigma^{2}} , & \textrm{for } | z | \leq \sigma , \\ 0 , & \textrm{for } | z | > \sigma . \end{array} \right.
\end{equation}

This very method is used in this paper to solve much more complicated models.

%%%%%%%%%%%%%%%%%%%%%%%%%%%%%%%%%%%%%%%%%%%%%%%%%%%%%%%%%%%%%%%%%%%%%%
%%%%%%%%%%%%%%%%%%%%%%%%%%%%%%%%%%%%%%%%%%%%%%%%%%%%%%%%%%%%%%%%%%%%%%
%%%%%%%%%%%%%%%%%%%%%%%%%%%%%%%%%%%%%%%%%%%%%%%%%%%%%%%%%%%%%%%%%%%%%%

\section{Mean spectral density of the equal-time covariance estimator for Toy Models 1, 2a, 3}
\label{a:TM12a3ETCE}

This article is devoted mainly to the TLCE, but an important conclusion may be drawn from comparing the MSD of the ETCE and TLCE for a given model which is that the former is a better tool for discerning true spatial covariances, while the latter for temporal correlations. Therefore, even though the MSD of the ETCE is already known for all the models considered here, it is helpful to present these results and their derivation for three simplest examples, i.e., Toy Models 1, 2a and 3, for Gaussian assets. The pertinent master equation in all these cases is (\ref{eq:Case2ETCEEq06}).

%%%%%%%%%%%%%%%%%%%%%%%%%%%%%%%%%%%%%%%%%%%%%%%%%%%%%%%%%%%%%%%%%%%%%%
%%%%%%%%%%%%%%%%%%%%%%%%%%%%%%%%%%%%%%%%%%%%%%%%%%%%%%%%%%%%%%%%%%%%%%

\subsection{Toy Model 1}
\label{aa:TM1ETCE}

For Toy Model 1 (\ref{eq:TM1DefinitionEq01}) (cf.~Sec.~\ref{ss:TM1}), the master equation becomes quadratic for \smash{$G \equiv G_{\mathbf{c}} ( z )$},
\begin{equation}\label{eq:TM1ETCEEq01}
r \sigma^{2} z G^{2} + \left( - z + \sigma^{2} ( 1 - r ) \right) G + 1 = 0 ,
\end{equation}
whose solution with the proper asymptotics at infinity [cf.~the discussion before eq.~(\ref{eq:HolomorphicGreenFunctionForGUE})] reads
\begin{equation}\label{eq:TM1ETCEEq02}
G = \frac{1}{2 r \sigma^{2} z} \left( z - \sigma^{2} ( 1 - r ) - \sqrt{z - x_{+}} \sqrt{z - x_{-}} \right) ,
\end{equation}
where \smash{$x_{\pm} \equiv \sigma^{2} ( 1 \pm \sqrt{r} )^{2}$}, and where the square roots are principal. Hence, the MSD follows from (\ref{eq:MSDFromHolomorphicGreenFunction}) to be
\begin{equation}
\begin{split}\label{eq:TM1ETCEEq03}
\rho_{\mathbf{c}} ( x ) &=\\
&= ( 1 - 1 / r ) \Theta ( r - 1 ) \delta ( x ) +\\
&+ \left\{ \begin{array}{ll} \frac{\sqrt{\left( x_{+} - x \right) \left( x - x_{-} \right)}}{2 \pi r \sigma^{2} x} , & \textrm{for } x \in \left[ x_{-} , x_{+} \right] , \\ 0 , & \textrm{otherwise} , \end{array} \right.
\end{split}
\end{equation}
which is the celebrated Mar\v{c}enko-Pastur distribution (cf.~Sec.~\ref{sss:MeasurementNoise} for its early appearance in econophysics). [The first term in (\ref{eq:TM1ETCEEq03}) represents the zero modes, which appear only for $r > 1$, and which are derived by expanding (\ref{eq:TM1ETCEEq02}) around $z = 0$, which yields \smash{$G = ( 1 - 1 / r ) \Theta ( r - 1 ) ( 1 / z ) + \textrm{O} ( z^{0} )$}, to which (\ref{eq:SokhotskyWeierstrassFormula}) needs to be applied. The normalization of the MSD essentially requires this contribution.]

%%%%%%%%%%%%%%%%%%%%%%%%%%%%%%%%%%%%%%%%%%%%%%%%%%%%%%%%%%%%%%%%%%%%%%
%%%%%%%%%%%%%%%%%%%%%%%%%%%%%%%%%%%%%%%%%%%%%%%%%%%%%%%%%%%%%%%%%%%%%%

\subsection{Toy Model 2a}
\label{aa:TM2aETCE}

\begin{figure}[t]
\includegraphics[width=\columnwidth]{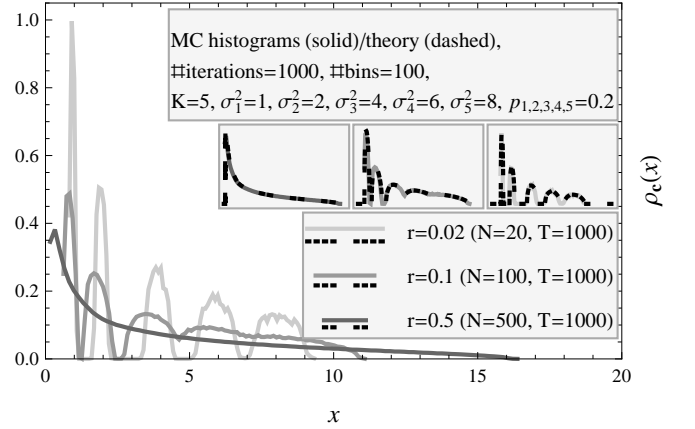}
\caption{Monte Carlo eigenvalues versus theory for the ETCE for Toy Model 2a. The graphs in the main figures are hardly distinguishable, hence the insets show them separately.}
\label{fig:TM2aETCE}
\end{figure}

For Toy Model 2a (\ref{eq:TM2aEq01}) (cf.~Sec.~\ref{sss:TM2a}), the master equation furnished with (\ref{eq:TM2aEq02}) and \smash{$N_{\mathbf{A}} ( z ) = 1 + 1 / z$} becomes
\begin{equation}\label{eq:TM2ETCEEq01}
M = \sum_{k = 1}^{K} \frac{p_{k} \sigma_{k}^{2}}{\frac{z}{r M + 1} - \sigma_{k}^{2}} ,
\end{equation}
which is a polynomial equation of order $( K + 1 )$ for the holomorphic $M$-transform \smash{$M \equiv M_{\mathbf{c}} ( z )$}, whose solution---with [(\ref{eq:HolomorphicMTransformDefinition}), (\ref{eq:MSDFromHolomorphicGreenFunction})] applied to it---yields the desired MSD.

A more detailed analysis of this model has been done elsewhere (cf.~e.g.~\cite{BurdaGorlichJaroszJurkiewicz2004}), but in Fig.~\ref{fig:TM2aETCE} histograms of Monte Carlo eigenvalues are compared to the solutions of (\ref{eq:TM2ETCEEq01}), for $K = 5$ with \smash{$\sigma_{1}^{2} = 1$}, \smash{$\sigma_{2}^{2} = 2$}, \smash{$\sigma_{3}^{2} = 4$}, \smash{$\sigma_{4}^{2} = 6$}, \smash{$\sigma_{5}^{2} = 8$} and equal relative multiplicities $0.2$, as well as three values of $r = 0.02 , 0.1 , 0.5$, discovering perfect agreement between the two. One recognizes how decreasing $r$ makes the peaks around the true variances narrower, allowing their determination from this MSD. This should be contrasted with Fig.~\ref{fig:TM2aTLCE} (c), which shows that even for very small $r$ the separate variances are not visible in the spectrum of the TLCE.

%%%%%%%%%%%%%%%%%%%%%%%%%%%%%%%%%%%%%%%%%%%%%%%%%%%%%%%%%%%%%%%%%%%%%%
%%%%%%%%%%%%%%%%%%%%%%%%%%%%%%%%%%%%%%%%%%%%%%%%%%%%%%%%%%%%%%%%%%%%%%

\subsection{Toy Model 3}
\label{aa:TM3ETCE}

\begin{figure}[t]
\includegraphics[width=\columnwidth]{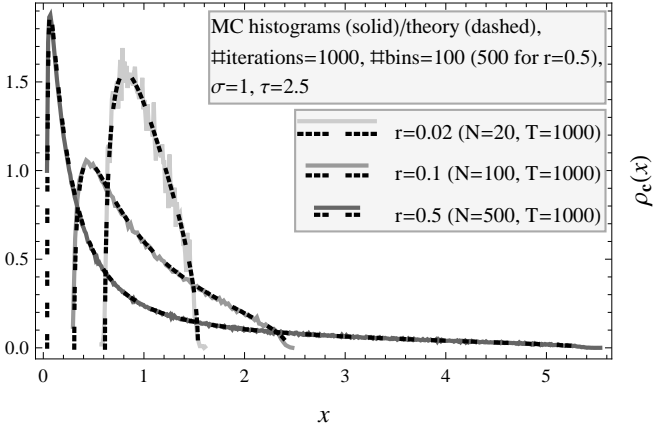}
\caption{Monte Carlo eigenvalues versus theory for the ETCE for Toy Model 3.}
\label{fig:TM3ETCE}
\end{figure}

%%%%%%%%%%%%%%%%%%%%%%%%%%%%%%%%%%%%%%%%%%%%%%%%%%%%%%%%%%%%%%%%%%%%%%

\subsubsection{Mean spectral density}
\label{aaa:TM3ETCEMSD}

For Toy Model 3 (\ref{eq:TM3DefinitionEq01}) (cf.~Sec.~\ref{ss:TM3}), the master equation with \smash{$N_{\mathbf{C}} ( z ) = 1 + 1 / z$} and
\begin{equation}
\begin{split}\label{eq:TM3ETCEEq01}
M_{\mathbf{A}} ( z ) &= \int_{- 1 / 2}^{1 / 2} \dd p \frac{1}{\frac{z}{\hat{A} ( p )} - 1} =\\
&= \pm \frac{1}{\sqrt{\frac{z}{\sigma^{2}} - \tanh \frac{1}{2 \tau}} \sqrt{\frac{z}{\sigma^{2}} - \coth \frac{1}{2 \tau}}}
\end{split}
\end{equation}
(the sign is irrelevant, the square roots are principal), is a quartic polynomial equation for \smash{$G \equiv G_{\mathbf{c}} ( z )$},
\begin{equation}
\begin{split}\label{eq:TM3ETCEEq02}
&r^{2} z_{\sigma}^{2} G_{\sigma}^{4} - 2 r z_{\sigma} \left( \chi z_{\sigma} + r \right) G_{\sigma}^{3} +\\
&+ \left( z_{\sigma}^{2} + 4 r \chi z_{\sigma} + r^{2} - 1 \right) G_{\sigma}^{2} -\\
&- 2 \left( z_{\sigma} + r \chi \right) G_{\sigma} + 1 = 0 ,
\end{split}
\end{equation}
where for short, \smash{$\chi \equiv \coth ( 1 / \tau )$}, where the argument and the holomorphic Green function are rescaled by the variance, \smash{$z_{\sigma} \equiv z / \sigma^{2}$} and \smash{$G_{\sigma} \equiv \sigma^{2} G$}.

This result has been found before~\cite{BurdaJaroszJurkiewiczNowakPappZahed2010}, nevertheless Fig.~\ref{fig:TM3ETCE} presents some tests of (\ref{eq:TM3ETCEEq02}) against Monte Carlo data, for $\tau = 2.5$ and three values of $r = 0.02 , 0.1 , 0.5$, with perfect concord. A crucial observation is that the MSD is ``MP-shaped,'' i.e., it may be impossible to distinguish from the case with no temporal correlations, which means that the ETCE is not an effective tool in discerning whether a given empirical set of data exhibits true temporal covariances or not. On the other hand, comparing Fig.~\ref{fig:TM1TLCE} with Figs.~[\ref{fig:TM3TLCEEIGt1}, \ref{fig:TM3TLCEEIGt5}, \ref{fig:TM3TLCEEIGt15}] reveals that the spectrum of the TLCE is highly sensitive to whether there are intrinsic temporal correlations or not, and of what strength.

%%%%%%%%%%%%%%%%%%%%%%%%%%%%%%%%%%%%%%%%%%%%%%%%%%%%%%%%%%%%%%%%%%%%%%

\subsubsection{Edges of the spectrum (new result)}
\label{aaa:TM3ETCEEdges}

It is a general principle~\cite{Zee1996,JanikNowakPappZahed1997-02} that for any Hermitian random matrix model $\mathbf{H}$, the edges \smash{$x_{\star}$} of its mean spectrum (in the thermodynamic limit) are given by the divergences of the Green function, \smash{$\dd G_{\mathbf{H}} ( z ) / \dd z |_{z = x_{\star}} = \infty$}. Combining this formula with (\ref{eq:TM3ETCEEq02}), one arrives at a sextic polynomial equation for the positions of the edges,
\begin{equation}
\begin{split}\label{eq:TM3ETCEEq03}
&\left( \chi^{2} - 1 \right) x_{\sigma \star}^{6} - 6 r \chi \left( \chi^{2} - 1 \right) x_{\sigma \star}^{5} +\\
&+ \Big( 3 \left( 1 - r^{2} \right) - \left( 2 + 9 r^{2} \right) \chi^{2} + 12 r^{2} \chi^{4} \Big) x_{\sigma \star}^{4} +\\
&+ 2 r \chi \Big( 3 \left( 1 + 2 r^{2} \right) - \left( 5 + 2 r^{2} \right) \chi^{2} - 4 r^{2} \chi^{4} \Big) x_{\sigma \star}^{3} +\\
&+ \Big( - 3 \left( 1 + 7 r^{2} + r^{4} \right) + \left( 1 + 26 r^{2} - 9 r^{4} \right) \chi^{2} +\\
&+ r^{2} \left( 1 + 12 r^{2} \right) \chi^{4} \Big) x_{\sigma \star}^{2} -\\
&- 2 r \chi \Big( 3 \left( 1 - r^{2} \right) \left( 2 + r^{2} \right) + r^{2} \left( 5 + 3 r^{2} \right) \chi^{2} \Big) x_{\sigma \star} +\\
&+ \left( 1 - r^{2} \right)^{2} \left( 1 - r^{2} + r^{2} \chi^{2} \right) = 0 ,
\end{split}
\end{equation}
where real and nonnegative solutions should be picked, and it is a numerical observation (to be proven) that there are two of them, as expected. Figure~\ref{fig:TM3TLCEEIGEdges} shows these edges as functions of $\tau$, compared with the analogous quantities for the TLCE.

%%%%%%%%%%%%%%%%%%%%%%%%%%%%%%%%%%%%%%%%%%%%%%%%%%%%%%%%%%%%%%%%%%%%%%
%%%%%%%%%%%%%%%%%%%%%%%%%%%%%%%%%%%%%%%%%%%%%%%%%%%%%%%%%%%%%%%%%%%%%%
%%%%%%%%%%%%%%%%%%%%%%%%%%%%%%%%%%%%%%%%%%%%%%%%%%%%%%%%%%%%%%%%%%%%%%

\section{Numerical evaluation of the mean spectral density of the time-lagged estimator for Toy Model 3}
\label{a:NumericalEvaluationOfTheMSDOfTheTLCEForToyModel3}

\begin{figure*}[t]
\includegraphics[width=\columnwidth]{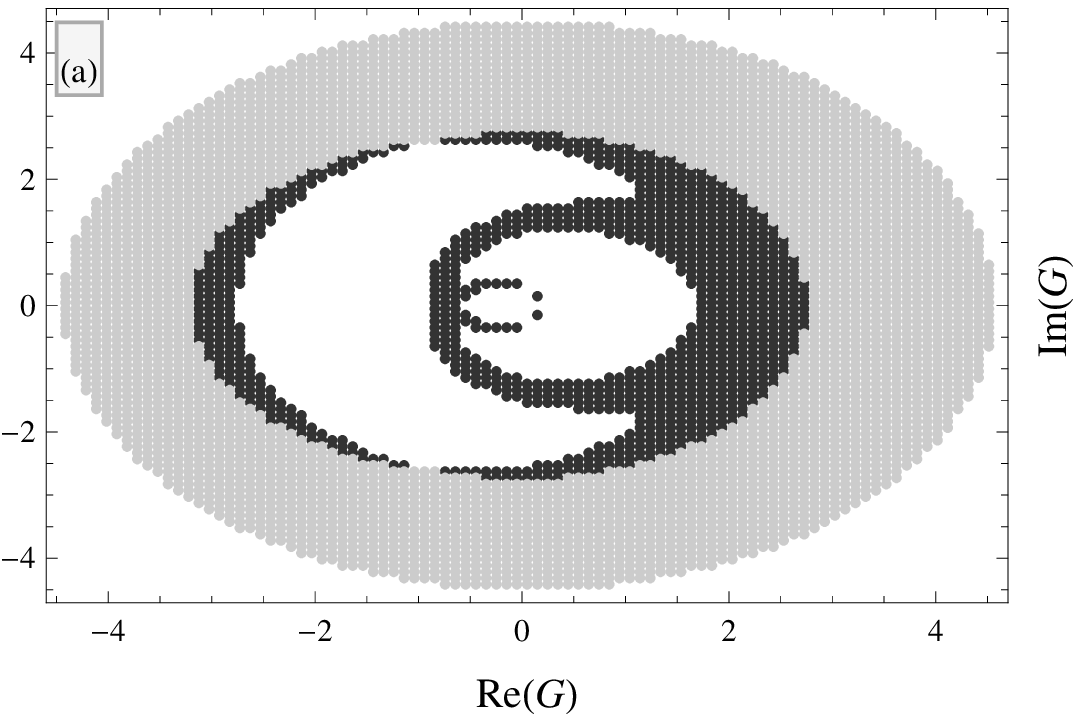}
\includegraphics[width=\columnwidth]{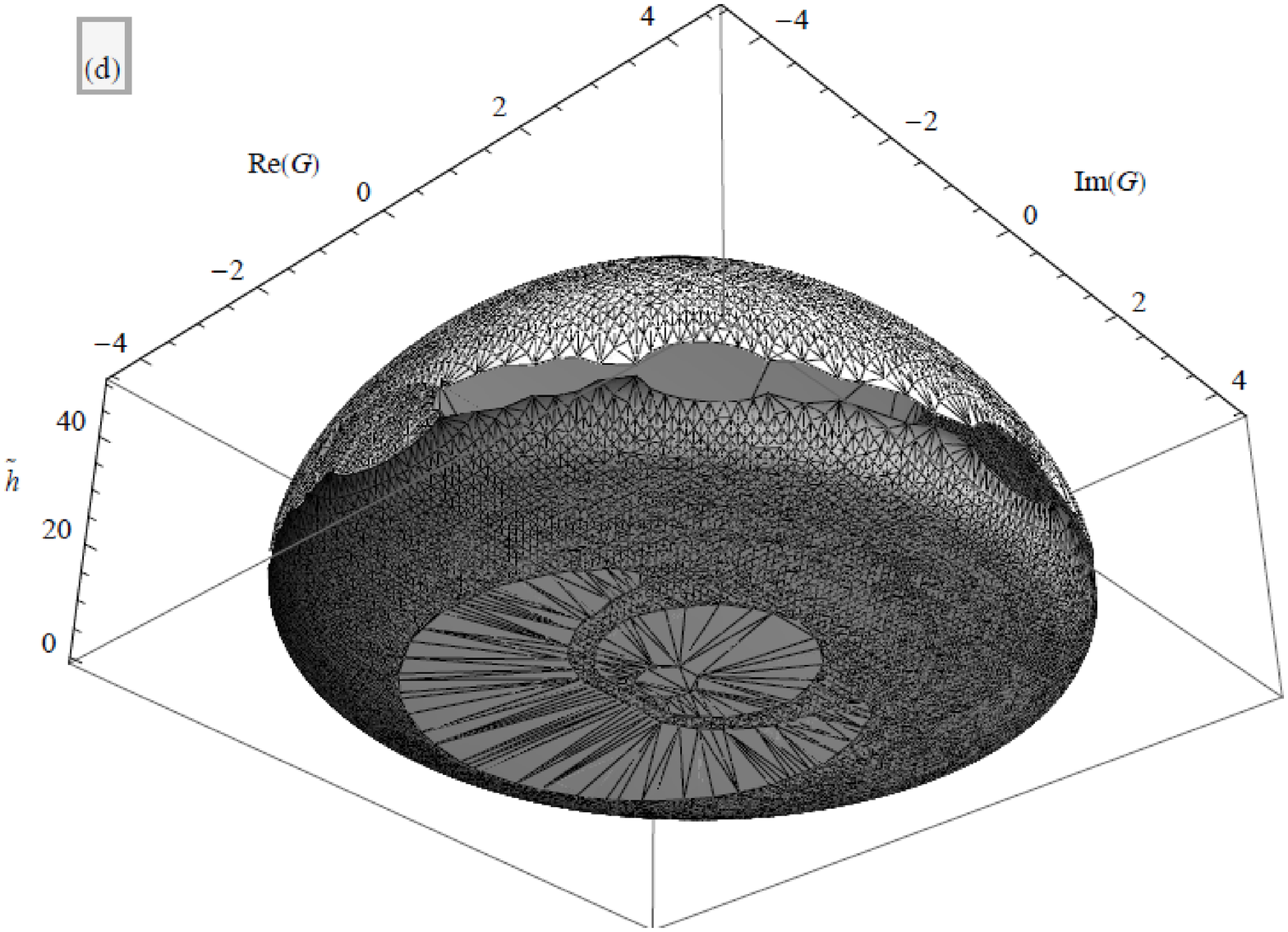}
\includegraphics[width=\columnwidth]{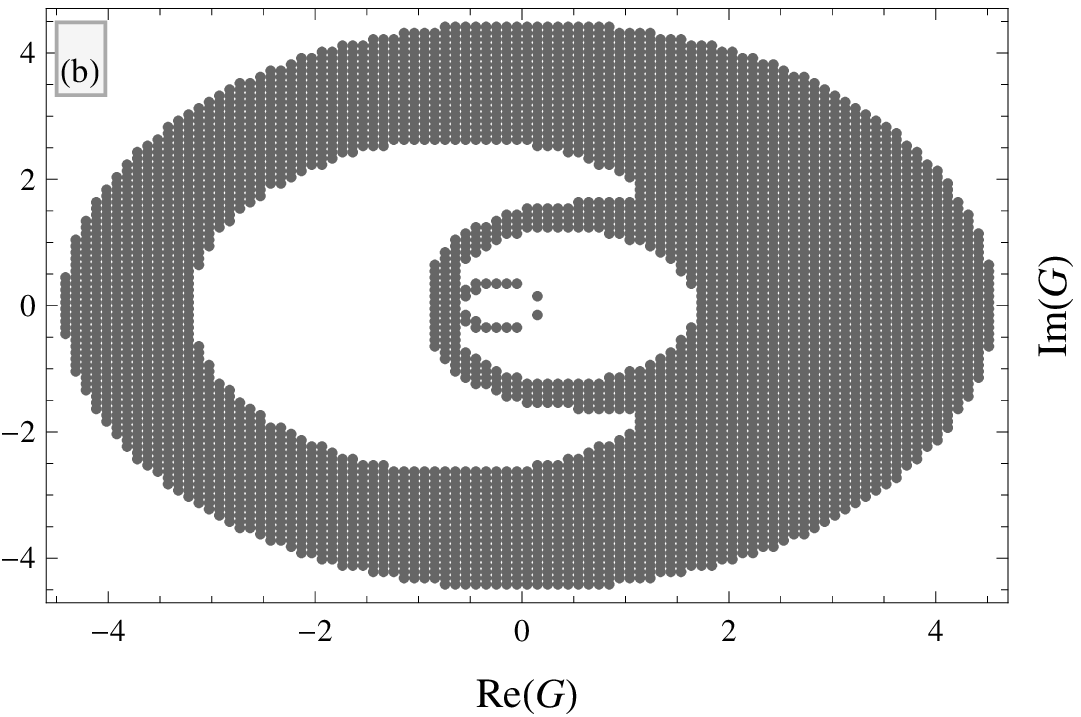}
\includegraphics[width=\columnwidth]{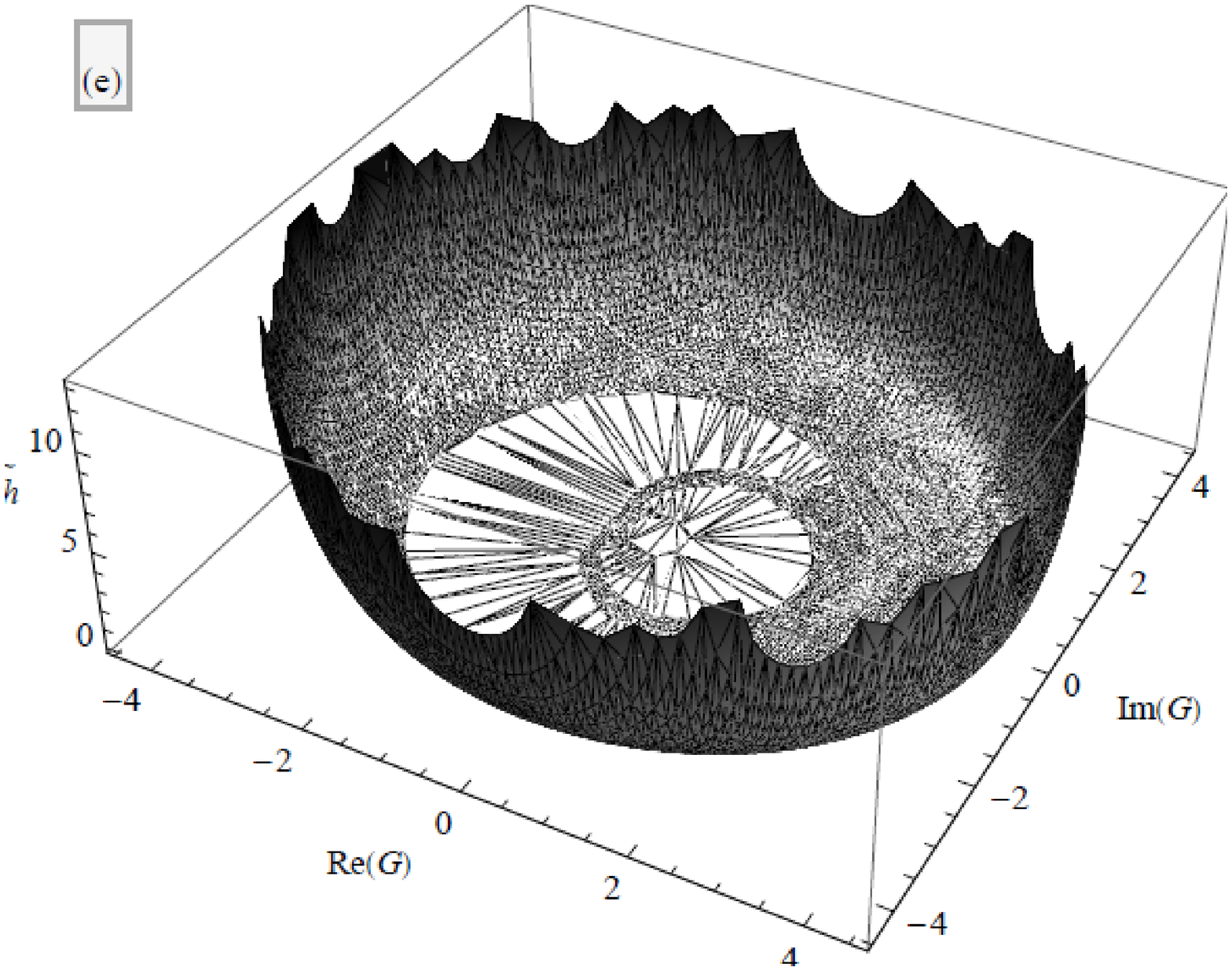}
\includegraphics[width=\columnwidth]{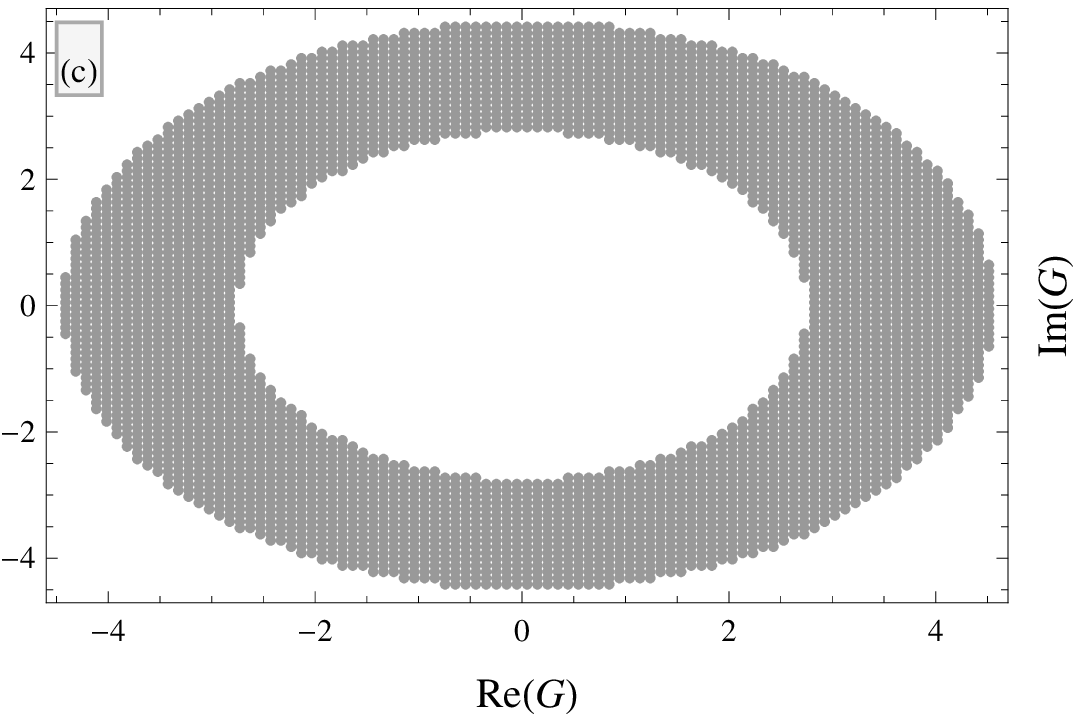}
\includegraphics[width=\columnwidth]{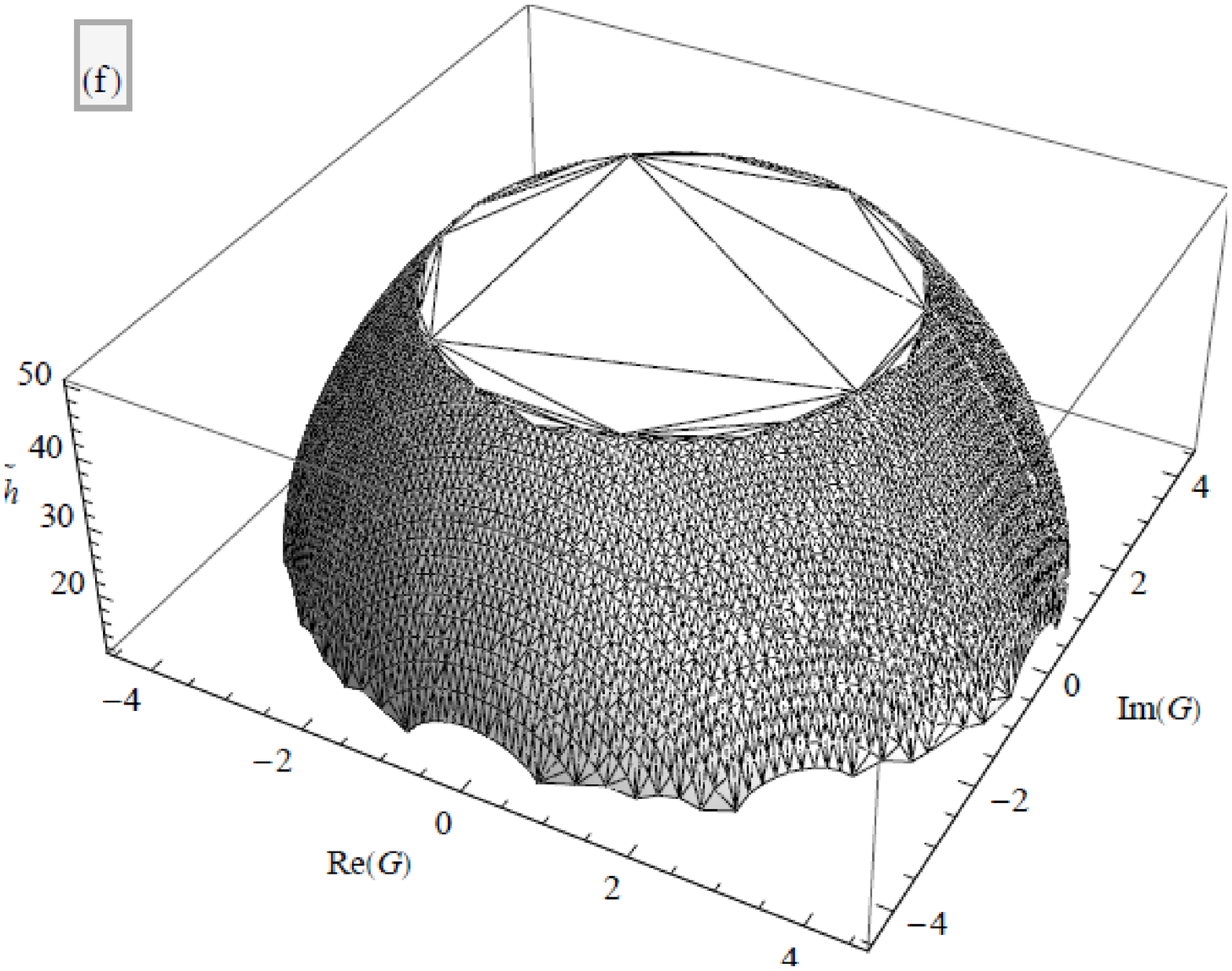}
\caption{Toy Model 3: Solving the first master equation (\ref{eq:TM3TLCEEq01a}) for \smash{$\tilde{h}$} (Sec.~\ref{aa:SolvingTheFirstMasterEquationForhtilde}). The initial lattice in the $G$-plane has the limits \smash{$X^{\min} = Y^{\min} = - 6.0$} and \smash{$X^{\max} = Y^{\max} = 6.0$} with \smash{$s^{\re} = s^{\im} = 121$} divisions.}
\label{fig:TM3TLCEMSDhtilde}
\end{figure*}

\begin{figure*}[t]
\includegraphics[width=\columnwidth]{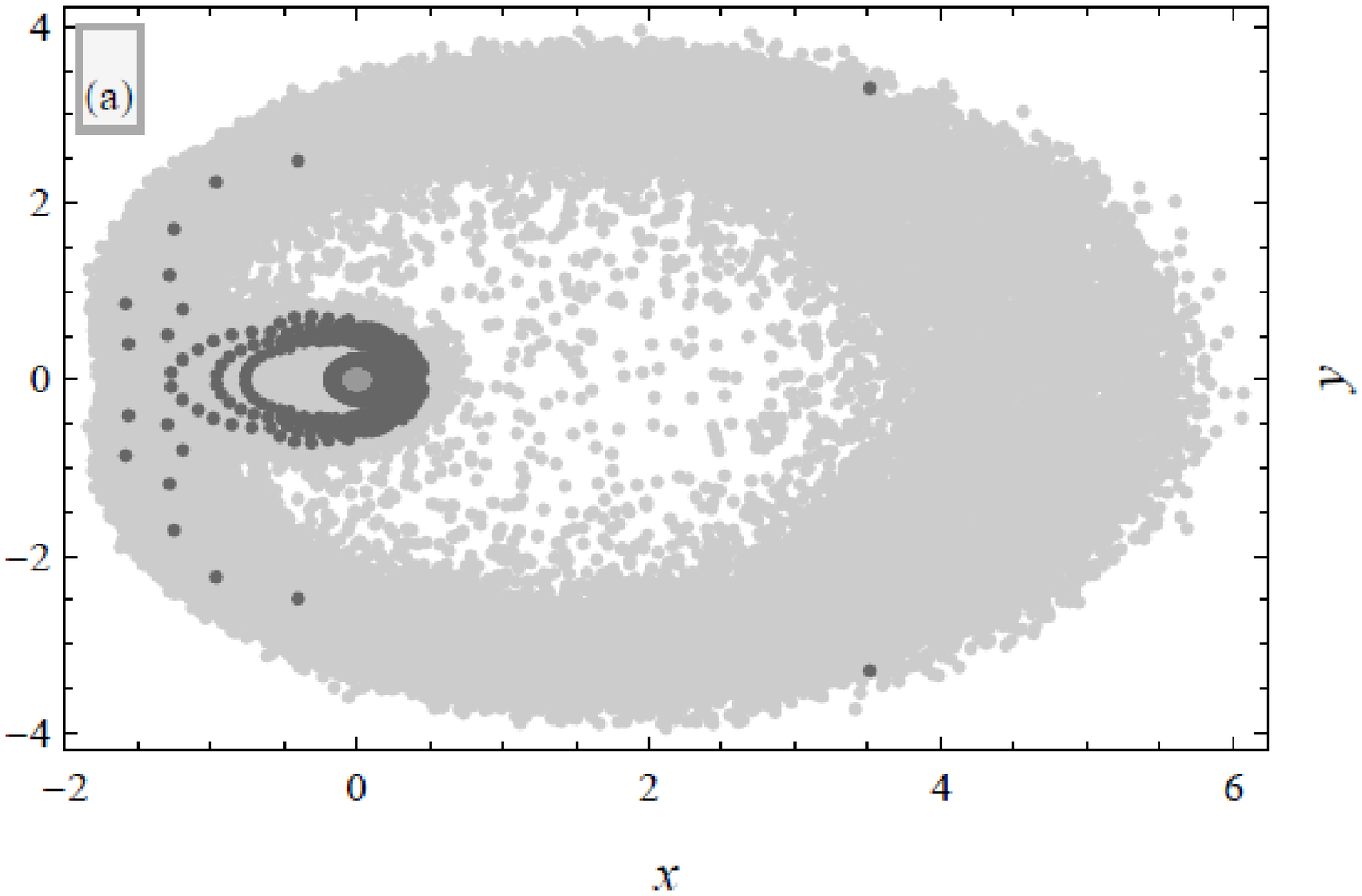}
\includegraphics[width=\columnwidth]{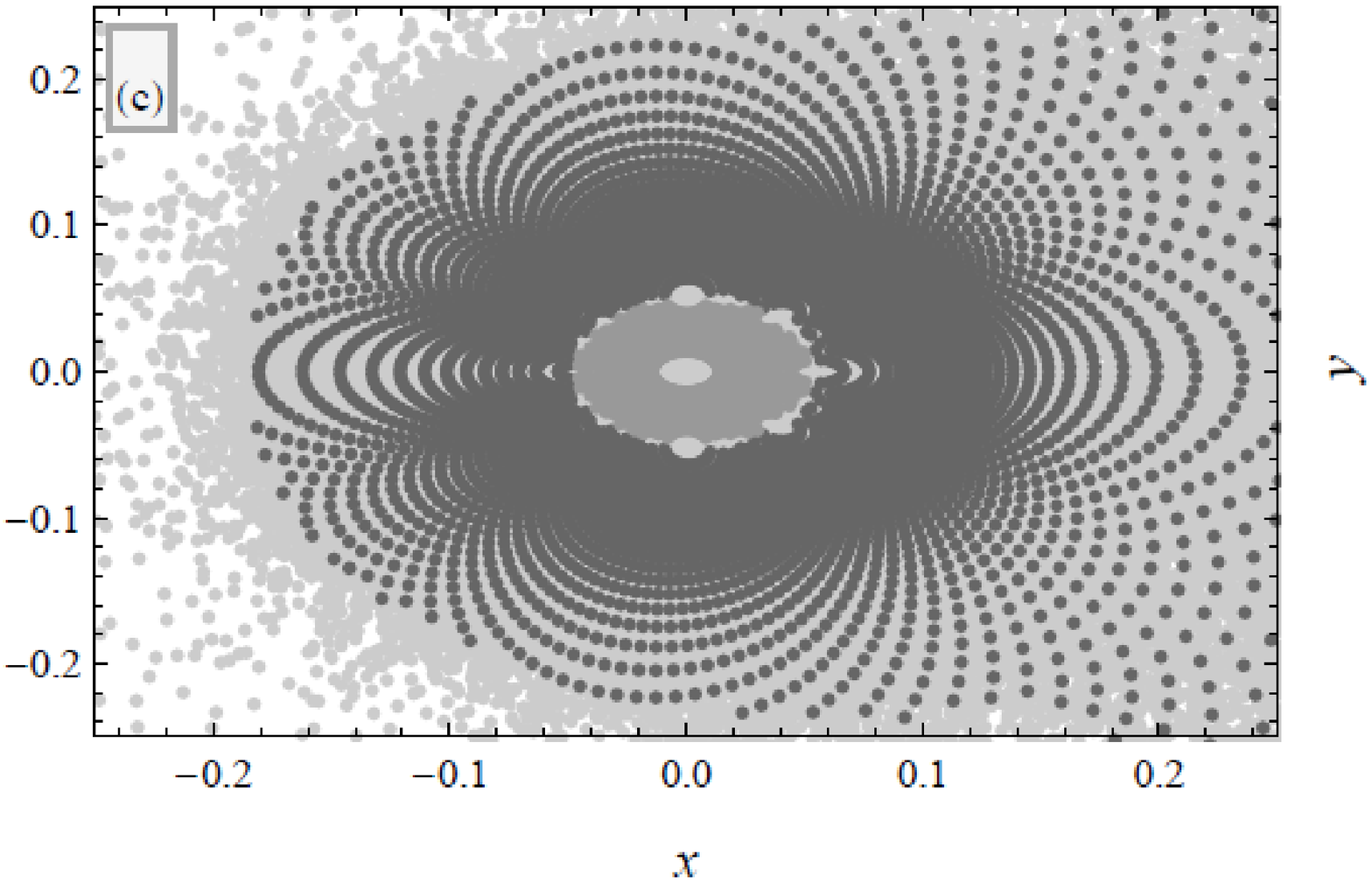}
\includegraphics[width=\columnwidth]{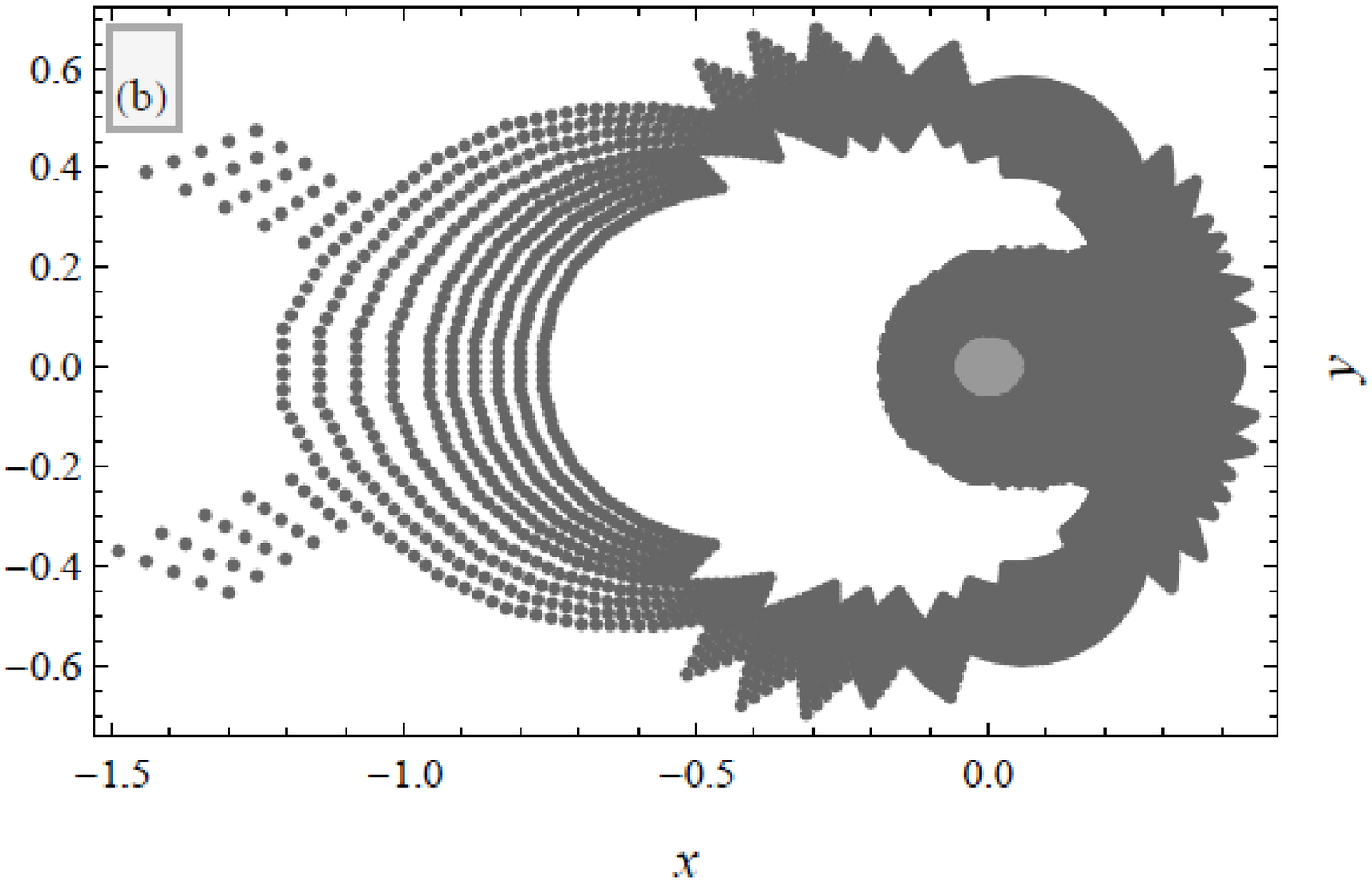}
\includegraphics[width=\columnwidth]{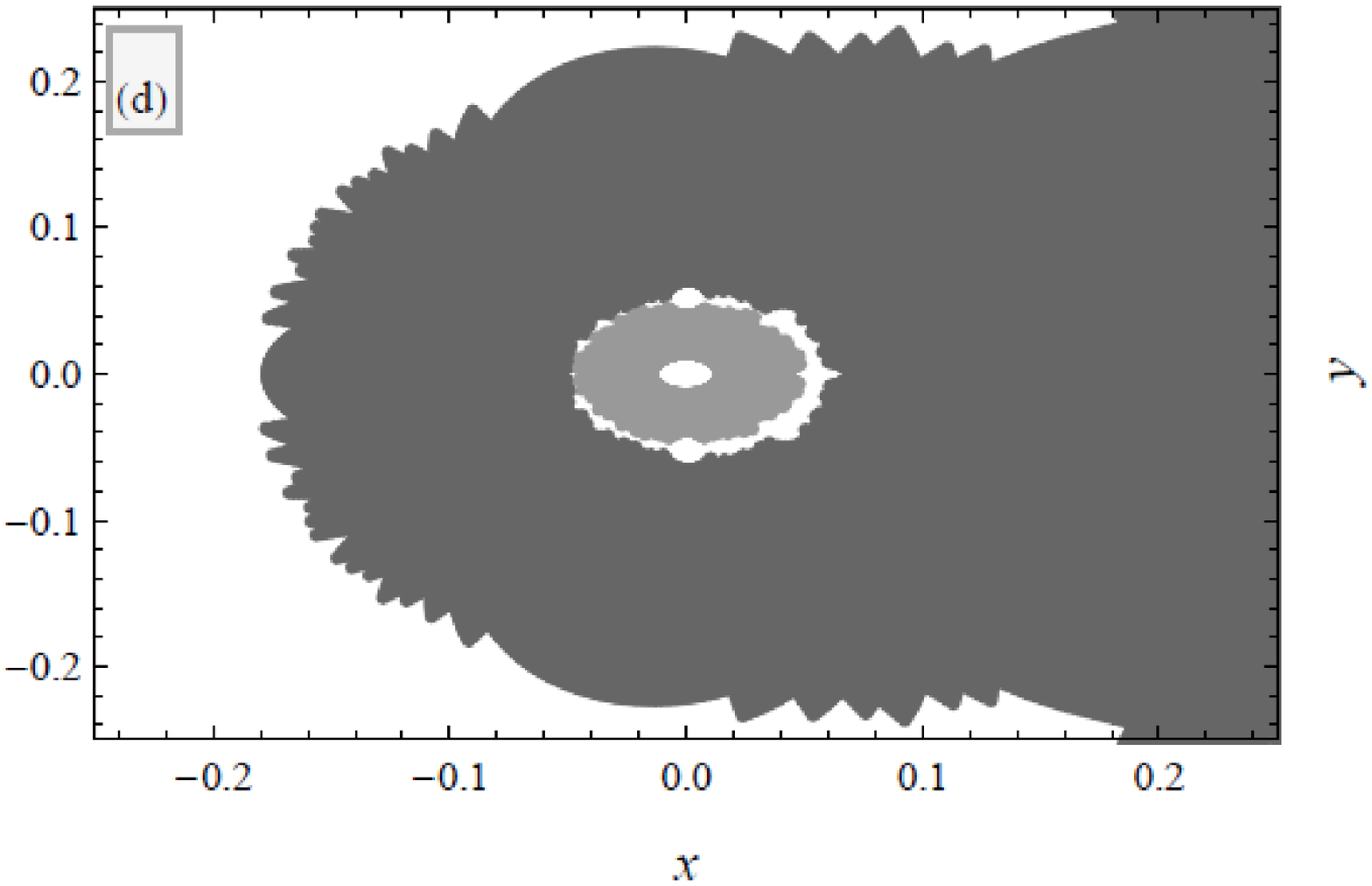}
\caption{Toy Model 3: Interpolation of the MSD (Sec.~\ref{aa:InterpolationAndCalculatingTheMSD}). The refining divisions \smash{$s^{\re}_{\textrm{int.}} = s^{\im}_{\textrm{int.}} = 5$}. Figs.~(c) and~(d) are zooms of~(a) and~(b), respectively, into $[ - 0.25 , 0.25 ] \times [ - 0.25 , 0.25 ]$.}
\label{fig:TM3TLCEMSDz}
\end{figure*}

This appendix briefly describes a simple numerical algorithm used to calculate the MSD, \smash{$\rho_{\mathbf{c} ( t )} ( x , y ) = \frac{1}{\pi} \partial_{\overline{z}} G = \frac{1}{2 \pi} ( \partial_{x} X - \partial_{y} Y + \ii ( \partial_{y} X + \partial_{x} Y ) )$} (\ref{eq:MSDFromNonHolomorphicGreenFunction}) of the TLCE for Toy Model 3 (\ref{eq:TM3DefinitionEq01}) (cf.~Sec.~\ref{ss:TM3}) from the master equations [(\ref{eq:TM3TLCEEq01a}), (\ref{eq:TM3TLCEEq01b}), (\ref{eq:TM3TLCEEq02a})-(\ref{eq:TM3TLCEEq03})]. The algorithm relies on certain observations, theorems and conjectures which are made for some special values of the parameters of the model (i.e., $r = 0.5$, $\sigma = 1$, $\tau = 5$, $t = 10$), but are supposed to hold in general. Even though Toy Model 3 is very simple, the algorithm captures certain generic properties any numerical approach to the MSD of more realistic financial models will probably share.

The basic observation is that the argument $z$ appears in the master equations only on the RHS of (\ref{eq:TM3TLCEEq01b}), hence the method will be to (i) pick a value of $G$ (from a certain regular lattice), and solve (\ref{eq:TM3TLCEEq01a}) for \smash{$\tilde{h} \geq 0$}, (ii) insert these $G$ and \smash{$\tilde{h}$} into (\ref{eq:TM3TLCEEq01b}), thereby obtaining the value of $z$ corresponding to the assumed value of $G$.

In this program, one is therefore faced with two problems: (i) How to solve (\ref{eq:TM3TLCEEq01a}) for \smash{$\tilde{h}$}, with given $G$? (Sec.~\ref{aa:SolvingTheFirstMasterEquationForhtilde}.) (ii) How to numerically compute the derivative \smash{$\partial_{\overline{z}} G$} from the set of values $z$ obtained from a regular lattice of the values of $G$? (Sec.~\ref{aa:InterpolationAndCalculatingTheMSD}.)

%%%%%%%%%%%%%%%%%%%%%%%%%%%%%%%%%%%%%%%%%%%%%%%%%%%%%%%%%%%%%%%%%%%%%%
%%%%%%%%%%%%%%%%%%%%%%%%%%%%%%%%%%%%%%%%%%%%%%%%%%%%%%%%%%%%%%%%%%%%%%

\subsection{Solving the first master equation (\ref{eq:TM3TLCEEq01a}) for \smash{$\tilde{h}$}}
\label{aa:SolvingTheFirstMasterEquationForhtilde}

An initial observation is that solving (\ref{eq:TM3TLCEEq01a}) for \smash{$\tilde{h}$} needs to be done in some non-iterative way, because it seems to behave chaotically upon iterating.

Before attempting a solution, one has to know how to calculate the integrals \smash{$F_{1 , 2} ( G , \tilde{h} )$} [(\ref{eq:TM3TLCEEq02a}), (\ref{eq:TM3TLCEEq02b})] for given arguments. For this, note that for \smash{$\tilde{h} > 0$} the roots of $W ( u )$ (\ref{eq:TM3TLCEEq03}) can never lie on the unit circle $C ( 0 , 1 )$; it is clear that if \smash{$u_{0}$} was a root satisfying \smash{$u_{0} = 1 / \overline{u_{0}}$}, then \smash{$W ( u_{0} ) = u_{0}^{t + 1}$} \smash{$( | 1 - A_{2} u_{0} + u_{0}^{2} + r A_{1} G u_{0}^{t + 1} |^{2} + r^{2} A_{1}^{2} \tilde{h} )$} \smash{$> 0$}, which is a contradiction. Hence, there is certainly no divergence in the contour integrals present in the master equations. Moreover, it also follows from (\ref{eq:TM3TLCEEq03}) that the roots of $W ( u )$ come in pairs, \smash{$( u_{n} , 1 / \overline{u_{n}} )$}, for $n = 1 , \ldots , t + 1$. These two properties mean that always half of the roots of $W ( u )$ lie inside $C ( 0 , 1 )$, and the other half outside; only the internal ones contribute to the contour integrals. Therefore, by expressing the integrals through the residues at these internal roots, one acquires precise control over the values of \smash{$F_{1 , 2} ( G , \tilde{h} )$}, and obtains a set of algebraic (i.e., no longer integral) master equations, much more suitable for numerical evaluation.

Solving (\ref{eq:TM3TLCEEq01a}) for \smash{$\tilde{h}$} is thus performed as follows: Form a regular lattice in the $G$-plane (for example, a rectangle \smash{$[ X^{\min} , X^{\max} ] \times [ Y^{\min} , Y^{\max} ]$}, divided into \smash{$s^{\re}$}, \smash{$s^{\im}$} stripes in each direction, respectively), and for each point $G$, scan an interval \smash{$[ 0 , \tilde{h}_{\max} ]$} searching for solutions \smash{$\tilde{h}$} of (\ref{eq:TM3TLCEEq01a}). The upper bound of solutions \smash{$\tilde{h}_{\max}$}, as well as the lattice parameters, are tuned in by trial and error.

Performed this program numerically for the selected parameters of the model, two important observations are made, which are conjectured to hold in general: (i) The points $G$ for which (\ref{eq:TM3TLCEEq01a}) has at least one solution \smash{$\tilde{h} \geq 0$} form a bounded area \smash{$\mathcal{D}_{G}$}. This is crucial from the point of view of numerical analysis, as one needs to search through only a restricted domain in the $G$-plane. (ii) For any \smash{$G \in \mathcal{D}_{G}$}, there can only be either one or two solutions \smash{$\tilde{h}$}. This is depicted in Fig.~\ref{fig:TM3TLCEMSDhtilde} (a), where the points are the subset of the initial $G$-lattice forming \smash{$\mathcal{D}_{G}$}, and the two shades of gray correspond to one or two solutions. These solutions form two continuous branches; Figs.~\ref{fig:TM3TLCEMSDhtilde} [(b), (c)] show the $G$-domains of these two branches, while their plots are presented in Figs.~\ref{fig:TM3TLCEMSDhtilde} (d) (both branches glued together to form a continuous surface $\mathcal{H}$) and [(e), (f)] (the branches separately).

Finally, one may substitute each $G$ from the lattice and its corresponding \smash{$\tilde{h}$} (from each of the two branches) into (\ref{eq:TM3TLCEEq01b}), obtaining the value of $z$. The points $z$ computed from both branches (reproducing theoretically the mean spectral domain $\mathcal{D}$) are marked by proper shades of gray in Figs.~\ref{fig:TM3TLCEMSDz} [(a), (c)], and shown on the background of the Monte Carlo eigenvalues of the TLCE. This describes a bijection $\mathcal{H} \to \mathcal{D}$.

%%%%%%%%%%%%%%%%%%%%%%%%%%%%%%%%%%%%%%%%%%%%%%%%%%%%%%%%%%%%%%%%%%%%%%
%%%%%%%%%%%%%%%%%%%%%%%%%%%%%%%%%%%%%%%%%%%%%%%%%%%%%%%%%%%%%%%%%%%%%%

\subsection{Interpolation and calculating the mean spectral density}
\label{aa:InterpolationAndCalculatingTheMSD}

Once the above numerical analysis is completed, one still faces another obstacle: Given the facts that one typically picks the initial $G$-lattice somewhat larger than the domain \smash{$\mathcal{D}_{G}$} plus that it is time-consuming to solve (\ref{eq:TM3TLCEEq01a}) at each lattice point---one generically obtains at the end too few points $z$ to calculate from them the MSD of the TLCE with a satisfactory accuracy.

For this reason, linear interpolation have been resorted to in order to create worthwhile MSD graphs: (i) For each of the two branches separately, select those lattice points from their $G$-domains which form the four corners of an elementary rectangle of the lattice which entirely lies within the respective domain. (ii) Create a sub-lattice by dividing the rectangle in question into \smash{$s^{\re}_{\textrm{int.}}$}, \smash{$s^{\im}_{\textrm{int.}}$} ($= 5$ here) stripes in each direction. (iii) Compute the value of $z$ at each point of this sub-lattice by linear interpolation from the values of $z$ at the four corners of the sub-lattice; this is presented in Figs.~\ref{fig:TM3TLCEMSDz} [(b), (d)]. In addition to $z$, it is straightforward to calculate the derivatives \smash{$\partial_{X} x$}, \smash{$\partial_{X} y$}, \smash{$\partial_{Y} x$}, \smash{$\partial_{Y} y$}, hence also the inverse derivatives \smash{$\partial_{x} X$}, \smash{$\partial_{y} X$}, \smash{$\partial_{x} Y$}, \smash{$\partial_{y} Y$} through an appropriate Jacobian---which is equivalent to knowing the MSD \smash{$\rho_{\mathbf{c} ( t )} ( x , y )$}, plotted in Fig.~\ref{fig:TM3TLCEMSD}.

%%%%%%%%%%%%%%%%%%%%%%%%%%%%%%%%%%%%%%%%%%%%%%%%%%%%%%%%%%%%%%%%%%%%%%
%%%%%%%%%%%%%%%%%%%%%%%%%%%%%%%%%%%%%%%%%%%%%%%%%%%%%%%%%%%%%%%%%%%%%%
%%%%%%%%%%%%%%%%%%%%%%%%%%%%%%%%%%%%%%%%%%%%%%%%%%%%%%%%%%%%%%%%%%%%%%

\end{document}